\documentclass[fleqn,usenatbib]{mnras}

\usepackage{newtxtext,newtxmath}

\usepackage[T1]{fontenc}

\DeclareRobustCommand{\VAN}[3]{#2}
\let\VANthebibliography\thebibliography
\def\thebibliography{\DeclareRobustCommand{\VAN}[3]{##3}\VANthebibliography}

\usepackage{graphicx}	
\usepackage{amsmath}	

\usepackage{amssymb}	
\usepackage{longtable}  
\usepackage{subfigure}  
\usepackage{caption}
\usepackage{multirow}

\newcommand{\kms}{\ensuremath{\mathrm{km}\,\mathrm{s}^{-1}}}
\newcommand{\Mpc}{\ensuremath{\mathrm{Mpc}^{-1}}}

\usepackage{graphicx}	
\usepackage{epstopdf}

\usepackage{amsmath}	
\usepackage{amssymb}	

\usepackage{lineno}

\usepackage{pdflscape}

\title[H{\sc i} gas kinematics of galaxy pairs in cluster environments]{WALLABY Pilot Survey: H{\sc i} gas kinematics of galaxy pairs in cluster environment}

\author[Shin-Jeong Kim et al.]
{Shin-Jeong Kim$^{1}$, 
Se-Heon Oh$^{1,2}$\thanks{E-mail: seheon.oh@sejong.ac.kr}, 
Jing Wang$^{3}$,
Lister Staveley-Smith$^{4}$,
B\"{a}rbel S. Koribalski$^{5,6}$,
\newauthor
Minsu Kim$^{2}$, 
Hye-Jin Park$^{7}$, 
Shinna Kim$^{1}$, 
Kristine Spekkens$^{8}$,
Tobias Westmeier$^{4}$,
O. Ivy Wong$^{9,4,10}$,
\newauthor
Gerhardt R. Meurer$^{4}$,
Peter Kamphuis.$^{11}$,
Barbara Catinella$^{4,10}$,
Kristen B.W. McQuinn$^{12}$,
Frank Bigiel$^{13}$,
\newauthor
Benne W. Holwerda$^{14}$,
Jonghwan Rhee$^{4,10}$,
Karen Lee-Waddell$^{4,9}$,
Nathan Deg$^{15}$,
\newauthor
Lourdes Verdes-Montenegro$^{16}$,
Bi-Qing For$^{4,10}$,
Juan P. Madrid$^{17}$,
Helga Dénes$^{18}$, 
and Ahmed Elagali$^{19}$ \\
$^{1}$Department of Physics and Astronomy, Sejong University, 209 Neungdong-ro, Gwangjin-gu, Seoul, Republic of Korea \\
$^{2}$Department of Astronomy and Space Science, Sejong University, 209 Neungdong-ro, Gwangjin-gu, Seoul, Republic of Korea \\
$^{3}$Kavli Institute for Astronomy and Astrophysics, Peking University, Beijing 100871, People’s Republic of China \\
$^{4}$ICRAR, The University of Western Australia, 35 Stirling Highway, Crawley WA 6009, Australia \\
$^{5}$CSIRO Astronomy and Space Science, Australia Telescope National Facility, P.O. Box 76, NSW 1710, Australia \\
$^{6}$School of Science, Western Sydney University, Locked Bag 1797, Penrith, NSW 2751, Australia \\
$^{7}$Research School of Astronomy and Astrophysics, Australian National University, Cotter Road, Weston Creek, ACT 2611, Australia\\
$^{8}$Royal Military College of Canada, P.O. Box 17000, Station Forces, Kingston, Ontario, K7K 7B4, Canada\\
$^{9}$CSIRO Space and Astronomy, PO Box 1130, Bentley, WA 6102, Australia \\
$^{10}$ARC Centre of Excellence for All Sky Astrophysics in 3 Dimensions (ASTRO 3D), Australia \\
$^{11}$Ruhr University Bochum, Faculty of Physics and Astronomy, Astronomical Institute, 44780 Bochum, Germany \\
$^{12}$Rutgers University, Department of Physics and Astronomy, 136 Frelinghuysen Road, Piscataway, NJ 08854, USA \\
$^{13}$Argelander-Institut für Astronomie, Universität Bonn, Auf dem Hügel 71, D-53121 Bonn, Germany\\
$^{14}$Department of Physics and Astronomy, University of Louisville, Natural Science Building 102, Louisville, KY 40292, USA \\
$^{15}$Queen's University, Department of Physics, Engineering Physics, and Astronomy, Queen's University, Kingston ON K7L 3N6, Canada \\
$^{16}$Instituto de Astrofísica de Andalucía (IAA-CSIC), Glorieta de la Astronomía, E-18008 Granada, Spain \\
$^{17}$Department of Physics and Astronomy, The University of Texas Rio Grande Valley, Brownsville, TX 78520, USA \\
$^{18}$ASTRON - The Netherlands Institute for Radio Astronomy, 7991 PD Dwingeloo, The Netherlands \\
$^{19}$Telethon Kids Institute, Perth Children’s Hospital, Perth, Australia }

\date{Accepted XXX. Received YYY; in original form ZZZ}

\pubyear{2022}

\begin{document}
\label{firstpage}
\pagerange{\pageref{firstpage}--\pageref{lastpage}}
\maketitle

\begin{abstract}
We examine the H{\sc i} gas kinematics of galaxy pairs in two clusters and a group using Australian Square Kilometre Array Pathfinder (ASKAP) WALLABY pilot survey observations. We compare the H{\sc i} properties of galaxy pair candidates in the Hydra I and Norma clusters, and the NGC 4636 group, with those of non-paired control galaxies selected in the same fields. We perform H{\sc i} profile decomposition of the sample galaxies using a tool, {\sc baygaud} which allows us to de-blend a line-of-sight velocity profile with an optimal number of Gaussian components. We construct H{\sc i} super-profiles of the sample galaxies via stacking of their line profiles after aligning the central velocities. We fit a double Gaussian model to the super-profiles and classify them as kinematically narrow and broad components with respect to their velocity dispersions. Additionally, we investigate the gravitational instability of H{\sc i} gas disks of the sample galaxies using Toomre Q parameters and H{\sc i} morphological disturbances. We investigate the effect of the cluster environment on the H{\sc i} properties of galaxy pairs by dividing the cluster environment into three subcluster regions (i.e., outskirts, infalling and central regions). We find that the denser cluster environment (i.e., infalling and central regions) is likely to impact the H{\sc i} gas properties of galaxies in a way of decreasing the amplitude of the kinematically narrow H{\sc i} gas ($M_{\rm{narrow}}^{\rm{HI}}$/$M_{\rm{total}}^{\rm{HI}}$), and increasing the Toomre Q values of the infalling and central galaxies. This tendency is likely to be more enhanced for galaxy pairs in the cluster environment.
\end{abstract}

\begin{keywords}
galaxies: evolution -- galaxies: clusters: general -- galaxies: groups: general -- galaxies: interactions -- ISM: kinematics and dynamics
\end{keywords}



\section{introduction}
According to the $\Lambda$ Cold Dark Matter ($\Lambda$CDM) hierarchical structure formation of galaxies, galaxy mergers and tidal interactions are critical for regulating the star formation (SF) histories of galaxies over cosmic time. Mergers and tidal interaction can lead to either enhancing or quenching of their star formation (\citealt{1996ApJ...464..641M}; \citealt{2007A&A...468...61D}; \citealt{2013MNRAS.433L..59P}; \citealt{2013MNRAS.435.3627E}; \citealt{2018MNRAS.478.3447E}; \citealt{2021MNRAS.504.1888Q}). The effective angular momentum transport outwards and the subsequent gas funneling inwards caused by the mergers and tidal interactions of galaxies can boost the creation of giant molecular gas clouds (GMCs) and thus star formation in galaxies (\citealt{1996ApJ...464..641M}; \citealt{2007A&A...468...61D}). On the other hand, the merging process can also give rise to turbulent gas motions which can suppress star formation (\citealt{2018MNRAS.476..122V}; \citealt{2018MNRAS.478.3447E}). It is important to examine how the merging process affects the gas properties of galaxies as this will lead us towards a better understanding of star formation and the resulting galaxy evolutionary path. 

Galaxy clusters are the largest known gravitationally bound systems and typically contain hundreds or thousands of galaxies, larger than galaxy groups which are also gravitaionally bound objects, but generally contain fewer galaxies. Galaxy clusters provide a high galaxy number density environment analogous to the early Universe where more frequent galaxy interactions are expected than in the local Universe. Many studies have discussed the effect of galaxy interactions in the cluster environment on their physical properties such as star formation rate, metallicity, colours, luminosity, morphology, and gas properties (\citealt{2004MNRAS.352.1081A}; \citealt{2009ApJ...699.1595P}; \citealt{2010MNRAS.407.1514E}; \citealt{2012A&A...539A..46A}; \citealt{2019MNRAS.488.4169O}; \citealt{2021PASA...38...35C}). 

An area of particular interest has been to examine how the merging process in cluster environments affects both molecular and atomic gas properties in and around galaxies: 1) How much of the cool gas reservoir is used up or replenished during galaxy mergers? (\citealt{2000MNRAS.318..124G}; \citealt{2017ApJ...844...96Y}), 2) How are the fractions of gas in different phases (atomic, molecular, and ionised) changed in the course of galaxy merger sequence, and how do they relate to star formation processes? (\citealt{2000MNRAS.318..124G}; \citealt{2021MNRAS.503.3113M}), 3) Which fuelling mechanism such as cool gas accretion from interacting galaxies or gas replenishment from the cooling of ionised gas in halos, is dominant during the galaxy merger? (\citealt{2011MNRAS.415.3750M}; \citealt{2015MNRAS.451L..35E}), 4) How can one distinguish between the effects of ram pressure stripping and tidal interaction on the gas properties of galaxies in a cluster environment, and which effect would prevail under what conditions? (\citealt{2003astro.ph..5512M}; \citealt{2017ApJ...838...81Y}). Recent studies of merging or interacting galaxies find an agreement between observed and simulated gas properties: the triggered nuclear star formation, enhanced molecular gas fractions (M$_{\rm{H2}}$/M$_{*}$), and higher star formation efficiency (SFE) compared to non-paired galaxies (\citealt{2013MNRAS.435.3627E,2016PASJ...68...96M, 2018ApJ...868..132P, 2018MNRAS.476.2591V}).

Close galaxy pairs, also known as pre\--merger galaxies, are in close proximity and have similar line-of-sight velocities. These close pairs provide crucial information about how galaxy interactions affect gas properties, and star formation, in the early phase of a merger sequence. High\--resolution, parsec-scale galaxy merger simulations show that the majority of close encounters experience centrally concentrated and enhanced SF that is mainly driven by cold gas inflow (\citealt{2020MNRAS.494.4969P}; \citealt{2021MNRAS.503.3113M}), whilst there are cases where SF in the central and outer regions of galaxies is suppressed by the reduced SFE. Hydrodynamical effects that drive turbulent gas motions such as stellar feedback and gravitational interactions can decrease SFE in galaxy pairs (\citealt{2021MNRAS.503.3113M}).

H{\sc i} in close galaxy pairs can be used for probing the environmental effect on the physical conditions of the cool gas reservoir. H{\sc i} is particularly useful for investigating the gas physics of galaxy pairs in the early phases of interaction as it is the most sensitive constituent of the system to any gravitational distortions. In addition, the dynamical time-scale on which H{\sc i} gas reacts to external influences is relatively shorter than that of other constituents in the galaxy pairs, like stars.

Several studies have performed H{\sc i} observations of galaxy pairs in order to investigate the effect of gravitational interactions between them on their cool gas properties (e.g., \citealt{2018ApJS..237....2Z}). These include the ALFALFA\footnote{The Arecibo Legacy Fast ALFA (ALFALFA) survey (\citealt{2005AJ....130.2598G}; \citealt{2018ApJ...861...49H}) } studies which discuss H{\sc i} gas fraction offsets of post-merger galaxies relative to isolated control galaxies  (\citealt{2015MNRAS.448..221E}). However, these observations lack spatial resolution due to the single dishes’ beam resolutions which are not high enough to resolve the internal gas kinematics and H{\sc i} distribution of close galaxy pairs.

H{\sc i} interferometric observations were carried out for a few tens of galaxy pairs but their beam resolutions were not high enough to spatially resolve individual galaxies given their distance (\citealt{2015MNRAS.449.3719S}). There was limited information about their spatially resolved kinematics (e.g., rotation curve analysis). To date, high-resolution H{\sc i} interferometric observations for a significant number of close galaxy pairs are limited (\citealt{1997AJ....114...77N}; \citealt{1997AJ....114..913N}; \citealt{2009MNRAS.400.1749K}). An important step towards a better understanding of the cool H{\sc i} gas properties of galaxy pairs is to obtain information about the spatially resolved H{\sc i} distribution and kinematics for a larger sample.

In this paper, we examine the resolved H{\sc i} gas properties of close galaxy pairs in two galaxy clusters (Hydra I and Norma), and a galaxy group (NGC 4636) from the Widefield ASKAP L-band Legacy All-sky Blind surveY (WALLABY) pilot observations undertaken with the Australian Square Kilometre Array Pathfinder (ASKAP). WALLABY is an ASKAP all-sky H{\sc i} galaxy survey that is expected to detect $\sim$210,000 galaxies out to $z$ $\sim$0.1 (\citealt{2020Ap&SS.365..118K}; Westmeier et al. accepted for publication in PASA). WALLABY pilot observations provide high-quality H{\sc i} 21\,cm data of resolved galaxies in the two galaxy clusters and the galaxy group with an angular resolution of 30\arcsec\ at an r.m.s of 2.0 mJy beam$^{-1}$ per 4 \kms\ channel spacing. The 30-square-degrees of ASKAP's wide field of view makes it possible to efficiently map the H{\sc i} emission over the large area of sky. Given the distances to these galaxy clusters and the group, H{\sc i} gas disks of more than 90$\%$ of member galaxies are marginally resolved, more than three beams across their major axes. 

We perform profile decomposition of H{\sc i} velocity profiles using a new tool, {\sc baygaud}. This new software allows us to de-blend a line-of-sight velocity profile with an optimal number of Gaussian components based on Bayesian analysis techniques (\citealt{2019MNRAS.485.5021O}; \citealt{2022ApJ...928..177O}; \citealt{2022arXiv220706698P}). We construct an H{\sc i} super-profile of each galaxy via stacking of individual line profiles after aligning their central velocities. We then fit a double Gaussian model to the H{\sc i} super-profile and classify them as the kinematically narrow and broad H{\sc i} gas components with respect to their velocity dispersions, respectively. From this, we aim to investigate how the H{\sc i} gas in different kinematic phases (i.e., the narrow and broad components) are affected by the merging process in different galaxy environments. Practically, we examine the variations of not only small-scale H{\sc i} kinematics but also the kinematically cool-to-total H{\sc i} mass ratios of galaxy pairs at their fixed total H{\sc i} masses in cluster environments. Additionally, we quantify the gravitational instability of the gas disks of galaxy pairs against gas collapse and star formation by deriving their Toomre Q parameters (\citealt{1964ApJ...139.1217T}). We also measure morphological H{\sc i} asymmetries to quantify the gas disks' disturbances. Galaxy pairs are classified as outskirts, infalling, and central ones based on the locations of the phase-space diagrams of the clusters and the group.

This paper is organized as follows. In Section~\ref{sec:2}, we present our data and perform a classification of galaxy pairs and non-paired galaxies.  In Section~\ref{sec:3}, we derive the H{\sc i} properties of the sample galaxies. In Section~\ref{sec:4}, we investigate the effect of the cluster environment on the H{\sc i} gas in the phase-space diagram of the galaxy clusters and group. Lastly, we summarize the main results of this paper in Section~\ref{sec:5}. Throughout this paper, we adopt the Hubble constant H$_0$ as 70 \kms\ ~\Mpc.

\section{Data}\label{sec:2}
\subsection{ASKAP H{\sc i} 21\,cm pilot observations}\label{sec:data}
In order to investigate the resolved H{\sc i} gas properties of galaxy pairs in the cluster and group environments, we use the H{\sc i} 21\,cm line data of two galaxy clusters, Hydra I and Norma, and a galaxy group, NGC 4636 taken from the ASKAP pilot observations. Prior to the start of the full survey, three 60-square-degree fields in the direction of the Hydra I cluster, Norma cluster, and NGC 4636 group were observed during the ASKAP pilot survey phase. ASKAP pilot observations with 36$\times$12-m dishes at a channel resolution of 4 \kms\ and a beam resolution of $\sim$30\arcsec\ at 1.4 GHz provide high-quality H{\sc i} data of the cluster and group fields with an average r.m.s. sensitivity level of $\sim$2.0 mJy beam$^{-1}$ (Westmeier et al. accepted for publication in PASA).

The raw visibility data were processed with ASKAPsoft, the ASKAP Science Data Processor pipeline \citep{2020ASPC..522..469W}. Data were flagged, calibrated, and imaged using the standard procedures described in \citet{2019MNRAS.482.3591R} (see also \citealt{2019MNRAS.487.2797E, 2019MNRAS.487.5248L, 2019MNRAS.488.5352K, 2019MNRAS.489.5723F}, for further details). In this paper, we use the second internal data release (DR2) of Hydra I (hereafter Hydra DR2) which covers the full Hydra field (60-square-degree) and contain more H{\sc i} detections than Hydra DR1. For the Norma and NGC 4636 fields, we use their first internal data release, DR1 as the preparation for their DR2 has been underway.

The Source Finding Application 2 (SoFiA2, \citealt{2021MNRAS.506.3962W}), the ASKAP source finding software package, has detected 272, 144, and 147  H{\sc i} sources in the Hydra I, Norma, and NGC 4636 fields, respectively. There are regions of the Norma and NGC 4636 fields that are significantly affected by continuum sources. Therefore, it was challenging to identify H{\sc i} sources that belong to these fields. Instead, SoFiA2 ran only on relatively clean parts of the Norma field without strong continuum. To reduce the effect of residual continuum emission on the source finding, the position of all continuum sources with a flux density greater than 150 mJy was flagged. Also, detections were visually inspected and obvious artifacts that were unlikely to be genuine H{\sc i} sources (e.g., ripple features) were removed. The entire western tile of the Norma cluster was discarded, while the entire eastern tile was kept. As the core of the Norma cluster itself is near the centre of the western tile, we did not include the central region of the Norma cluster in the end, but just its lower-density structures (i.e., infalling and outskirts regions). For the NGC 4636 field, SoFiA2 extracted sources from small regions around the position and redshift of the identified galaxies from optical (SDSS\footnote{The Sloan Digital Sky Survey (SDSS) \citep{2000AJ....120.1579Y}}, Cosmic Flows\footnote{Cosmic Flows \citep{2011MNRAS.414.2005C}}, 6dFGS\footnote{The 6dF Galaxy Survey (6dFGS) \citep{2011MNRAS.416.3017B}}) and H{\sc i} (ALFALFA, HIPASS\footnote{The HI Parkes All Sky Survey (HIPASS) \citep{2001MNRAS.322..486B}}) databases. Only H{\sc i} emission coinciding with the pre-selected optical/H{\sc i} source was kept even if additional emission, e.g., from a satellite without known redshift information, was picked up by SoFiA2 within the processed region. We refer to Westmeier et al. (accepted for publication in PASA) for the full description of the SoFiA2 source finding for the fields.

\begin{figure*}
    \centering
    \includegraphics[width = \textwidth]{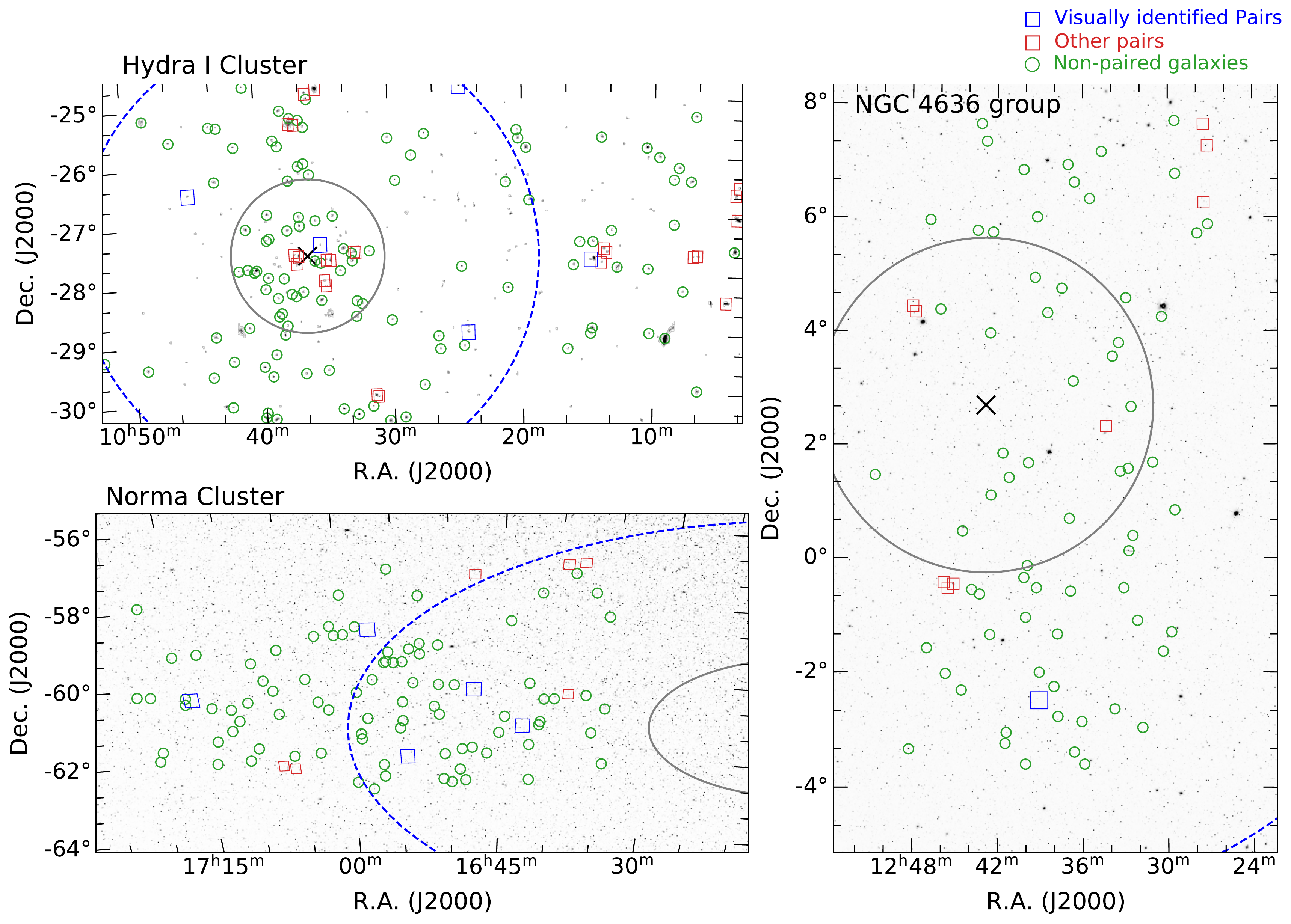}
    \caption[Image of clusters and group]{Integrated H{\sc i} intensity map of the Hydra I cluster and WISE images of the Norma cluster, and the NGC 4636 fields. ASKAP H{\sc i} detections from the SoFiA2 source finding are shown by the open red and blue rectangles (galaxy pairs) and green circles (non-paired control galaxies). The red rectangle hosts each member of a galaxy pair, and the larger blue one indicates a visually identified galaxy pair. The black cross symbols indicate the kinematic centres of the Hydra I and NGC 4636, and the references are given in Table~\ref{tab:cluster_info}. The grey solid and blue dashed circles indicate the virial radius, and a radius of three times the virial radius. See Section~\ref{sec:data} for details.}
    \label{fig_mosaic_cluster}
\end{figure*}

\setlength{\tabcolsep}{10pt}
\renewcommand{\arraystretch}{1.3}

Fig.~\ref{fig_mosaic_cluster} shows the projected H{\sc i} detections of galaxies in the fields of the two clusters (Hydra I and Norma) and the group (NGC 4636). The green circles, red rectangles, and the larger blue ones indicate the non-paired galaxies, each companion of a galaxy pair, and visually identified galaxy pairs (See Section~\ref{sec:pair} for a description). The integrated H{\sc i} intensities of the galaxies shown in the figures are derived from the SoFiA2 {\sc moment0} maps. 

The black cross mark, grey solid and blue dashed lines indicate the centres, virial radius, and a radius of three times the virial radius of the Hydra I and Norma clusters, and the NGC 4636 group. For the Norma field, only the eastern part is released due to strong continuum sources. The ASKAP pilot observations cover the Hydra I and Norma cluster fields out to approximately three times their virial radii within which the resolved H{\sc i} gas distribution and kinematics of member galaxies can be studied. Kinematic properties and distances of the two galaxy clusters and the group are given in Table~\ref{tab:cluster_info}.

\begin{table*}
    \centering
    {\small
    \begin{tabular}{c|c|c|c|c|c|c|c}
    \hline
    Name & Centre & $D$ & $\langle V \rangle$ & $\sigma$ & $R_{\rm{vir}}$ & $M_{\rm{vir}}$ & Ref.\\
    & R.A., Dec. (J2000) & (Mpc) & (\kms)\ & (\kms)\ & (Mpc) & ($10^{14} \, \rm{M}_{\odot}$) & \\
    (1) & (2) & (3) & (4) & (5) & (6) & (7) & (8)\\
    \hline
    Hydra I cluster & 10$^{\rm{h}}$36$^{\rm{m}}$41$^{\rm{s}}_.$8, -27$^{\rm{d}}$31$^{\rm{m}}$28$^{\rm{s}}$ & 61.0 & 3,777 & 676 $\pm$ 35 & 1.44 & 3.13 & a\\ 
    Norma cluster & 16$^{\rm{h}}$15$^{\rm{m}}$32$^{\rm{s}}_.$8, -60$^{\rm{d}}$54$^{\rm{m}}$30$^{\rm{s}}$ & 69.6 & 4,871 & 925 & 2.20 & 11.0 & b, c, d\\ 
    NGC 4636 group & 12$^{\rm{h}}$42$^{\rm{m}}$49$^{\rm{s}}_.$8, +02$^{\rm{d}}$41$^{\rm{m}}$16$^{\rm{s}}$ & 13.6 & 1,696 & 284 $\pm$ 73 & 0.70 & 0.39 & b, e, f, g\\
    \hline

    \end{tabular}
    }
\caption[Dynamical properties of two clusters (Hydra I and Norma) and NGC 4636 group]{Properties of the Hydra I and Norma clusters and NGC 4636 group: (1) cluster or group name; (2) R.A. and Dec. of the centre position (J2000); (3) the distance to the cluster or group (Mpc); (4) the systemic velocity (\kms); (5) the velocity dispersion of the cluster or group (\kms); (6) the virial radius (Mpc); (7) the virial mass ($10^{14}\mathrm{M}_{\odot}$); (8) references for columns (2)\--(7) \-- (a) \citet{2021MNRAS.505.1891R}, (b) NED = NASA/IPAC Extragalactic Database, (c) \citet{2008MNRAS.383..445W}, (d) \citet{2014ApJ...792...11J}, (e) \citet{2009MNRAS.400.1962K}, (f) \citet{2004MNRAS.350.1511O}, (g) \citet{2022ApJS..262...31L}}
\label{tab:cluster_info}
\end{table*}

\subsection{Galaxy classification: central, infalling and outskirts galaxies}
\label{sec:classification}

We classifiy member galaxies in Hydra I, Norma and NGC 4636 based on their spatial and spectroscopic information. In the fields toward the clusters and galaxy group, there are background and foreground H{\sc i} sources which are detected by SoFiA2. Given the systemic velocities of Hydra I ($\sim$3,777 \kms; \citealt{1999ApJS..125...35S}), Norma ($\sim$4,871 \kms; \citealt{2008MNRAS.383..445W} ), and NGC 4636 ($\sim$1,696 \kms; \citealt{2009MNRAS.400.1962K}), we only select H{\sc i} sources whose systemic velocities are within the velocity ranges of 500 $\sim$ 7,000 \kms (c$z$ $<$ 7,000 \kms), 500 $\sim$ 9,500 \kms ($<$ 9,500 \kms), and 500 $\sim$ 3,100 \kms ($<$ 3,100 \kms), respectively. These are five times the velocity dispersions of the clusters and the group with respect to their systemic velocities. This excludes the possibility of including any H{\sc i} rich massive galaxies at high redshifts. The numbers of the identified galaxies according to the velocity cuts are 156, 109, and 83 in the Hydra I, Norma, and NGC 4636 fields, respectively.

\begin{figure*}
    \centering
    \includegraphics[width = 0.99\columnwidth]{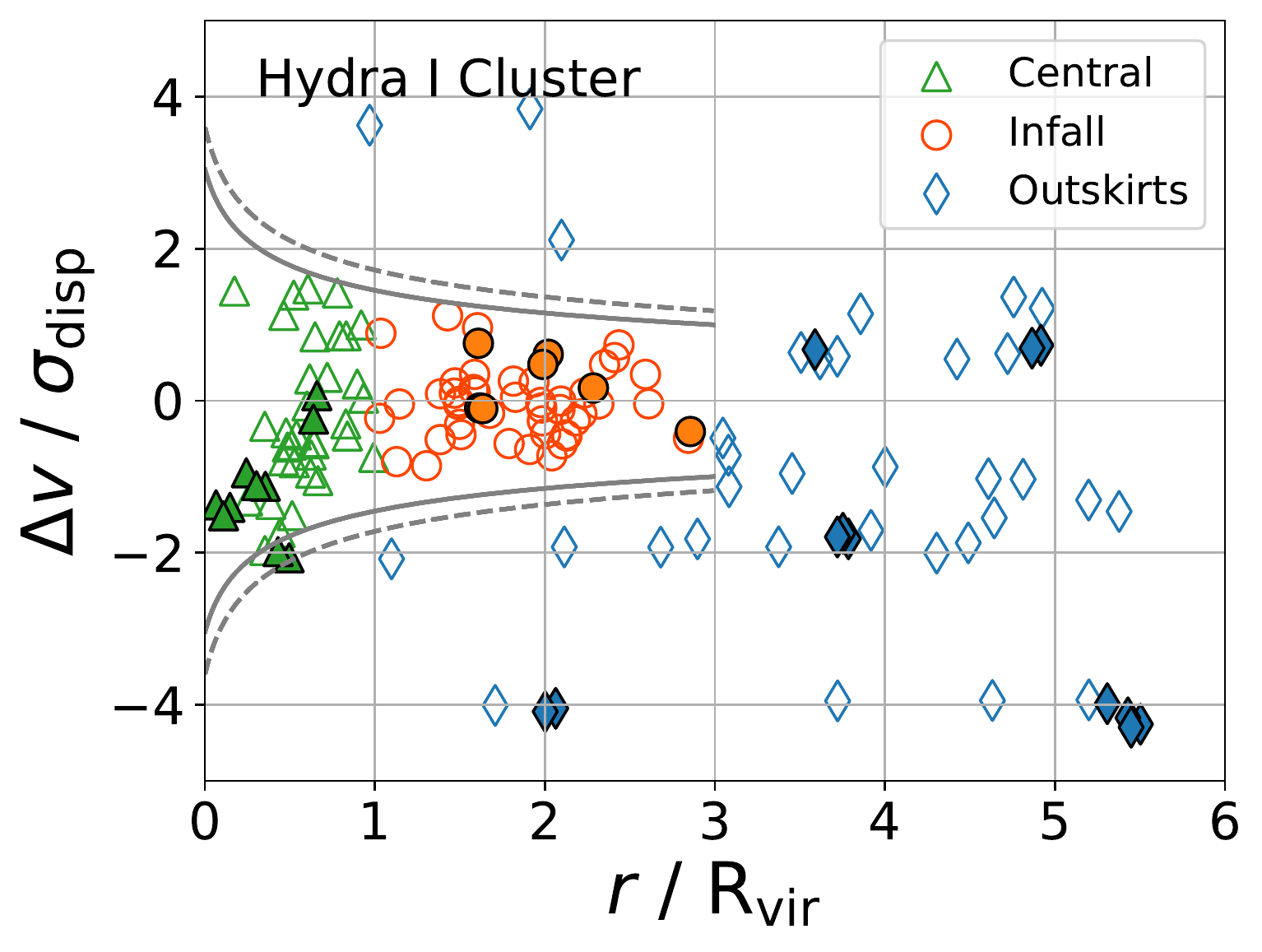}
    \includegraphics[width = 0.99\columnwidth]{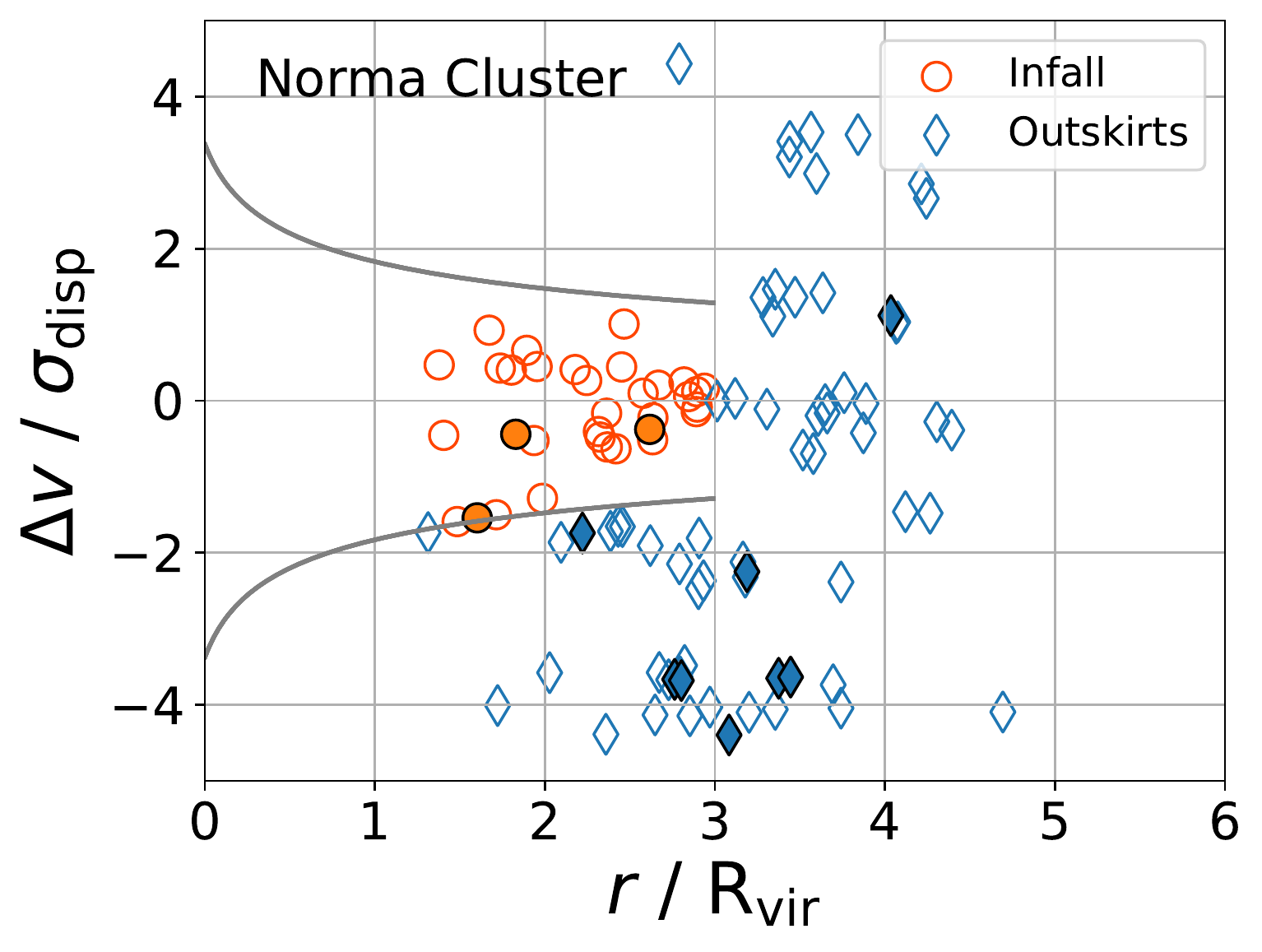}
    \includegraphics[width = 0.99\columnwidth]{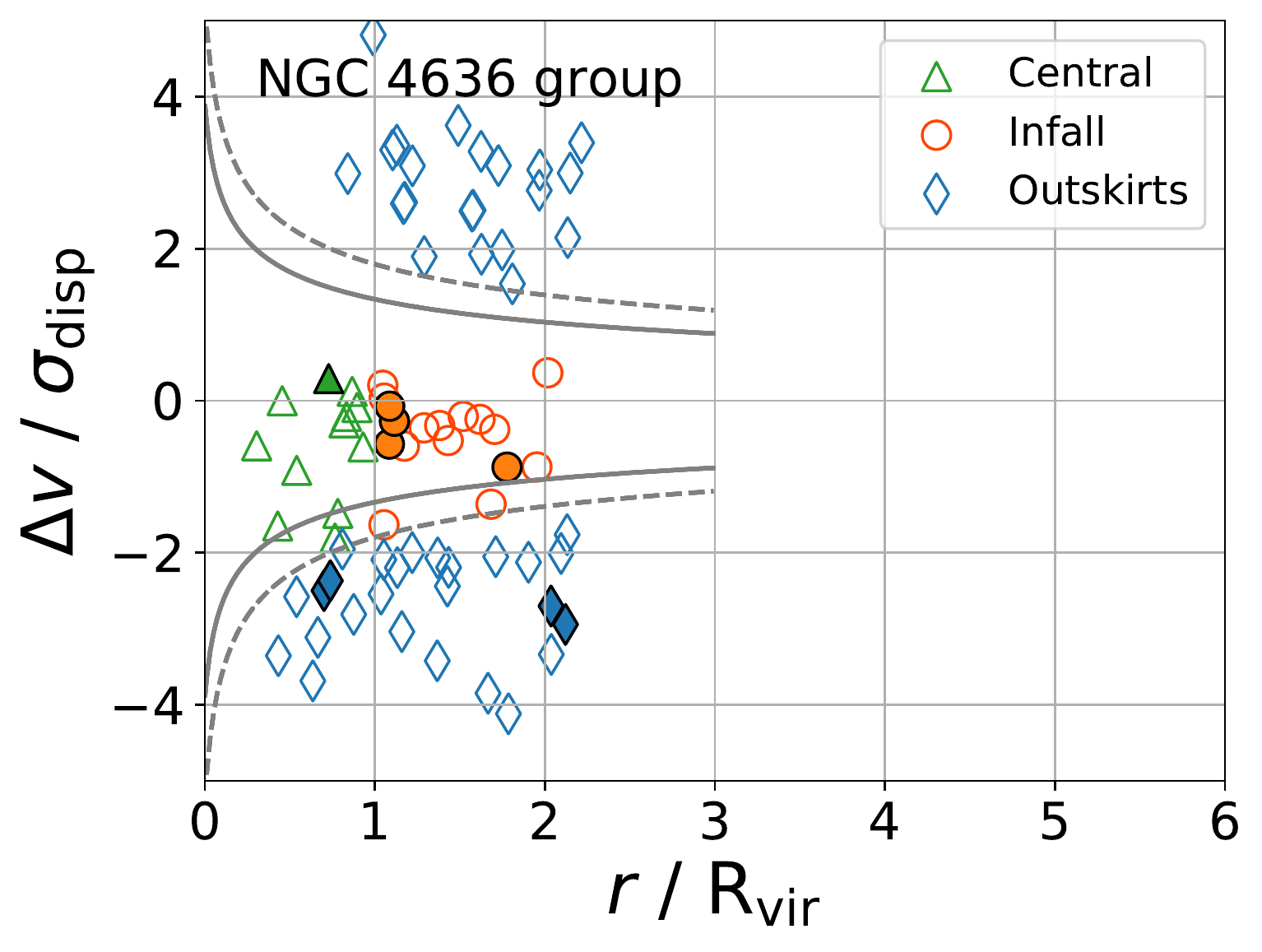}
    \caption[Phase-space diagram of clusters and group]{Phase-space diagrams of the Hydra I and Norma clusters, and NGC 4636 group. Galaxies are classified by their location on the phase-space diagram (see Section~\ref{sec:classification} for a description): central (green open triangles), infalling (orange open circles), and outskirts (blue open diamond) galaxies. Filled symbols indicate the corresponding galaxy pairs. The solid lines indicate the escape velocity curves of the clusters and the group which are derived using the Eqs. 1$\sim$4 in \cite{2017ApJ...843..128R}. The dashed curves in the Hydra I cluster and the NGC 4636 group indicate the escape velocity curves to which 3\,$\sigma_\mathrm{esc}$ and 1\,$\sigma_\mathrm{esc}$ are added, respectively.}
    \label{fig:phase_space_cluster_infall_field}
\end{figure*}

Secondly, we classify the sample galaxies as three different groups: 1) central, 2) infalling, and 3) outskirts galaxies based on their locations in the phase-space diagrams of the galaxy clusters and group. For this, we use the projected phase-space diagrams of the Hydra I and Norma clusters and the NGC 4636 group. As shown in Fig.~\ref{fig:phase_space_cluster_infall_field}, the $x$-axis ($r$/$R_{\rm{vir}}$) indicates the member galaxies' projected distances from the centres which are normalised by their virial radii $R_{\rm{vir}}$. The $y$-axis ($\Delta V$/$\sigma_{\rm{disp}}$) indicates their relative line-of-sight velocities ($\Delta V$) to the systemic velocities at the centres which are normalised by the velocity dispersions ($\sigma_{\rm{disp}}$) of the clusters and group. In many other works, this projected phase-space diagram has been widely used as a tool for classifying member galaxies as different groups with time since they infall into a galaxy cluster or group environment (\citealt{2015MNRAS.448.1715J}; \citealt{2017ApJ...838...81Y};  \citealt{2017ApJ...843..128R}). Numerical simulations have shown that galaxies tend to follow a known path in the phase-space diagram as they fall into the cluster or group potential (\citealt{2017ApJ...843..128R}).

We define a galaxy as a central galaxy if its projected distance ($r$) from the cluster or group centre is $<$ $R_{\rm{vir}}$, and its line-of-sight systemic velocity with respect to the cluster or group systemic velocity divided by the cluster or group velocity dispersion, $\Delta V$/$\sigma_{\rm{disp}}$, is less than an escape velocity limit of the cluster or group divided by the cluster or group velocity dispersion. The escape velocity ($v_{\rm{esc}}$) is derived using Eqs. 1$\sim$4 in \citet{2017ApJ...843..128R} which adopts the virial mass $M_{\rm{vir}}$ given in Table~\ref{tab:cluster_info}. We adopt ($v_{\rm{esc}}$ $\pm$ 3$\sigma_{\rm{esc}}$)/$\sigma_{\rm{disp}}$ and ($v_{\rm{esc}}$ $\pm$ 1$\sigma_{\rm{esc}}$)/$\sigma_{\rm{disp}}$ as the escape velocity limits for the Hydra I cluster and NGC 4636 group as in \cite{2022MNRAS.510.1716R} and \cite{2022ApJS..262...31L}. These are shown as grey dashed lines in Fig.~\ref{fig:phase_space_cluster_infall_field}. For the Norma cluster, we use $v_{\rm{esc}}$/$\sigma_{\rm{disp}}$ as the limit since the cluster's $\sigma_{\rm{esc}}$ is not available. If the projected distance ($r$) and normalised relative velocity ($\Delta V$/$\sigma_{\rm{disp}}$) of a galaxy satisfy the following criteria, $R_{\rm{vir}}$ $<$ $r$ $<$ 3$R_{\rm{vir}}$ and $\Delta V$/$\sigma_{\rm{disp}}$ $<$ $v_{\rm{esc}}$/$\sigma_{\rm{disp}}$, we classify it as an infalling galaxy. Other galaxies with either $\Delta V$/$\sigma_{\rm{disp}}$ $>$ $v_{\rm{esc}}$/$\sigma_{\rm{disp}}$ or $r$ $>$ 3$R_{\rm{vir}}$ are classified as outskirts galaxies. The virial radii and masses of Hydra I and Norma clusters and the NGC 4636 group adopted in our work are derived assuming hydrostatic equilibrium (Hydra I), dynamical equilibrium (NGC 4636), or NFW halo (Norma). The values derived using the different assumptions are not very different within a range of up to a factor of 1.5. For more details, we refer to \cite{2002ApJ...567..716R}, \cite{2014ApJ...792...11J}, and \cite{2022ApJS..262...31L} for Hydra I cluster, Norma cluster, and NGC 4636 group, respectively.

We show the locations of the central, infalling and outskirts galaxies in the Hydra I, Norma, and NGC 4636 fields classified according to the above criteria on the phase-space diagrams in Fig.~\ref{fig:phase_space_cluster_infall_field}. They are colour-coded as green (triangle), orange (circle), and blue (diamond) colours, respectively in the figure. As discussed above, no central galaxies are classified in the Norma field due to the strong continuum sources in the cluster centre (see Fig.~\ref{fig_mosaic_cluster}). The effect of continuum sources might have resulted in the asymmetric distribution of infall (orange) and central (green) galaxies relative to the systemic velocity for the Norma and NGC 4636 fields. Of the ASKAP H{\sc i} detections in the three fields, we exclude galaxies whose H{\sc i} emission is significantly affected by strong continuum sources or only partly detected. In our study, we focus on galaxy pairs at pre-merger stages, not the post-merger ones at coalescence. Hence, galaxies that appear to be in a post-merger stage are excluded. We identified post-merger galaxies both spatially and spectrally using their H{\sc i} intensity and velocity maps, respectively. These are overlapped with each other in their H{\sc i} intensity and velocity maps. On the other hand, the companions of a close galaxy pair are visually separated with each other in their H{\sc i} intensity and velocity maps.

\subsection{Sample selection of galaxy pairs and control galaxies}
\label{sec:pair}

Firstly, we compile galaxy pair candidates from the galaxies which are detected in H{\sc i} and classified as the outskirts, infalling, or central ones in the three fields, the Hydra I, Norma, and NGC 4636 in Section~\ref{sec:classification}. For this, we use the selection criteria described in \citet{2019A&A...631A..87V} which allows us to select galaxy pair candidates based on their spectral and spatial information. Based on cosmological simulations, \citet{2019A&A...631A..87V}'s selection criteria for closed pairs take into account 3D projection effects. If a system with two galaxies satisfies the following criteria, it is classified as a galaxy pair: the projected angular separation ($R_{\rm{p}}$) and the absolute difference of the systemic velocities ($\Delta V$) between the two galaxy encounters are within either 1) $R_{\rm{p}}$ $\leq$ 50 kpc and $\Delta V$ $\leq$ 300 \kms\ or 2) 50 kpc $\leq$ $R_{\rm{p}}$ $\leq$ 100 kpc and $\Delta V$ $\leq$ 100 \kms. This is compared with the ones typically adopted in other works including \cite{2008AJ....135.1877E} and \cite{2015MNRAS.449.3719S} where a more relaxed criterion of $\Delta V$ $\leq$ 300 - 500 \kms\ is used. In addition, enhanced SFR in galaxy pairs separated by $R_{\rm{p}}$ of $\sim$150 kpc has been found as in \cite{2013MNRAS.433L..59P}. The galaxy pairs classified using our selection criteria are likely to be gravitationally bound. 
Together, their H{\sc i} gas disks and systemic velocities should be resolved and separated by the ASKAP’s 30\arcsec\ beam ($\sim$\,8.9 kpc, 9.8 kpc, and 2.0 kpc for the Hydra I and Norma clusters and NGC 4636, respectively) and 4 \kms\ channel resolution, respectively. It is noted that we are targeting wet interactions as the galaxy pair candidates in this work are H{\sc i} selected samples. There are 38 galaxy pair candidates selected by the selection criteria. Through a visual inspection of their H{\sc i} data cubes, we find that most of them are well separated, either spectrally or spatially. These galaxy pair candidates can be in a stage of merging before the first encounter or after the first pericenter passage.

\begin{figure}
    \centering
    \includegraphics[width = 0.5\textwidth]{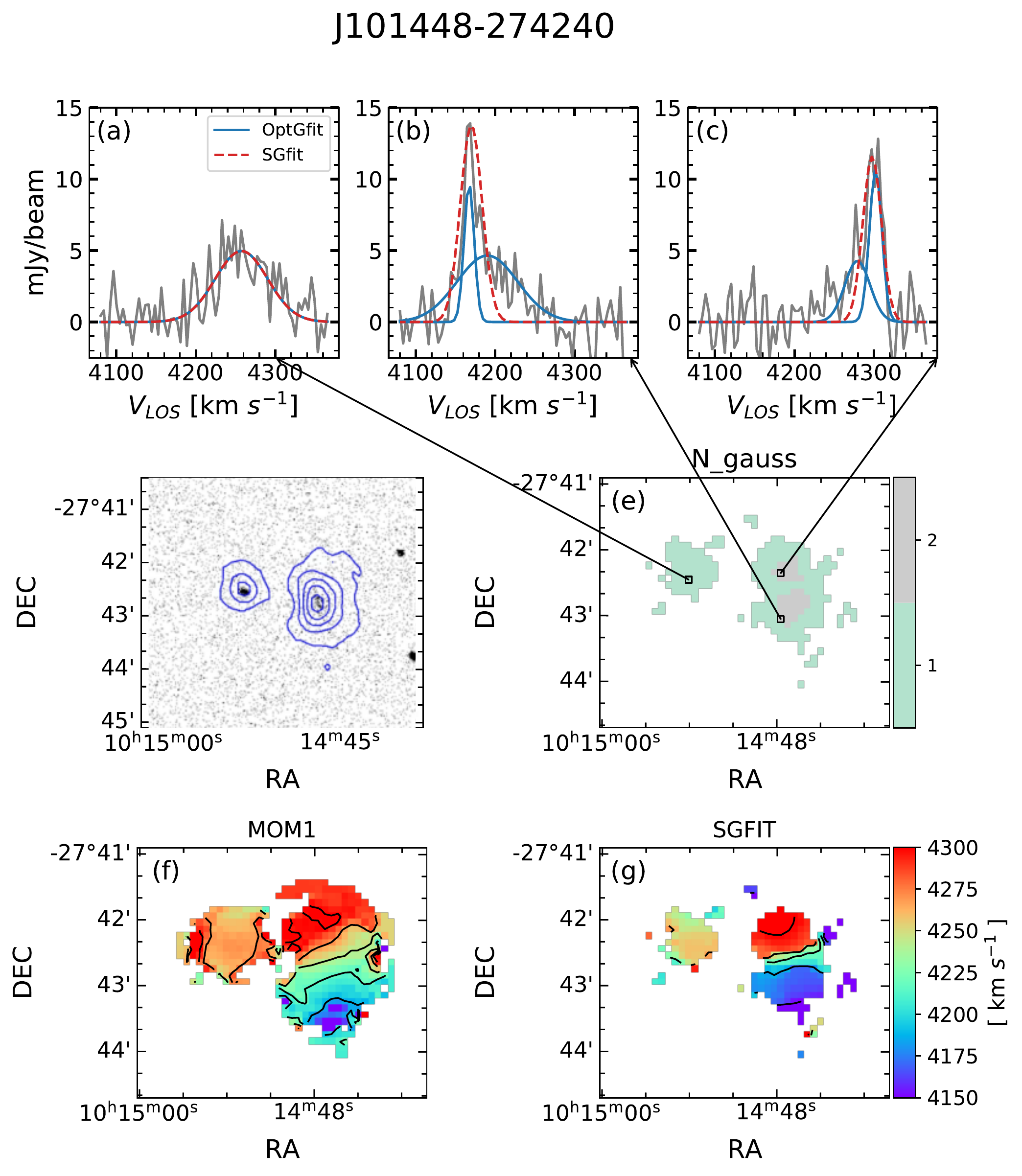}
    \caption[An example image of galaxy pairs in Hydra I cluster]{Velocity profiles and H{\sc i} maps of a galaxy pair J101448$-$274240 in the Hydra I cluster. (a,b,c): the blue solid and red dashed lines indicate the optimal Gaussian fitting (OptGfit) and single Gaussian fitting (SGfit) to the observed profiles (grey solid lines) at three locations. (d): WISE images overlaid with the contours of the H{\sc i} integrated intensity ({\sc moment0}) maps; the contours start from 0.075 Jy beam$^{-1}$ in steps of 0.125 Jy beam$^{-1}$, (e): 2D map of the optimal number of Gaussian components, (f): {\sc moment1} map (MOM1), and (g): single Gaussian fitting velocity field (SGFIT).}
    \label{fig2}
\end{figure}

Additionally, we visually identify 10 galaxy pair candidates which are close to each other so that they are not separable in the H{\sc i} emission detected by SoFiA2. These close galaxy pair candidates could be in a late stage of merging where the two interacting galaxies are not coalescent yet but overlapped spatially and/or spectrally. This could make their systemic velocities comparable with each other. These galaxy pair candidates are visually verified and indicated by the symbol, ‘*’ in Tables~\ref{long_hydra_pair} and \ref{long_norma_pair}. An example of a visually identified close galaxy pair candidate which can be in a late stage of the galaxy merger sequence is presented in Fig.~\ref{fig2}. Of all the galaxies detected at H{\sc i} in the three fields, there are 48 galaxy pair candidates selected by either the above selection criteria (38 galaxies) or our visual inspection (10 galaxies) (see Figs.~\ref{figA1} $\sim$ \ref{figA3} in Appendix. The full tables and figures are available as supplementary online material). These are subsequently classified as the outskirts, infalling, and central galaxy pairs based on their locations in the phase-space diagrams of Fig.~\ref{fig:phase_space_cluster_infall_field}. 

H{\sc i} gas mass ratio values for the galaxy pairs range from 1:1.6 $\sim$ 1:66.6 with the peak value being between 1:1 $\sim$ 1:6. Simulations predict that minor mergers with dynamical mass ratios 1:10 or less do not usually produce significant morphological disturbance in their gas and stellar distributions (e.g., \citealt{2005AJ....129..682H}; \citealt{2006ApJ...638..686C}). In general, such morphological disturbances are more enhanced in major mergers with mass ratios of 1:4 or closer to unity. In this respect, our $f_{\rm{narrow}}$ and Toomre-Q parameter measurements might better extract any kinematic disturbances (if present) even in minor mergers even though the H{\sc i} mass ratio may not directly scale with the dynamical or stellar mass ratio.

Secondly, as a control sample we select a set of galaxies in the Hydra I, Norma, and NGC 4636 fields which do not satisfy the selection criteria above and thus have no close companions in their spatial and spectral vicinity. These control galaxies are relatively less affected by the gravitational effects of any close companions compared to the galaxy pair candidates selected above. From these, 266 control sample galaxies are selected in the three fields. Likewise, these are classified as the outskirts, infalling, and central galaxies according to their locations in the phase-space diagrams. H{\sc i} properties of these control galaxies will be compared with those of the galaxy pairs.

We make additional cuts to both the galaxy pairs and the control sample galaxies above in order to match the distribution of their H{\sc i} masses and the galaxy number ratios (outskirts/infalling, outskirts/central) between the different environments in the three fields. This is to reduce potential bias due to differences in the galaxy number ratios for the clusters and group or in galaxy properties. Since we examine how the small-scale kinematics and kinematically cool-to-total H{\sc i} mass ratio of galaxies are affected in galaxy pairs and cluster environments at fixed total H{\sc i} mass, we match the sample galaxies by H{\sc i} mass. Here, we assume that the sample galaxies on average have not yet been significantly affected by the cluster environment.

\begin{figure*}
    \centering
    \includegraphics[width = 0.99\textwidth]{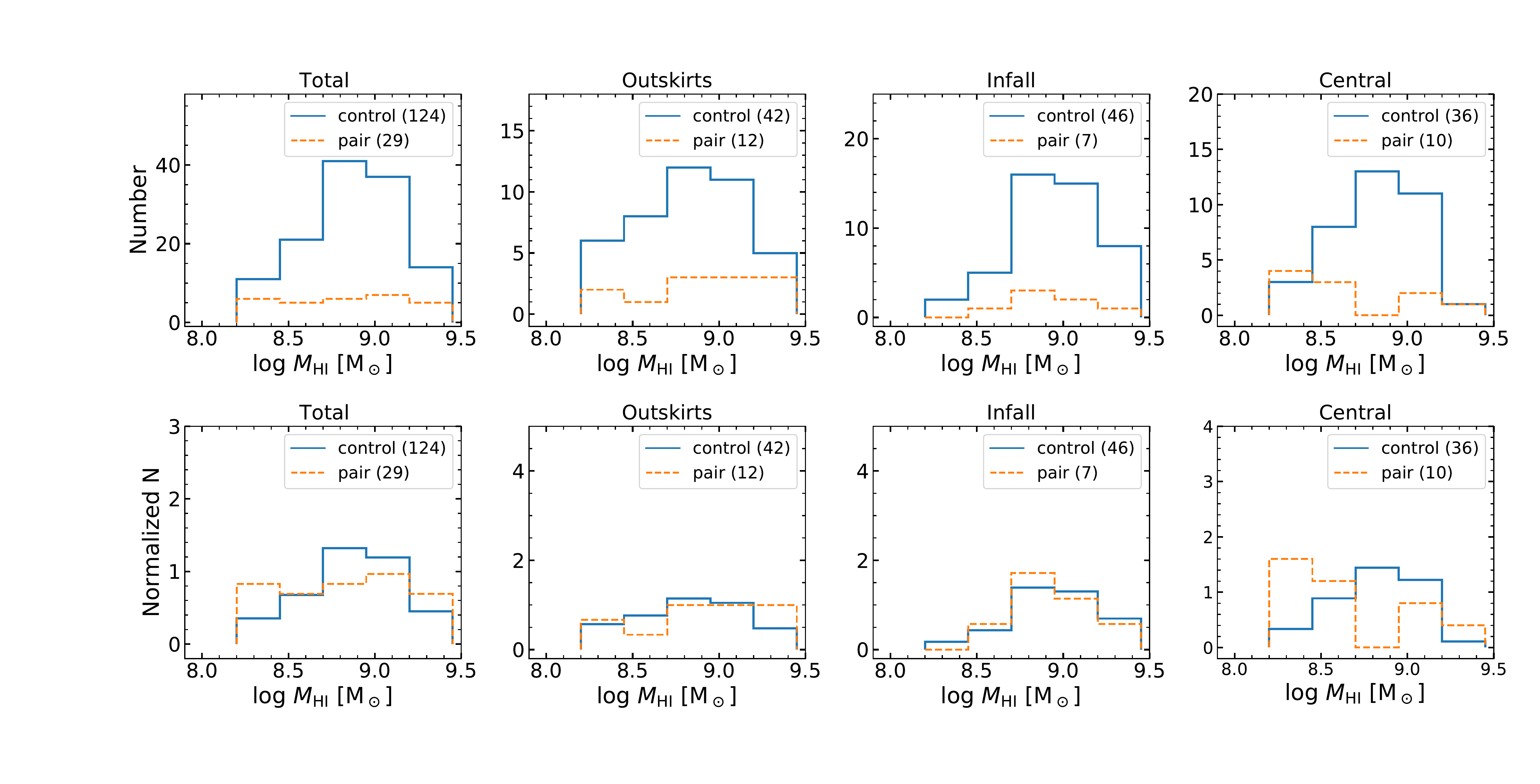}
    \caption{The H{\sc i} mass distribution of the galaxy pairs and control sample galaxies of the Hydra I and Norma clusters and NGC 4636 group in the outskirts, infalling, and central environments as denoted on the top of each panel. The  mass distribution of all the sample galaxies are shown in the left-most panels. The orange dashed and blue solid lines show the galaxy pairs and control sample galaxies, respectively. The number of the control galaxies and galaxy pairs are denoted.}
    \label{fig:HI_mass_distribution}
\end{figure*}

The H{\sc i} mass and galaxy number ratio cuts are made based on the galaxies in the Hydra I field as both the Norma and NGC 4636 fields are affected by continuum sources as discussed in Section~\ref{sec:data}. When matching the H{\sc i} mass distribution and galaxy number ratios for the galaxy pairs and control galaxies selected above, we preferentially remove the ones which are relatively less resolved by the ASKAP beam and with low S/N. We exclude sources which are less resolved (across their major axes) than two, four, and five synthesized beams in the Hydra, Norma, and NGC 4636 fields, respectively. The corresponding S/N of individual H{\sc i} profiles of the selected galaxies is greater than three. The resulting H{\sc i} mass distribution for the final sample galaxies is shown in Fig.~\ref{fig:HI_mass_distribution}. The H{\sc i} mass distribution of the infalling galaxy pairs is narrower than that of the control galaxies due to the lack of galaxy pairs in the H{\sc i} mass bins ($<$ 8.45). The corresponding galaxy number ratios (Outskirts/Infall, Outskirts/Central) for the galaxy pairs and control galaxies in the Hydra I, Norma and NGC 4636 fields are given in Table~\ref{tab:galaxy_num_ratio}.

\begin{table*}
    \centering
    
    \begin{tabular}{c|c|c|c|c|c|}
    \hline
    & N(Outskirts) & N(Infall) & N(Central) & N(Outskirts)/N(Infall) & N(Outskirts)/N(Central) \\
    \hline
    Hydra I & 24 & 30 & 30 & 0.80 & 0.80 \\ 
          & 9* & 5* & 10* & 1.80* & 0.90* \\    
          \\
    Norma & 13 & 12 & - & 1.08 & - \\ 
          & 1* & 1* & -* & 1.0* & -* \\
          \\
    NGC 4636 & 5 & 4 & 6 & 1.25 & 0.83 \\
             & 2* & 1* & -* & 2.0* & -*\\
    \hline
    \end{tabular}
\caption[galaxy_number_ratio]{The number of galaxy pairs (indicated by ‘*’ symbol) and control sample galaxies, and the galaxy number ratios, N(outskirts)/N(infall) and N(outskirts)/N(central) in the Hydra I, Norma and NGC 4636 fields.}\label{tab:galaxy_num_ratio}
\end{table*}

In summary, 29 galaxy pair candidates (Hydra I: 24, Norma: 2, and NGC 4636: 3) and 124 control sample galaxies (Hydra I: 84, Norma: 25, and NGC 4636: 15) are selected by the selection criteria adopted in this work. The H{\sc i} images (moment maps, single Gaussian fitting, and model velocity fields) of the galaxy pairs are shown in Figs.~\ref{figA1} $\sim$ \ref{figA3} in Appendix (The full figures are available as supplementary online material). The observational properties of the sample galaxies including their H{\sc i} properties derived in this work are presented in Table~\ref{long_hydra_control} $\sim$ \ref{long_norma_pair} in Appendix (The full tables are available as supplementary online material). 

\section[H{\sc i} data analysis]{H{\sc i} data analysis}\label{sec:3}
\subsection{Decomposition of H{\sc i} velocity profiles}\label{sec:baygaud}

We perform profile decompositions of individual line-of-sight line profiles of the ASKAP H{\sc i} data cubes. For this, we use a newly developed profile decomposition tool, {\sc baygaud} (\citealt{2019MNRAS.485.5021O}). For each line profile of an H{\sc i} data cube, {\sc baygaud} performs profile decomposition based on Bayesian nested sampling techniques, and finds an optimal number of Gaussian components with which the profile is best described.

In practice, {\sc baygaud} fits a Gaussian model which consists of a number of Gaussian components to each line profile,
\begin{equation}
    \label{eq:baygaud}
    G(x) = \sum_{i=1}^{N} \frac{A_{i}}{\sqrt{2\pi}} \exp\bigg(\frac{-(x-\mu_{i})^{2}}{2\sigma_{i}^{2}}\bigg) + \sum_{j=0}^{n} b_{j}x^{j}\,,
\end{equation}
where {\it G($x$)} is the Gaussian model, {\it N} is the maximum number of Gaussian components, and $A_{i}$, $\sigma_{i}$, and $\mu_{i}$ are the parameters of integrated intensity, velocity dispersion, and central velocity of the {\it i}-th Gaussian component, respectively. $b_{j}$ are the coefficients of the $n^{th}$ order polynomial function which is used for the baseline fitting. In this work, we fit $b_{j}$ with a constant value. Among the {\it N} different Gaussian models fitted to the line profile, {\sc baygaud} determines an optimal number of Gaussian components based on a Bayesian model selection where a model with the largest bayes factor, the ratio of Bayesian evidences of any two competing models, is chosen as the most appropriate model.

From the performance test discussed in \citet{2019MNRAS.485.5021O}, {\sc baygaud} is found to be robust for parameter estimation and model selection against any local minima in the course of the fitting as long as the integrated signal-to-noise (S/N) of the line profile is high enough for the analysis, for example, larger than three or so. We refer to \citet{2019MNRAS.485.5021O} for the full description of the fitting algorithm and its performance test (see also \citealt{2022ApJ...928..177O} and \citealt{2022arXiv220706698P}).

We run {\sc baygaud} for the ASKAP H{\sc i} data cubes, letting it fit the individual velocity profiles with the models in Eq.~(\ref{eq:baygaud}) consisting of different numbers of Gaussian components starting from one up to a user-defined maxim number. From a visual inspection of line profiles of the cubes, we find that modelling them with up to three Gaussian components appears to be enough to take their non-Gaussian profile shape into account. In this work, we adopt the maximum number of Gaussian components as three for the {\sc baygaud} analysis. In addition, {\sc baygaud} is only ran over the line profiles of the cubelets within the SoFiA2 masks whose S/N is greater than three. As described in \citet{2019MNRAS.485.5021O}, we select the most appropriate model for each H{\sc i} velocity profile among the competing Gaussian models with one, two, and three Gaussian components based on their calculated Bayesian evidence. In this work, we adopt the `substantial' model whose Bayes factor against the second-best model is at least larger than 3.2. In the rest of this paper, we designate the best model as `optimal Gaussian fitting'. If the data is best described by a single Gaussian component, we call it `single Gaussian fitting'. 

Of the optimally decomposed Gaussian components for each line profile, we filter out the ones whose `integrated S/N' is lower than 3. For this, we derive an integrated S/N for each Gaussian component as follows,
\begin{equation}
    \mathrm{RMS}_{\mathrm{int}} = \sqrt{n_{\mathrm{chan}}}\delta V \times \mathrm{RMS}_{\mathrm{bg}},
    \label{eq:integratedsn}
\end{equation}
\noindent where $n_\mathrm{chan}$ is the number of available channels for the line profile over which the Gaussian component is well covered, and $\delta V$ is the channel resolution of the cube. Typically, a velocity range of $6 \times v_{\rm{disp}}$, where $v_{\rm{disp}}$ is the velocity dispersion of a line profile, is wide enough over which $\sim$99.7\% of the total flux of the profile is contained. We therefore derive the corresponding $n_{chan}$ from $6 \times$ $v_{\rm{disp}}$ / $\delta V$. ${\rm RMS_{bg}}$ is the estimated rms of the line-free channels of the profile. For a given line profile, unless all of its decomposed Gaussian components satisfy the S/N cut adopted in this work, i.e., the integrated S/N limit of three, we replace its optimal Gaussian fitting with the single Gaussian fitting result.

In Fig.~\ref{fig2}, examples of line profiles are given by the grey solid lines, and the blue solid and red dashed lines represent their optimal Gaussian fitting and single Gaussian fitting results, respectively. The single Gaussian fitting loses part of the flux, which is unsuitable for accounting for the non-Gaussianity of the profiles. On the other hand, the optimal Gaussian fitting better models the non-Gaussian profiles, retrieving more flux than the single Gaussian fitting. For example, the total integrated intensity of J101448$-$274240 derived from the single Gaussian fitting analysis is 63.3 Jy beam$^{-1}$ which is approximately 7$\%$ lower than that derived from the optimal Gaussian fitting analysis with 68.0 Jy beam$^{-1}$. The number of optimally decomposed Gaussian components is mapped in the panel (e) of Fig.~\ref{fig2}. This map visualizes the non-Gaussian feature of the line profiles in a quantitative manner.

\subsection{H{\sc i} rotation curves}\label{sec:rotationcurve}
We derive the H{\sc i} rotation curves of the sample galaxies whose number of resolved elements across the major axis by the ASKAP's 30\arcsec\ beam is at least larger than five ($\geq$ 6) on their single Gaussian fitting velocity fields. There are 77 galaxies available for the H{\sc i} rotation curve analysis. As described in Section~\ref{sec:baygaud}, we mask the pixels of the velocity fields which do not satisfy the integrated S/N limit of three for their line profiles. An example single Gaussian fitting velocity field of a galaxy pair classified as in Section~\ref{sec:pair} is presented in panel (g) of Fig.~\ref{fig2}. We note that these single Gaussian fitting velocity fields which are based on the line profile fitting analysis are less affected by any spike-like noise in the line profiles, and thus better estimate their representative centroid velocities than the conventional moment analysis ({\sc moment1}) in most cases.

However, the single Gaussian fitting method has limited success in modelling non-Gaussian profile shapes (e.g., \citealt{2022ApJ...928..177O}). Alternative fitting forms, on the other hand, such as the Hermite polynomial and multiple Gaussian functions have worked in other cases (\citealt{2008AJ....136.2648D, 2008AJ....136.2761O, 2011AJ....141..193O, 2015AJ....149..180O, 2022ApJ...928..177O}). Since the main purpose of the rotation curve analysis in this work is to derive rotation velocities at a given radius in order to calculate the gravitational instabilities of the gas disk, and their geometrical parameters which are used to measure H{\sc i} morphology, the use of single Gaussian fitting velocity fields is adequate. However, there are two galaxies whose single Gaussian fitting velocity field maps have large blank areas due to low S/N. For these galaxies, we instead use their SoFiA2 {\sc moment1} maps for the rotation curve anlaysis. These galaxies are indicated by '*' symbol in Table~\ref{long_hydra_control} $\sim$ \ref{long_norma_pair}.

We perform a 2D tilted-ring analysis of the galaxy pairs and the control sample galaxies by fitting a model line-of-sight velocity field given in Eq.~(\ref{eq:12}) below. The 2D tilted-ring parameters of the galaxy pairs and the control sample galaxies selected in Section~\ref{sec:pair}  are estimated by maximizing the following log-likelihood function.

\begin{align}\label{eq:12}
{\rm log}\,L &= \sum_{\rho}^{} \sum_{\Phi}^{{}}\,\Biggl[{\rm log}\,\biggl(\frac{\Gamma(\frac{\nu + 1}{2})}{\sqrt{\pi(\nu-2)}\,\Gamma(\frac{\nu}{2})}\biggr) \\ \nonumber
&- \frac{1}{2}\,{\rm log}\,\epsilon^{2}  \\ \nonumber
&- \frac{\nu+1}{2}\,{\rm log}\,\biggl(1+\frac{\biggl({\rm v}^{\rm LOS}(\rho, \Phi) - {\rm v}^{\rm MODEL}(\rho, \Phi)\biggr)^{2}}{\epsilon^{2}\,(\nu-2)}\biggr)\Biggr],
\end{align}
\noindent where log\,L is the log-likelihood, ${\rm v}^{\rm LOS}(\rho, \Phi)$ is the single Gaussian fitting velocity field at a sky position of ($\rho$, $\Phi$), ${\rm v}^{\rm MODEL}(\rho, \Phi)$ is a model line-of-sight velocity field, and $\nu$ is the normality parameter which is set to three in this analysis. As discussed in \citealt{2018MNRAS.473.3256O}, the normality parameter of three is the smallest value for the Student-t distribution, which makes the fit as insensitive to any outliers as possible. $\Gamma$ is the gamma function given as
\begin{align}\label{eq:13}
\Gamma \left( x \right) = \int\limits_0^\infty {s^{x - 1} e^{ - s} ds}.
\end{align}
The distribution of the velocity residual between the observed and model velocity field is scaled by a free parameter, $\epsilon$ in Eq.~(\ref{eq:12}). This is largely proportional to the standard deviation of the residual map between the observed and model velocity fields.

To derive the rotation curves of the sample galaxies, we use 2D Bayesian Automated Tilted-ring fitter ({\sc 2dbat}; \citealt{2018MNRAS.473.3256O}) which allows us to fit 2D tilted-ring models to a velocity field in a fully automated manner based on Bayesian analysis techniques. We refer to \citet{2018MNRAS.473.3256O} for the full description of {\sc 2dbat} which describes its detailed fitting algorithm and performance test using simulated and observed H{\sc i} data cubes.

In the course of the non-parametric 2D tilted-ring analysis which fits concentric tilted-rings to the observed velocity field, the kinematic centre (XPOS, YPOS) and systemic velocity (VSYS) of a galaxy can be modelled with single values over all the rings. On the other hand, we adopt a cubic spline curve to model the rotation velocity parameter (v$_{\rm ROT}$) which varies with galaxy radius in the analysis. 

For the position angle ($\phi$) and the inclination ($i$) which tend to vary radially depending on the kinematic features of the gaseous disk of a galaxy such as bars, spiral arms, gas inflow or outflow, and warps, we regularise their radial variations using B-spline functions to suppress any sudden change or unphysical small-scale fluctuations so as to derive their representative radial gradients. The model $\phi$ and $i$ are given as, 

\begin{align}\label{eq:9}
\phi(R) = \sum_{l=1}^{P}c^{\phi}_{p}B^{\phi}_{l}(r), \\ \nonumber
i(R) = \sum_{m=1}^{Q}c^{i}_{q}B^{i}_{m}(r)
\end{align}

\noindent where $P$ and $Q$ are the numbers of B-splines, and $c^{\phi}_{p}B^{\phi}_{l,k}$ and $c^{i}_{q}B^{i}_{m,k}$ are the coefficients of the B-splines for $\phi$ and
$i$, respectively. In this work, we use two sets of B-spline orders, constant and cubic for $\phi(R)$ and $i(R)$ in Eqs.~(\ref{eq:9}). This is for checking the effect of different B-spline orders on the resulting rotation curves. 

\begin{figure*}
    \centering
    \includegraphics[width = 0.95\textwidth]{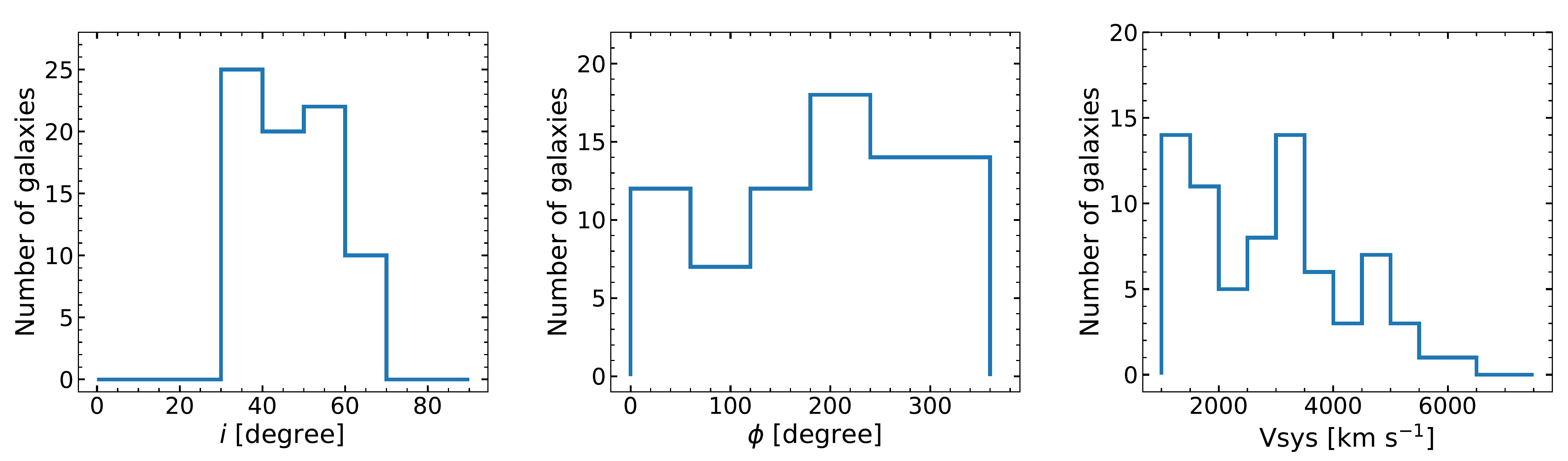}
    \caption{Distributions of inclination ($i$), position angle ($\phi$), and systemic velocities (VSYS) values of 77 sample galaxies from the rotation curve analysis.}
    \label{fig:INCL_PA_VSYS}
\end{figure*}

Given the regularization setup of the ring parameters, we derive two rotation curves ($\phi$: cubic spline, $i$: constant) for each galaxy.
The {\sc 2dbat} analysis was performed on all the galaxy pairs (48) and control galaxies (266) which are not matched in H{\sc i} mass and galaxy number ratios, and performed adequately in 77 cases ($\sim$25\%). The  exceptions, 237 cases, are significantly disturbed by non-circular motion or not well-resolved and hence were excluded from further rotation curve analysis. Most galaxies show no significant radial variations of $\phi$ and $i$ in the cubic-spline fits, providing comparable best-fit measurements of $\phi$ and $i$ between the two regularization setup (i.e., constant and cubic spline) within their uncertainties. No evidence of over-fitting of $\phi$ and $i$ using the cubic splines is seen in the analysis from visual inspection. In Fig.~\ref{fig:INCL_PA_VSYS}, we show the distributions of {\it i}, $\phi$ and VSYS of the 77 sample galaxies, derived from the tilted-ring analysis.

\subsection{H{\sc i} super-profiles of the sample galaxies}\label{sec:super-profile}
For each galaxy in our sample, we stack individual line profiles of the data cube after aligning their centroid velocities to derive the so-called `H{\sc i} super-profile' as described in \citet{2012AJ....144...96I}. In general, the S/N of the super-profile stacked using independent profiles increases with $\sqrt{N}$ where $N$ is the total number of the co-added profiles since the noise decreases as $1/\sqrt{N}$. Several studies have used super-profiles of galaxies, constructed using H{\sc i} or CO data cubes, in order to investigate correlations between the super-profile velocity dispersion and physical properties of galaxies (e.g., metallicity, FUV-NUV colours, and H$\alpha$ luminosities; see \citealt{2012AJ....144...96I}; \citealt{2013ApJ...765..136S}; \citealt{2022AJ....163..132H}). In addition, a super-profile can be decomposed into double Gaussian components which can be then classified as kinematically narrow and broad Gaussian components representing narrower and broader velocity dispersions, respectively. Based on the analysis of THINGS\footnote{The HI Nearby Galaxy Survey \citep{2008AJ....136.2563W}} galaxies, \citet{2012AJ....144...96I} show that the narrower (kinematically cool) H{\sc i} component of the super-profiles are likely to be associated with star formation and possibly with molecular hydrogen gas.

\begin{figure}
    \centering
    \includegraphics[width = 0.99\columnwidth]{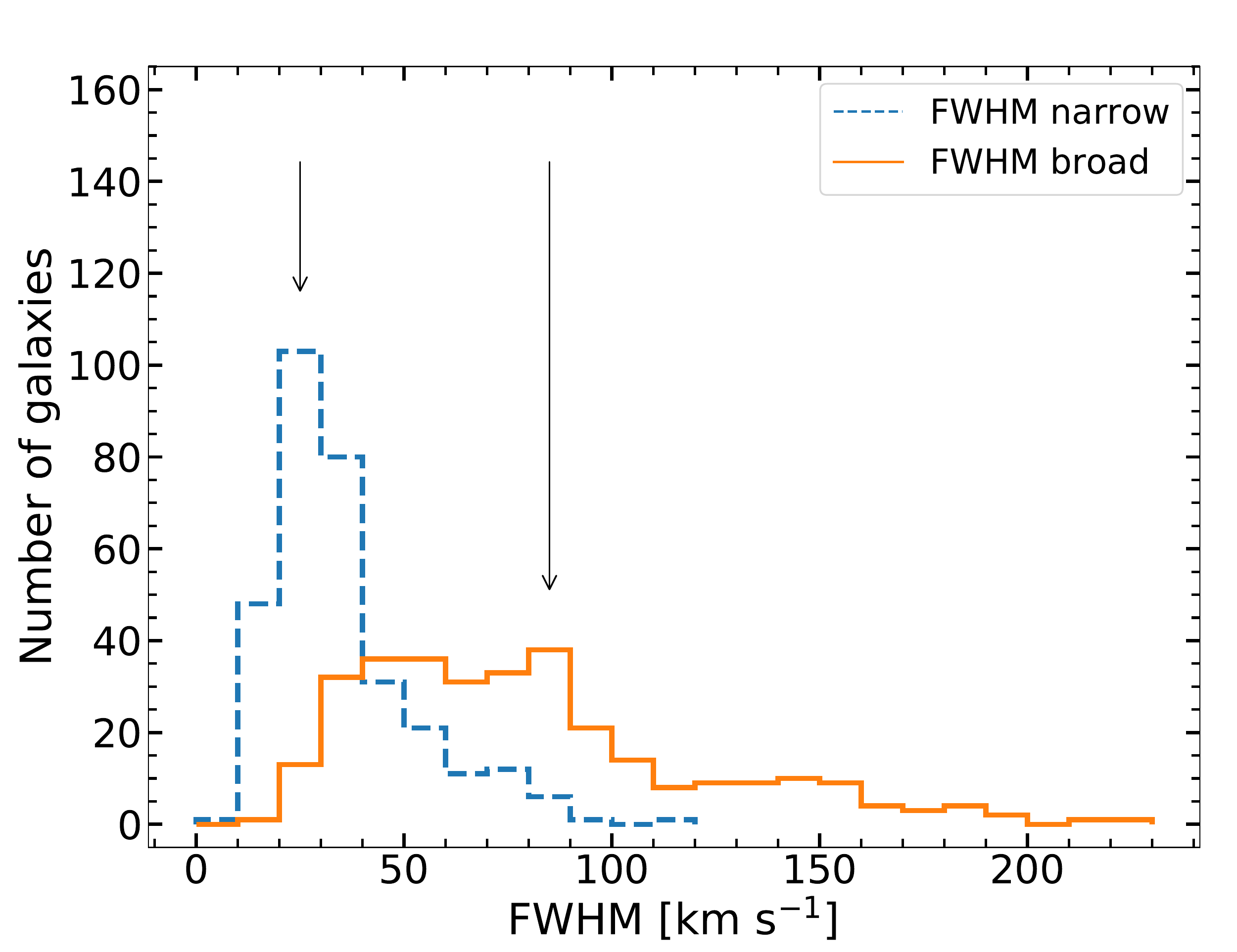}
    \caption{Distributions of FWHM values of narrow and broad components for all sample galaxies. The blue dashed and orange solid lines indicate the FWHM values of the narrow and broad components, respectively. The arrows indicate the mode value of each component.}
    \label{fig:FWHM}
\end{figure}

\begin{figure*}
    \centering
    \includegraphics[width = 0.9\textwidth]{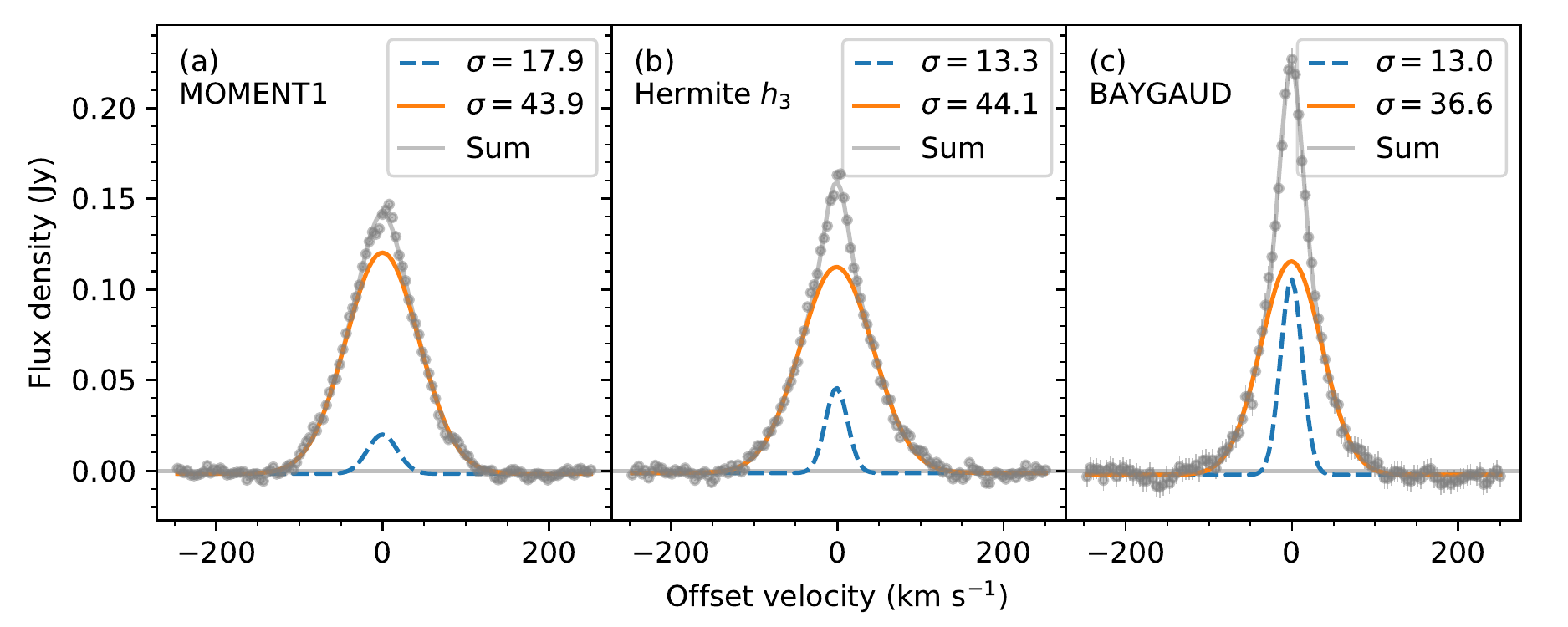}
    \caption[An example of H{\sc i} super-profile]{H{\sc i} super-profiles of J101247-275028 derived from the three profile analysis methods: (a) moment analysis, (b) Gauss-Hermite {\it h3} polynomial fitting, and (c) {\sc baygaud} analysis with double Gaussian fitting. Grey dots show the corresponding H{\sc i} super-profiles, and the blue dashed and orange solid lines indicate the narrow and broad components decomposed from the double Gaussian fitting. The grey solid lines are the sum of the narrow and broad components. The derived velocity dispersions of the narrow and broad components are denoted on the top-right corner of the panels.}
    \label{fig4}
\end{figure*}

In this work, we construct super-profiles of the sample galaxies using their decomposed H{\sc i} gas components from {\sc baygaud} analysis in Section~\ref{sec:baygaud}. The super-profile stacking analysis that we adopt is as follows:

\noindent 1) Decompose individual H{\sc i} line profiles into optimal numbers of Gaussian components using {\sc baygaud}; determine their central velocities from the Gaussian fitting.
\newline
2) For the extraction of each Gaussian component of a line profile, we subtract the other Gaussian model components from the original line profile. If the profile is best described by a single Gaussian function, use the input velocity profile whose central velocity is determined from the single Gaussian fitting.
\newline
3) Align all the residual line profiles from which the other Gaussian models are subtracted with respect to their centroid velocities; Use the original line profiles for the alignment if they are best modelled by a single Gaussian function.
\newline
4) Construct the H{\sc i} super-profile by co-adding all the aligned line profiles.
\newline
5) Fit a double Gaussian model to the super-profile; use the velocity channels whose velocities are within $\pm$ three times velocity dispersion with respect to the central velocity of the super-profile; the velocity dispersion and the central velocity are derived by fitting a single Gaussian function to the super-profile; this prevents us fitting unreasonably broad velocity dispersions, which might be caused by background ripples that result from inaccurate baseline subtraction.
\newline
6) Derive the amplitudes, velocity dispersions, and central velocity of the primary and secondary Gaussian components and constant background as a means to quantify the super-profile.

The Gaussian model parameters are estimated using the python module {\sc emcee} with a prior range of $\delta V$ $<$ velocity dispersion ($\sigma$) $<$ 150 \kms, and $-3\times\delta V$ $<$ central velocity $<$ $+3\times\delta V$, where $\delta V$ is the channel resolution.
In this work, we do not include line profiles whose integrated S/N is less than three when constructing the super-profiles. We note that the optimally decomposed line profiles whose central velocities are determined from the Gaussian fitting are stacked to produce the super-profiles. This is compared with the previous stacking methods which align the line profiles to their central velocities determined from the moment analysis (intensity-weighted mean velocity) or Hermite polynomial fit.

In Fig.~\ref{fig:FWHM}, we show the ranges of full width at half maximum (FWHM) values of the narrow and broad components in the H{\sc i} super-profile analysis. The blue dashed and orange solid lines indicate the FWHM values of narrow, and broad components, respectively. Approximately, they range from 10 \kms to 119 \kms and 20 \kms to 226 \kms\ , respectively. The mode values of the distributions for the narrow and broad components indicated by the solid vertical lines are $\sim$25 and $\sim$85 \kms, respectively. For this, we also derive additional H{\sc i} super-profiles of a sample galaxy, J101247$-$275028, using the centroid velocities of line profiles determined from the moment analysis ({\sc moment0}) and the Hermite {\it h3} polynomial fitting. We then compare them with the {\sc baygaud}-based H{\sc i} super-profile as shown in Fig.~\ref{fig4}. Turbulent gas motions caused by hydrodynamical processes in galaxies like star formation or supernova (SN) explosions can result in larger velocity dispersions and asymmetric tails to H{\sc i} velocity profiles. A super-profile constructed using these asymmetric non-Gaussian line profiles with larger velocity dispersions will have broader wings and lower peaks than the one constructed using the decomposed Gaussian profiles which tend to have smaller velocity dispersions. We refer to \citet{2022arXiv220900390K} for the full description of the super-profile stacking analysis adopted in this work and its comparison with the previous methods using THINGS and LITTLE THINGS\footnote{Local Irregulars That Trace Luminosity Extremes, The HI Nearby Galaxy Survey (LITTLE THINGS) \citep{2012AJ....144..134H}} sample galaxies.

The influence of galaxy\--galaxy interactions, or merging processes, on the gas distribution and kinematics of galaxies that experience close encounters varies for in different parts of a galaxy; i.e. core and outer regions (\citealt{1990ApJ...361..426O}; \citealt{1991ApJ...370L..65B}; \citealt{2017ApJ...844...96Y}; \citealt{2021MNRAS.503.3113M}). Accordingly, this can lead to different spatial distribution and star formation rates on a range of time scales in the galaxies (\citealt{2012ApJ...758...73S}; \citealt{2022ApJ...927...66W}). Many post-merger galaxies show enhanced star formation in their central kpc region (\citealt{2013MNRAS.435.3627E}). This is mainly attributed to the inflow of interstellar medium into the central regions of the galaxies, which results in the formerly atomic medium being compressed and thus converted into molecular gas due to the higher hydrostatic pressures (\citealt{2021MNRAS.503.3113M}).

In this regard, we also split the inner and outer gas disk regions of the sample galaxies at an H{\sc i} break radius, $R_{\rm{HI-break}}$. For each sample galaxy, we derive two additional super-profiles for the central and outer regions as well as the one for the entire gas disk region. As a $R_{\rm{HI-break}}$, we adopt 0.5$R_{\rm{HI}}$ based on which the H{\sc i} properties of the super-profiles for the central and outer regions show the most significant change with respect to the galaxy environments. $R_{\rm{HI}}$ is the radius of H{\sc i} disk defined at a surface density ($\sum_{\rm{HI}}$) of 1 $\rm{M_{\odot} \, pc^{-2}}$ in the $\sum_{\rm{HI}}$ radial profile. To derive $R_{\rm{HI}}$, we use the H{\sc i} mass-size relation (\citealt{1997A&A...324..877B}; \citealt{2016MNRAS.460.2143W}). These regional super-profiles will be useful for examining how the H{\sc i} gas properties react with the gravitational impact of galaxies in different galaxy environments.

The H{\sc i} super-profiles of galaxy pair candidates in the Hydra I, Norma clusters and NGC 4636 group are presented in Figs.~\ref{figA4} $\sim$ \ref{figA6} in Appendix (The full super-profile figures are available as supplementary online material). We classify the decomposed Gaussian profiles of an H{\sc i} super-profile as kinematically narrow and broad components with respect to their velocity dispersions ($\sigma$). Relatively, the kinematically narrow component has a smaller $\sigma$ than the broad one.

We derive ratios of the H{\sc i} properties of the kinematically narrow and broad components, such as the H{\sc i} mass ratios, $M_{\rm{narrow}}^{\rm{HI}}$/$M_{\rm{total}}^{\rm{HI}}$ and $M_{\rm{broad}}^{\rm{HI}}$/$M_{\rm{total}}^{\rm{HI}}$. These are used for comparing the different super-profiles derived. The H{\sc i} gas mass of a galaxy is derived using the following equation:
\begin{equation}
    \label{eq:MHI}
    \frac{M_{\rm{HI}}}{\rm{M_{\odot}}} = 49.7  \left(\frac{D}{\rm{Mpc}}\right)^2 \left(\frac{S_{\rm{int}}}{\rm{Jy} \ Hz}\right),
\end{equation}
where $D$ is the distance to the galaxy in $\rm{Mpc}$, and $S_{\rm{int}}$ is the integrated H{\sc i} flux in $\rm{Jy} \ Hz$. We note that the sum of the $M_{\rm{narrow}}^{\rm{HI}}$ and $M_{\rm{broad}}^{\rm{HI}}$ equals the total H{\sc i} gas mass of a galaxy. We list the $M_{\rm{narrow}}^{\rm{HI}}$ and $M_{\rm{broad}}^{\rm{HI}}$ of the sample galaxies derived from their H{\sc i} super-profiles in Table~\ref{long_hydra_control} $\sim$ \ref{long_norma_pair} in Appendix.

\subsection{Toomre Q parameter}\label{sec:toomre}
We derive the Toomre Q parameter (\citealt{1964ApJ...139.1217T}) of the galaxies in our sample to examine how they are correlated with the galaxy environments. The Toomre-Q parameter has been widely used for measuring the degree of gravitational instability in the rotating gaseous thin disk of a galaxy (\citealt{1989ApJ...344..685K}; \citealt{2001ApJ...555..301M}; \citealt{2006AJ....131..363D}; \citealt{2008AJ....136.2782L}; \citealt{2010AJ....140.1194B}; \citealt{2017MNRAS.472.3761N}; \citealt{2018MNRAS.476..122V}). The Toomre-Q parameter for gas instability can be derived as follows:
\begin{equation}
    \label{eq:Q}
    Q_{\rm{gas}} = \alpha \frac{\sigma_g \kappa}{\pi G \Sigma_{\rm{gas}}}\,,
\end{equation}
where $\alpha$ is a dimensionless constant value, $\sigma_g$ is the gas velocity dispersion in units of \kms, {\it G} is the gravitational constant, and $\Sigma_{\rm{gas}}$ is the gas surface density in units of $\rm{M_{\odot}}$\,pc$^{-2}$. $\kappa$ is an epicyclic frequency of the galaxy disk in unit of s$^{-1}$ varying with radius by its definition:
\begin{equation}
    \label{eq:kappa}
    \kappa^{2} = 2 \left(\frac{V}{R}\right)^{2} \left[1+\left(\frac{R}{V}\right)\left(\frac{dV}{dR}\right) \right]\,,
\end{equation}
where {\it V} is the rotation velocity in units of \kms\ and {\it R} is the radius in pc. Following  \citet{1989ApJ...344..685K}, we adopt the value $\alpha$ = 0.63. Given the above, \citet{1989ApJ...344..685K} defined a critical surface density ($\Sigma_{\rm{crit}}$) over which gas clouds become gravitationally unstable: 
\begin{equation}
    \label{eq:crit}
    \Sigma_{\rm{crit}} = \alpha \frac{\sigma_g \kappa}{\pi G}\,.
\end{equation}

Gas instability leads to gravitational collapse and star formation is expected in regions where the observed gas surface density $\Sigma_{\rm{gas}}$ is higher than the critical surface density $\Sigma_{\rm{crit}}$, that is, $\Sigma_{\rm{gas}} / \Sigma_{\rm{crit}} >$ 1, which means $Q_{\rm{gas}}$ $<$ 1. Below this so-called star formation threshold, gas pressure stabilizes the gas cloud against gravitational collapse. This Toomre Q criterion is analogous to the Jeans criterion, but it includes the kinematic effect of the disk's rotation. 

In Eq.~(\ref{eq:Q}), we use the gas surface density summed over the optimally decomposed Gaussian components. We then multiply by a factor of 1.4 to the H{\sc i} gas surface density derived in order to take the amount of helium and metal abundances of the gas into account. To estimate $\kappa$ in Eq.~(\ref{eq:kappa}), we adopt the rotation curves derived from the {\sc 2dbat} analysis in Section~\ref{sec:rotationcurve}.
As $\kappa$ is derived from the rotation curve analysis, we only select galaxies which show a clear rotation pattern in their H{\sc i} velocity fields and that have intermediate values of inclination (30$^\circ$ $<$ $i$ $<$ 70$^\circ$), and are resolved with $\geq$ 6 beams across their major axes. 

For the gas velocity dispersion in Eq.~(\ref{eq:Q}), we use the velocity dispersion values derived from the single Gaussian fitting to the H{\sc i} velocity profiles. As for the H{\sc i} super-profiles derived in Section~\ref{sec:super-profile} for each galaxy, we derive Toomre Q parameters over the central ($0 < r < 0.5R_{\rm{HI}}$), outer ($0.5R_{\rm{HI}}$ $< r < R_{\rm{HI}}$), and entire ($0 < r < R_{\rm{HI}}$) disk regions. 

In theory, if $Q_{\rm{gas}}$ $<$ 1, the disk is expected to become gravitationally unstable and trigger star formation. However, \citet{2008AJ....136.2782L} and \citet{2017MNRAS.469..286R} found that star-forming galaxies often have $Q_{\rm{gas}}$ values above unity. This is because the Toomre Q criterion assumes a simple disk model of symmetric perturbations and we assume a pure gas disk. Therefore, we should keep in mind that the Toomre Q analysis is limited in identifying the local disk gravitational instability in a disturbed gas disk.

\begin{figure}
    \centering
    \includegraphics[width = 0.5\textwidth]{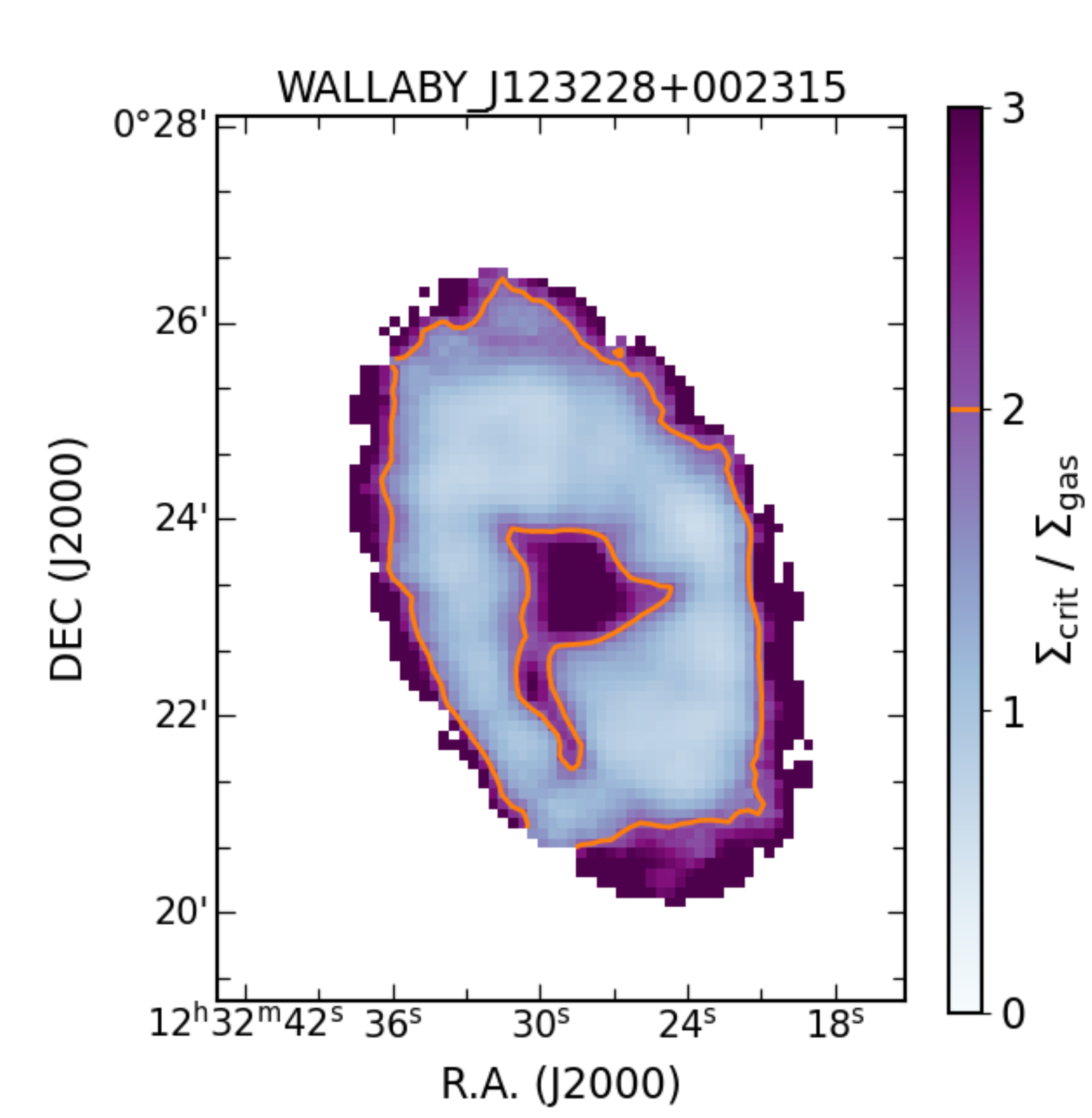}
    \caption[An example of 2D map of Toomre Q values]{A 2D map of the Toomre Q parameter values as indicated in the colourbar for J123228$+$002315. The regions with $Q_{\rm{gas}}$ $<$ 2 are enclosed by the orange contours.}
    \label{fig:toomre_entire_inner_outer}
\end{figure}

In Fig.~\ref{fig:toomre_entire_inner_outer}, we present an example 2D map of the Toomre Q values of a galaxy, J123228$+$002315, overlaid with the orange contours of $Q_{\rm{gas}}$ $<$ 2 which indicate gravitationally unstable regions (\citealt{2013MNRAS.434.3389Z}, \citealt{2016MNRAS.460.1106W}). From this, we are able to identify gravitationally unstable candidate regions where $Q_{\rm{gas}}$ $<$ 2 corresponding to $\Sigma_{\rm{gas}} / \Sigma_{\rm{crit}} >$ 2. As shown in Fig.~\ref{fig:toomre_entire_inner_outer}, the $Q_{\rm{gas}}$ values in the central and outer regions are higher. However, this might be due to an overestimated gas velocity dispersion of H{\sc i} velocity profiles which are presumably caused by the effect of the beam resolution (so-called beam smearing). Given this, the estimated $Q_{\rm{gas}}$ values can be considered as upper limits for the gravitational instability of the gas disk.

\subsection{H{\sc i} asymmetry}\label{sec:mor}
As an additional parameter for quantifying the H{\sc i} morphological disturbance of the sample galaxies, we measure the morphological asymmetries of their integrated H{\sc i} intensity maps ({\sc moment0}). For this, we use the following formula as described in \citet{2020MNRAS.493.5089R},
\begin{equation}
    \label{eq:1}
    A_{\rm{map}}^{\rm{HI}} = \frac{\Sigma_{i,j}|I(i,j) - I_{180}(i,j)|}{2\Sigma_{i,j}|I(i,j)|}\,
\end{equation} where $I(i,j)$ is the integrated intensity map at a sky position of ($i, j$), and $I_{180}(i,j)$ is the same map but rotated by 180$^\circ$.

\begin{figure*}
    \centering
    \includegraphics[width = 0.99\textwidth]{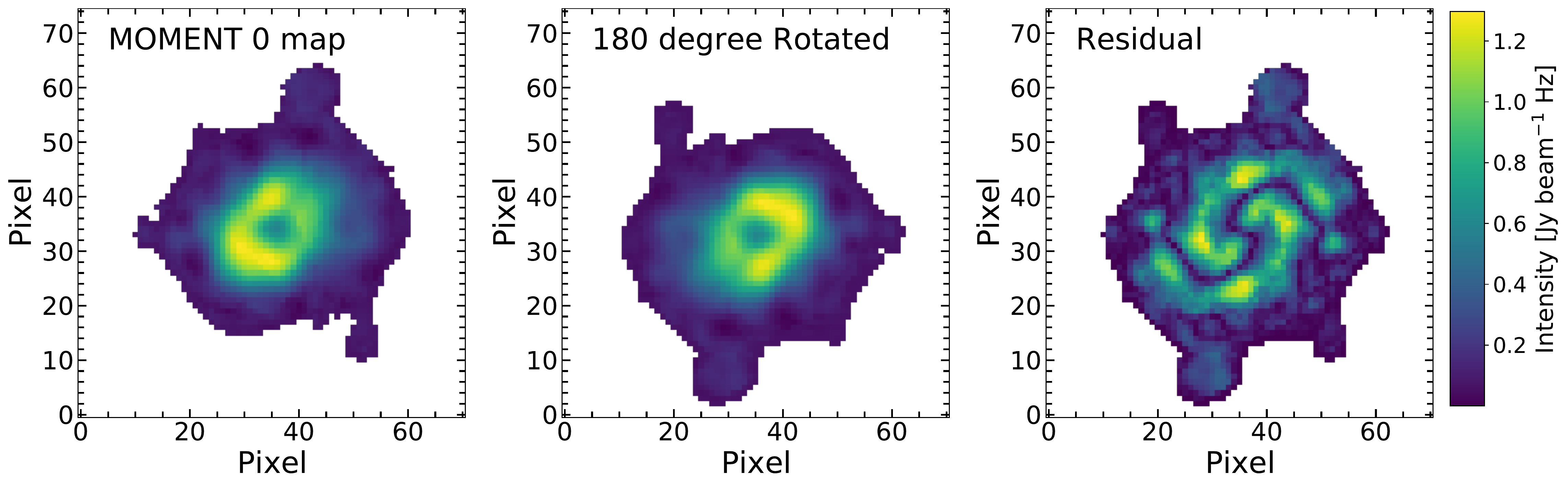}
    \caption[Illustration of integrated intensity asymmetry]{The H{\sc i} integrated-intensity maps of J104016$-$274630 used for deriving the galaxy’s morphological H{\sc i} asymmetry. Left: H{\sc i} {\sc moment0} map, middle: H{\sc i} {\sc moment0} map rotated 180$^{\circ}$, and right: the absolute residual map as indicated in the colourbar. See Section~\ref{sec:mor} for more details.}
    \label{fig:amap}
\end{figure*}

The morphological asymmetry of the {\sc moment0} can be measured by summing the absolute residuals between the two maps, $I$ and $I_{180}$. We rotate the SoFiA2 {\sc moment0} map of a galaxy with respect to its kinematic centre derived from the {\sc 2bdat} analysis. We refer to \citet{2020MNRAS.493.5089R} for the full description of measuring the H{\sc i} morphological asymmetry of a galaxy. As an example, Fig.~\ref{fig:amap} shows the H{\sc i} intensity maps (the original, 180$^\circ$ rotated, and residual ones) of a galaxy, J104016$-$274630. The $A_{\rm{map}}^{\rm{HI}}$ value for this galaxy derived using  Eq.~(\ref{eq:1}) is 0.19. The larger value of $A_{\rm{map}}^{\rm{HI}}$, the more asymmetric H{\sc i} gas distribution is.

Both internal and external hydrodynamical effects of galaxies such as gas outflows driven by star formation or SN explosions, ram pressure stripping, and tidal interactions between galaxies can make the gas disk of galaxies disturbed, particularly in the outer regions (\citealt{1972ApJ...176....1G}; \citealt{1996ApJ...471..115B}; \citealt{2005ApJ...621..227M}). The analysis of the morphological H{\sc i} asymmetry in different environments is useful for investigating how environments affect the H{\sc i} distribution of galaxies. This is discussed in Section~\ref{sec:4}.

\section{Environmental effects on the resolved H{\sc i} properties of galaxy pairs}
\label{sec:4}

In this section, we examine the effects of galaxy environment (outskirts, infalling, and central) on the resolved H{\sc i} properties of galaxy pairs in the two clusters (Hydra I and Norma) and the galaxy group (NGC 4636). The following analyses are carried out: 1) First, we investigate how the fractions of the kinematically decomposed narrow H{\sc i} gas components in the total H{\sc i} gas of galaxy pairs and control sample galaxies change with the infall stage (i.e., central, infalling, and outskirts). 2) Second, we assess whether the infall stage has an impact on the gravitational instability of gas disks using the Toomre Q parameters. 3) Third, we inspect the morphological H{\sc i} asymmetries of galaxies to see whether there is any systematic difference between the pairs and control galaxies in the environments. Using these analyses, we examine whether the resolved H{\sc i} properties of the sample galaxies are affected by the galaxy environments in a quantitative manner.

\subsection{Kinematically narrow H{\sc i} gas}\label{sec:fnarrow}

It has been shown that galaxy mergers and strong tidal interactions of gas-rich galaxies can trigger star formation (\citealt{1996ApJ...464..641M}; \citealt{2007A&A...468...61D}; \citealt{2013MNRAS.433L..59P}; \citealt{2013MNRAS.435.3627E}). On the other hand, galaxy cluster and group environments are more likely to quench star formation as galaxies reach close to the cluster or group centre (\citealt{2011MNRAS.415.1797C}; \citealt{2016A&A...596A..11B}; \citealt{2021PASA...38...35C}).

Molecular hydrogen in galaxies serves as a gas reservoir for star formation, which is sustained and/or fed from cool H{\sc i} gas (e.g. \citealt{2008AJ....136.2846B}; \citealt{2022arXiv220200690S}). In the course of forming the molecular hydrogen gas, atomic hydrogen should have passed the kinematically narrow (cool) H{\sc i} gas before turning into H$_2$. Thus, the kinematically narrow H{\sc i} gas components decomposed from the stacked profiles in Section~\ref{sec:super-profile} may be associated with the H$_2$ component (e.g., \citealt{2012AJ....144...96I}). 

\begin{figure*}
    \centering
    {
    \includegraphics[width = 0.33\textwidth]{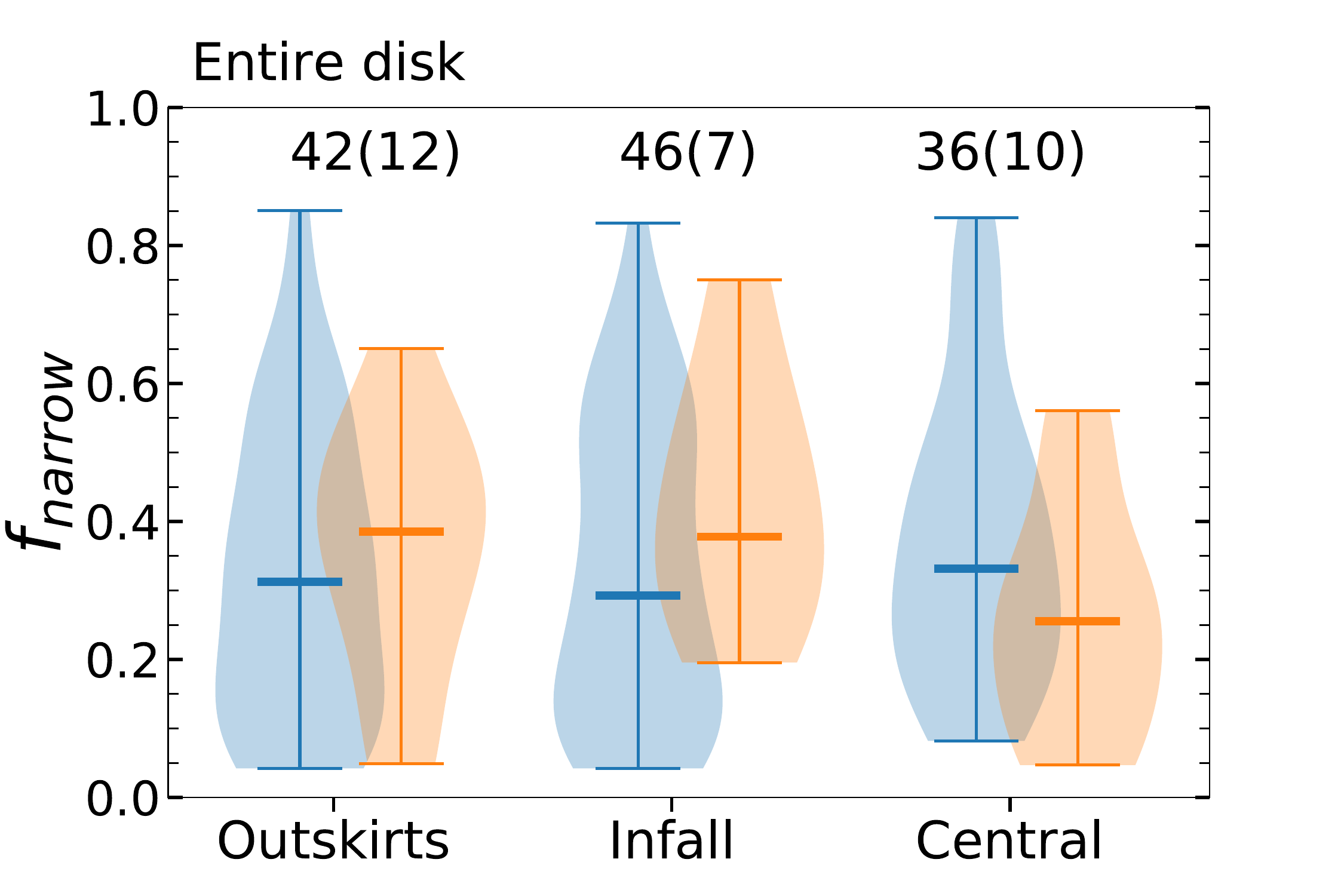}
    \includegraphics[width = 0.33\textwidth]{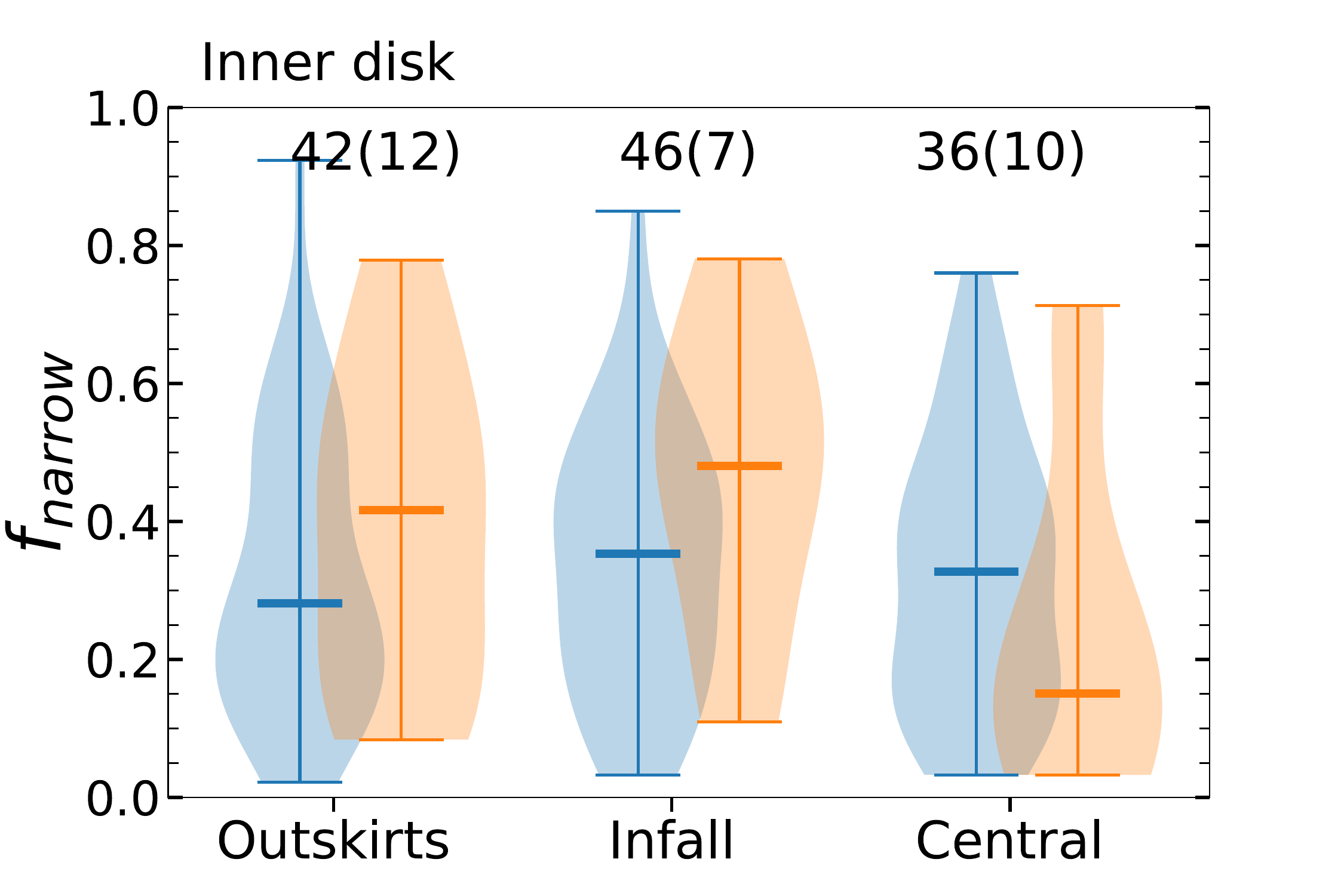}
    \includegraphics[width = 0.33\textwidth]{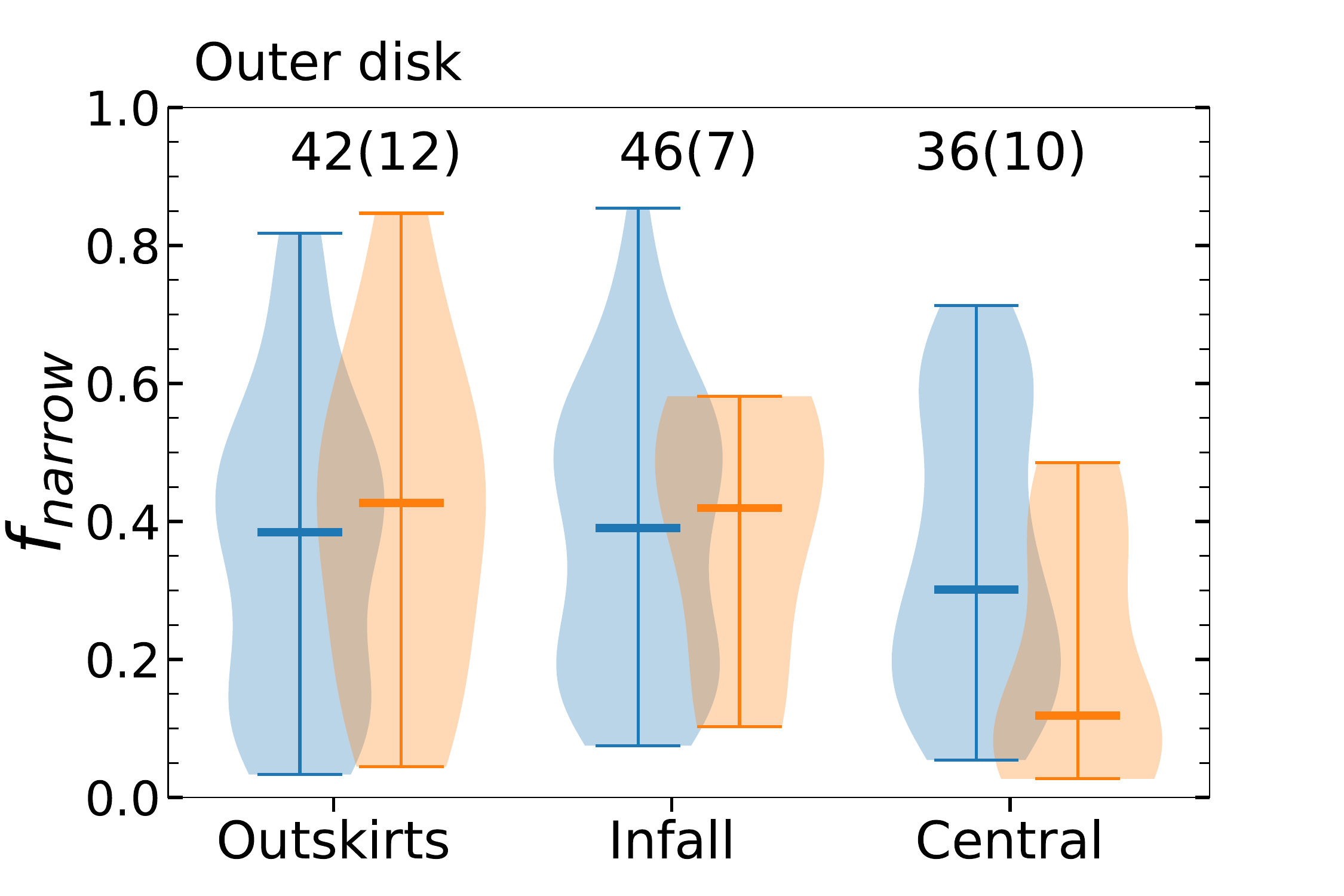}
    }
    \caption[Distributions of the kinematically narrow H{\sc i} gas fraction at different environments]{Distributions of the kinematically narrow H{\sc i} gas fractions $f_{\rm{narrow}}$ =  $M_{\rm{narrow}}^{\rm{HI}}$ / $M_{\rm{total}}^{\rm{HI}}$ of the three different regions (‘Entire’, ‘Inner’, and ‘Outer’) of the sample galaxies, in the outskirts, infalling, and central environments. The control galaxies and galaxy pairs are shown with blue and orange colours, respectively. The numbers of the control and pair (bracket) galaxies are shown in each panel. The border thickness represents the relative frequency of the sample galaxies with the corresponding $f_{\rm{narrow}}$ values. The small horizontal lines in each panel indicate the median values.}
    \label{fig:cold}
\end{figure*}

Here, we investigate the influence of galaxy environments associated with the infall stage on the kinematically narrow H{\sc i} gas components of the sample galaxies.
Fig.~\ref{fig:cold} shows the mass fractions of the kinematically narrow H{\sc i} gas in the total H{\sc i} masses, $f_{\rm{narrow}}$ = $M_{\rm{narrow}}^{\rm{HI}}$ / $M_{\rm{total}}^{\rm{HI}}$, derived for the three different disk regions of the sample galaxies. In addition, these are shown for the different galaxy environments, outskirts, infalling, and central as classified in Section~\ref{sec:classification}. Here, we use all the sample galaxies in the three fields, the Hydra I and Norma clusters, and NGC 4636. The mass fractions for the galaxy pairs and control galaxies are displayed with different colours, galaxy pair (orange) and control galaxy (blue). The numbers of the control and pair (bracket) galaxies in the galaxy environment bins are denoted on each panel in Fig.~\ref{fig:cold}.

\begin{table*}
\tiny{
\centering
\begin{tabular}{lccccccccccccccc}
\hline
\hline
 \multicolumn{2}{c}{} & \multicolumn{4}{c}{Outskirts} & \multicolumn{4}{c}{Infalling} & \multicolumn{4}{c}{Central} \\
 \\

        & & $\langle f_{\rm{narrow}} \rangle^{\rm{median}}$  & 1$\sigma$  & CI & N & $\langle f_{\rm{narrow}} \rangle^{\rm{median}}$ & 1$\sigma$  & CI & N & $\langle f_{\rm{narrow}} \rangle^{\rm{median}}$  & 1$\sigma$  & CI & N\\
        
\hline

Entire & Control & 0.31 & 0.21 & (0.22, 0.39) & 42 & 0.29 & 0.22 & (0.21, 0.46) & 46 & 0.33 & 0.21 & (0.26, 0.45) & 36 \\
& Pair & 0.39 & 0.18 & (0.27, 0.54) & 12 & 0.38 & 0.17 & - & 7 & 0.26 & 0.16 & (0.09, 0.56) & 10 \\ \\

Inner & Control & 0.28 & 0.20 & (0.20, 0.45) & 42 & 0.35 & 0.19 & (0.25, 0.42) & 46 & 0.33 & 0.19 & (0.21, 0.40) & 36 \\
& Pair    & 0.42 & 0.22 & (0.20, 0.69) & 12 & 0.48 & 0.21 & - & 7 & 0.15 & 0.23 & (0.08, 0.71) & 10 \\ \\

Outer & Control & 0.38 & 0.22 & (0.25, 0.47) & 42 & 0.39 & 0.20 & (0.30, 0.49) & 46 & 0.30 & 0.20 & (0.24, 0.47) & 36 \\
& Pair    & 0.43 & 0.23 & (0.23, 0.68) & 12 & 0.42 & 0.18 & - & 7 & 0.12 & 0.17 & (0.05, 0.49) & 10 \\

\hline
\end{tabular}}
\caption[Median and standard deviation of kinematically narrow H{\sc i} gas fractions at different environments]{The median values ($\langle f_{\rm{narrow}} \rangle^{\rm{median}}$) and standard deviations ($\sigma_{f_{\rm{narrow}}}$) of the kinematically narrow H{\sc i} gas fractions for the three disk regions (entire, inner and outer) of the galaxy pairs and control sample galaxies in the outskirts, infalling, and central environments. N is the number of samples, and CI denotes the 95$\%$ confidence interval for median value.}
\label{tab:cold}

\end{table*}

As shown in Fig.~\ref{fig:cold}, the median $f_{\rm{narrow}}$ values indicated by the thick horizontal lines for the entire disk do not appear to vary much over the environments for the control galaxies (blue). The median $f_{\rm{narrow}}$ values of the infalling galaxies are slightly higher in both the inner and outer disk regions than the others in the outskirts and central environment.
Despite the wide range of the $f_{\rm{narrow}}$ values for the outer disk region of the control galaxies, the median value is the lowest in the central environment although it is not significant. More frequent galaxy interactions are expected in the central environment with higher local densities, which could give rise to more turbulent gas motions with higher velocity dispersions. 

The ram pressure caused by the intergalactic medium (IGM) could also have influenced the $f_{\rm{narrow}}$ values of galaxies in the central environment. It could strip part of the kinematically narrow H{\sc i} gas in denser environments (\citealt{2009AJ....138.1741C}; \citealt{2011MNRAS.415.1797C}; \citealt{2013MNRAS.436...34C}; \citealt{2016MNRAS.461.1202J}; \citealt{2017MNRAS.466.1275B}; \citealt{2021ApJ...915...70W}). Molecular hydrogen (H$_2$) gas in galaxies traced by CO emission, particularly in the outer regions, can also be stripped by ram pressure in a cluster environment (\citealt{2014A&A...564A..67B}; \citealt{2017ApJ...843...50C}; \citealt{2018A&A...617A.103C}; \citealt{2021MNRAS.502.3158S}; \citealt{2021PASA...38...35C}). 

The removal of cool H{\sc i} gas by ram pressure stripping will in turn lead to star formation quenching (\citealt{2004ApJ...613..866K}; \citealt{2006PASP..118..517B}; \citealt{2015MNRAS.448.1715J}). Additionally, numerous high-speed galaxy encounters, i.e. “galaxy harassment” can increase the gas velocity dispersion of galaxies in the denser environment (\citealt{1996Natur.379..613M}). This in turn will decrease their $f_{\rm{narrow}}$ values. In the course of galaxy harassment, the cool (kinematically narrow) H{\sc i} component can be heated, which results in a higher velocity dispersion.

As shown in Fig.~\ref{fig:cold}, the median $f_{\rm{narrow}}$ values for the three disk regions (entire, inner and outer) of galaxy pairs (orange) are the lowest in the central environment. The gravitational effect of the close galaxy pairs may play a role in inducing the trend.  As discussed in \cite{2010MNRAS.407.1514E}, star formation triggered by galaxy\--galaxy interaction is more enhanced in the outskirts (i.e. field) than in dense environments. This is likely to be associated with the higher $f_{\rm{narrow}}$ values of galaxy pairs in low-density environments as found in this work (see also \citealt{2012A&A...543A..33V}, \citealt{2017MNRAS.466.1382L}, and \citealt{2015ApJ...805....2S}). \citealt{2016MNRAS.461.2630M} concluded that the satellite encounters are more effective at stripping H{\sc i} gas than ram pressure stripping in the dense environment using the EAGLE simulations.
We note that the difference in H{\sc i} mass distribution between the environments (outskirts, infalling and central) may impact the $f_{\rm{narrow}}$ trend found in this work. As shown in Fig.~\ref{fig:HI_mass_distribution}, the fraction of galaxies in low $M_{\rm{HI}}$ mass bins (${\rm log\,M_{HI}}$ $<$ 8.7) is relatively lower in the infalling and central regions than in the outskirts. In these denser infalling and central environments, galaxies may have already experienced moderate gas loss which is below the current WALLABY’s sensitivity. However, it is difficult to predict how exactly this process affects the $f_{\rm{narrow}}$ trend in the subcluster environments due to the low number statistics. A larger sample of galaxies is needed to further improve the analysis. In addition, we also note that there are a number of sample galaxies ($\sim$11 and 7$\%$ of the control and paired galaxies) that have similar velocity widths (velocity dispersion difference $<$ ASKAP channel resolution of 4 \kms) of the narrow and broad components in the H{\sc i} super-profile analysis. These galaxies are denoted with a † symbol in Table~\ref{long_hydra_control} $\sim$ \ref{long_norma_pair}. This could be attributed to factors like low S/N, fractions of cool or warm gas components over the total gas etc. A scaled single Gaussian model fit would be appropriate for measuring global velocity dispersions of these super-profiles as described in \cite{2013ApJ...765..136S}. In this work, since we need to decompose the super-profiles into kinematically narrow and broad components, a double Gaussian model fit is made to the profiles. We tabulate the median and standard deviations of the $f_{\rm{narrow}}$ values derived for the entire, inner and outer gas disks of the sample galaxies in Table~\ref{tab:toomre}.

\begin{figure*}
    \centering
    \includegraphics[width=0.98\textwidth]{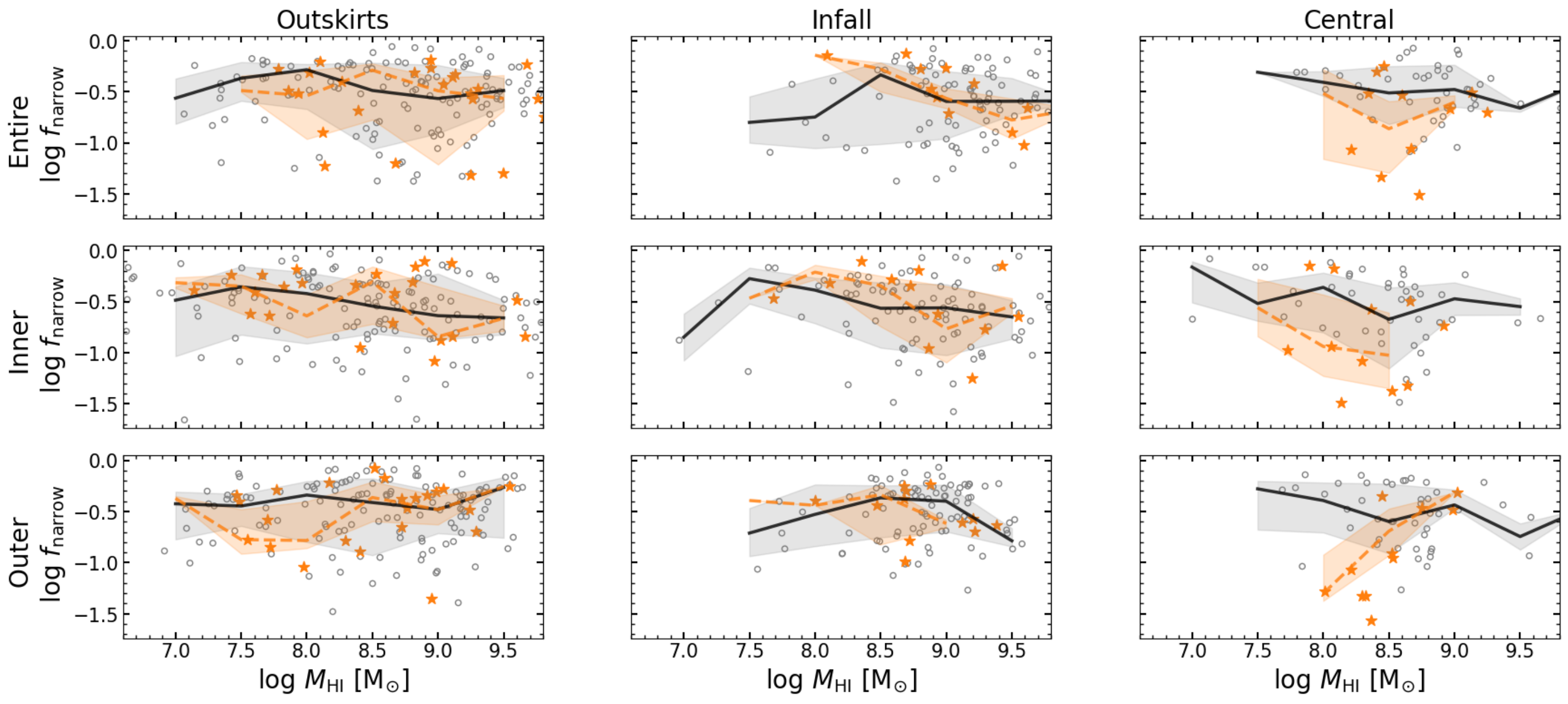}
    \caption{The kinematically narrow H{\sc i} gas fractions for the three disk regions (entire, inner and outer) of sample galaxies with their H{\sc i} masses in the outskirts (left column panels), infall (middle column panels) and central (right column panels) environments. Grey open circles and orange stars indicate non-paired and paired galaxies. The black solid and orange dashed lines indicate the median values for the non-paired and paired galaxies with 1$\sigma$ bounds (grey and orange shaded), respectively. See Section~\ref{sec:fnarrow} for more details.}
    \label{fig:corr_cold}
\end{figure*}

\cite{2018MNRAS.478.3447E} argue that merger-induced star formation quenching is not via exhaustion or expulsion of the gas reservoir. Instead, it could be due to enhanced turbulent gas motions which make the gas reservoir less capable of forming stars. The decreasing trend of the median $f_{\rm{narrow}}$ values for the galaxy pairs in the central environment seen in our work seems to support such a scenario. In Fig.~\ref{fig:corr_cold}, we show the $f_{\rm{narrow}}$ values for the three disk regions (entire, inner and outer) of galaxy pairs with their H{\sc i} masses, and compare them with those of non-paired galaxies in the outskirts, infall and central environments. For this, we use 48 galaxy pairs and 266 non-paired galaxies detected at H{\sc i} in the Hydra I, Norma and NGC 4636 fields from the ASKAP pilot observations.

As shown in Fig.~\ref{fig:corr_cold}, despite the large scatter, the median $f_{\rm{narrow}}$ values of the galaxy pairs (the orange dashed lines) are likely to be lower than those of the non-paired galaxies (the black solid lines) in the central environments. In particular, this trend seems to be more enhanced in the outer disk regions of the galaxy pairs. Although the H{\sc i} mass range of the galaxy pairs in the central environment is not wide ($8 < {\rm log\,M_{HI}} < 9$), this may be an indicative of the enhanced gas velocity dispersions in galaxy pairs as proposed in \citet{2018MNRAS.478.3447E} (see Fig. 2 in their work).

\subsection{The effect of projected separations between galaxy pairs on $f_{\rm{narrow}}$}\label{kinamatic}

\begin{figure}
    \centering
    {
    \includegraphics[width = 0.5\textwidth]{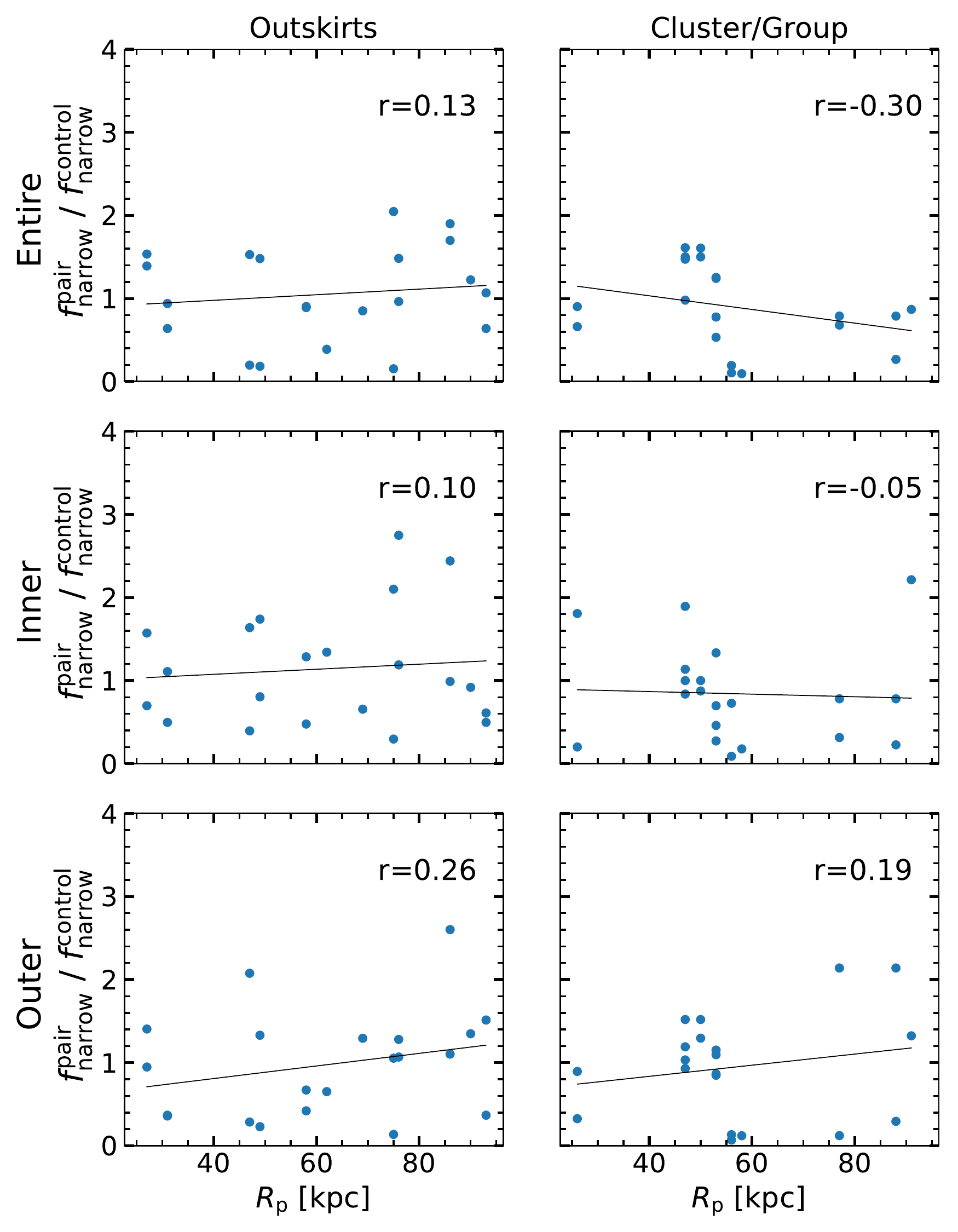}
    }
    \caption[values as a function of projected separations]{The kinematically narrow H{\sc i} gas fractions ($f_{\rm{narrow}}$) of the galaxy pairs normalised by $f_{\rm{narrow}}$ of the control galaxies at similar H{\sc i} mass ($f_{\rm{narrow}}^{\rm{pair}}$/$f_{\rm{narrow}}^{\rm{control}}$)  against the projected separation between the pairs for the entire, inner, and outer disk regions in the outskirts (left panel) and cluster/group environments (right panel). Linear regressions are indicated by the solid lines. The Pearson correlation coefficients $r$ are shown in the top right corner of each panel.}
    \label{fig:rp}
\end{figure}

We examine whether there is a correlation between projected separations $R_{\rm{p}}$ of galaxy pairs and their normalised $f_{\rm{narrow}}$ values (ratio of $f_{\rm{narrow}}^{\rm{pair}}$ to $f_{\rm{narrow}}^{\rm{control}}$) in different environments (i.e., outskirts and cluster/group). For normalised $f_{\rm{narrow}}$ values, we use the median values for the control galaxies which have similar H{\sc i} masses derived in Fig.~\ref{fig:corr_cold} (black solid lines). The effect of tidal interactions between the galaxies in a galaxy pair is expected to become significant when they approach each other. We explore this possibility using the $f_{\rm{narrow}}$ values derived for the ‘Entire’, ‘Inner’, and ‘Outer’ regions of galaxies in Section~\ref{sec:super-profile}. These are shown in Fig.~\ref{fig:rp} with the best fit linear regression lines over-plotted. The corresponding Pearson correlation coefficients $r$ are given in the top right corner of each panel.

As shown in Fig.~\ref{fig:rp}, no clear correlations between $R_{\rm{p}}$ and normalised $f_{\rm{narrow}}$ are found for the galaxy pairs in the environments given their Pearson correlation coefficients. However, the relations could be also affected by other galaxy properties like the stellar mass ratio between pairs. \citet{2021arXiv210805874D} show that SFRs tend to be enhanced with decreasing $R_{\rm{p}}$ for SDSS galaxy pairs which have stellar mass ratio of 1$<$ $M_1/M_2$ $<$3 where $M_1$ and $M_2$ are the stellar masses of the primary and secondary galaxies. On the other hand, the SFRs are likely to be suppressed in the ones with stellar mass ratios of 3$<$ $M_1/M_2$ $<$10. No sign of correlations between normalised $f_{\rm{narrow}}$ and $R_{\rm{p}}$ and their large scatter shown in Fig.~\ref{fig:rp} for the paired galaxies could be attributed to the effect of their stellar mass ratios.

\begin{figure*}
    \centering
    \includegraphics[width=0.33\textwidth]{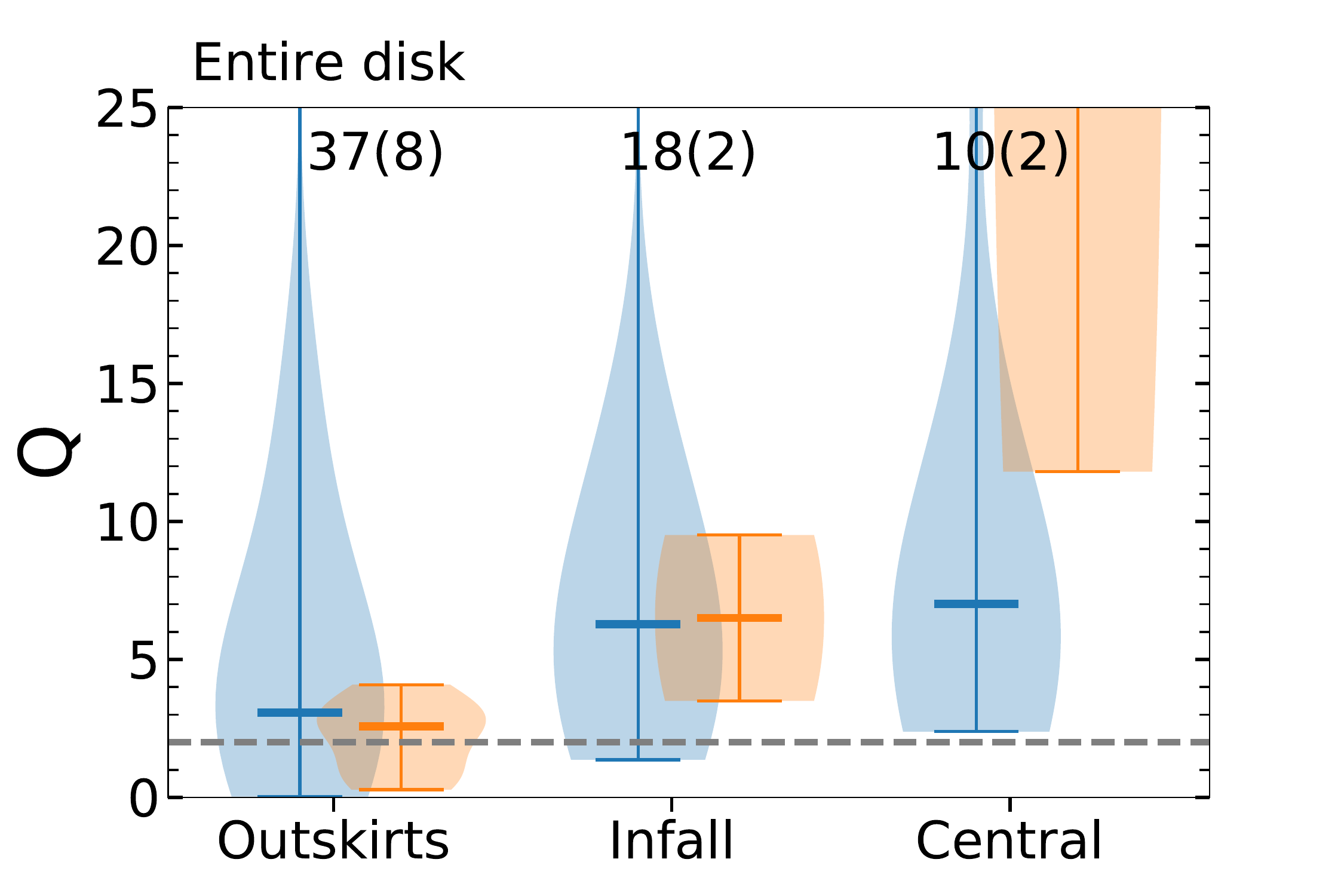}
    \includegraphics[width=0.33\textwidth]{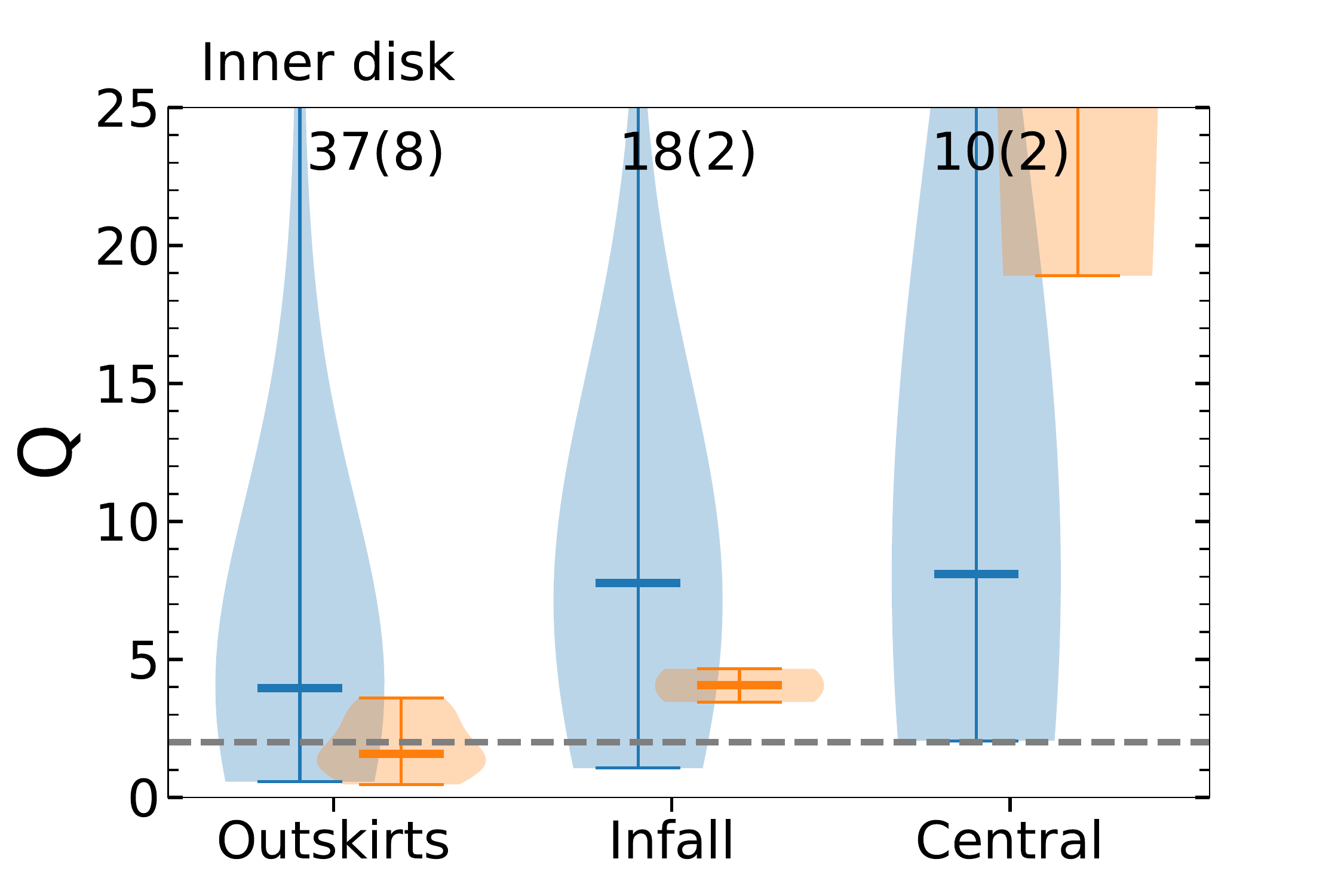}
    \includegraphics[width=0.33\textwidth]{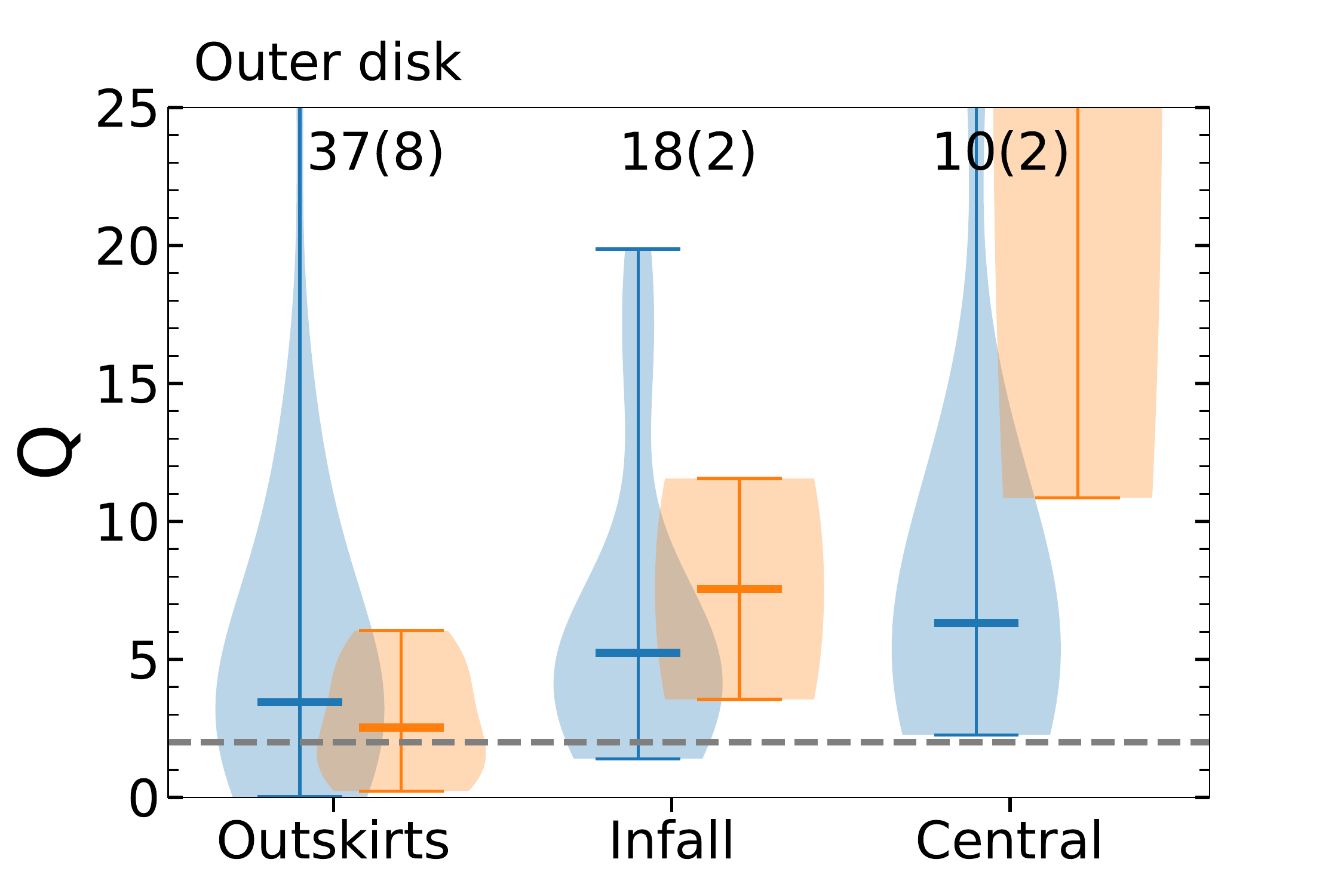}
    \vspace{0.2cm}
    \hrule
    \vspace{0.2cm}
    \includegraphics[width=0.33\textwidth]{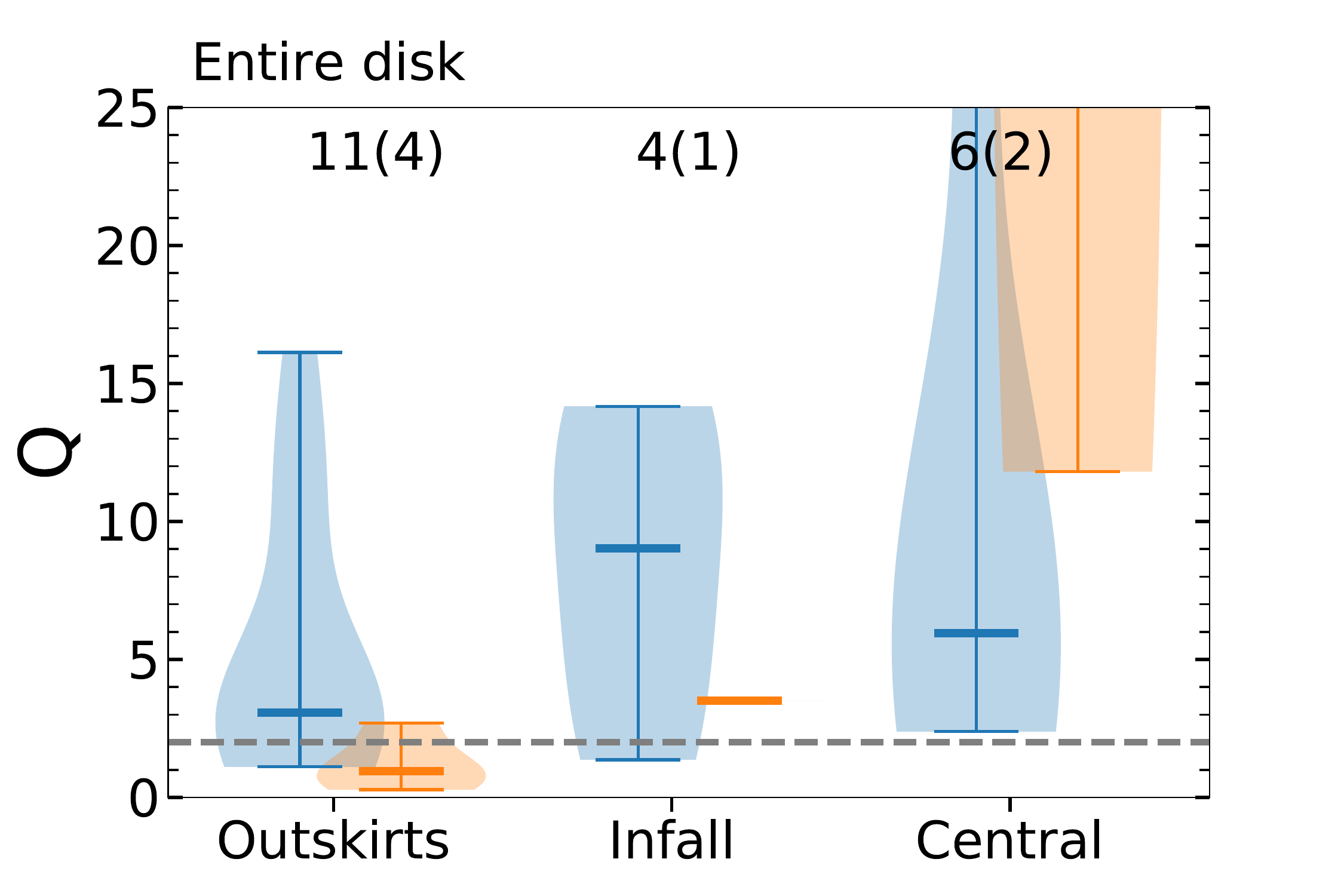}
    \includegraphics[width=0.33\textwidth]{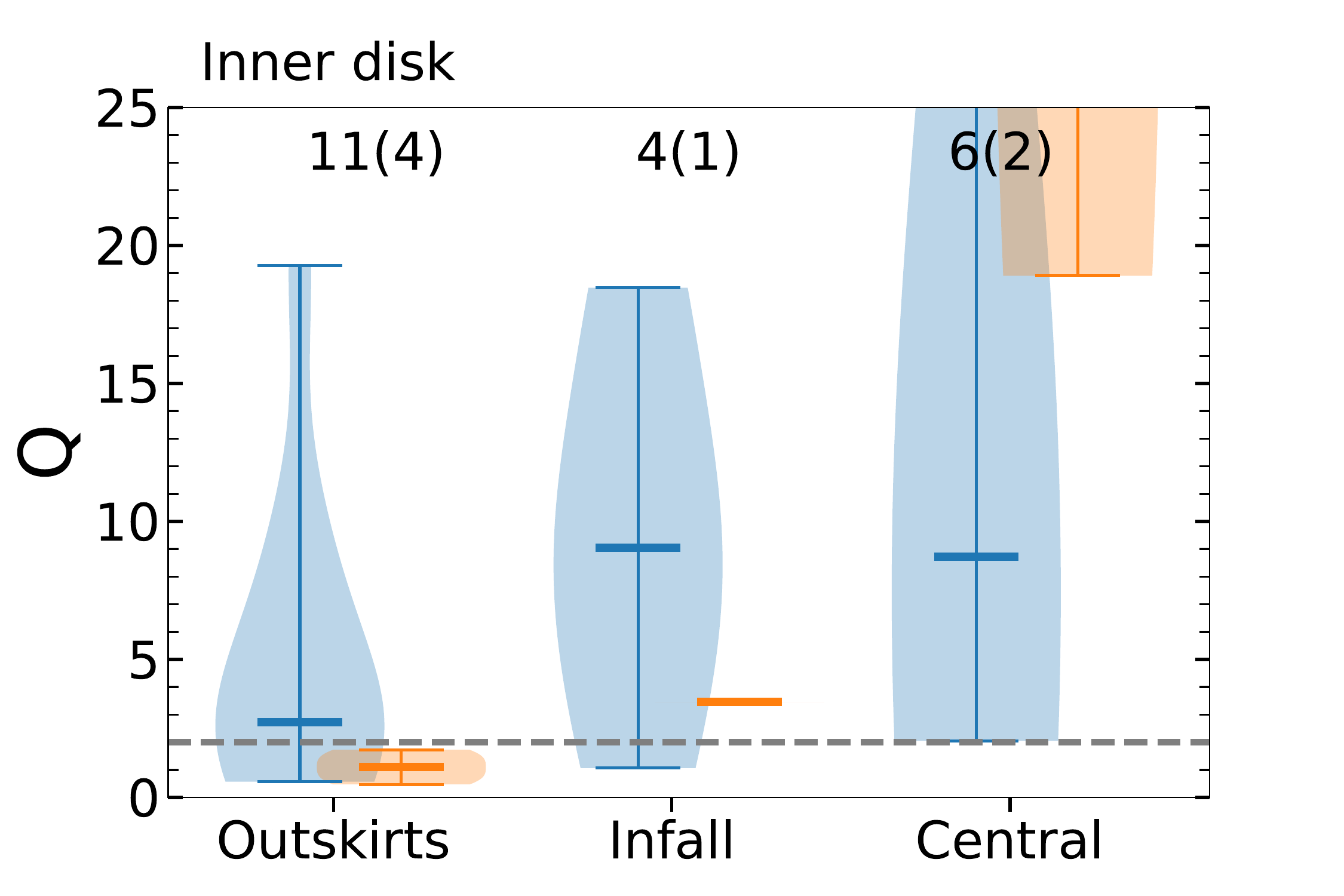}
    \includegraphics[width=0.33\textwidth]{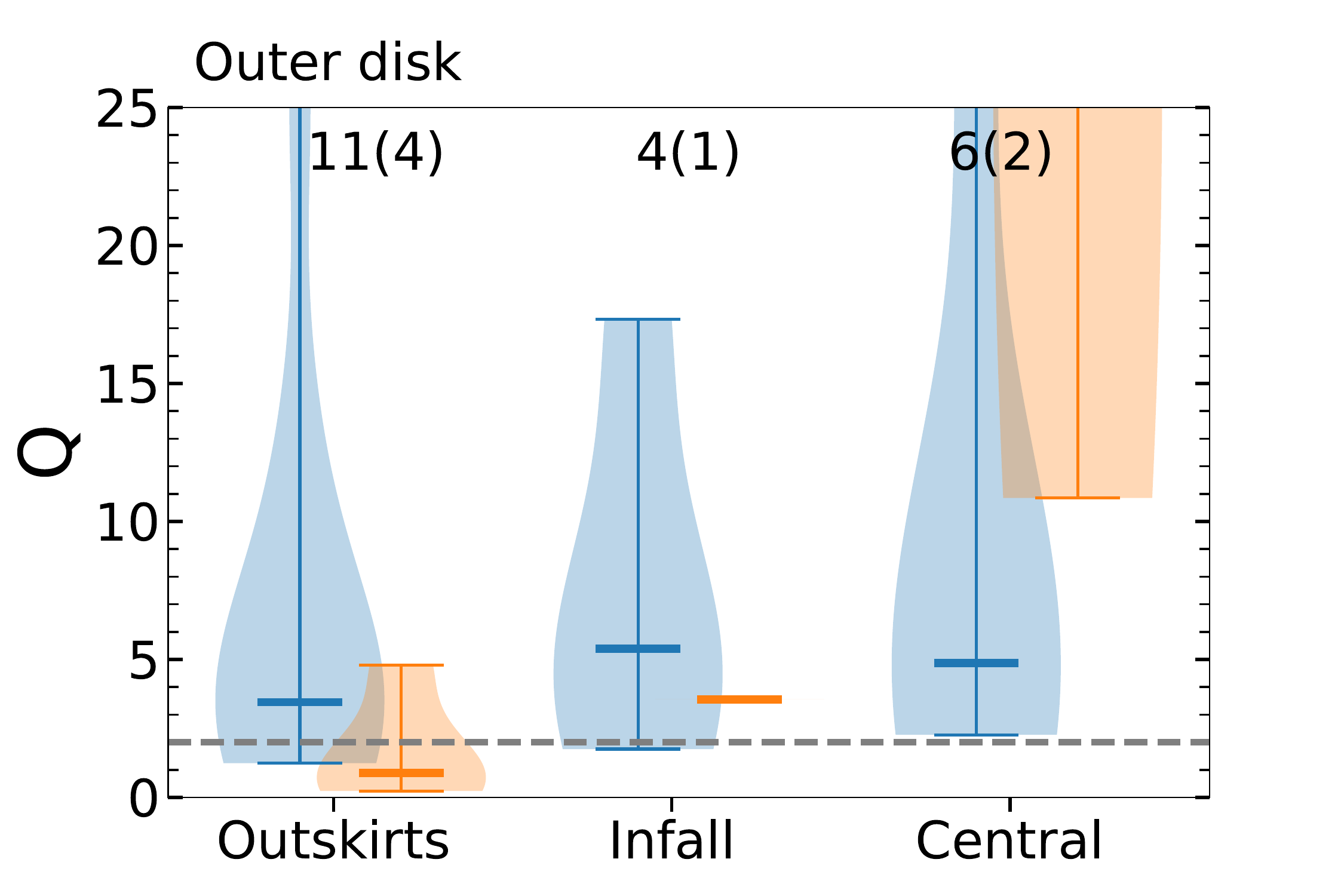}
    \caption[Distributions of the Toomre Q values at different environments]{The Toomre Q parameter median values for the three different regions (‘Entire’, ‘Inner’, and ‘Outer’) of the sample galaxies which are not matched (upper panels) and matched (lower panels) in H{\sc i} mass and galaxy number ratios, respectively, in the outskirts, infalling, and central environments. The control galaxies and galaxy pairs are shown with blue and orange colours, respectively. The numbers of the control and pair (bracket) galaxies are shown in each panel. The border thickness represents the relative frequency of the sample galaxies with the corresponding Toomre Q parameter median values. The small horizontal lines in each panel indicate the median values. The grey horizontal dashed lines indicate the Q = 2 values. See Section~\ref{sec:gravitational} for more details.}
    \label{fig:toomre}
\end{figure*}

\subsection{Gravitational instability of the gas disk}\label{sec:gravitational}

In this Section, we examine the influence of galaxy environments associated with the infall stage of galaxies (i.e., clusters, groups, pairs, and outskirts) on the gravitational instability of their gas disks. To quantify the degree of gravitational instability in the sample galaxies’ gas disks, we use the Toomre Q parameter value which allows us to measure gas instability against gravitational collapse and star formation. This is described in Section~\ref{sec:toomre}.

The estimation of the Toomre Q parameter is only valid for galaxies for which rotation curves are available. We therefore restrict our Toomre Q parameter analysis to the sample galaxies which show a clear rotation pattern in their single Gaussian fitting velocity fields and that have inclination values of 30$^\circ$ $<$ $i$ $<$ 70$^\circ$ and resolved with more than $\geq$ 6 beams across their major axes. In addition, as mentioned in Section~\ref{sec:rotationcurve}, for the galaxies whose single Gaussian fitting velocity field maps have large blank areas, we use their SoFiA2 {\sc moment1} maps. Typically, the percentage of masked pixels is less than 40$\%$ but with two exceptions, over about 50$\%$ are masked. 
These are indicated by ‘*’ in Table~\ref{long_hydra_control} $\sim$ \ref{long_norma_pair}.

There are 65 control galaxies and 12 galaxy pairs in the Hydra I, Norma and NGC 4636 fields whose Toomre Q parameter values can be measured from the rotation curve analysis. For each galaxy, we derive median values of the Toomre Q parameters measured in the inner and outer gas disk regions as well as in the entire one as described in Section~\ref{sec:toomre}. The distribution of the Toomre Q parameter median values for the disk regions (inner, outer and entire) of the sample galaxies in the different environments (outskirts, infall and central) are presented in the upper panels of Fig.~\ref{fig:toomre}. The additional cuts for H{\sc i} masses of galaxies and galaxy number ratios described in Section~\ref{sec:pair} result in 21 control galaxies and 7 paired galaxies.  Their Toomre Q parameter distributions are shown in the lower panels of Fig.~\ref{fig:toomre}.

\begin{table*}
\scriptsize{
\centering
\begin{tabular}{lccccccccccccccc}
\hline
\hline
\multicolumn{2}{c}{} & \multicolumn{4}{c}{Outskirts} & \multicolumn{4}{c}{Infalling} & \multicolumn{4}{c}{Central} \\ \\
        & & $ \langle \rm{Q} \rangle^{\rm{median}}$   & 1$\sigma$  & CI & N & $ \langle \rm{Q} \rangle^{\rm{median}}$   & 1$\sigma$  & CI & N & $ \langle \rm{Q} \rangle^{\rm{median}}$   & 1$\sigma$  & CI & N \\
\hline \\
Entire & Control & 3.08 & 8.87 & (2.24, 6.30) & 37 & 6.28 & 7.87 & (3.14, 11.9) & 18 & 7.01 & 8.56 & (3.70, 32.68) & 10 \\
& Pair & 2.58 & 1.24 & - & 8 & 6.50 & 3.01 & - & 2 & 31.36 & 19.55 & - & 2 \\ \\
Inner & Control & 3.96 & 12.56 & (2.74, 6.05) & 37 & 7.77 & 12.12 & (3.56, 11.05) & 18 & 8.10 & 21.54 & (3.82, 78.2) & 10 \\
& Pair    & 1.57 & 1.04 & - & 8 & 4.06 & 0.60 & - & 2 & 36.98 & 18.07 & - & 2 \\ \\
Outer & Control & 3.45 & 8.20 & (2.53, 6.32) & 37 & 5.23 & 5.26 & (3.27, 8.12) & 18 & 6.32 & 7.65 & (3.15, 29.14) & 10 \\
& Pair    & 2.54 & 1.98 & - & 8 & 7.56 & 4.00 & - & 2 & 26.99 & 16.14 & - & 2 \\ \\ 
\hline \\
Entire & Control & 3.08 & 4.77 & (2.20, 11.8) & 11 & 9.02 & 5.01 & - & 4 & 5.95 & 10.65 & - & 6 \\
& Pair & 0.96 & 0.91 & - & 4 & 3.50 & 0.00 & - & 1 & 31.36 & 19.55 & - & 2 \\ \\
Inner & Control & 2.74 & 5.31 & (1.80, 9.86) & 11 & 9.04 & 6.23 & - & 4 & 8.72 & 26.80 & - & 6 \\
& Pair    & 1.10 & 0.49 & - & 4 & 3.46 & 0.00 & - & 1 & 36.98 & 18.07 & - & 2 \\ \\
Outer & Control & 3.45 & 6.94 & (2.58, 13.47) & 11 & 5.39 & 5.88 & - & 4 & 4.88 & 9.54 & - & 6 \\
& Pair    & 0.89 & 1.81 & - & 4 & 3.55 & 0.00 & - & 1 & 26.99 & 16.14 & - & 2 \\ \\
\hline
\end{tabular}}
\caption[Median and standard deviation of Toomre Q values at different environments]{The median values ($ \langle \rm{Q} \rangle^{\rm{median}}$) and standard deviations ($\sigma_\mathrm{Q}$) of the Toomre Q parameter values for the three disk regions (entire, inner and outer) of the galaxy pairs and control sample galaxies galaxies which are not matched (upper) and matched (lower) in H{\sc i} mass and galaxy number ratios, respectively, in the outskirts, infalling, and central environments. N is the number of samples, and CI denotes the 95$\%$ confidence interval for median value.}
\label{tab:toomre}
\end{table*}

As shown in the upper panels of Fig.~\ref{fig:toomre}, the median values of the Toomre Q parameters of {\it all} the control galaxies are the lowest (more unstable) and highest (more stable) in the outskirts and central environments, respectively. Largely, this trend holds for the three gas disk regions (inner, outer and entire) although the distribution tails for outskirts and central galaxies are further extended towards higher Q values up to 60.4. The central environment with a higher local density of galaxies and IGM could result in such higher Q values for the central galaxies in a way of increasing their gas velocity dispersions (See Eq.~\ref{eq:Q}). Galaxies in central environments are likely to be more affected by gravitational and/or hydrodynamical influences including ram pressure stripping, galaxy\--galaxy interactions, and gas accretion than infalling and outskirts galaxies. These all might have given rise to turbulent gas motions in their gas disks (\citealt{1972ApJ...176....1G}; \citealt{1996Natur.379..613M}; \citealt{2009ApJ...699.1595P}; \citealt{2014ApJ...783..109W}; \citealt{2017ApJ...838...81Y}).

Despite the small number of statistics, the trend of the higher Toomre Q values for the galaxies in the central environment is also seen for the galaxy pairs. The higher Q values of the galaxy pairs in the central environment could be also attributed by the aforementioned dynamical effects like ram pressure stripping and tidal interactions among galaxies. External gravitational and/or hydrodynamical effects in the central environment might be associated with the higher Toomre Q values. Alternatively, internal processes (e.g., star formation and feedback) of the galaxies could also have disturbed the gas motions, resulting in the higher Toomre Q values. However, we note that the H{\sc i} masses of the galaxies and the galaxy number ratios in the environments are not matched. The trend of increasing Toomre Q parameter values towards the central environment could be biased by the unmatched sample.

The distributions of the Toomre Q parameters of the matched control galaxies are shown in the lower panels of Fig.~\ref{fig:toomre}. The median values for the three regions (entire, inner and outer disks) of the control galaxies (blue) in the infall and central environments are higher than those of the galaxies in the outskirts environment. However, it is difficult to see whether this trend is present for the matched galaxy pairs (orange) due to the limited number of galaxies.

We point out that the Toomre Q analysis of the sample galaxies could be affected by resolution effects. If the beam resolution was not high enough to resolve the inner regions of our sample galaxies where velocity gradients are high, their gas velocity dispersions could be overestimated. In addition, rotation velocities of galaxies with steeply rising rotation curves are prone to be underestimated at low angular resolutions. 2D tilted-ring analysis is usually affected by this so-called beam smearing effect. Inaccurate measurements (i.e., $R/V$ and $dV/dR$) of the rotation curve shape of a galaxy, particularly in the inner region, will lead to uncertainties in the Toomre Q parameter analysis. This could result in the wider distribution of the Toomre Q parameter values for the inner disk region than those for the outer disk of the control galaxies in the infalling and central environments in Fig.~\ref{fig:toomre}. This demonstrates the need for higher angular resolution H{\sc i} observations not only to minimize beam smearing effects but also to improve the small number of statistics in the Toomre Q parameter analysis of this work. We tabulate the values for the median and standard deviation of the Toomre Q parameter values derived for the entire, inner and outer gas disks of the sample galaxies in Table~\ref{tab:toomre}.

\begin{figure}
    \centering
    \includegraphics[width = 0.45\textwidth]{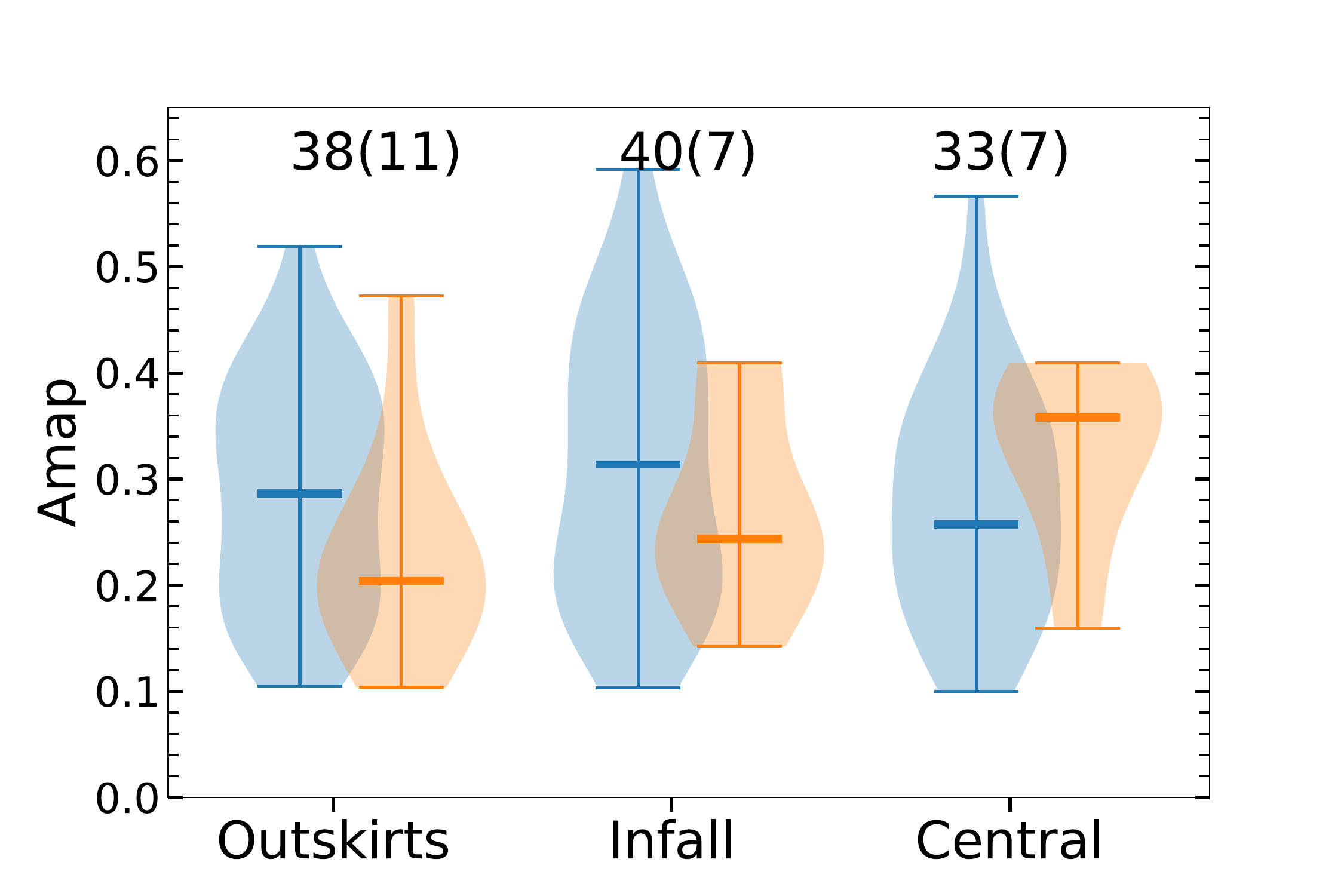}
    \caption[Distributions of the H{\sc i} morphology asymmetries at different environments]{The H{\sc i} morphological asymmetries of the sample galaxies, in the outskirts, infalling, and central environments. The control galaxies and galaxy pairs are shown with blue and orange colours, respectively. The numbers of the control and pair (bracket) galaxies are shown in each panel. The border thickness represents the relative frequency of the sample galaxies with the corresponding H{\sc i} morphological asymmetries. The small horizontal lines in each panel indicate the median values. See Section~\ref{sec:mordistribution} for more details.}
    \label{fig:asymmetry}
\end{figure}

\begin{table*}
\scriptsize{
\centering
\begin{tabular}{lcccccccccccccc}
\hline
\hline
 & \multicolumn{4}{c}{Outskirts} & \multicolumn{4}{c}{Infalling} & \multicolumn{4}{c}{Central} \\ \\
        & $\langle A_{\rm{map}}^{\rm{HI}} \rangle^{\rm{median}}$   & 1$\sigma$  &  CI & N & $\langle A_{\rm{map}}^{\rm{HI}} \rangle^{\rm{median}}$   & 1$\sigma$  & CI & N & $\langle A_{\rm{map}}^{\rm{HI}} \rangle^{\rm{median}}$ & 1$\sigma$  & CI & N\\
\hline

Control & 0.29 & 0.11 & (0.23, 0.35) & 38 & 0.31 & 0.13 & (0.24, 0.39) & 40 & 0.26 & 0.11 & (0.21, 0.34) & 33 \\
Pair    & 0.20 & 0.10 & (0.18, 0.33) & 11 & 0.24 & 0.09 & - & 7 & 0.36 & 0.08 & - & 7 \\

\hline
\end{tabular}}
\caption[Median and standard deviation of asymmetries at different environments]{The median values ($\langle A_{\rm{map}}^{\rm{HI}} \rangle^{\rm{median}}$) and standard deviations ($\sigma_{A_{\rm{map}}^{\rm{HI}}}$) of the morphological H{\sc i} asymmetries of the galaxy pairs and control sample galaxies in the outskirts, infalling, and central environments. N is the number of samples, and CI denotes the 95$\%$ confidence interval for median value.}
\label{tab:asym}

\end{table*}

\subsection{Morphological H{\sc i} asymmetries}\label{sec:mordistribution}

In this Section, we examine morphological disturbances of the gas distribution of the sample galaxies in different galaxy environments. H{\sc i} gas in galaxies usually extends further than the readily discernible stellar disk and thus is a useful probe of environmental effects (\citealt{2011MNRAS.410.2217W}; \citealt{2016ApJ...824..110O}; \citealt{2017ApJ...838...81Y}). Several hydrodynamical processes in and around galaxies (internal: spiral arms, and bars; external: ram pressure stripping, minor or major mergers, tidal interactions, and gas accretion) are able to disturb the gas distribution, particularly in the outer region, which increases morphological H{\sc i} asymmetries of the gas disks (\citealt{1972ApJ...176....1G}; \citealt{2011MNRAS.416.2415H}; \citealt{2008MNRAS.388..697M}; \citealt{2020MNRAS.492.3672W}; \citealt{2020MNRAS.493.5089R}; \citealt{2020MNRAS.499.5205W}). As described in Section~\ref{sec:mor}, to quantify the morphological H{\sc i} disturbances of galaxies in our sample, we measure H{\sc i} asymmetries using their integrated intensity maps ({\sc moment0}). We exclude four galaxy pairs and 13 control galaxies which are not well-resolved spatially and thus whose kinematic centres are not reliably derived. Fig.~\ref{fig:asymmetry} shows the derived H{\sc i} asymmetries of the selected galaxies in the three galaxy environments.

As discussed in \cite{2021MNRAS.505.1891R} and \cite{2021ApJ...915...70W}, the effect of ram pressure stripping on the galaxies in the Hydra I cluster could affect their H{\sc i} asymmetries in a way of that gives rise to morphological disturbances. Additionally, external tidal forces from other galaxies in the central environment might have disturbed their H{\sc i} gas distribution, resulting in complex H{\sc i} gas structures (e.g., H{\sc i} tails or streams) and kinematics, particularly in their outer regions where the gravitational potential is shallow. This can also apply to the Norma cluster and the NGC 4636 galaxies in dense environments. In this respect, environmental effects, including ram pressure stripping, on the H{\sc i} asymmetry of a galaxy are expected to be minimal in the outskirts environment.

In Fig.~\ref{fig:asymmetry}, no clear trend in the median values of the H{\sc i} asymmetries ($A_{\rm{map}}^{\rm{HI}}$) of the control sample galaxies (blue) is found with respect to the environments (outskirts to central). On the other hand, the median $A_{\rm{map}}^{\rm{HI}}$ value (0.20) of the galaxy pairs (orange) in the outskirts is slightly smaller than those (0.24 and 0.36) of the pairs in the infalling and central environments. \cite{2021MNRAS.505.1891R} derive the $A_{\rm{map}}^{\rm{HI}}$ value of a galaxy in a way of minimizing it by varying the mass centre of the {\sc moment0} map within one ASKAP beam size. However, we fix the centre in Eq.~(\ref{eq:1}) with the value determined from the {\sc 2dbat} analysis when deriving $A_{\rm{map}}^{\rm{HI}}$. So the H{\sc i} asymmetry values of the galaxy pairs in the infalling and central environments could potentially be increased by the way of asymmetry measurement adopted in this work.

More importantly, as discussed in \cite{2016MNRAS.461.1656G} (see also \citealt{2020MNRAS.493.5089R, 2021MNRAS.505.1891R}), the measurement of morphological H{\sc i} asymmetries of galaxies using Eq.~(\ref{eq:1}) is not reliable for H{\sc i} observations at low spatial resolutions as well as with low S/N. Additionally, the current sensitivity of ASKAP pilot observations may not be high enough to detect, if any, signatures of H{\sc i} morphological disturbances in the outer regions of the galaxy pairs in the outskirts (see also \citealt{2021MNRAS.505.1891R}). This demonstrates the need for deeper H{\sc i} observations in different environments at high sensitivities as well as at high angular resolutions in order to better examine the environmental effects on the morphological H{\sc i} asymmetries of galaxies. We list the median values and standard deviations of the derived H{\sc i} asymmetries of the sample galaxies in Table~\ref{tab:asym}.

\section{Summary} \label{sec:5}

In this paper, we examine the galaxy environmental effects on the resolved H{\sc i} properties of galaxy pairs in the two clusters (Hydra I and Norma) and the galaxy group (NGC 4636) from the ASKAP pilot observations. We perform profile decomposition of the H{\sc i} data cubes using a new tool, {\sc baygaud} which allows us to de-blend a line-of-sight velocity profile with an optimal number of Gaussian components based on Bayesian nested sampling. Then, we construct H{\sc i} super-profiles via stacking individual line profiles after aligning their centroid velocities. We fit the H{\sc i} super-profiles with a double Gaussian model and classify them as kinematically narrow and broad H{\sc i} gas components with respect to their velocity dispersions, i.e., having narrower and broader $\sigma$, respectively. The kinematically narrow H{\sc i} gas components can be associated with cool H{\sc i} gas (\citealt{2012AJ....144...96I}). Additionally, we quantify the gravitational instability of the H{\sc i} gas disk of galaxy pairs using their Toomre Q parameters, and H{\sc i} morphological disturbances. For this, we use their {\sc moment0} maps produced by SoFiA2. These H{\sc i} properties are then compared with those of a set of control galaxies which are classified as non-paired systems in the same fields.

We investigate the cluster environmental effects on the H{\sc i} gas properties of galaxy pairs by comparing H{\sc i} super-profiles, Toomre Q parameter values, and morphological asymmetries in three different environments, 1) outskirts, 2) infalling and 3) central which are defined using the phase-space diagrams of the galaxy clusters and group. The main results of this paper are as follows:\\

\noindent 1) The median value of $f_{\rm{narrow}}$ ($M_{\rm{narrow}}^{\rm{HI}}$ / $M_{\rm{total}}^{\rm{HI}}$) for the outer gas disk regions of the control sample galaxies appears to be lower in the central environment than outskirts and infalling environments. This could be attributed to the enhanced local densities of galaxies in the central environment. Tidal interactions among galaxies in denser environments can give rise to more turbulent gas motions in the galaxies with higher velocity dispersions. Together, the ram pressure induced by the IGM in the clusters may also contribute to decrease $f_{\rm{narrow}}$ in a way of stripping part of the kinematically narrow H{\sc i} gas in the outer regions of the galaxies (\citealt{2009AJ....138.1741C}; \citealt{2011MNRAS.415.1797C}; \citealt{2016MNRAS.461.1202J}; \citealt{2017MNRAS.466.1275B}). Despite the small number statistics, a decreasing trend of$f_{\rm{narrow}}$ values for the galaxy pairs is seen with the infall stage, outskirts to central, particularly in their outer gas disk regions. This tendency is also seen in the inner gas disk regions but is less significant, possibly due to the higher gravitational potential. Tidal interactions of galaxies in the close galaxy pairs in denser environments could have induced such a tendency, giving rise to more turbulent gas motions. The lower $f_{\rm{narrow}}$ values of the galaxy pairs in denser environments are consistent with the results from other works that galaxy pairs in low-density environments tend to have high star formation rates caused by higher gas fractions (\citealt{2010MNRAS.407.1514E}).\\

\noindent 2) Despite a wide range of the Toomre Q parameter values and the small number of statistics, the median values for the control galaxies are likely to increase (i.e., being more stable) in denser environments. This tendency is further enhanced for the galaxy pairs. The increased kinetic energy which is possibly caused by the ram pressure stripping and/or tidal interactions in the central environment might have stirred up the gas in the galaxies, making the gas disks of the galaxies more stable against gravitational collapse.\\

\noindent 3) The median value of the H{\sc i} morphological asymmetries ($A_{\rm{map}}^{\rm{HI}}$) of the control galaxies is the lowest in the central environment but it is not significant. On the other hand, the median values of $A_{\rm{map}}^{\rm{HI}}$ of the galaxy pairs increase towards the denser environments (outskirts to central). External processes like tidal interactions and ram pressure stripping in the central environment might have led to disturbed H{\sc i} gas structures (e.g., H{\sc i} tails or streams) in the outer gas disks of the galaxies. However, the current sensitivity and beam resolution of the ASKAP pilot observations may not be high enough to detect such disturbed H{\sc i} features in the outer regions of the galaxies (\citealt{2021ApJ...915...70W}; \citealt{2022MNRAS.510.1716R}). The $A_{\rm{map}}^{\rm{HI}}$ values derived in our work could suffer from resolution effects as discussed in \cite{2016MNRAS.461.1656G} and \cite{2020MNRAS.493.5089R, 2021MNRAS.505.1891R}. Deeper H{\sc i} observations of the cluster and group galaxies to better sensitivity are needed in order to examine the environmental effects on their H{\sc i} morphological disturbances.


\section*{Acknowledgements}
SHOH acknowledges a support from the National Research Foundation of Korea (NRF) grant funded by the Korea government (Ministry of Science and ICT: MSIT) (No. NRF-2020R1A2C1008706). PK is partially supported by the BMBF project 05A17PC2 for D-MeerKAT. Parts of this research were conducted by the Australian Research Council Centre of Excellence for All Sky Astrophysics in 3 Dimensions (ASTRO 3D), through project number CE170100013. FB acknowledges funding from the European Research Council (ERC) under the
European Union’s Horizon 2020 research and innovation programme (grant agreement
No.726384/Empire).


\section*{Data availability}
The galaxy properties used in this work are available as online supplementary material. We show the first four rows of the tables for the H{\sc i} properties and distances and representative two images and super-profiles of sample galaxies in Appendix~\ref{appendixA}. The 30 arcsec H{\sc i} data cubes and associated moment maps are available from the CSIRO ASKAP Science Data Archive (CASDA, \citealt{2015Champman}; \citealt{2020Huynh}) with the DOI \url{https://doi.org/10.25919/5f7bde37c20b5}.



\bibliographystyle{mnras}
\bibliography{askap-galaxy-pairs} 

\providecommand{\noopsort}[1]{}
\begin{thebibliography}{}
\makeatletter
\relax
\def\mn@urlcharsother{\let\do\@makeother \do\$\do\&\do\#\do\^\do\_\do\%\do\~}
\def\mn@doi{\begingroup\mn@urlcharsother \@ifnextchar [ {\mn@doi@}
  {\mn@doi@[]}}
\def\mn@doi@[#1]#2{\def\@tempa{#1}\ifx\@tempa\@empty \href
  {http://dx.doi.org/#2} {doi:#2}\else \href {http://dx.doi.org/#2} {#1}\fi
  \endgroup}
\def\mn@eprint#1#2{\mn@eprint@#1:#2::\@nil}
\def\mn@eprint@arXiv#1{\href {http://arxiv.org/abs/#1} {{\tt arXiv:#1}}}
\def\mn@eprint@dblp#1{\href {http://dblp.uni-trier.de/rec/bibtex/#1.xml}
  {dblp:#1}}
\def\mn@eprint@#1:#2:#3:#4\@nil{\def\@tempa {#1}\def\@tempb {#2}\def\@tempc
  {#3}\ifx \@tempc \@empty \let \@tempc \@tempb \let \@tempb \@tempa \fi \ifx
  \@tempb \@empty \def\@tempb {arXiv}\fi \@ifundefined
  {mn@eprint@\@tempb}{\@tempb:\@tempc}{\expandafter \expandafter \csname
  mn@eprint@\@tempb\endcsname \expandafter{\@tempc}}}

\bibitem[\protect\citeauthoryear{{Alonso}, {Tissera}, {Coldwell}  \&
  {Lambas}}{{Alonso} et~al.}{2004}]{2004MNRAS.352.1081A}
{Alonso} M.~S.,  {Tissera} P.~B.,  {Coldwell} G.,   {Lambas} D.~G.,  2004,
  \mn@doi [\mnras] {10.1111/j.1365-2966.2004.08002.x}, \href
  {https://ui.adsabs.harvard.edu/abs/2004MNRAS.352.1081A} {352, 1081}

\bibitem[\protect\citeauthoryear{{Alonso}, {Mesa}, {Padilla}  \&
  {Lambas}}{{Alonso} et~al.}{2012}]{2012A&A...539A..46A}
{Alonso} S.,  {Mesa} V.,  {Padilla} N.,   {Lambas} D.~G.,  2012, \mn@doi [\aap]
  {10.1051/0004-6361/201117901}, \href
  {https://ui.adsabs.harvard.edu/abs/2012A&A...539A..46A} {539, A46}

\bibitem[\protect\citeauthoryear{{Barnes} \& {Hernquist}}{{Barnes} \&
  {Hernquist}}{1991}]{1991ApJ...370L..65B}
{Barnes} J.~E.,  {Hernquist} L.~E.,  1991, \mn@doi [\apjl] {10.1086/185978},
  \href {https://ui.adsabs.harvard.edu/abs/1991ApJ...370L..65B} {370, L65}

\bibitem[\protect\citeauthoryear{{Barnes} \& {Hernquist}}{{Barnes} \&
  {Hernquist}}{1996}]{1996ApJ...471..115B}
{Barnes} J.~E.,  {Hernquist} L.,  1996, \mn@doi [\apj] {10.1086/177957}, \href
  {https://ui.adsabs.harvard.edu/abs/1996ApJ...471..115B} {471, 115}

\bibitem[\protect\citeauthoryear{{Barnes} et~al.,}{{Barnes}
  et~al.}{2001}]{2001MNRAS.322..486B}
{Barnes} D.~G.,  et~al., 2001, \mn@doi [\mnras]
  {10.1046/j.1365-8711.2001.04102.x}, \href
  {https://ui.adsabs.harvard.edu/abs/2001MNRAS.322..486B} {322, 486}

\bibitem[\protect\citeauthoryear{{Beutler} et~al.,}{{Beutler}
  et~al.}{2011}]{2011MNRAS.416.3017B}
{Beutler} F.,  et~al., 2011, \mn@doi [\mnras]
  {10.1111/j.1365-2966.2011.19250.x}, \href
  {https://ui.adsabs.harvard.edu/abs/2011MNRAS.416.3017B} {416, 3017}

\bibitem[\protect\citeauthoryear{{Bigiel}, {Leroy}, {Walter}, {Brinks}, {de
  Blok}, {Madore}  \& {Thornley}}{{Bigiel} et~al.}{2008}]{2008AJ....136.2846B}
{Bigiel} F.,  {Leroy} A.,  {Walter} F.,  {Brinks} E.,  {de Blok} W.~J.~G.,
  {Madore} B.,   {Thornley} M.~D.,  2008, \mn@doi [\aj]
  {10.1088/0004-6256/136/6/2846}, \href
  {https://ui.adsabs.harvard.edu/abs/2008AJ....136.2846B} {136, 2846}

\bibitem[\protect\citeauthoryear{{Bigiel}, {Leroy}, {Walter}, {Blitz},
  {Brinks}, {de Blok}  \& {Madore}}{{Bigiel}
  et~al.}{2010}]{2010AJ....140.1194B}
{Bigiel} F.,  {Leroy} A.,  {Walter} F.,  {Blitz} L.,  {Brinks} E.,  {de Blok}
  W.~J.~G.,   {Madore} B.,  2010, \mn@doi [\aj] {10.1088/0004-6256/140/5/1194},
  \href {https://ui.adsabs.harvard.edu/abs/2010AJ....140.1194B} {140, 1194}

\bibitem[\protect\citeauthoryear{{Boselli} \& {Gavazzi}}{{Boselli} \&
  {Gavazzi}}{2006}]{2006PASP..118..517B}
{Boselli} A.,  {Gavazzi} G.,  2006, \mn@doi [\pasp] {10.1086/500691}, \href
  {https://ui.adsabs.harvard.edu/abs/2006PASP..118..517B} {118, 517}

\bibitem[\protect\citeauthoryear{{Boselli}, {Cortese}, {Boquien}, {Boissier},
  {Catinella}, {Gavazzi}, {Lagos}  \& {Saintonge}}{{Boselli}
  et~al.}{2014}]{2014A&A...564A..67B}
{Boselli} A.,  {Cortese} L.,  {Boquien} M.,  {Boissier} S.,  {Catinella} B.,
  {Gavazzi} G.,  {Lagos} C.,   {Saintonge} A.,  2014, \mn@doi [\aap]
  {10.1051/0004-6361/201322313}, \href
  {https://ui.adsabs.harvard.edu/abs/2014A&A...564A..67B} {564, A67}

\bibitem[\protect\citeauthoryear{{Boselli} et~al.,}{{Boselli}
  et~al.}{2016}]{2016A&A...596A..11B}
{Boselli} A.,  et~al., 2016, \mn@doi [\aap] {10.1051/0004-6361/201629221},
  \href {https://ui.adsabs.harvard.edu/abs/2016A&A...596A..11B} {596, A11}

\bibitem[\protect\citeauthoryear{{Broeils} \& {Rhee}}{{Broeils} \&
  {Rhee}}{1997}]{1997A&A...324..877B}
{Broeils} A.~H.,  {Rhee} M.~H.,  1997, \aap, \href
  {https://ui.adsabs.harvard.edu/abs/1997A&A...324..877B} {324, 877}

\bibitem[\protect\citeauthoryear{{Brown} et~al.,}{{Brown}
  et~al.}{2017}]{2017MNRAS.466.1275B}
{Brown} T.,  et~al., 2017, \mn@doi [\mnras] {10.1093/mnras/stw2991}, \href
  {https://ui.adsabs.harvard.edu/abs/2017MNRAS.466.1275B} {466, 1275}

\bibitem[\protect\citeauthoryear{{Castignani} et~al.,}{{Castignani}
  et~al.}{2018}]{2018A&A...617A.103C}
{Castignani} G.,  et~al., 2018, \mn@doi [\aap] {10.1051/0004-6361/201832887},
  \href {https://ui.adsabs.harvard.edu/abs/2018A&A...617A.103C} {617, A103}

\bibitem[\protect\citeauthoryear{{Catinella} et~al.,}{{Catinella}
  et~al.}{2013}]{2013MNRAS.436...34C}
{Catinella} B.,  et~al., 2013, \mn@doi [\mnras] {10.1093/mnras/stt1417}, \href
  {https://ui.adsabs.harvard.edu/abs/2013MNRAS.436...34C} {436, 34}

\bibitem[\protect\citeauthoryear{{Chapman}}{{Chapman}}{2015}]{2015Champman}
{Chapman} J.~M.,  2015, in IAU General Assembly. p. 2232458

\bibitem[\protect\citeauthoryear{{Chung}, {van Gorkom}, {Kenney}, {Crowl}  \&
  {Vollmer}}{{Chung} et~al.}{2009}]{2009AJ....138.1741C}
{Chung} A.,  {van Gorkom} J.~H.,  {Kenney} J. D.~P.,  {Crowl} H.,   {Vollmer}
  B.,  2009, \mn@doi [\aj] {10.1088/0004-6256/138/6/1741}, \href
  {https://ui.adsabs.harvard.edu/abs/2009AJ....138.1741C} {138, 1741}

\bibitem[\protect\citeauthoryear{{Chung}, {Yun}, {Verheijen}  \&
  {Chung}}{{Chung} et~al.}{2017}]{2017ApJ...843...50C}
{Chung} E.~J.,  {Yun} M.~S.,  {Verheijen} M. A.~W.,   {Chung} A.,  2017,
  \mn@doi [\apj] {10.3847/1538-4357/aa756b}, \href
  {https://ui.adsabs.harvard.edu/abs/2017ApJ...843...50C} {843, 50}

\bibitem[\protect\citeauthoryear{{Conselice}}{{Conselice}}{2006}]{2006ApJ...638..686C}
{Conselice} C.~J.,  2006, \mn@doi [\apj] {10.1086/499067}, \href
  {https://ui.adsabs.harvard.edu/abs/2006ApJ...638..686C} {638, 686}

\bibitem[\protect\citeauthoryear{{Cortese}, {Catinella}, {Boissier}, {Boselli}
  \& {Heinis}}{{Cortese} et~al.}{2011}]{2011MNRAS.415.1797C}
{Cortese} L.,  {Catinella} B.,  {Boissier} S.,  {Boselli} A.,   {Heinis} S.,
  2011, \mn@doi [\mnras] {10.1111/j.1365-2966.2011.18822.x}, \href
  {https://ui.adsabs.harvard.edu/abs/2011MNRAS.415.1797C} {415, 1797}

\bibitem[\protect\citeauthoryear{{Cortese}, {Catinella}  \& {Smith}}{{Cortese}
  et~al.}{2021}]{2021PASA...38...35C}
{Cortese} L.,  {Catinella} B.,   {Smith} R.,  2021, \mn@doi [\pasa]
  {10.1017/pasa.2021.18}, \href
  {https://ui.adsabs.harvard.edu/abs/2021PASA...38...35C} {38, e035}

\bibitem[\protect\citeauthoryear{{Courtois}, {Tully}, {Makarov}, {Mitronova},
  {Koribalski}, {Karachentsev}  \& {Fisher}}{{Courtois}
  et~al.}{2011}]{2011MNRAS.414.2005C}
{Courtois} H.~M.,  {Tully} R.~B.,  {Makarov} D.~I.,  {Mitronova} S.,
  {Koribalski} B.,  {Karachentsev} I.~D.,   {Fisher} J.~R.,  2011, \mn@doi
  [\mnras] {10.1111/j.1365-2966.2011.18515.x}, \href
  {https://ui.adsabs.harvard.edu/abs/2011MNRAS.414.2005C} {414, 2005}

\bibitem[\protect\citeauthoryear{{Das}, {Pandey}, {Sarkar}  \& {Dutta}}{{Das}
  et~al.}{2021}]{2021arXiv210805874D}
{Das} A.,  {Pandey} B.,  {Sarkar} S.,   {Dutta} A.,  2021, arXiv e-prints,
  \href {https://ui.adsabs.harvard.edu/abs/2021arXiv210805874D} {p.
  arXiv:2108.05874}

\bibitem[\protect\citeauthoryear{{\noopsort{De}{de Blok}} \&
  {Walter}}{{\noopsort{De}{de Blok}} \& {Walter}}{2006}]{2006AJ....131..363D}
{\noopsort{De}{de Blok}} W.~J.~G.,  {Walter} F.,  2006, \mn@doi [\aj]
  {10.1086/497828}, \href
  {https://ui.adsabs.harvard.edu/abs/2006AJ....131..363D} {131, 363}

\bibitem[\protect\citeauthoryear{{\noopsort{De}{de Blok}}, {Walter}, {Brinks},
  {Trachternach}, {Oh}  \& {Kennicutt}}{{\noopsort{De}{de Blok}}
  et~al.}{2008}]{2008AJ....136.2648D}
{\noopsort{De}{de Blok}} W.~J.~G.,  {Walter} F.,  {Brinks} E.,  {Trachternach}
  C.,  {Oh} S.~H.,   {Kennicutt} R.~C. J.,  2008, \mn@doi [\aj]
  {10.1088/0004-6256/136/6/2648}, \href
  {https://ui.adsabs.harvard.edu/abs/2008AJ....136.2648D} {136, 2648}

\bibitem[\protect\citeauthoryear{{Di Matteo}, {Combes}, {Melchior}  \&
  {Semelin}}{{Di Matteo} et~al.}{2007}]{2007A&A...468...61D}
{Di Matteo} P.,  {Combes} F.,  {Melchior} A.~L.,   {Semelin} B.,  2007, \mn@doi
  [\aap] {10.1051/0004-6361:20066959}, \href
  {https://ui.adsabs.harvard.edu/abs/2007A&A...468...61D} {468, 61}

\bibitem[\protect\citeauthoryear{{Elagali} et~al.,}{{Elagali}
  et~al.}{2019}]{2019MNRAS.487.2797E}
{Elagali} A.,  et~al., 2019, \mn@doi [\mnras] {10.1093/mnras/stz1448}, \href
  {https://ui.adsabs.harvard.edu/abs/2019MNRAS.487.2797E} {487, 2797}

\bibitem[\protect\citeauthoryear{{Ellison}, {Patton}, {Simard}  \&
  {McConnachie}}{{Ellison} et~al.}{2008}]{2008AJ....135.1877E}
{Ellison} S.~L.,  {Patton} D.~R.,  {Simard} L.,   {McConnachie} A.~W.,  2008,
  \mn@doi [\aj] {10.1088/0004-6256/135/5/1877}, \href
  {https://ui.adsabs.harvard.edu/abs/2008AJ....135.1877E} {135, 1877}

\bibitem[\protect\citeauthoryear{{Ellison}, {Patton}, {Simard}, {McConnachie},
  {Baldry}  \& {Mendel}}{{Ellison} et~al.}{2010}]{2010MNRAS.407.1514E}
{Ellison} S.~L.,  {Patton} D.~R.,  {Simard} L.,  {McConnachie} A.~W.,  {Baldry}
  I.~K.,   {Mendel} J.~T.,  2010, \mn@doi [\mnras]
  {10.1111/j.1365-2966.2010.17076.x}, \href
  {https://ui.adsabs.harvard.edu/abs/2010MNRAS.407.1514E} {407, 1514}

\bibitem[\protect\citeauthoryear{{Ellison}, {Mendel}, {Patton}  \&
  {Scudder}}{{Ellison} et~al.}{2013}]{2013MNRAS.435.3627E}
{Ellison} S.~L.,  {Mendel} J.~T.,  {Patton} D.~R.,   {Scudder} J.~M.,  2013,
  \mn@doi [\mnras] {10.1093/mnras/stt1562}, \href
  {https://ui.adsabs.harvard.edu/abs/2013MNRAS.435.3627E} {435, 3627}

\bibitem[\protect\citeauthoryear{{Ellison}, {Fertig}, {Rosenberg}, {Nair},
  {Simard}, {Torrey}  \& {Patton}}{{Ellison}
  et~al.}{2015a}]{2015MNRAS.448..221E}
{Ellison} S.~L.,  {Fertig} D.,  {Rosenberg} J.~L.,  {Nair} P.,  {Simard} L.,
  {Torrey} P.,   {Patton} D.~R.,  2015a, \mn@doi [\mnras]
  {10.1093/mnras/stu2744}, \href
  {https://ui.adsabs.harvard.edu/abs/2015MNRAS.448..221E} {448, 221}

\bibitem[\protect\citeauthoryear{{Ellison}, {Patton}  \& {Hickox}}{{Ellison}
  et~al.}{2015b}]{2015MNRAS.451L..35E}
{Ellison} S.~L.,  {Patton} D.~R.,   {Hickox} R.~C.,  2015b, \mn@doi [\mnras]
  {10.1093/mnrasl/slv061}, \href
  {https://ui.adsabs.harvard.edu/abs/2015MNRAS.451L..35E} {451, L35}

\bibitem[\protect\citeauthoryear{{Ellison}, {Catinella}  \&
  {Cortese}}{{Ellison} et~al.}{2018}]{2018MNRAS.478.3447E}
{Ellison} S.~L.,  {Catinella} B.,   {Cortese} L.,  2018, \mn@doi [\mnras]
  {10.1093/mnras/sty1247}, \href
  {https://ui.adsabs.harvard.edu/abs/2018MNRAS.478.3447E} {478, 3447}

\bibitem[\protect\citeauthoryear{{For} et~al.,}{{For}
  et~al.}{2019}]{2019MNRAS.489.5723F}
{For} B.~Q.,  et~al., 2019, \mn@doi [\mnras] {10.1093/mnras/stz2501}, \href
  {https://ui.adsabs.harvard.edu/abs/2019MNRAS.489.5723F} {489, 5723}

\bibitem[\protect\citeauthoryear{{Georgakakis}, {Forbes}  \&
  {Norris}}{{Georgakakis} et~al.}{2000}]{2000MNRAS.318..124G}
{Georgakakis} A.,  {Forbes} D.~A.,   {Norris} R.~P.,  2000, \mn@doi [\mnras]
  {10.1046/j.1365-8711.2000.03709.x}, \href
  {https://ui.adsabs.harvard.edu/abs/2000MNRAS.318..124G} {318, 124}

\bibitem[\protect\citeauthoryear{{Giese}, {van der Hulst}, {Serra}  \&
  {Oosterloo}}{{Giese} et~al.}{2016}]{2016MNRAS.461.1656G}
{Giese} N.,  {van der Hulst} T.,  {Serra} P.,   {Oosterloo} T.,  2016, \mn@doi
  [\mnras] {10.1093/mnras/stw1426}, \href
  {https://ui.adsabs.harvard.edu/abs/2016MNRAS.461.1656G} {461, 1656}

\bibitem[\protect\citeauthoryear{{Giovanelli} et~al.,}{{Giovanelli}
  et~al.}{2005}]{2005AJ....130.2598G}
{Giovanelli} R.,  et~al., 2005, \mn@doi [\aj] {10.1086/497431}, \href
  {https://ui.adsabs.harvard.edu/abs/2005AJ....130.2598G} {130, 2598}

\bibitem[\protect\citeauthoryear{{Gunn} \& {Gott}}{{Gunn} \&
  {Gott}}{1972}]{1972ApJ...176....1G}
{Gunn} J.~E.,  {Gott} J.~Richard I.,  1972, \mn@doi [\apj] {10.1086/151605},
  \href {https://ui.adsabs.harvard.edu/abs/1972ApJ...176....1G} {176, 1}

\bibitem[\protect\citeauthoryear{{Haynes} et~al.,}{{Haynes}
  et~al.}{2018}]{2018ApJ...861...49H}
{Haynes} M.~P.,  et~al., 2018, \mn@doi [\apj] {10.3847/1538-4357/aac956}, \href
  {https://ui.adsabs.harvard.edu/abs/2018ApJ...861...49H} {861, 49}

\bibitem[\protect\citeauthoryear{{Hern{\'a}ndez-Toledo}, {Avila-Reese},
  {Conselice}  \& {Puerari}}{{Hern{\'a}ndez-Toledo}
  et~al.}{2005}]{2005AJ....129..682H}
{Hern{\'a}ndez-Toledo} H.~M.,  {Avila-Reese} V.,  {Conselice} C.~J.,
  {Puerari} I.,  2005, \mn@doi [\aj] {10.1086/427134}, \href
  {https://ui.adsabs.harvard.edu/abs/2005AJ....129..682H} {129, 682}

\bibitem[\protect\citeauthoryear{{Holwerda}, {Pirzkal}, {de Blok}, {Bouchard},
  {Blyth}, {van der Heyden}  \& {Elson}}{{Holwerda}
  et~al.}{2011}]{2011MNRAS.416.2415H}
{Holwerda} B.~W.,  {Pirzkal} N.,  {de Blok} W.~J.~G.,  {Bouchard} A.,  {Blyth}
  S.~L.,  {van der Heyden} K.~J.,   {Elson} E.~C.,  2011, \mn@doi [\mnras]
  {10.1111/j.1365-2966.2011.17683.x}, \href
  {https://ui.adsabs.harvard.edu/abs/2011MNRAS.416.2415H} {416, 2415}

\bibitem[\protect\citeauthoryear{{Hunter} et~al.,}{{Hunter}
  et~al.}{2012}]{2012AJ....144..134H}
{Hunter} D.~A.,  et~al., 2012, \mn@doi [\aj] {10.1088/0004-6256/144/5/134},
  \href {https://ui.adsabs.harvard.edu/abs/2012AJ....144..134H} {144, 134}

\bibitem[\protect\citeauthoryear{{Hunter}, {van Zee}, {McQuinn}, {Garner}  \&
  {Dolphin}}{{Hunter} et~al.}{2022}]{2022AJ....163..132H}
{Hunter} L.~C.,  {van Zee} L.,  {McQuinn} K. B.~W.,  {Garner} R.,   {Dolphin}
  A.~E.,  2022, \mn@doi [\aj] {10.3847/1538-3881/ac4d2c}, \href
  {https://ui.adsabs.harvard.edu/abs/2022AJ....163..132H} {163, 132}

\bibitem[\protect\citeauthoryear{{Huynh}, {Dempsey}, Whiting~{M.}  \&
  {Ophel}}{{Huynh} et~al.}{2020}]{2020Huynh}
{Huynh} M.,  {Dempsey} J.,  Whiting~{M.} T.,   {Ophel} M.,  2020, in Ballester
  P., Ibsen J., Solar M., Shortridge K., eds, ASP Conf. Ser. Vol. 522,
  Astronomical Data Analysis Software and Systems XXVII. Astron. Soc. Pac., San
  Francisco, p. 263

\bibitem[\protect\citeauthoryear{{Ianjamasimanana}, {de Blok}, {Walter}  \&
  {Heald}}{{Ianjamasimanana} et~al.}{2012}]{2012AJ....144...96I}
{Ianjamasimanana} R.,  {de Blok} W.~J.~G.,  {Walter} F.,   {Heald} G.~H.,
  2012, \mn@doi [\aj] {10.1088/0004-6256/144/4/96}, \href
  {https://ui.adsabs.harvard.edu/abs/2012AJ....144...96I} {144, 96}

\bibitem[\protect\citeauthoryear{{J{\'a}chym}, {Combes}, {Cortese}, {Sun}  \&
  {Kenney}}{{J{\'a}chym} et~al.}{2014}]{2014ApJ...792...11J}
{J{\'a}chym} P.,  {Combes} F.,  {Cortese} L.,  {Sun} M.,   {Kenney} J. D.~P.,
  2014, \mn@doi [\apj] {10.1088/0004-637X/792/1/11}, \href
  {https://ui.adsabs.harvard.edu/abs/2014ApJ...792...11J} {792, 11}

\bibitem[\protect\citeauthoryear{{Jaff{\'e}}, {Smith}, {Candlish}, {Poggianti},
  {Sheen}  \& {Verheijen}}{{Jaff{\'e}} et~al.}{2015}]{2015MNRAS.448.1715J}
{Jaff{\'e}} Y.~L.,  {Smith} R.,  {Candlish} G.~N.,  {Poggianti} B.~M.,  {Sheen}
  Y.-K.,   {Verheijen} M. A.~W.,  2015, \mn@doi [\mnras]
  {10.1093/mnras/stv100}, \href
  {https://ui.adsabs.harvard.edu/abs/2015MNRAS.448.1715J} {448, 1715}

\bibitem[\protect\citeauthoryear{{Jaff{\'e}} et~al.,}{{Jaff{\'e}}
  et~al.}{2016}]{2016MNRAS.461.1202J}
{Jaff{\'e}} Y.~L.,  et~al., 2016, \mn@doi [\mnras] {10.1093/mnras/stw984},
  \href {https://ui.adsabs.harvard.edu/abs/2016MNRAS.461.1202J} {461, 1202}

\bibitem[\protect\citeauthoryear{{Kennicutt}}{{Kennicutt}}{1989}]{1989ApJ...344..685K}
{Kennicutt} Robert~C. J.,  1989, \mn@doi [\apj] {10.1086/167834}, \href
  {https://ui.adsabs.harvard.edu/abs/1989ApJ...344..685K} {344, 685}

\bibitem[\protect\citeauthoryear{{Kilborn}, {Forbes}, {Barnes}, {Koribalski},
  {Brough}  \& {Kern}}{{Kilborn} et~al.}{2009}]{2009MNRAS.400.1962K}
{Kilborn} V.~A.,  {Forbes} D.~A.,  {Barnes} D.~G.,  {Koribalski} B.~S.,
  {Brough} S.,   {Kern} K.,  2009, \mn@doi [\mnras]
  {10.1111/j.1365-2966.2009.15587.x}, \href
  {https://ui.adsabs.harvard.edu/abs/2009MNRAS.400.1962K} {400, 1962}

\bibitem[\protect\citeauthoryear{{Kim} \& {Oh}}{{Kim} \&
  {Oh}}{2022}]{2022arXiv220900390K}
{Kim} M.,  {Oh} S.-H.,  2022, arXiv e-prints, \href
  {https://ui.adsabs.harvard.edu/abs/2022arXiv220900390K} {p. arXiv:2209.00390}

\bibitem[\protect\citeauthoryear{{Kleiner} et~al.,}{{Kleiner}
  et~al.}{2019}]{2019MNRAS.488.5352K}
{Kleiner} D.,  et~al., 2019, \mn@doi [\mnras] {10.1093/mnras/stz2063}, \href
  {https://ui.adsabs.harvard.edu/abs/2019MNRAS.488.5352K} {488, 5352}

\bibitem[\protect\citeauthoryear{{Koopmann} \& {Kenney}}{{Koopmann} \&
  {Kenney}}{2004}]{2004ApJ...613..866K}
{Koopmann} R.~A.,  {Kenney} J. D.~P.,  2004, \mn@doi [\apj] {10.1086/423191},
  \href {https://ui.adsabs.harvard.edu/abs/2004ApJ...613..866K} {613, 866}

\bibitem[\protect\citeauthoryear{{Koribalski} \&
  {L{\'o}pez-S{\'a}nchez}}{{Koribalski} \&
  {L{\'o}pez-S{\'a}nchez}}{2009}]{2009MNRAS.400.1749K}
{Koribalski} B.~S.,  {L{\'o}pez-S{\'a}nchez} {\'A}.~R.,  2009, \mn@doi [\mnras]
  {10.1111/j.1365-2966.2009.15610.x}, \href
  {https://ui.adsabs.harvard.edu/abs/2009MNRAS.400.1749K} {400, 1749}

\bibitem[\protect\citeauthoryear{{Koribalski} et~al.,}{{Koribalski}
  et~al.}{2020}]{2020Ap&SS.365..118K}
{Koribalski} B.~S.,  et~al., 2020, \mn@doi [\apss]
  {10.1007/s10509-020-03831-4}, \href
  {https://ui.adsabs.harvard.edu/abs/2020Ap&SS.365..118K} {365, 118}

\bibitem[\protect\citeauthoryear{{Lee-Waddell} et~al.,}{{Lee-Waddell}
  et~al.}{2019}]{2019MNRAS.487.5248L}
{Lee-Waddell} K.,  et~al., 2019, \mn@doi [\mnras] {10.1093/mnras/stz017}, \href
  {https://ui.adsabs.harvard.edu/abs/2019MNRAS.487.5248L} {487, 5248}

\bibitem[\protect\citeauthoryear{{Lee} et~al.,}{{Lee}
  et~al.}{2017}]{2017MNRAS.466.1382L}
{Lee} B.,  et~al., 2017, \mn@doi [\mnras] {10.1093/mnras/stw3162}, \href
  {https://ui.adsabs.harvard.edu/abs/2017MNRAS.466.1382L} {466, 1382}

\bibitem[\protect\citeauthoryear{{Lee} et~al.,}{{Lee}
  et~al.}{2022}]{2022ApJS..262...31L}
{Lee} B.,  et~al., 2022, \mn@doi [\apjs] {10.3847/1538-4365/ac7eba}, \href
  {https://ui.adsabs.harvard.edu/abs/2022ApJS..262...31L} {262, 31}

\bibitem[\protect\citeauthoryear{{Leroy}, {Walter}, {Brinks}, {Bigiel}, {de
  Blok}, {Madore}  \& {Thornley}}{{Leroy} et~al.}{2008}]{2008AJ....136.2782L}
{Leroy} A.~K.,  {Walter} F.,  {Brinks} E.,  {Bigiel} F.,  {de Blok} W.~J.~G.,
  {Madore} B.,   {Thornley} M.~D.,  2008, \mn@doi [\aj]
  {10.1088/0004-6256/136/6/2782}, \href
  {https://ui.adsabs.harvard.edu/abs/2008AJ....136.2782L} {136, 2782}

\bibitem[\protect\citeauthoryear{{Mapelli}, {Moore}  \&
  {Bland-Hawthorn}}{{Mapelli} et~al.}{2008}]{2008MNRAS.388..697M}
{Mapelli} M.,  {Moore} B.,   {Bland-Hawthorn} J.,  2008, \mn@doi [\mnras]
  {10.1111/j.1365-2966.2008.13421.x}, \href
  {https://ui.adsabs.harvard.edu/abs/2008MNRAS.388..697M} {388, 697}

\bibitem[\protect\citeauthoryear{{Marasco}, {Crain}, {Schaye}, {Bah{\'e}}, {van
  der Hulst}, {Theuns}  \& {Bower}}{{Marasco}
  et~al.}{2016}]{2016MNRAS.461.2630M}
{Marasco} A.,  {Crain} R.~A.,  {Schaye} J.,  {Bah{\'e}} Y.~M.,  {van der Hulst}
  T.,  {Theuns} T.,   {Bower} R.~G.,  2016, \mn@doi [\mnras]
  {10.1093/mnras/stw1498}, \href
  {https://ui.adsabs.harvard.edu/abs/2016MNRAS.461.2630M} {461, 2630}

\bibitem[\protect\citeauthoryear{{Martin}}{{Martin}}{2005}]{2005ApJ...621..227M}
{Martin} C.~L.,  2005, \mn@doi [\apj] {10.1086/427277}, \href
  {https://ui.adsabs.harvard.edu/abs/2005ApJ...621..227M} {621, 227}

\bibitem[\protect\citeauthoryear{{Martin} \& {Kennicutt}}{{Martin} \&
  {Kennicutt}}{2001}]{2001ApJ...555..301M}
{Martin} C.~L.,  {Kennicutt} Robert~C. J.,  2001, \mn@doi [\apj]
  {10.1086/321452}, \href
  {https://ui.adsabs.harvard.edu/abs/2001ApJ...555..301M} {555, 301}

\bibitem[\protect\citeauthoryear{{Michiyama} et~al.,}{{Michiyama}
  et~al.}{2016}]{2016PASJ...68...96M}
{Michiyama} T.,  et~al., 2016, \mn@doi [\pasj] {10.1093/pasj/psw087}, \href
  {https://ui.adsabs.harvard.edu/abs/2016PASJ...68...96M} {68, 96}

\bibitem[\protect\citeauthoryear{{Mihos}}{{Mihos}}{2003}]{2003astro.ph..5512M}
{Mihos} C.,  2003, arXiv e-prints, \href
  {https://ui.adsabs.harvard.edu/abs/2003astro.ph..5512M} {pp
  astro--ph/0305512}

\bibitem[\protect\citeauthoryear{{Mihos} \& {Hernquist}}{{Mihos} \&
  {Hernquist}}{1996}]{1996ApJ...464..641M}
{Mihos} J.~C.,  {Hernquist} L.,  1996, \mn@doi [\apj] {10.1086/177353}, \href
  {https://ui.adsabs.harvard.edu/abs/1996ApJ...464..641M} {464, 641}

\bibitem[\protect\citeauthoryear{{Moore}, {Katz}, {Lake}, {Dressler}  \&
  {Oemler}}{{Moore} et~al.}{1996}]{1996Natur.379..613M}
{Moore} B.,  {Katz} N.,  {Lake} G.,  {Dressler} A.,   {Oemler} A.,  1996,
  \mn@doi [\nat] {10.1038/379613a0}, \href
  {https://ui.adsabs.harvard.edu/abs/1996Natur.379..613M} {379, 613}

\bibitem[\protect\citeauthoryear{{Moreno} et~al.,}{{Moreno}
  et~al.}{2021}]{2021MNRAS.503.3113M}
{Moreno} J.,  et~al., 2021, \mn@doi [\mnras] {10.1093/mnras/staa2952}, \href
  {https://ui.adsabs.harvard.edu/abs/2021MNRAS.503.3113M} {503, 3113}

\bibitem[\protect\citeauthoryear{{Moster}, {Macci{\`o}}, {Somerville}, {Naab}
  \& {Cox}}{{Moster} et~al.}{2011}]{2011MNRAS.415.3750M}
{Moster} B.~P.,  {Macci{\`o}} A.~V.,  {Somerville} R.~S.,  {Naab} T.,   {Cox}
  T.~J.,  2011, \mn@doi [\mnras] {10.1111/j.1365-2966.2011.18984.x}, \href
  {https://ui.adsabs.harvard.edu/abs/2011MNRAS.415.3750M} {415, 3750}

\bibitem[\protect\citeauthoryear{{Namumba}, {Carignan}, {Passmoor}  \& {de
  Blok}}{{Namumba} et~al.}{2017}]{2017MNRAS.472.3761N}
{Namumba} B.,  {Carignan} C.,  {Passmoor} S.,   {de Blok} W.~J.~G.,  2017,
  \mn@doi [\mnras] {10.1093/mnras/stx2256}, \href
  {https://ui.adsabs.harvard.edu/abs/2017MNRAS.472.3761N} {472, 3761}

\bibitem[\protect\citeauthoryear{{Nordgren}, {Chengalur}, {Salpeter}  \&
  {Terzian}}{{Nordgren} et~al.}{1997a}]{1997AJ....114...77N}
{Nordgren} T.~E.,  {Chengalur} J.~N.,  {Salpeter} E.~E.,   {Terzian} Y.,
  1997a, \mn@doi [\aj] {10.1086/118454}, \href
  {https://ui.adsabs.harvard.edu/abs/1997AJ....114...77N} {114, 77}

\bibitem[\protect\citeauthoryear{{Nordgren}, {Chengalur}, {Salpeter}  \&
  {Terzian}}{{Nordgren} et~al.}{1997b}]{1997AJ....114..913N}
{Nordgren} T.~E.,  {Chengalur} J.~N.,  {Salpeter} E.~E.,   {Terzian} Y.,
  1997b, \mn@doi [\aj] {10.1086/118523}, \href
  {https://ui.adsabs.harvard.edu/abs/1997AJ....114..913N} {114, 913}

\bibitem[\protect\citeauthoryear{{Odekon} et~al.,}{{Odekon}
  et~al.}{2016}]{2016ApJ...824..110O}
{Odekon} M.~C.,  et~al., 2016, \mn@doi [\apj] {10.3847/0004-637X/824/2/110},
  \href {https://ui.adsabs.harvard.edu/abs/2016ApJ...824..110O} {824, 110}

\bibitem[\protect\citeauthoryear{{Oh}, {de Blok}, {Walter}, {Brinks}  \&
  {Kennicutt}}{{Oh} et~al.}{2008}]{2008AJ....136.2761O}
{Oh} S.-H.,  {de Blok} W.~J.~G.,  {Walter} F.,  {Brinks} E.,   {Kennicutt}
  Robert~C. J.,  2008, \mn@doi [\aj] {10.1088/0004-6256/136/6/2761}, \href
  {https://ui.adsabs.harvard.edu/abs/2008AJ....136.2761O} {136, 2761}

\bibitem[\protect\citeauthoryear{{Oh}, {de Blok}, {Brinks}, {Walter}  \&
  {Kennicutt}}{{Oh} et~al.}{2011}]{2011AJ....141..193O}
{Oh} S.-H.,  {de Blok} W.~J.~G.,  {Brinks} E.,  {Walter} F.,   {Kennicutt}
  Robert~C. J.,  2011, \mn@doi [\aj] {10.1088/0004-6256/141/6/193}, \href
  {https://ui.adsabs.harvard.edu/abs/2011AJ....141..193O} {141, 193}

\bibitem[\protect\citeauthoryear{{Oh} et~al.,}{{Oh}
  et~al.}{2015}]{2015AJ....149..180O}
{Oh} S.-H.,  et~al., 2015, \mn@doi [\aj] {10.1088/0004-6256/149/6/180}, \href
  {https://ui.adsabs.harvard.edu/abs/2015AJ....149..180O} {149, 180}

\bibitem[\protect\citeauthoryear{{Oh}, {Staveley-Smith}, {Spekkens}, {Kamphuis}
   \& {Koribalski}}{{Oh} et~al.}{2018}]{2018MNRAS.473.3256O}
{Oh} S.-H.,  {Staveley-Smith} L.,  {Spekkens} K.,  {Kamphuis} P.,
  {Koribalski} B.~S.,  2018, \mn@doi [\mnras] {10.1093/mnras/stx2304}, \href
  {https://ui.adsabs.harvard.edu/abs/2018MNRAS.473.3256O} {473, 3256}

\bibitem[\protect\citeauthoryear{{Oh}, {Staveley-Smith}  \& {For}}{{Oh}
  et~al.}{2019a}]{2019MNRAS.485.5021O}
{Oh} S.-H.,  {Staveley-Smith} L.,   {For} B.-Q.,  2019a, \mn@doi [\mnras]
  {10.1093/mnras/stz710}, \href
  {https://ui.adsabs.harvard.edu/abs/2019MNRAS.485.5021O} {485, 5021}

\bibitem[\protect\citeauthoryear{{Oh} et~al.,}{{Oh}
  et~al.}{2019b}]{2019MNRAS.488.4169O}
{Oh} S.,  et~al., 2019b, \mn@doi [\mnras] {10.1093/mnras/stz1920}, \href
  {https://ui.adsabs.harvard.edu/abs/2019MNRAS.488.4169O} {488, 4169}

\bibitem[\protect\citeauthoryear{{Oh}, {Kim}, {For}  \& {Staveley-Smith}}{{Oh}
  et~al.}{2022}]{2022ApJ...928..177O}
{Oh} S.-H.,  {Kim} S.,  {For} B.-Q.,   {Staveley-Smith} L.,  2022, \mn@doi
  [\apj] {10.3847/1538-4357/ac5905}, \href
  {https://ui.adsabs.harvard.edu/abs/2022ApJ...928..177O} {928, 177}

\bibitem[\protect\citeauthoryear{{Olson} \& {Kwan}}{{Olson} \&
  {Kwan}}{1990}]{1990ApJ...361..426O}
{Olson} K.~M.,  {Kwan} J.,  1990, \mn@doi [\apj] {10.1086/169208}, \href
  {https://ui.adsabs.harvard.edu/abs/1990ApJ...361..426O} {361, 426}

\bibitem[\protect\citeauthoryear{{Osmond} \& {Ponman}}{{Osmond} \&
  {Ponman}}{2004}]{2004MNRAS.350.1511O}
{Osmond} J. P.~F.,  {Ponman} T.~J.,  2004, \mn@doi [\mnras]
  {10.1111/j.1365-2966.2004.07742.x}, \href
  {https://ui.adsabs.harvard.edu/abs/2004MNRAS.350.1511O} {350, 1511}

\bibitem[\protect\citeauthoryear{{Pan} et~al.,}{{Pan}
  et~al.}{2018}]{2018ApJ...868..132P}
{Pan} H.-A.,  et~al., 2018, \mn@doi [\apj] {10.3847/1538-4357/aaeb92}, \href
  {https://ui.adsabs.harvard.edu/abs/2018ApJ...868..132P} {868, 132}

\bibitem[\protect\citeauthoryear{{Park} \& {Hwang}}{{Park} \&
  {Hwang}}{2009}]{2009ApJ...699.1595P}
{Park} C.,  {Hwang} H.~S.,  2009, \mn@doi [\apj]
  {10.1088/0004-637X/699/2/1595}, \href
  {https://ui.adsabs.harvard.edu/abs/2009ApJ...699.1595P} {699, 1595}

\bibitem[\protect\citeauthoryear{{Park}, {Oh}, {Wang}, {Zheng}, {Zhang}  \& {de
  Blok}}{{Park} et~al.}{2022}]{2022arXiv220706698P}
{Park} H.-J.,  {Oh} S.-H.,  {Wang} J.,  {Zheng} Y.,  {Zhang} H.-X.,   {de Blok}
  W.~J.~G.,  2022, arXiv e-prints, \href
  {https://ui.adsabs.harvard.edu/abs/2022arXiv220706698P} {p. arXiv:2207.06698}

\bibitem[\protect\citeauthoryear{{Patton}, {Torrey}, {Ellison}, {Mendel}  \&
  {Scudder}}{{Patton} et~al.}{2013}]{2013MNRAS.433L..59P}
{Patton} D.~R.,  {Torrey} P.,  {Ellison} S.~L.,  {Mendel} J.~T.,   {Scudder}
  J.~M.,  2013, \mn@doi [\mnras] {10.1093/mnrasl/slt058}, \href
  {https://ui.adsabs.harvard.edu/abs/2013MNRAS.433L..59P} {433, L59}

\bibitem[\protect\citeauthoryear{{Patton} et~al.,}{{Patton}
  et~al.}{2020}]{2020MNRAS.494.4969P}
{Patton} D.~R.,  et~al., 2020, \mn@doi [\mnras] {10.1093/mnras/staa913}, \href
  {https://ui.adsabs.harvard.edu/abs/2020MNRAS.494.4969P} {494, 4969}

\bibitem[\protect\citeauthoryear{{Quai}, {Hani}, {Ellison}, {Patton}  \&
  {Woo}}{{Quai} et~al.}{2021}]{2021MNRAS.504.1888Q}
{Quai} S.,  {Hani} M.~H.,  {Ellison} S.~L.,  {Patton} D.~R.,   {Woo} J.,  2021,
  \mn@doi [\mnras] {10.1093/mnras/stab988}, \href
  {https://ui.adsabs.harvard.edu/abs/2021MNRAS.504.1888Q} {504, 1888}

\bibitem[\protect\citeauthoryear{{Reiprich} \& {B{\"o}hringer}}{{Reiprich} \&
  {B{\"o}hringer}}{2002}]{2002ApJ...567..716R}
{Reiprich} T.~H.,  {B{\"o}hringer} H.,  2002, \mn@doi [\apj] {10.1086/338753},
  \href {https://ui.adsabs.harvard.edu/abs/2002ApJ...567..716R} {567, 716}

\bibitem[\protect\citeauthoryear{{Reynolds} et~al.,}{{Reynolds}
  et~al.}{2019}]{2019MNRAS.482.3591R}
{Reynolds} T.~N.,  et~al., 2019, \mn@doi [\mnras] {10.1093/mnras/sty2930},
  \href {https://ui.adsabs.harvard.edu/abs/2019MNRAS.482.3591R} {482, 3591}

\bibitem[\protect\citeauthoryear{{Reynolds}, {Westmeier}, {Staveley-Smith},
  {Chauhan}  \& {Lagos}}{{Reynolds} et~al.}{2020}]{2020MNRAS.493.5089R}
{Reynolds} T.~N.,  {Westmeier} T.,  {Staveley-Smith} L.,  {Chauhan} G.,
  {Lagos} C.~D.~P.,  2020, \mn@doi [\mnras] {10.1093/mnras/staa597}, \href
  {https://ui.adsabs.harvard.edu/abs/2020MNRAS.493.5089R} {493, 5089}

\bibitem[\protect\citeauthoryear{{Reynolds} et~al.,}{{Reynolds}
  et~al.}{2021}]{2021MNRAS.505.1891R}
{Reynolds} T.~N.,  et~al., 2021, \mn@doi [\mnras] {10.1093/mnras/stab1371},
  \href {https://ui.adsabs.harvard.edu/abs/2021MNRAS.505.1891R} {505, 1891}

\bibitem[\protect\citeauthoryear{{Reynolds} et~al.,}{{Reynolds}
  et~al.}{2022}]{2022MNRAS.510.1716R}
{Reynolds} T.~N.,  et~al., 2022, \mn@doi [\mnras] {10.1093/mnras/stab3522},
  \href {https://ui.adsabs.harvard.edu/abs/2022MNRAS.510.1716R} {510, 1716}

\bibitem[\protect\citeauthoryear{{Rhee}, {Smith}, {Choi}, {Yi}, {Jaff{\'e}},
  {Candlish}  \& {S{\'a}nchez-J{\'a}nssen}}{{Rhee}
  et~al.}{2017}]{2017ApJ...843..128R}
{Rhee} J.,  {Smith} R.,  {Choi} H.,  {Yi} S.~K.,  {Jaff{\'e}} Y.,  {Candlish}
  G.,   {S{\'a}nchez-J{\'a}nssen} R.,  2017, \mn@doi [\apj]
  {10.3847/1538-4357/aa6d6c}, \href
  {https://ui.adsabs.harvard.edu/abs/2017ApJ...843..128R} {843, 128}

\bibitem[\protect\citeauthoryear{{Romeo} \& {Mogotsi}}{{Romeo} \&
  {Mogotsi}}{2017}]{2017MNRAS.469..286R}
{Romeo} A.~B.,  {Mogotsi} K.~M.,  2017, \mn@doi [\mnras]
  {10.1093/mnras/stx844}, \href
  {https://ui.adsabs.harvard.edu/abs/2017MNRAS.469..286R} {469, 286}

\bibitem[\protect\citeauthoryear{{Saintonge} \& {Catinella}}{{Saintonge} \&
  {Catinella}}{2022}]{2022arXiv220200690S}
{Saintonge} A.,  {Catinella} B.,  2022, arXiv e-prints, \href
  {https://ui.adsabs.harvard.edu/abs/2022arXiv220200690S} {p. arXiv:2202.00690}

\bibitem[\protect\citeauthoryear{{Saintonge} et~al.,}{{Saintonge}
  et~al.}{2012}]{2012ApJ...758...73S}
{Saintonge} A.,  et~al., 2012, \mn@doi [\apj] {10.1088/0004-637X/758/2/73},
  \href {https://ui.adsabs.harvard.edu/abs/2012ApJ...758...73S} {758, 73}

\bibitem[\protect\citeauthoryear{{Scudder}, {Ellison}, {Momjian}, {Rosenberg},
  {Torrey}, {Patton}, {Fertig}  \& {Mendel}}{{Scudder}
  et~al.}{2015}]{2015MNRAS.449.3719S}
{Scudder} J.~M.,  {Ellison} S.~L.,  {Momjian} E.,  {Rosenberg} J.~L.,  {Torrey}
  P.,  {Patton} D.~R.,  {Fertig} D.,   {Mendel} J.~T.,  2015, \mn@doi [\mnras]
  {10.1093/mnras/stv588}, \href
  {https://ui.adsabs.harvard.edu/abs/2015MNRAS.449.3719S} {449, 3719}

\bibitem[\protect\citeauthoryear{{Stevens} et~al.,}{{Stevens}
  et~al.}{2021}]{2021MNRAS.502.3158S}
{Stevens} A. R.~H.,  et~al., 2021, \mn@doi [\mnras] {10.1093/mnras/staa3662},
  \href {https://ui.adsabs.harvard.edu/abs/2021MNRAS.502.3158S} {502, 3158}

\bibitem[\protect\citeauthoryear{{Stierwalt}, {Besla}, {Patton}, {Johnson},
  {Kallivayalil}, {Putman}, {Privon}  \& {Ross}}{{Stierwalt}
  et~al.}{2015}]{2015ApJ...805....2S}
{Stierwalt} S.,  {Besla} G.,  {Patton} D.,  {Johnson} K.,  {Kallivayalil} N.,
  {Putman} M.,  {Privon} G.,   {Ross} G.,  2015, \mn@doi [\apj]
  {10.1088/0004-637X/805/1/2}, \href
  {https://ui.adsabs.harvard.edu/abs/2015ApJ...805....2S} {805, 2}

\bibitem[\protect\citeauthoryear{{Stilp}, {Dalcanton}, {Warren}, {Skillman},
  {Ott}  \& {Koribalski}}{{Stilp} et~al.}{2013}]{2013ApJ...765..136S}
{Stilp} A.~M.,  {Dalcanton} J.~J.,  {Warren} S.~R.,  {Skillman} E.,  {Ott} J.,
   {Koribalski} B.,  2013, \mn@doi [\apj] {10.1088/0004-637X/765/2/136}, \href
  {https://ui.adsabs.harvard.edu/abs/2013ApJ...765..136S} {765, 136}

\bibitem[\protect\citeauthoryear{{Struble} \& {Rood}}{{Struble} \&
  {Rood}}{1999}]{1999ApJS..125...35S}
{Struble} M.~F.,  {Rood} H.~J.,  1999, \mn@doi [\apjs] {10.1086/313274}, \href
  {https://ui.adsabs.harvard.edu/abs/1999ApJS..125...35S} {125, 35}

\bibitem[\protect\citeauthoryear{{Toomre}}{{Toomre}}{1964}]{1964ApJ...139.1217T}
{Toomre} A.,  1964, \mn@doi [\apj] {10.1086/147861}, \href
  {https://ui.adsabs.harvard.edu/abs/1964ApJ...139.1217T} {139, 1217}

\bibitem[\protect\citeauthoryear{{\noopsort{Van}{van de Voort}}
  et~al.,}{{\noopsort{Van}{van de Voort}} et~al.}{2018}]{2018MNRAS.476..122V}
{\noopsort{Van}{van de Voort}} F.,  et~al., 2018, \mn@doi [\mnras]
  {10.1093/mnras/sty228}, \href
  {https://ui.adsabs.harvard.edu/abs/2018MNRAS.476..122V} {476, 122}

\bibitem[\protect\citeauthoryear{{Ventou} et~al.,}{{Ventou}
  et~al.}{2019}]{2019A&A...631A..87V}
{Ventou} E.,  et~al., 2019, \mn@doi [\aap] {10.1051/0004-6361/201935597}, \href
  {https://ui.adsabs.harvard.edu/abs/2019A&A...631A..87V} {631, A87}

\bibitem[\protect\citeauthoryear{{Violino}, {Ellison}, {Sargent}, {Coppin},
  {Scudder}, {Mendel}  \& {Saintonge}}{{Violino}
  et~al.}{2018}]{2018MNRAS.476.2591V}
{Violino} G.,  {Ellison} S.~L.,  {Sargent} M.,  {Coppin} K. E.~K.,  {Scudder}
  J.~M.,  {Mendel} T.~J.,   {Saintonge} A.,  2018, \mn@doi [\mnras]
  {10.1093/mnras/sty345}, \href
  {https://ui.adsabs.harvard.edu/abs/2018MNRAS.476.2591V} {476, 2591}

\bibitem[\protect\citeauthoryear{{Vollmer}, {Wong}, {Braine}, {Chung}  \&
  {Kenney}}{{Vollmer} et~al.}{2012}]{2012A&A...543A..33V}
{Vollmer} B.,  {Wong} O.~I.,  {Braine} J.,  {Chung} A.,   {Kenney} J.~D.~P.,
  2012, \mn@doi [\aap] {10.1051/0004-6361/201118690}, \href
  {https://ui.adsabs.harvard.edu/abs/2012A&A...543A..33V} {543, A33}

\bibitem[\protect\citeauthoryear{{Walter}, {Brinks}, {de Blok}, {Bigiel},
  {Kennicutt}, {Thornley}  \& {Leroy}}{{Walter}
  et~al.}{2008}]{2008AJ....136.2563W}
{Walter} F.,  {Brinks} E.,  {de Blok} W.~J.~G.,  {Bigiel} F.,  {Kennicutt}
  Robert~C. J.,  {Thornley} M.~D.,   {Leroy} A.,  2008, \mn@doi [\aj]
  {10.1088/0004-6256/136/6/2563}, \href
  {https://ui.adsabs.harvard.edu/abs/2008AJ....136.2563W} {136, 2563}

\bibitem[\protect\citeauthoryear{{Wang}, {Koribalski}, {Serra}, {van der
  Hulst}, {Roychowdhury}, {Kamphuis}  \& {Chengalur}}{{Wang}
  et~al.}{2016}]{2016MNRAS.460.2143W}
{Wang} J.,  {Koribalski} B.~S.,  {Serra} P.,  {van der Hulst} T.,
  {Roychowdhury} S.,  {Kamphuis} P.,   {Chengalur} J.~N.,  2016, \mn@doi
  [\mnras] {10.1093/mnras/stw1099}, \href
  {https://ui.adsabs.harvard.edu/abs/2016MNRAS.460.2143W} {460, 2143}

\bibitem[\protect\citeauthoryear{{Wang} et~al.,}{{Wang}
  et~al.}{2021}]{2021ApJ...915...70W}
{Wang} J.,  et~al., 2021, \mn@doi [\apj] {10.3847/1538-4357/abfc52}, \href
  {https://ui.adsabs.harvard.edu/abs/2021ApJ...915...70W} {915, 70}

\bibitem[\protect\citeauthoryear{{Wang} et~al.,}{{Wang}
  et~al.}{2022}]{2022ApJ...927...66W}
{Wang} S.,  et~al., 2022, \mn@doi [\apj] {10.3847/1538-4357/ac4270}, \href
  {https://ui.adsabs.harvard.edu/abs/2022ApJ...927...66W} {927, 66}

\bibitem[\protect\citeauthoryear{{Watts}, {Catinella}, {Cortese}  \&
  {Power}}{{Watts} et~al.}{2020a}]{2020MNRAS.492.3672W}
{Watts} A.~B.,  {Catinella} B.,  {Cortese} L.,   {Power} C.,  2020a, \mn@doi
  [\mnras] {10.1093/mnras/staa094}, \href
  {https://ui.adsabs.harvard.edu/abs/2020MNRAS.492.3672W} {492, 3672}

\bibitem[\protect\citeauthoryear{{Watts}, {Power}, {Catinella}, {Cortese}  \&
  {Stevens}}{{Watts} et~al.}{2020b}]{2020MNRAS.499.5205W}
{Watts} A.~B.,  {Power} C.,  {Catinella} B.,  {Cortese} L.,   {Stevens} A.
  R.~H.,  2020b, \mn@doi [\mnras] {10.1093/mnras/staa3200}, \href
  {https://ui.adsabs.harvard.edu/abs/2020MNRAS.499.5205W} {499, 5205}

\bibitem[\protect\citeauthoryear{{Westmeier}, {Braun}  \&
  {Koribalski}}{{Westmeier} et~al.}{2011}]{2011MNRAS.410.2217W}
{Westmeier} T.,  {Braun} R.,   {Koribalski} B.~S.,  2011, \mn@doi [\mnras]
  {10.1111/j.1365-2966.2010.17596.x}, \href
  {https://ui.adsabs.harvard.edu/abs/2011MNRAS.410.2217W} {410, 2217}

\bibitem[\protect\citeauthoryear{{Westmeier} et~al.,}{{Westmeier}
  et~al.}{2021}]{2021MNRAS.506.3962W}
{Westmeier} T.,  et~al., 2021, \mn@doi [\mnras] {10.1093/mnras/stab1881}, \href
  {https://ui.adsabs.harvard.edu/abs/2021MNRAS.506.3962W} {506, 3962}

\bibitem[\protect\citeauthoryear{{Whiting}}{{Whiting}}{2020}]{2020ASPC..522..469W}
{Whiting} M.~T.,  2020, in {Ballester} P.,  {Ibsen} J.,  {Solar} M.,
  {Shortridge} K.,  eds,  Astronomical Society of the Pacific Conference Series
  Vol. 522, Astronomical Data Analysis Software and Systems XXVII. p.~469

\bibitem[\protect\citeauthoryear{{Wong}, {Kenney}, {Murphy}  \& {Helou}}{{Wong}
  et~al.}{2014}]{2014ApJ...783..109W}
{Wong} O.~I.,  {Kenney} J. D.~P.,  {Murphy} E.~J.,   {Helou} G.,  2014, \mn@doi
  [\apj] {10.1088/0004-637X/783/2/109}, \href
  {https://ui.adsabs.harvard.edu/abs/2014ApJ...783..109W} {783, 109}

\bibitem[\protect\citeauthoryear{{Wong}, {Meurer}, {Zheng}, {Heckman},
  {Thilker}  \& {Zwaan}}{{Wong} et~al.}{2016}]{2016MNRAS.460.1106W}
{Wong} O.~I.,  {Meurer} G.~R.,  {Zheng} Z.,  {Heckman} T.~M.,  {Thilker} D.~A.,
    {Zwaan} M.~A.,  2016, \mn@doi [\mnras] {10.1093/mnras/stw993}, \href
  {https://ui.adsabs.harvard.edu/abs/2016MNRAS.460.1106W} {460, 1106}

\bibitem[\protect\citeauthoryear{{Woudt}, {Kraan-Korteweg}, {Lucey}, {Fairall}
  \& {Moore}}{{Woudt} et~al.}{2008}]{2008MNRAS.383..445W}
{Woudt} P.~A.,  {Kraan-Korteweg} R.~C.,  {Lucey} J.,  {Fairall} A.~P.,
  {Moore} S.~A.~W.,  2008, \mn@doi [\mnras] {10.1111/j.1365-2966.2007.12571.x},
  \href {https://ui.adsabs.harvard.edu/abs/2008MNRAS.383..445W} {383, 445}

\bibitem[\protect\citeauthoryear{{Yamashita} et~al.,}{{Yamashita}
  et~al.}{2017}]{2017ApJ...844...96Y}
{Yamashita} T.,  et~al., 2017, \mn@doi [\apj] {10.3847/1538-4357/aa7af1}, \href
  {https://ui.adsabs.harvard.edu/abs/2017ApJ...844...96Y} {844, 96}

\bibitem[\protect\citeauthoryear{{Yoon}, {Chung}, {Smith}  \&
  {Jaff{\'e}}}{{Yoon} et~al.}{2017}]{2017ApJ...838...81Y}
{Yoon} H.,  {Chung} A.,  {Smith} R.,   {Jaff{\'e}} Y.~L.,  2017, \mn@doi [\apj]
  {10.3847/1538-4357/aa6579}, \href
  {https://ui.adsabs.harvard.edu/abs/2017ApJ...838...81Y} {838, 81}

\bibitem[\protect\citeauthoryear{{York} et~al.,}{{York}
  et~al.}{2000}]{2000AJ....120.1579Y}
{York} D.~G.,  et~al., 2000, \mn@doi [\aj] {10.1086/301513}, \href
  {https://ui.adsabs.harvard.edu/abs/2000AJ....120.1579Y} {120, 1579}

\bibitem[\protect\citeauthoryear{{Zheng}, {Meurer}, {Heckman}, {Thilker}  \&
  {Zwaan}}{{Zheng} et~al.}{2013}]{2013MNRAS.434.3389Z}
{Zheng} Z.,  {Meurer} G.~R.,  {Heckman} T.~M.,  {Thilker} D.~A.,   {Zwaan}
  M.~A.,  2013, \mn@doi [\mnras] {10.1093/mnras/stt1242}, \href
  {https://ui.adsabs.harvard.edu/abs/2013MNRAS.434.3389Z} {434, 3389}

\bibitem[\protect\citeauthoryear{{Zuo}, {Xu}, {Yun}, {Lisenfeld}, {Li}  \&
  {Cao}}{{Zuo} et~al.}{2018}]{2018ApJS..237....2Z}
{Zuo} P.,  {Xu} C.~K.,  {Yun} M.~S.,  {Lisenfeld} U.,  {Li} D.,   {Cao} C.,
  2018, \mn@doi [\apjs] {10.3847/1538-4365/aabd30}, \href
  {https://ui.adsabs.harvard.edu/abs/2018ApJS..237....2Z} {237, 2}

\makeatother
\end{thebibliography}





\appendix
\section{H{\sc i} properties, images and super-profiles of the sample galaxies.}\label{appendixA}


\begin{landscape}

\begin{table}
\centering
\scriptsize
\caption{H{\sc i} properties and distances of the control galaxies in the ASKAP Hydra I cluster field. The full table is provided with supplementary material.}
\label{long_hydra_control}

\resizebox{1.3\textwidth}{!}{
\begin{tabular}{cccccccccccc}
 \hline
 Name & R.A. (J2000) & Dec. (J2000) & $D$ & $V_{\rm{sys}}$ & $R_{\rm{HI}}$ & $S_{\rm{HI}}$ & $M_{\rm{total}}^{\rm{HI}}$ &  $M_{\rm{narrow}}^{\rm{HI}}$ & $M_{\rm{broad}}^{\rm{HI}}$ & $A_{\rm{map}}^{\rm{HI}}$ & $Q$\\
 &   &  & (Mpc) & (\kms)\ & (kpc) & (Jy km s$^{-1}$) & ($10^8 \ \rm{M_{\odot}}$) & ($10^8 \ \rm{M_{\odot}}$) & ($10^8 \ \rm{M_{\odot}}$) \\
    (1) & (2) & (3) & (4) & (5) & (6) & (7) & (8) & (9) & (10) & (11) & (12) \\
 \hline
 
J100351-273417  & 10$^{\rm{h}}$03$^{\rm{m}}$52$^{\rm{s}}_.$0 & -27$^{\rm{d}}$34$^{\rm{m}}$15$^{\rm{s}}_.$3 & 39.9 & 2,792 & 30.5 &  12.6 &  47.2 & 12.4 &  34.8 & 0.23 &  4.22\\ 
J100634-295615  & 10$^{\rm{h}}$06$^{\rm{m}}$34$^{\rm{s}}_.$2 & -29$^{\rm{d}}$56$^{\rm{m}}$12$^{\rm{s}}_.$2 & 15.9 & 1,117 &  9.6 &   9.2 &   5.5 &  1.3 &   4.2 & 0.23 & 10.21\\ 
J100656-251731†  & 10$^{\rm{h}}$06$^{\rm{m}}$55$^{\rm{s}}_.$9 & -25$^{\rm{d}}$17$^{\rm{m}}$32$^{\rm{s}}_.$5 & 41.3 & 2,894 & 10.5 &   1.6 &   6.5 &  3.0 &   3.5 & 0.27 &   ...\\ 
J100713-262336  & 10$^{\rm{h}}$07$^{\rm{m}}$13$^{\rm{s}}_.$8 & -26$^{\rm{d}}$23$^{\rm{m}}$31$^{\rm{s}}_.$4 & 65.7 & 4,599 & 17.1 &   2.0 &  20.5 & 16.8 &   3.7 & 0.21 &   ...\\ 
 
 \hline
 
\end{tabular}}
\flushleft
 \textbf{Notes.} The columns are (1) source name; (2) kinematic centre in R.A. units from {\sc 2dbat}; (3) kinematic centre Dec. units from {\sc 2dbat}; (4) distance to the galaxy from redshift (Mpc); (5) systemic velocity from redshift (\kms)\,; (6) radius of H{\sc i} disk from the {\sc baygaud} analysis, $R_{\rm{HI}}$ (kpc);  (7) integrated H{\sc i} intensities derived from super-profiles in Section~\ref{sec:3} (Jy km s$^{-1}$); (8) H{\sc i} mass derived from super-profiles in Section~\ref{sec:3} ($10^8 \ \rm{M_{\odot}}$); (9) narrow H{\sc i} mass derived from super-profiles in Section~\ref{sec:3} ($10^8 \ \rm{M_{\odot}}$); (10) broad H{\sc i} mass derived from super-profiles in Section~\ref{sec:3} ($10^8 \ \rm{M_{\odot}}$); (11) H{\sc i} morphological asymmetry of the galaxy ; (12) median values of the Toomre Q parameter values of the galaxy. '†' indicates the galaxy that have similar velocity widths of the narrow and broad components in the H{\sc i} super-profile analysis.
 
\end{table}

\begin{table}

\scriptsize
\caption{H{\sc i} properties and distances of the paired galaxies in the ASKAP Hydra I cluster field. The full table is provided with supplementary material.}
\label{long_hydra_pair}

\resizebox{1.3\textwidth}{!}{
\begin{tabular}{cccccccccccccc}
 \hline
 Name & R.A. (J2000) & Dec. (J2000) & $D$ & $V_{\rm{sys}}$ & $R_{\rm{HI}}$ & $S_{\rm{HI}}$ & $M_{\rm{total}}^{\rm{HI}}$ &  $M_{\rm{narrow}}^{\rm{HI}}$ & $M_{\rm{broad}}^{\rm{HI}}$ & $R_{\rm{p}}$ & $\Delta V$ & $A_{\rm{map}}^{\rm{HI}}$ & $Q$\\
 &   &  & (Mpc) & (\kms)\ & (kpc) & (Jy km s$^{-1}$) & ($10^8 \ \rm{M_{\odot}}$) & ($10^8 \ \rm{M_{\odot}}$) & ($10^8 \ \rm{M_{\odot}}$) & (kpc) & (\kms)\ \\
    (1) & (2) & (3) & (4) & (5) & (6) & (7) & (8) & (9) & (10) & (11) & (12) & (13) & (14)\\
 \hline
 
J100336-262923  & 10$^{\rm{h}}$03$^{\rm{m}}$37$^{\rm{s}}_.$4 & -26$^{\rm{d}}$29$^{\rm{m}}$39$^{\rm{s}}_.$1 & 12.9 &   901 &  4.3 &  1.8 &   0.7 &  0.2 &   0.5 &    31 &    26 &  ... &   ...\\ 
J100342-270137  & 10$^{\rm{h}}$03$^{\rm{m}}$41$^{\rm{s}}_.$9 & -27$^{\rm{d}}$01$^{\rm{m}}$39$^{\rm{s}}_.$5 & 13.7 &   957 & 15.5 & 45.7 &  20.2 &  6.8 &  13.3 &    93 &    83 & 0.09 &  2.46\\ 
J100351-263707  & 10$^{\rm{h}}$03$^{\rm{m}}$51$^{\rm{s}}_.$9 & -26$^{\rm{d}}$36$^{\rm{m}}$35$^{\rm{s}}_.$5 & 12.5 &   874 &  6.6 &  6.7 &   2.4 &  0.5 &   1.9 & 31/93 & 26/83 &  ... &   ...\\ 
J100426-282638  & 10$^{\rm{h}}$04$^{\rm{m}}$26$^{\rm{s}}_.$3 & -28$^{\rm{d}}$26$^{\rm{m}}$40$^{\rm{s}}_.$8 & 15.5 & 1,083 & 15.6 & 32.4 &  18.3 &  4.9 &  13.3 &    69 &    41 & 0.10 &  2.70\\ 
\hline

\end{tabular}
}
\flushleft
\textbf{Notes.} The columns are (1) source name; (2) kinematic centre in R.A. units from {\sc 2dbat}; (3) kinematic centre Dec. units from {\sc 2dbat}; (4) distance to the galaxy from redshift (Mpc); (5) systemic velocity from redshift (\kms)\,; (6) radius of H{\sc i} disk from the {\sc baygaud} analysis, $R_{\rm{HI}}$ (kpc);  (7) integrated H{\sc i} intensities derived from super-profiles in Section~\ref{sec:3} (Jy km s$^{-1}$); (8) H{\sc i} mass derived from super-profiles in Section~\ref{sec:3} ($10^8 \ \rm{M_{\odot}}$); (9) narrow H{\sc i} mass derived from super-profiles in Section~\ref{sec:3} ($10^8 \ \rm{M_{\odot}}$); (10) broad H{\sc i} mass derived from super-profiles in Section~\ref{sec:3} ($10^8 \ \rm{M_{\odot}}$); (11) projected distance to the companion galaxy (kpc). For galaxies with two nearby companions, we provide two $R_{\rm{p}}$ values separated by '/'; (12) relative line-of-sight velocity to the companion galaxy (\kms)\,. For galaxies with two nearby companions, we provide two $\Delta V$ values separated by '/'; (13) H{\sc i} morphological asymmetry of the galaxy, (14) median values of the Toomre Q parameter values of the galaxy. '*' indicates the visually identified galaxy pair. '†' indicates the galaxy that have similar velocity widths of the narrow and broad components in the H{\sc i} super-profile analysis.
\end{table}
\end{landscape}


\begin{landscape}
\begin{table}
\scriptsize
\caption{H{\sc i} properties and distances of the control galaxies in the ASKAP NGC 4636 field. The full table is provided with supplementary material.}
\label{long_ngc_control}

\resizebox{1.3\textwidth}{!}{
\begin{tabular}{cccccccccccc}
\hline
 Name & R.A. (J2000) & Dec. (J2000) & $D$ & $V_{\rm{sys}}$ & $R_{\rm{HI}}$ & $S_{\rm{HI}}$ & $M_{\rm{total}}^{\rm{HI}}$ &  $M_{\rm{narrow}}^{\rm{HI}}$ & $M_{\rm{broad}}^{\rm{HI}}$ & $A_{\rm{map}}^{\rm{HI}}$ & $Q$\\
  &   &  & (Mpc) & (\kms)\ & (kpc) & (Jy km s$^{-1}$) & ($10^8 \ \rm{M_{\odot}}$) & ($10^8 \ \rm{M_{\odot}}$) & ($10^8 \ \rm{M_{\odot}}$) \\
    (1) & (2) & (3) & (4) & (5) & (6) & (7) & (8) & (9) & (10) & (11) & (12) \\
 \hline
 
J122708+055255  &  12$^{\rm{h}}$27$^{\rm{m}}$09$^{\rm{s}}_.$3 & +05$^{\rm{d}}$52$^{\rm{m}}$50$^{\rm{s}}_.$6 & 15.9 & 1,113 &  9.9 &  11.2 &   6.7 &  5.7 &   1.0 & 0.40 &  4.00\\ 
J122755+054312  & 12$^{\rm{h}}$27$^{\rm{m}}$55$^{\rm{s}}_.$5 & +05$^{\rm{d}}$43$^{\rm{m}}$14$^{\rm{s}}_.$8 & 32.0 & 2,243 &  5.3 &   1.2 &   2.8 &  0.4 &   2.4 & 0.19 &   ...\\ 
J122928+064623  & 12$^{\rm{h}}$29$^{\rm{m}}$28$^{\rm{s}}_.$6 & +06$^{\rm{d}}$46$^{\rm{m}}$12$^{\rm{s}}_.$0 &  7.5 &   527 &  2.3 &   3.6 &   0.5 &  0.1 &   0.4 & 0.41 &   ...\\ 
J122930+074142  & 12$^{\rm{h}}$29$^{\rm{m}}$30$^{\rm{s}}_.$5 & +07$^{\rm{d}}$41$^{\rm{m}}$46$^{\rm{s}}_.$8 & 10.7 &   748 &  3.0 &   1.5 &   0.4 &  0.1 &   0.3 & 0.22 &   ...\\ 
 \hline
 
\end{tabular}}
\flushleft
 \textbf{Notes.} The columns are (1) source name; (2) kinematic centre in R.A. units from {\sc 2dbat}; (3) kinematic centre Dec. units from {\sc 2dbat}; (4) distance to the galaxy from redshift (Mpc); (5) systemic velocity from redshift (\kms)\,; (6) radius of H{\sc i} disk from the {\sc baygaud} analysis, $R_{\rm{HI}}$ (kpc);  (7) integrated H{\sc i} intensities derived from super-profiles in Section~\ref{sec:3} (Jy km s$^{-1}$); (8) H{\sc i} mass derived from super-profiles in Section~\ref{sec:3} ($10^8 \ \rm{M_{\odot}}$); (9) narrow H{\sc i} mass derived from super-profiles in Section~\ref{sec:3} ($10^8 \ \rm{M_{\odot}}$); (10) broad H{\sc i} mass derived from super-profiles in Section~\ref{sec:3} ($10^8 \ \rm{M_{\odot}}$); (11) H{\sc i} morphological asymmetry of the galaxy, (12) median values of the Toomre Q parameter values of the galaxy ('*': {\sc moment1} velocity field is used for the rotation curve analysis).
\end{table}

\begin{table}
\scriptsize
\caption{H{\sc i} properties and distances of the paired galaxies in the ASKAP NGC 4636 field. The full table is provided with supplementary material.}
\label{long_ngc_pair}

\resizebox{1.3\textwidth}{!}{
\begin{tabular}{cccccccccccccc}
 \hline
 Name & R.A. (J2000) & Dec. (J2000) & $D$ & $V_{\rm{sys}}$ & $R_{\rm{HI}}$ & $S_{\rm{HI}}$ & $M_{\rm{total}}^{\rm{HI}}$ &  $M_{\rm{narrow}}^{\rm{HI}}$ & $M_{\rm{broad}}^{\rm{HI}}$ & $R_{\rm{p}}$ & $\Delta V$ & $A_{\rm{map}}^{\rm{HI}}$ & $Q$\\
  &   &  & (Mpc) & (\kms)\ & (kpc) & (Jy km s$^{-1}$) & ($10^8 \ \rm{M_{\odot}}$) & ($10^8 \ \rm{M_{\odot}}$) & ($10^8 \ \rm{M_{\odot}}$) & (kpc) & (\kms)\ \\
    (1) & (2) & (3) & (4) & (5) & (6) & (7) & (8) & (9) & (10) & (11) & (12) & (13) & (14)\\
 \hline
 
J122710+071549  & 12$^{\rm{h}}$27$^{\rm{m}}$11$^{\rm{s}}_.$3 & +07$^{\rm{d}}$15$^{\rm{m}}$49$^{\rm{s}}_.$2 & 13.3 &   929 & 11.3 &  21.1 &   8.8 &  4.7 &   4.0 & 86 & 68 & 0.20 &   ...\\ 
J122729+073841  & 12$^{\rm{h}}$27$^{\rm{m}}$29$^{\rm{s}}_.$4 & +07$^{\rm{d}}$38$^{\rm{m}}$48$^{\rm{s}}_.$4 & 12.3 &   860 &  4.6 &   3.5 &   1.3 &  0.8 &   0.5 & 86 & 68 & 0.36 &   ...\\ 
J123422+021914  & 12$^{\rm{h}}$34$^{\rm{m}}$21$^{\rm{s}}_.$8 & +02$^{\rm{d}}$19$^{\rm{m}}$25$^{\rm{s}}_.$3 & 25.4 & 1,779 &  7.5 &   3.5 &   5.4 &  0.2 &   5.2 & 58 & 19 &  ... &   ...\\ 
J124508-002747  &  12$^{\rm{h}}$45$^{\rm{m}}$08$^{\rm{s}}_.$4 & -00$^{\rm{d}}$27$^{\rm{m}}$47$^{\rm{s}}_.$4 & 21.9 & 1,534 & 34.3 &  80.5 &  91.2 & 30.0 &  61.2 & 47 & 84 & 0.17 &  9.51\\ 

\hline

\end{tabular}}

\flushleft
\textbf{Notes.} The columns are (1) source name; (2) kinematic centre in R.A. units from {\sc 2dbat}; (3) kinematic centre Dec. units from {\sc 2dbat}; (4) distance to the galaxy from redshift (Mpc); (5) systemic velocity from redshift (\kms)\,; (6) radius of H{\sc i} disk from the {\sc baygaud} analysis, $R_{\rm{HI}}$ (kpc);  (7) integrated H{\sc i} intensities derived from super-profiles in Section~\ref{sec:3} (Jy km s$^{-1}$); (8) H{\sc i} mass derived from super-profiles in Section~\ref{sec:3} ($10^8 \ \rm{M_{\odot}}$); (9) narrow H{\sc i} mass derived from super-profiles in Section~\ref{sec:3} ($10^8 \ \rm{M_{\odot}}$); (10) broad H{\sc i} mass derived from super-profiles in Section~\ref{sec:3} ($10^8 \ \rm{M_{\odot}}$); (11) projected distance to the companion galaxy (kpc). For galaxies with two nearby companions, we provide two $R_{\rm{p}}$ values separated by '/'; (12) relative line-of-sight velocity to the companion galaxy (\kms)\,. For galaxies with two nearby companions, we provide two $\Delta V$ values separated by '/'; (13) H{\sc i} morphological asymmetry of the galaxy; (14) median values of the Toomre Q parameter of the galaxy.

\end{table}
\end{landscape}
\clearpage


\begin{landscape}

\begin{table}
\scriptsize
\caption{{H{\sc i} properties and distances of the control galaxies in the ASKAP Norma cluster field. The full table is provided with supplementary material.}}
\label{long_norma_control}

\resizebox{1.3\textwidth}{!}{
\begin{tabular}{cccccccccccc}
 \hline
 Name & R.A. (J2000) & Dec. (J2000) & $D$ & $V_{\rm{sys}}$ & $R_{\rm{HI}}$ & $S_{\rm{HI}}$ & $M_{\rm{total}}^{\rm{HI}}$ &  $M_{\rm{narrow}}^{\rm{HI}}$ & $M_{\rm{broad}}^{\rm{HI}}$ & $A_{\rm{map}}^{\rm{HI}}$ & $Q$\\
 &   &  & (Mpc) & (\kms)\ & (kpc) & (Jy km s$^{-1}$) & ($10^8 \ \rm{M_{\odot}}$) & ($10^8 \ \rm{M_{\odot}}$) & ($10^8 \ \rm{M_{\odot}}$) \\
    (1) & (2) & (3) & (4) & (5) & (6) & (7) & (8) & (9) & (10) & (11) & (12) \\
 \hline
 
J163435-620248  & 16$^{\rm{h}}$34$^{\rm{m}}$34$^{\rm{s}}_.$7 & -62$^{\rm{d}}$02$^{\rm{m}}$45$^{\rm{s}}_.$3 &  63.6 & 4,450 & 28.4 &   5.2 &  49.4 & 10.8 &  38.5 & 0.14 &  3.14\\ 
J163452-603705  & 16$^{\rm{h}}$34$^{\rm{m}}$52$^{\rm{s}}_.$8 &  -60$^{\rm{d}}$37$^{\rm{m}}$07$^{\rm{s}}_.$8 &  46.7 & 3,266 & 32.5 &  15.1 &  77.5 & 18.3 &  59.2 & 0.15 &  0.03\\ 
J163518-581311  & 16$^{\rm{h}}$35$^{\rm{m}}$19$^{\rm{s}}_.$1 &  -58$^{\rm{d}}$13$^{\rm{m}}$09$^{\rm{s}}_.$0 &  22.3 & 1,562 & 13.4 &  10.6 &  12.4 &  1.3 &  11.2 & 0.24 &  4.55\\ 
J163600-611510  &  16$^{\rm{h}}$36$^{\rm{m}}$00$^{\rm{s}}_.$4 & -61$^{\rm{d}}$15$^{\rm{m}}$11$^{\rm{s}}_.$2 &  75.8 & 5,307 & 16.3 &   1.3 &  17.4 & 11.9 &   5.5 & 0.32 &   ...\\ 

 \hline
 
\end{tabular}}
\flushleft
 \textbf{Notes.} The columns are (1) source name; (2) kinematic centre in R.A. units from {\sc 2dbat}; (3) kinematic centre Dec. units from {\sc 2dbat}; (4) distance to the galaxy from redshift (Mpc); (5) systemic velocity from redshift (\kms)\,; (6) radius of H{\sc i} disk from the {\sc baygaud} analysis, $R_{\rm{HI}}$ (kpc);  (7) integrated H{\sc i} intensities derived from super-profiles in Section~\ref{sec:3} (Jy km s$^{-1}$); (8) H{\sc i} mass derived from super-profiles in Section~\ref{sec:3} ($10^8 \ \rm{M_{\odot}}$); (9) narrow H{\sc i} mass derived from super-profiles in Section~\ref{sec:3} ($10^8 \ \rm{M_{\odot}}$); (10) broad H{\sc i} mass derived from super-profiles in Section~\ref{sec:3} ($10^8 \ \rm{M_{\odot}}$); (11) H{\sc i} morphological asymmetry of the galaxy; (12) median values of the Toomre Q parameter values of the galaxy ('*': {\sc moment1} velocity field is used for the rotation curve analysis). '†' indicates the galaxy that have similar velocity widths of the narrow and broad components in the H{\sc i} super-profile analysis.

\end{table}

\begin{table}
\scriptsize
\caption{H{\sc i} properties and distances of the paired galaxies in the ASKAP Norma cluster field. The full table is provided with supplementary material.}
\label{long_norma_pair}

\resizebox{1.3\textwidth}{!}
{
\begin{tabular}{cccccccccccccc}
 \hline
 Name & R.A. (J2000) & Dec. (J2000) & $D$ & $V_{\rm{sys}}$ & $R_{\rm{HI}}$ & $S_{\rm{HI}}$ & $M_{\rm{total}}^{\rm{HI}}$ &  $M_{\rm{narrow}}^{\rm{HI}}$ & $M_{\rm{broad}}^{\rm{HI}}$ & $R_{\rm{p}}$ & $\Delta V$ & $A_{\rm{map}}^{\rm{HI}}$ & $Q$\\
 &   &  & (Mpc) & (\kms)\ & (kpc) & (Jy km s$^{-1}$) & ($10^8 \ \rm{M_{\odot}}$) & ($10^8 \ \rm{M_{\odot}}$) & ($10^8 \ \rm{M_{\odot}}$) & (kpc) & (\kms)\ \\
    (1) & (2) & (3) & (4) & (5) & (6) & (7) & (8) & (9) & (10) & (11) & (12) & (13) & (14)\\
 \hline
 
J163754-564907  & 16$^{\rm{h}}$37$^{\rm{m}}$55$^{\rm{s}}_.$7 & -56$^{\rm{d}}$48$^{\rm{m}}$59$^{\rm{s}}_.$2 &  21.1 & 1,481 &  4.8 &   1.0 &   1.0 &  0.5 &   0.5 &  76 &  14 &  ... &   ...\\ 
J163834-601517  & 16$^{\rm{h}}$38$^{\rm{m}}$34$^{\rm{s}}_.$2 & -60$^{\rm{d}}$15$^{\rm{m}}$14$^{\rm{s}}_.$2 &  49.2 & 3,446 &  9.7 &   1.5 &   8.4 &  2.4 &   6.0 &  91 &  91 & 0.23 &   ...\\ 
J163924-565221  & 16$^{\rm{h}}$39$^{\rm{m}}$24$^{\rm{s}}_.$7 & -56$^{\rm{d}}$52$^{\rm{m}}$18$^{\rm{s}}_.$4 &  20.9 & 1,467 & 15.0 &  17.1 &  17.7 &  5.4 &  12.3 &  76 &  14 & 0.17 &  1.15\\ 
J164249-610527* & 16$^{\rm{h}}$42$^{\rm{m}}$49$^{\rm{s}}_.$7 & -61$^{\rm{d}}$05$^{\rm{m}}$24$^{\rm{s}}_.$9 &  63.8 & 4,463 & 22.3 &   4.1 &  39.0 &  3.7 &  35.3 & ... & ... & 0.13 &   ...\\ 

\hline

\end{tabular}}

\flushleft
\textbf{Notes.} The columns are (1) source name; (2) kinematic centre in R.A. units from {\sc 2dbat}; (3) kinematic centre Dec. units from {\sc 2dbat}; (4) distance to the galaxy from redshift (Mpc); (5) systemic velocity from redshift (\kms)\,; (6) radius of H{\sc i} disk from the {\sc baygaud} analysis, $R_{\rm{HI}}$ (kpc);  (7) integrated H{\sc i} intensities derived from super-profiles in Section~\ref{sec:3} (Jy km s$^{-1}$); (8) H{\sc i} mass derived from super-profiles in Section~\ref{sec:3} ($10^8 \ \rm{M_{\odot}}$); (9) narrow H{\sc i} mass derived from super-profiles in Section~\ref{sec:3} ($10^8 \ \rm{M_{\odot}}$); (10) broad H{\sc i} mass derived from super-profiles in Section~\ref{sec:3} ($10^8 \ \rm{M_{\odot}}$); (11) projected distance to the companion galaxy (kpc); (12) relative line-of-sight velocity to the companion galaxy (\kms)\,; (13) H{\sc i} morphological asymmetry of the galaxy; (14) median values of the Toomre Q parameter of the galaxy. '*' indicates the visually identified galaxy pair.

\end{table}
\end{landscape}



\begin{figure*}
    \centering
    \includegraphics[width=8.5cm]{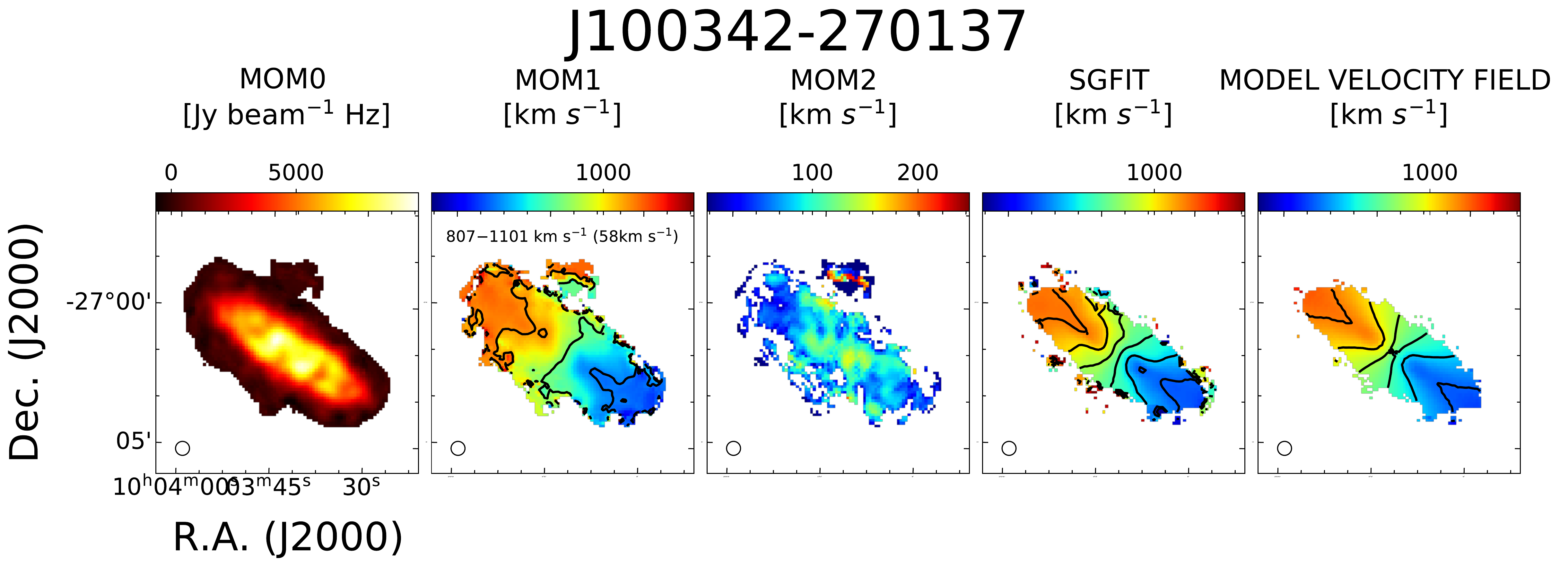}
    \vspace{0.5cm}
    \hspace{0.5cm}
    \includegraphics[width=8.5cm]{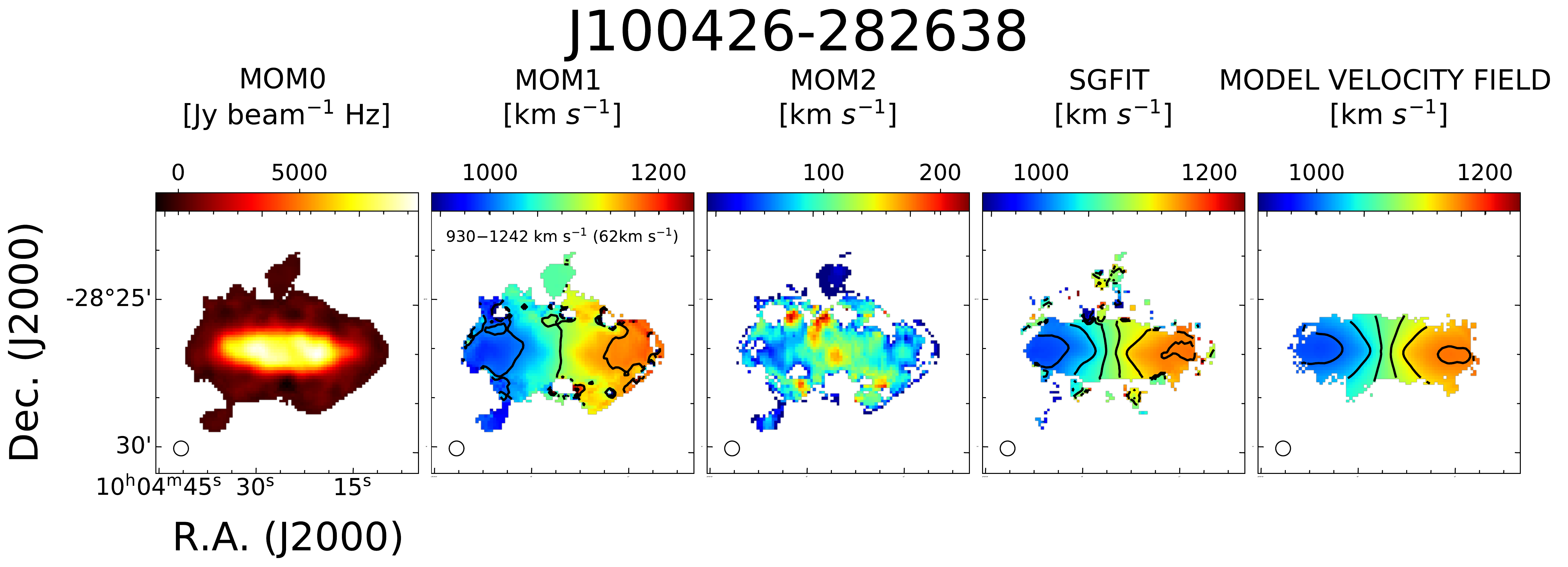}
    \caption{H{\sc i} images of the galaxy pairs (ordered by the WALLABY ID) in the ASKAP Hydra I cluster field: ‘MOM0’ (SoFiA2 integrated intensity map, {\sc moment0}), ‘MOM1’ (SoFiA2 velocity field map , {\sc moment1}), ‘MOM2’ (SoFiA2, velocity dispersion map, {\sc moment2}), SGFIT ({\sc baygaud} Single Gaussian fitting (SGfit) velocity field map), and ‘MODEL VELOCITY FIELD’ ({\sc 2dbat} model velocity field map). The contour levels for each velocity field are denoted in the second panel of each figure. The ASKAP beam is shown as an ellipse on the bottom-left corner of each panel. The full figure is provided with supplementary material.}\label{figA1}
   
\end{figure*}

\begin{figure*}
    \includegraphics[width=8.5cm]{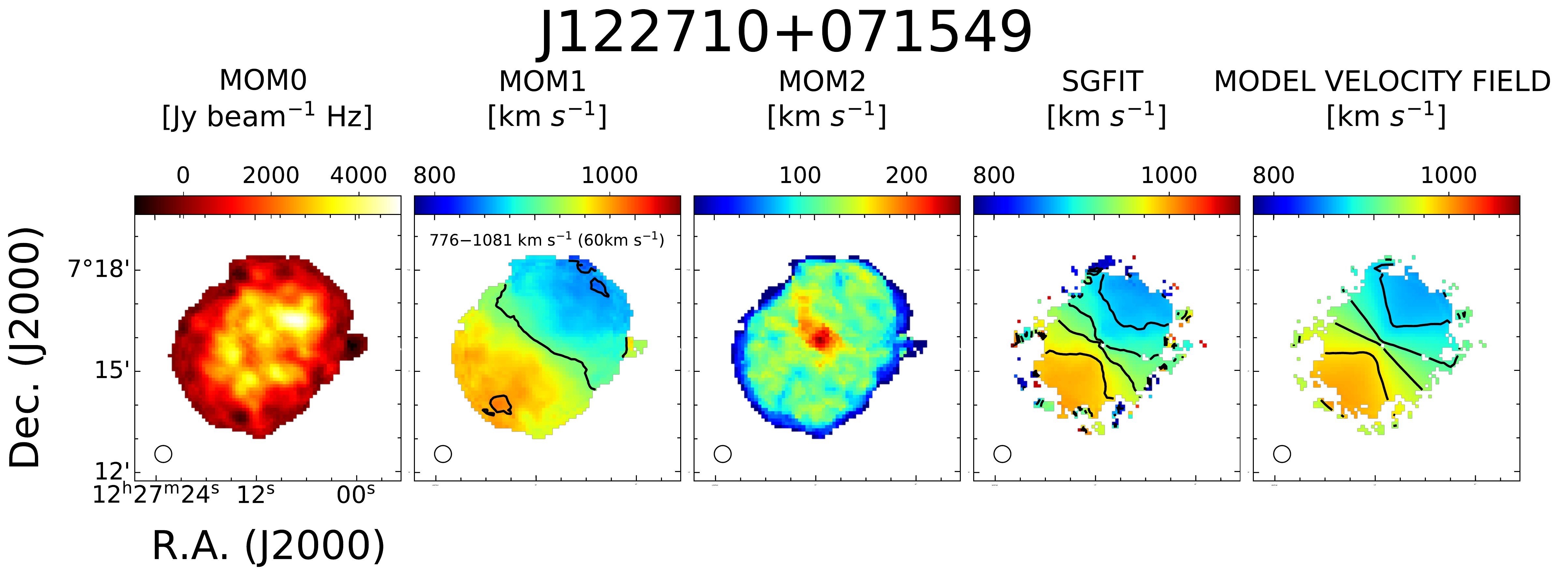}
    \vspace{0.5cm}
    \hspace{0.5cm}
    \includegraphics[width=8.5cm]{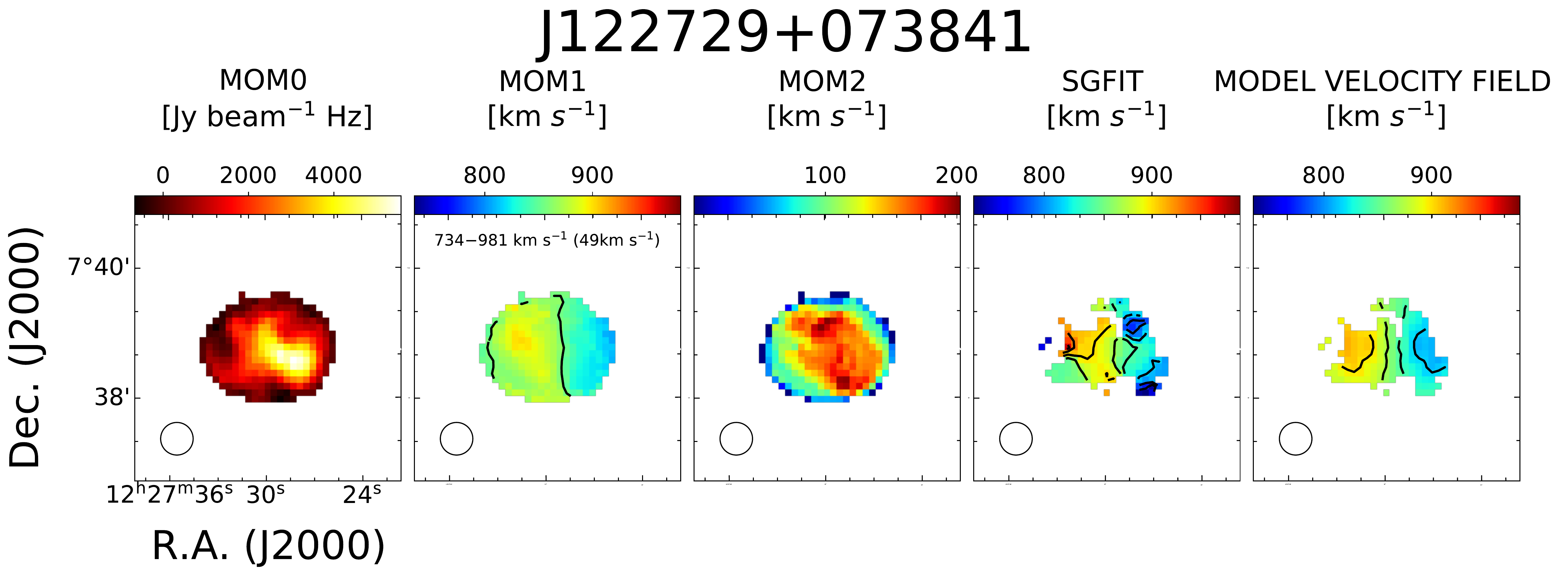}
    \caption{H{\sc i} images of the galaxy pairs (ordered by the WALLABY ID) in the ASKAP NGC 4636 field: ‘MOM0’ (SoFiA2 integrated intensity map, {\sc moment0}), ‘MOM1’ (SoFiA2 velocity field map , {\sc moment1}), ‘MOM2’ (SoFiA2, velocity dispersion map, {\sc moment2}), SGFIT ({\sc baygaud} Single Gaussian fitting (SGfit) velocity field map), and ‘MODEL VELOCITY FIELD’ ({\sc 2dbat} model velocity field map). The contour levels for each velocity field are denoted in the second panel of each figure. The ASKAP beam is shown as an ellipse on the bottom-left corner of each panel. The full figure is provided with supplementary material.}
    \label{fig:my_label}\label{figA2}
\end{figure*}

\begin{figure*}
    \includegraphics[width=8.5cm]{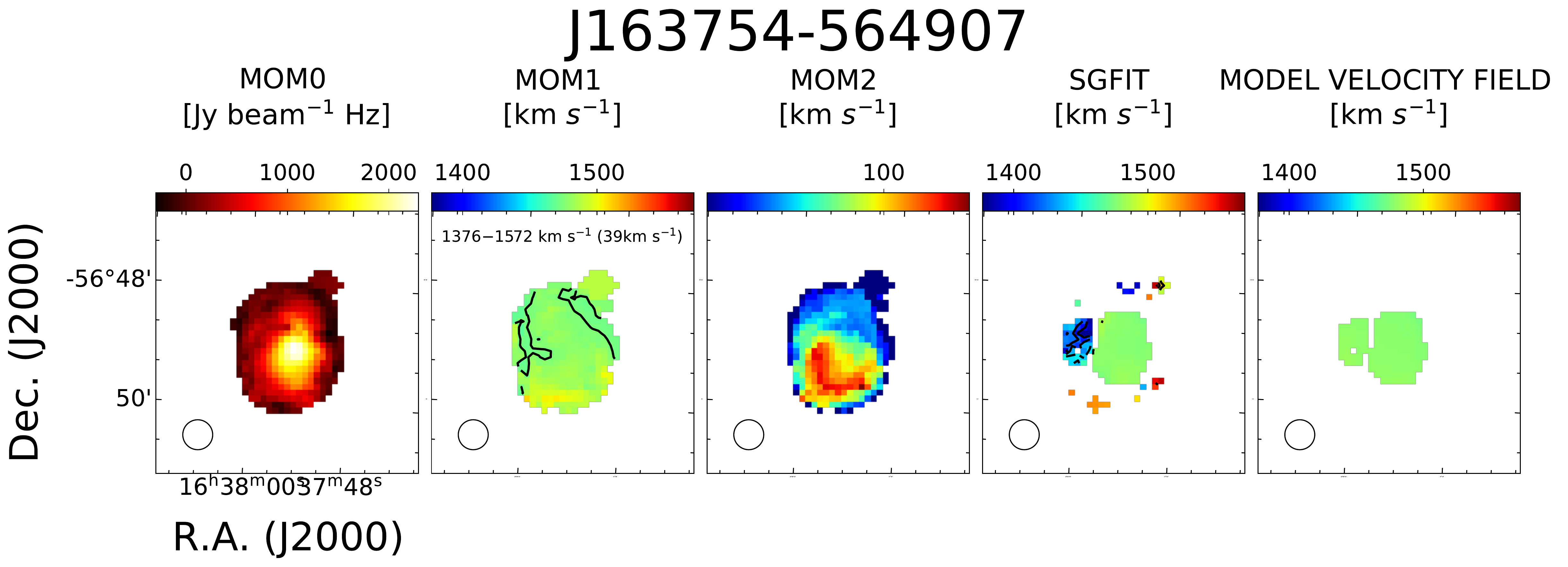}
    \vspace{0.5cm}
    \hspace{0.5cm}
    \includegraphics[width=8.5cm]{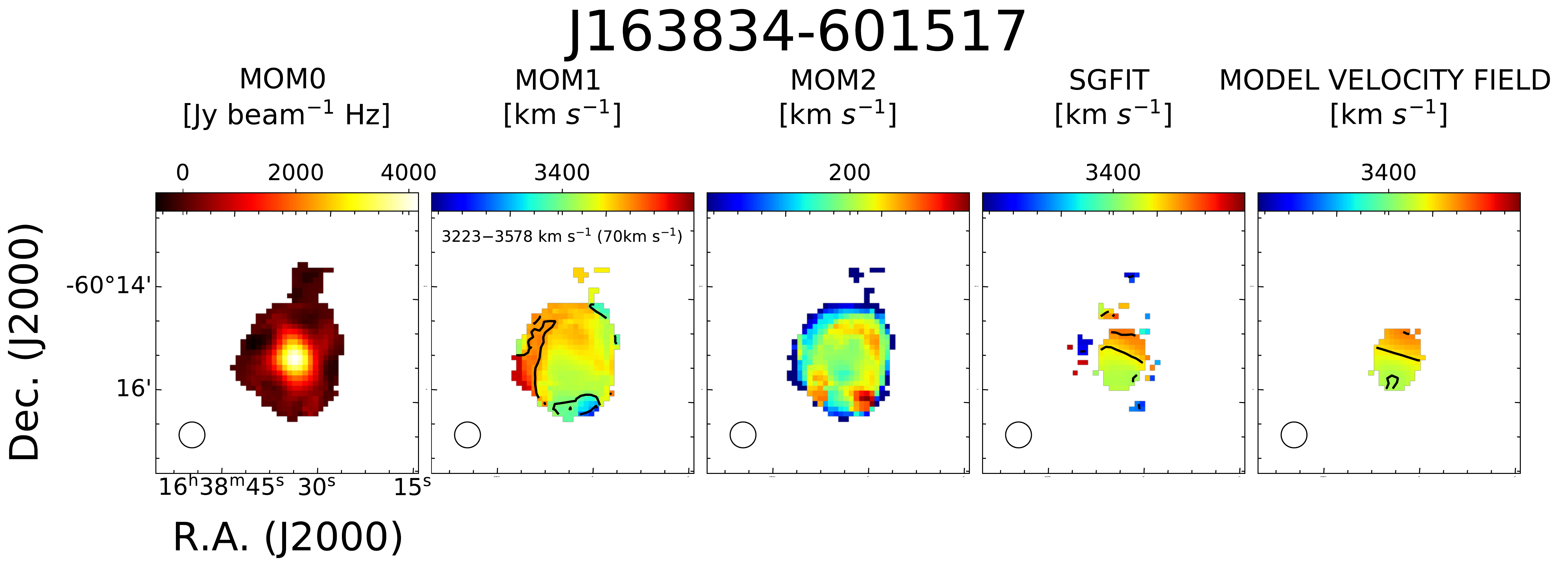}
    \caption{H{\sc i} images of the galaxy pairs (ordered by the WALLABY ID) in the ASKAP Norma cluster field: ‘MOM0’ (SoFiA2 integrated intensity map, {\sc moment0}), ‘MOM1’ (SoFiA2 velocity field map , {\sc moment1}), ‘MOM2’ (SoFiA2, velocity dispersion map, {\sc moment2}), SGFIT ({\sc baygaud} Single Gaussian fitting (SGfit) velocity field map), and ‘MODEL VELOCITY FIELD’ ({\sc 2dbat} model velocity field map). The contour levels for each velocity field are denoted in the second panel of each figure. The ASKAP beam is shown as an ellipse on the bottom-left corner of each panel. The full figure is provided with supplementary material.}\label{figA3}
    \end{figure*}


\begin{figure*}
    \centering
    \includegraphics[width=0.45\textwidth]{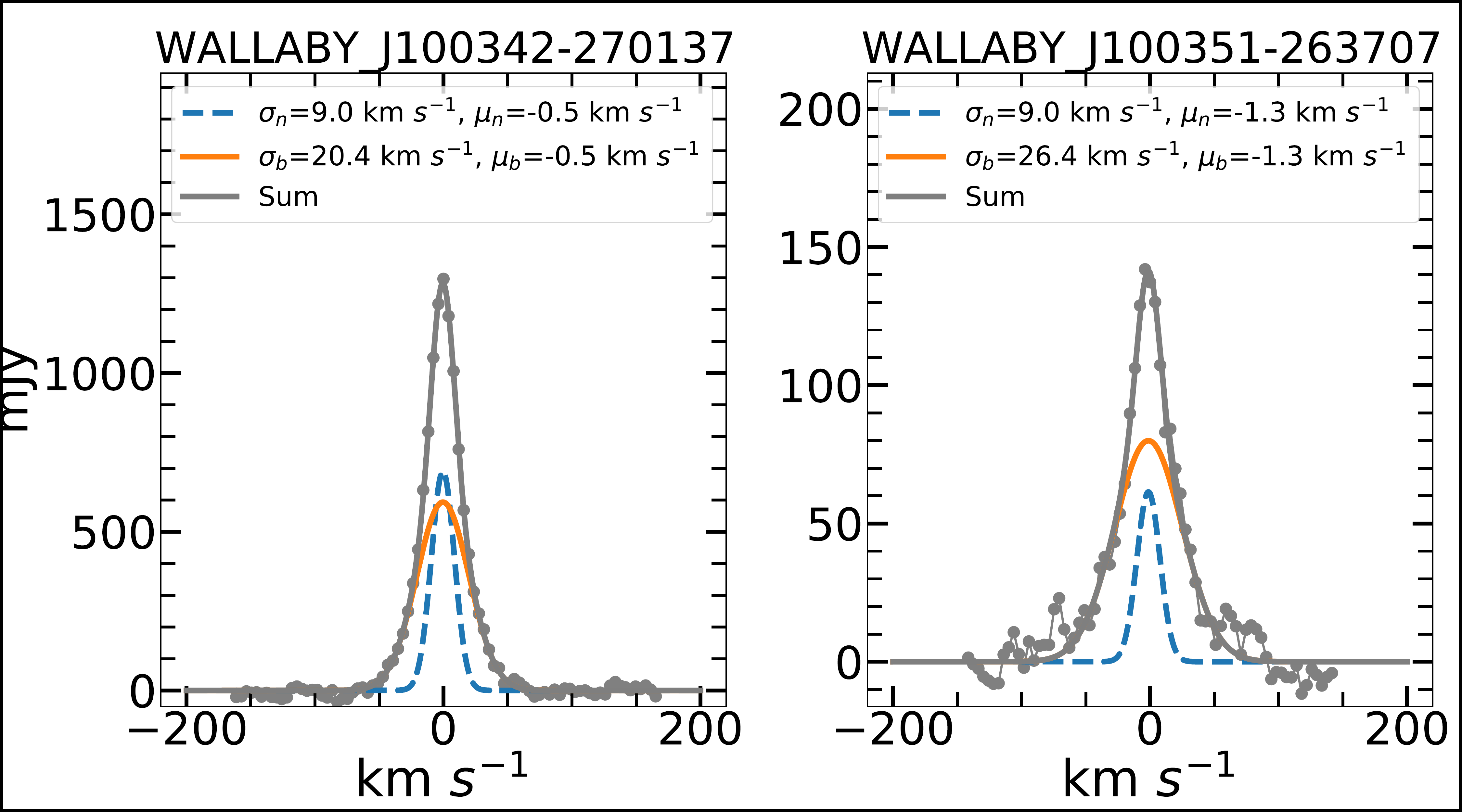}
    \hspace{0.3cm}
    \vspace{0.5cm}
    \includegraphics[width=0.45\textwidth]{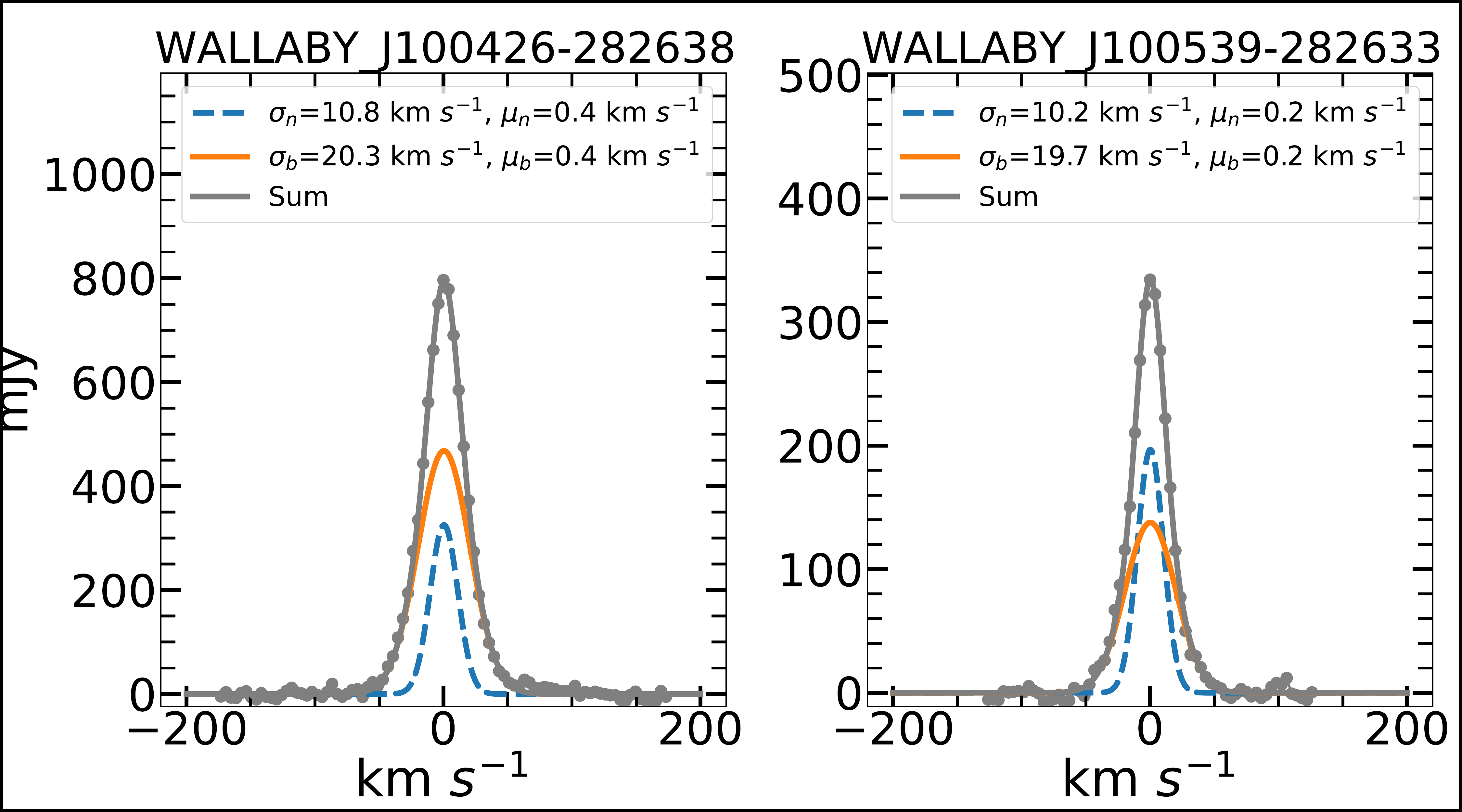}
    \caption{{\sc baygaud} based H{\sc i} super-profiles of the galaxy pairs (ordered by the WALLABY ID) in the ASKAP Hydra I cluster field. The grey circles show the stacked fluxes at the corresponding channels. The blue dashed and orange solid lines represent the narrow and broad components decomposed from the double Gaussian fitting to the super-profiles. Their velocity dispersions ($\sigma$) and centroid velocities ($\mu$) are labeled on the top-right corner of each panel. H{\sc i} super-profiles of galaxies which are poorly resolved are blanked. The full figure is provided with supplementary material.}
    \label{figA4}
\end{figure*}

\begin{figure*}  
    \includegraphics[width=0.45\textwidth]{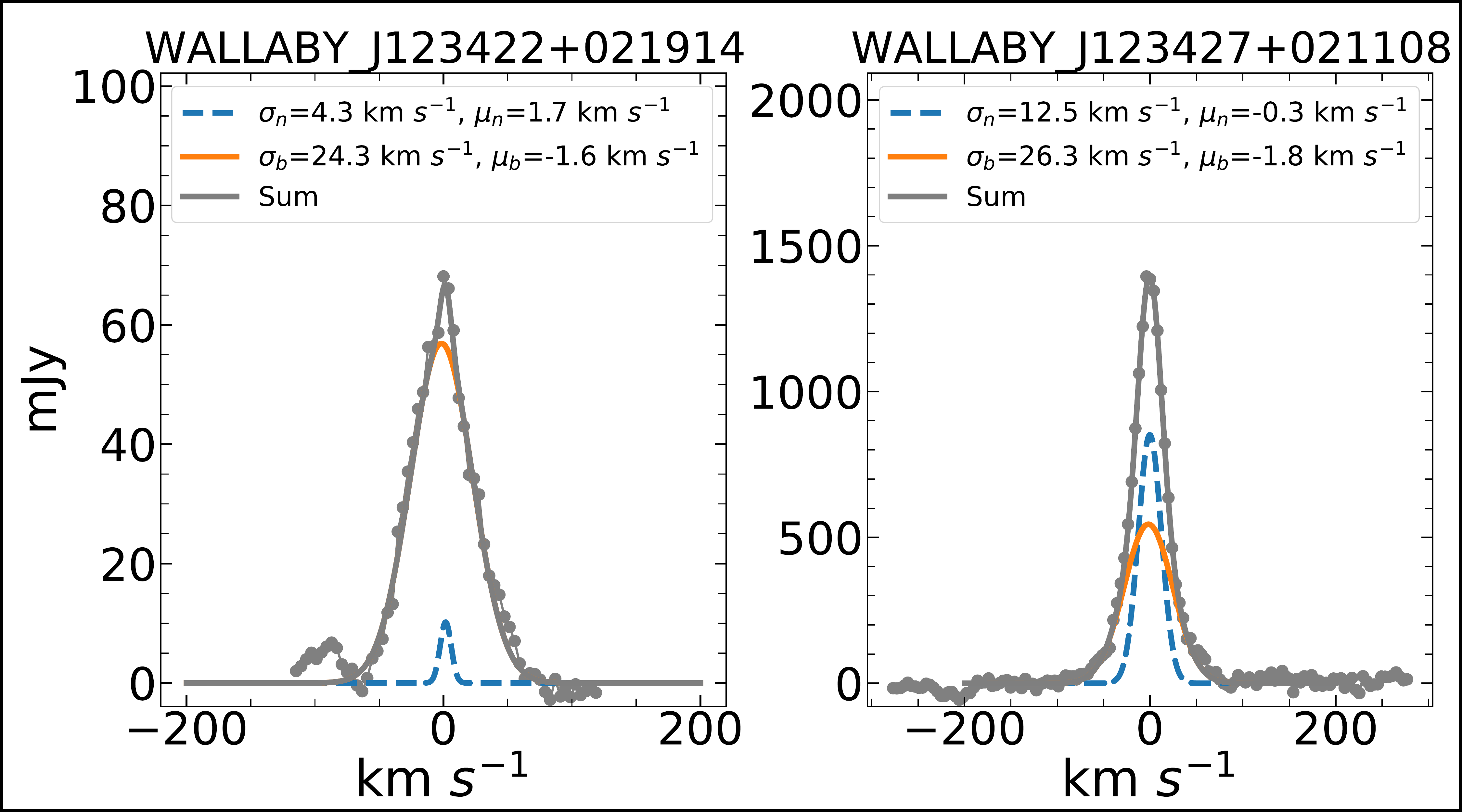}
    \hspace{0.3cm}
    \vspace{0.5cm}
    \includegraphics[width=0.45\textwidth]{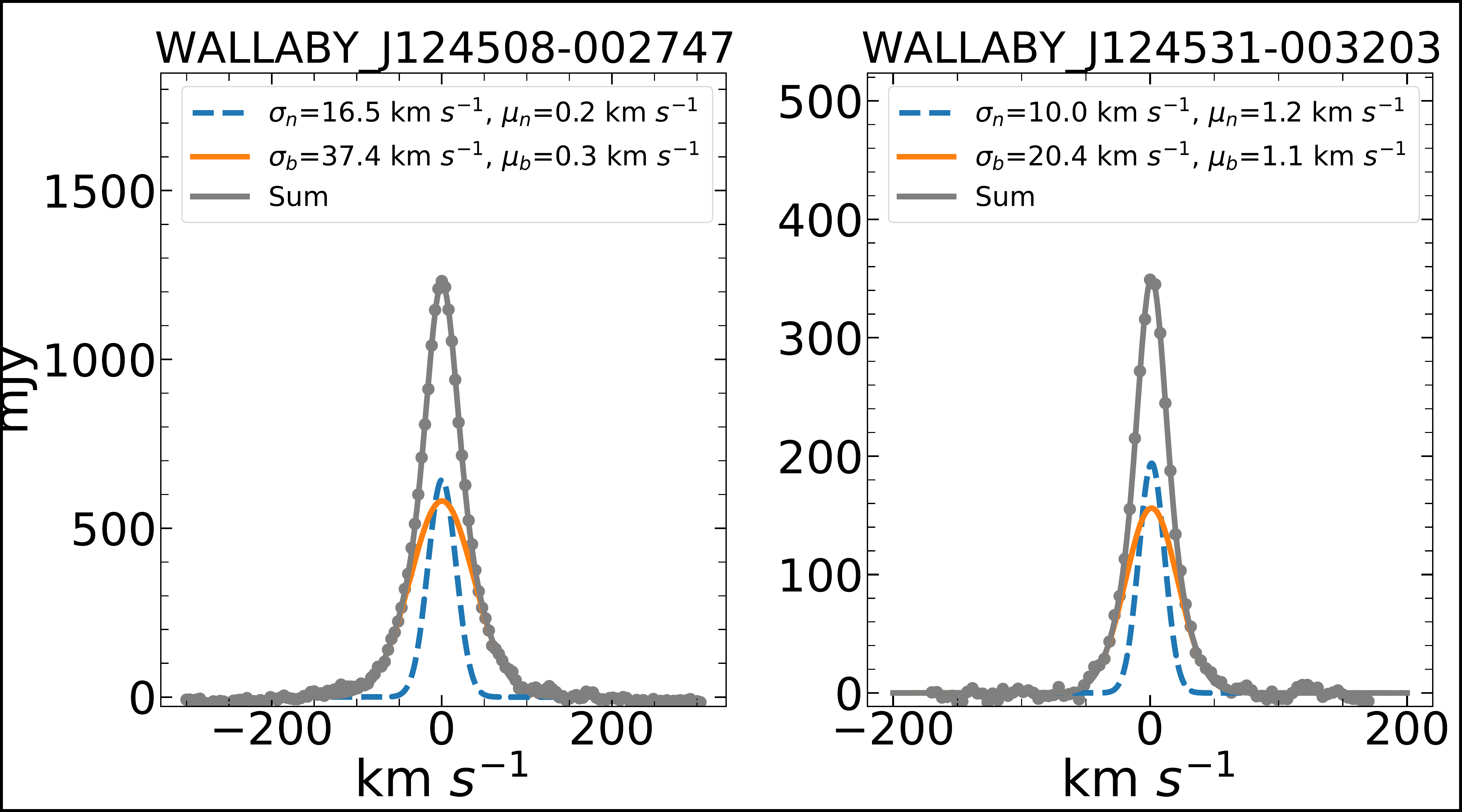}
   
    \caption{{\sc baygaud} based H{\sc i} super-profiles of the galaxy pairs (ordered by the WALLABY ID) in the ASKAP NGC 4636 group field. The grey circles show the stacked fluxes at the corresponding channels. The blue dashed and orange solid lines represent the narrow and broad components decomposed from the double Gaussian fitting to the super-profiles. Their velocity dispersions ($\sigma$) and centroid velocities ($\mu$) are labeled on the top-right corner of each panel. H{\sc i} super-profiles of galaxies which are poorly resolved are blanked. The full figure is provided with supplementary material.}
    \label{figA5}
    \end{figure*}
 
\begin{figure*}
    \includegraphics[width=0.45\textwidth]{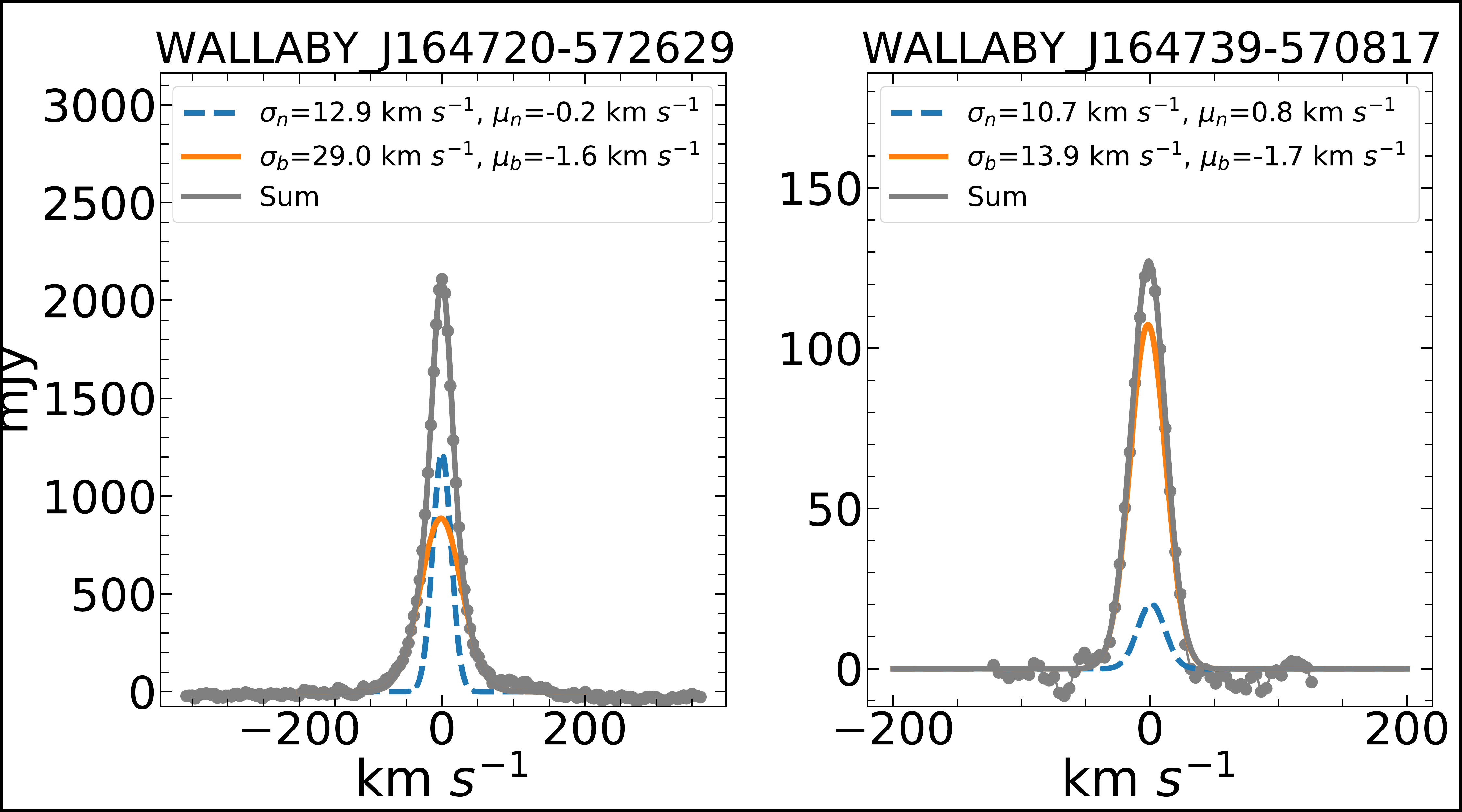}
    \hspace{0.3cm}
    \vspace{0.5cm}
    \includegraphics[width=0.45\textwidth]{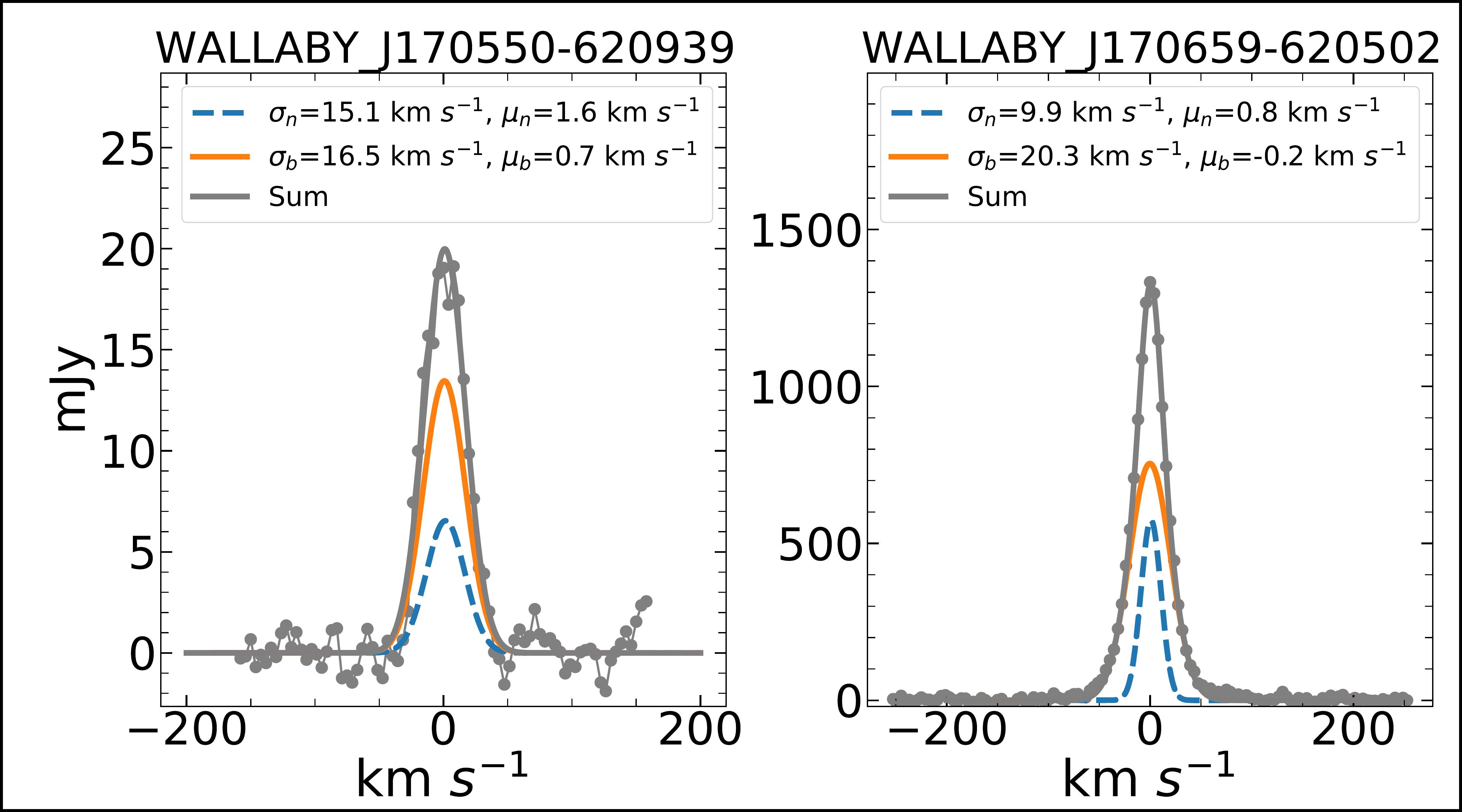}
    \caption{{\sc baygaud} based H{\sc i} super-profiles of the galaxy pairs (ordered by the WALLABY ID) in the ASKAP Norma cluster field. The grey circles show the stacked fluxes at the corresponding channels. The blue dashed and orange solid lines represent the narrow and broad components decomposed from the double Gaussian fitting to the super-profiles. Their velocity dispersions ($\sigma$) and centroid velocities ($\mu$) are labeled on the top-right corner of each panel. H{\sc i} super-profiles of galaxies which are poorly resolved are blanked. The full figure is provided with supplementary material.}
    \label{figA6}
\end{figure*}

\label{lastpage}

\renewcommand{\thetable}{A\arabic{table}} 
\renewcommand{\thefigure}{A\arabic{figure}}
\setcounter{figure}{0}
\setcounter{table}{0}


\maketitle

\begin{titlepage}
   \begin{center}
       \vspace*{1cm}

       \Large
       \textbf{Supplementary materials: H{\sc i} properties, images and super-profiles of the sample galaxies}

       \vspace{0.5cm}
        \vspace{0.8cm}
                        
   \end{center}
\end{titlepage}

\title{Supplementary materials}









\begin{table*}
\centering
\tiny
\caption{H{\sc i} properties and distances of the control galaxies in the ASKAP Hydra I cluster field.}
\label{long_hydra_control}

\begin{tabular}{cccccccccccc}
 \hline
 Name & R.A. (J2000) & Dec. (J2000) & $D$ & $V_{\rm{sys}}$ & $R_{\rm{HI}}$ & $S_{\rm{HI}}$ & $M_{\rm{total}}^{\rm{HI}}$ &  $M_{\rm{narrow}}^{\rm{HI}}$ & $M_{\rm{broad}}^{\rm{HI}}$ & $A_{\rm{map}}^{\rm{HI}}$ & $Q$\\
 &   &  & (Mpc) & (\kms)\ & (kpc) & (Jy km s$^{-1}$) & ($10^8 \ \rm{M_{\odot}}$) & ($10^8 \ \rm{M_{\odot}}$) & ($10^8 \ \rm{M_{\odot}}$) \\
    (1) & (2) & (3) & (4) & (5) & (6) & (7) & (8) & (9) & (10) & (11) & (12) \\
 \hline
 
J100351-273417  & 10$^{\rm{h}}$03$^{\rm{m}}$52$^{\rm{s}}_.$0 & -27$^{\rm{d}}$34$^{\rm{m}}$15$^{\rm{s}}_.$3 & 39.9 & 2,792 & 30.5 &  12.6 &  47.2 & 12.4 &  34.8 & 0.23 &  4.22\\ 
J100634-295615  & 10$^{\rm{h}}$06$^{\rm{m}}$34$^{\rm{s}}_.$2 & -29$^{\rm{d}}$56$^{\rm{m}}$12$^{\rm{s}}_.$2 & 15.9 & 1,117 &  9.6 &   9.2 &   5.5 &  1.3 &   4.2 & 0.23 & 10.21\\ 
J100656-251731†  & 10$^{\rm{h}}$06$^{\rm{m}}$55$^{\rm{s}}_.$9 & -25$^{\rm{d}}$17$^{\rm{m}}$32$^{\rm{s}}_.$5 & 41.3 & 2,894 & 10.5 &   1.6 &   6.5 &  3.0 &   3.5 & 0.27 &   ...\\ 
J100713-262336  & 10$^{\rm{h}}$07$^{\rm{m}}$13$^{\rm{s}}_.$8 & -26$^{\rm{d}}$23$^{\rm{m}}$31$^{\rm{s}}_.$4 & 65.7 & 4,599 & 17.1 &   2.0 &  20.5 & 16.8 &   3.7 & 0.21 &   ...\\ 
J100746-281451  & 10$^{\rm{h}}$07$^{\rm{m}}$45$^{\rm{s}}_.$8 & -28$^{\rm{d}}$14$^{\rm{m}}$40$^{\rm{s}}_.$9 & 67.1 & 4,699 &  9.0 &   0.3 &   3.6 &  0.3 &   3.3 & 0.46 &   ...\\ 
J100808-260942†  &  10$^{\rm{h}}$08$^{\rm{m}}$08$^{\rm{s}}_.$7 & -26$^{\rm{d}}$09$^{\rm{m}}$38$^{\rm{s}}_.$5 & 44.0 & 3,078 & 10.1 &   1.8 &   8.4 &  3.6 &   4.8 & 0.24 &   ...\\ 
J100827-270707  & 10$^{\rm{h}}$08$^{\rm{m}}$27$^{\rm{s}}_.$3 & -27$^{\rm{d}}$07$^{\rm{m}}$11$^{\rm{s}}_.$9 & 39.1 & 2,735 &  6.8 &   0.9 &   3.2 &  0.8 &   2.5 & 0.37 &   ...\\ 
J100830-262140  & 10$^{\rm{h}}$08$^{\rm{m}}$29$^{\rm{s}}_.$9 & -26$^{\rm{d}}$21$^{\rm{m}}$38$^{\rm{s}}_.$7 & 59.9 & 4,196 &  9.4 &   0.8 &   6.9 &  0.3 &   6.6 & 0.18 &   ...\\ 
J100903-290239  &  10$^{\rm{h}}$09$^{\rm{m}}$07$^{\rm{s}}_.$0 & -29$^{\rm{d}}$03$^{\rm{m}}$47$^{\rm{s}}_.$5 & 15.9 & 1,112 & 27.7 & 118.7 &  70.6 & 31.9 &  38.7 &  ... &  1.84\\ 
J100938-255850  & 10$^{\rm{h}}$09$^{\rm{m}}$38$^{\rm{s}}_.$4 & -25$^{\rm{d}}$58$^{\rm{m}}$52$^{\rm{s}}_.$4 & 44.0 & 3,084 & 15.5 &   3.9 &  17.7 &  0.9 &  16.8 & 0.17 &   ...\\ 
J101018-285748  & 10$^{\rm{h}}$10$^{\rm{m}}$18$^{\rm{s}}_.$0 & -28$^{\rm{d}}$57$^{\rm{m}}$53$^{\rm{s}}_.$7 & 59.2 & 4,147 & 11.1 &   0.9 &   7.6 &  3.0 &   4.6 & 0.40 &   ...\\ 
J101025-275214  & 10$^{\rm{h}}$10$^{\rm{m}}$25$^{\rm{s}}_.$5 & -27$^{\rm{d}}$52$^{\rm{m}}$11$^{\rm{s}}_.$1 & 34.6 & 2,422 &  7.2 &   1.4 &   3.8 &  1.3 &   2.5 & 0.15 &   ...\\ 
J101035-254920  & 10$^{\rm{h}}$10$^{\rm{m}}$35$^{\rm{s}}_.$0 & -25$^{\rm{d}}$49$^{\rm{m}}$27$^{\rm{s}}_.$3 & 35.9 & 2,513 & 15.3 &   7.1 &  21.5 &  4.8 &  16.7 &  ... &  0.63\\ 
J101247-275028  & 10$^{\rm{h}}$12$^{\rm{m}}$47$^{\rm{s}}_.$7 & -27$^{\rm{d}}$50$^{\rm{m}}$18$^{\rm{s}}_.$9 & 37.5 & 2,624 & 24.0 &  13.7 &  45.3 &  8.7 &  36.6 &  ... &  7.36\\ 
J101314-271308  & 10$^{\rm{h}}$13$^{\rm{m}}$14$^{\rm{s}}_.$5 & -27$^{\rm{d}}$12$^{\rm{m}}$57$^{\rm{s}}_.$3 & 65.0 & 4,550 & 12.6 &   1.1 &  10.5 &  1.3 &   9.2 & 0.44 &   ...\\ 
J101359-253824†  & 10$^{\rm{h}}$13$^{\rm{m}}$59$^{\rm{s}}_.$3 & -25$^{\rm{d}}$38$^{\rm{m}}$21$^{\rm{s}}_.$5 & 45.5 & 3,188 & 18.3 &   5.3 &  26.1 &  8.6 &  17.4 & 0.12 &  4.56\\ 
J101438-272431  & 10$^{\rm{h}}$14$^{\rm{m}}$39$^{\rm{s}}_.$0 & -27$^{\rm{d}}$24$^{\rm{m}}$36$^{\rm{s}}_.$1 & 59.3 & 4,148 & 16.5 &   2.1 &  17.4 &  5.4 &  12.1 & 0.29 &   ...\\ 
J101441-285221  & 10$^{\rm{h}}$14$^{\rm{m}}$41$^{\rm{s}}_.$9 & -28$^{\rm{d}}$52$^{\rm{m}}$19$^{\rm{s}}_.$0 & 15.8 & 1,108 &  9.1 &  11.0 &   6.5 &  0.3 &   6.2 & 0.12 & 16.13\\ 
J101448-285723  & 10$^{\rm{h}}$14$^{\rm{m}}$48$^{\rm{s}}_.$8 & -28$^{\rm{d}}$57$^{\rm{m}}$26$^{\rm{s}}_.$4 & 59.6 & 4,172 & 15.6 &   1.7 &  13.9 &  2.6 &  11.3 & 0.35 &   ...\\ 
J101537-272427  & 10$^{\rm{h}}$15$^{\rm{m}}$37$^{\rm{s}}_.$2 & -27$^{\rm{d}}$24$^{\rm{m}}$24$^{\rm{s}}_.$5 & 44.7 & 3,130 &  5.7 &   0.7 &   3.1 &  1.7 &   1.5 & 0.33 &   ...\\ 
J101606-274758  &  10$^{\rm{h}}$16$^{\rm{m}}$06$^{\rm{s}}_.$2 &  -27$^{\rm{d}}$48$^{\rm{m}}$01$^{\rm{s}}_.$1 & 35.4 & 2,477 &  6.5 &   0.8 &   2.3 &  1.3 &   1.0 & 0.28 &   ...\\ 
J101632-291258  & 10$^{\rm{h}}$16$^{\rm{m}}$32$^{\rm{s}}_.$8 & -29$^{\rm{d}}$12$^{\rm{m}}$56$^{\rm{s}}_.$8 & 60.1 & 4,205 & 12.6 &   1.5 &  13.0 &  8.0 &   5.0 & 0.14 &   ...\\ 
J101927-264159†  & 10$^{\rm{h}}$19$^{\rm{m}}$28$^{\rm{s}}_.$3 & -26$^{\rm{d}}$41$^{\rm{m}}$57$^{\rm{s}}_.$6 & 36.4 & 2,548 &  5.9 &   0.7 &   2.1 &  1.2 &   0.9 & 0.34 &   ...\\ 
J101941-254852  & 10$^{\rm{h}}$19$^{\rm{m}}$41$^{\rm{s}}_.$3 & -25$^{\rm{d}}$48$^{\rm{m}}$51$^{\rm{s}}_.$4 & 43.0 & 3,012 & 18.4 &   5.9 &  25.7 &  4.0 &  21.8 & 0.16 &  2.91\\ 
J102017-253913  & 10$^{\rm{h}}$20$^{\rm{m}}$18$^{\rm{s}}_.$1 & -25$^{\rm{d}}$39$^{\rm{m}}$12$^{\rm{s}}_.$8 & 49.2 & 3,445 & 16.9 &   4.4 &  25.2 & 17.5 &   7.7 & 0.14 &  2.21\\ 
J102023-253050  & 10$^{\rm{h}}$20$^{\rm{m}}$23$^{\rm{s}}_.$7 & -25$^{\rm{d}}$30$^{\rm{m}}$50$^{\rm{s}}_.$5 & 47.0 & 3,293 &  6.7 &   0.7 &   3.6 &  0.6 &   3.0 & 0.26 &   ...\\ 
J102107-281054  &  10$^{\rm{h}}$21$^{\rm{m}}$07$^{\rm{s}}_.$2 & -28$^{\rm{d}}$10$^{\rm{m}}$49$^{\rm{s}}_.$6 & 57.3 & 4,012 & 14.8 &   2.3 &  18.2 &  4.2 &  14.0 & 0.20 &  6.12\\ 
J102113-262325  & 10$^{\rm{h}}$21$^{\rm{m}}$13$^{\rm{s}}_.$5 & -26$^{\rm{d}}$23$^{\rm{m}}$23$^{\rm{s}}_.$7 & 35.4 & 2,475 &  4.1 &   0.6 &   1.7 &  0.6 &   1.2 & 0.14 &   ...\\ 
J102430-290904  & 10$^{\rm{h}}$24$^{\rm{m}}$30$^{\rm{s}}_.$6 & -29$^{\rm{d}}$08$^{\rm{m}}$55$^{\rm{s}}_.$5 & 53.5 & 3,748 &  8.7 &   0.8 &   5.5 &  2.8 &   2.8 & 0.59 &   ...\\ 
J102439-274841  & 10$^{\rm{h}}$24$^{\rm{m}}$39$^{\rm{s}}_.$4 & -27$^{\rm{d}}$48$^{\rm{m}}$42$^{\rm{s}}_.$3 & 51.4 & 3,598 &  8.1 &   1.0 &   6.4 &  2.3 &   4.1 & 0.20 &   ...\\ 
J102621-291150  & 10$^{\rm{h}}$26$^{\rm{m}}$21$^{\rm{s}}_.$9 & -29$^{\rm{d}}$11$^{\rm{m}}$57$^{\rm{s}}_.$3 & 52.8 & 3,699 & 12.8 &   1.7 &  11.1 &  1.0 &  10.0 & 0.46 &   ...\\ 
J102629-285851  & 10$^{\rm{h}}$26$^{\rm{m}}$29$^{\rm{s}}_.$8 & -28$^{\rm{d}}$58$^{\rm{m}}$56$^{\rm{s}}_.$3 & 53.0 & 3,713 &  7.1 &   0.6 &   4.0 &  2.5 &   1.5 & 0.33 &   ...\\ 
J102719-253256  & 10$^{\rm{h}}$27$^{\rm{m}}$19$^{\rm{s}}_.$4 & -25$^{\rm{d}}$32$^{\rm{m}}$52$^{\rm{s}}_.$4 & 49.5 & 3,464 & 10.9 &   1.4 &   7.9 &  5.2 &   2.6 & 0.24 &   ...\\ 
J102738-294752  & 10$^{\rm{h}}$27$^{\rm{m}}$38$^{\rm{s}}_.$9 & -29$^{\rm{d}}$47$^{\rm{m}}$52$^{\rm{s}}_.$6 & 55.0 & 3,846 & 14.6 &   2.9 &  20.8 &  9.8 &  11.0 & 0.13 &   ...\\ 
J102818-255446†  & 10$^{\rm{h}}$28$^{\rm{m}}$17$^{\rm{s}}_.$8 & -25$^{\rm{d}}$54$^{\rm{m}}$39$^{\rm{s}}_.$9 & 54.4 & 3,808 &  9.4 &   0.4 &   3.0 &  1.5 &   1.5 &  ... &   ...\\ 
J102911-302031  & 10$^{\rm{h}}$29$^{\rm{m}}$11$^{\rm{s}}_.$3 & -30$^{\rm{d}}$20$^{\rm{m}}$37$^{\rm{s}}_.$3 & 59.4 & 4,160 & 18.0 &   3.5 &  29.1 & 18.8 &  10.3 & 0.25 &   ...\\ 
J102934-261937†  & 10$^{\rm{h}}$29$^{\rm{m}}$34$^{\rm{s}}_.$6 & -26$^{\rm{d}}$19$^{\rm{m}}$36$^{\rm{s}}_.$1 & 56.2 & 3,935 & 12.1 &   1.0 &   7.1 &  2.2 &   5.0 & 0.20 &   ...\\ 
J103002-284116  &  10$^{\rm{h}}$30$^{\rm{m}}$3$^{\rm{s}}_.$1 & -28$^{\rm{d}}$41$^{\rm{m}}$23$^{\rm{s}}_.$0 & 54.8 & 3,839 &  9.6 &   0.7 &   5.2 &  0.5 &   4.7 &  ... &   ...\\ 
J103004-253630  &  10$^{\rm{h}}$30$^{\rm{m}}$05$^{\rm{s}}_.$3 & -25$^{\rm{d}}$36$^{\rm{m}}$34$^{\rm{s}}_.$9 & 48.5 & 3,396 &  7.0 &   0.6 &   3.1 &  0.9 &   2.2 &  ... &   ...\\ 
J103023-302336  & 10$^{\rm{h}}$30$^{\rm{m}}$23$^{\rm{s}}_.$2 & -30$^{\rm{d}}$23$^{\rm{m}}$35$^{\rm{s}}_.$0 & 58.5 & 4,096 & 25.9 &   6.1 &  49.6 & 38.4 &  11.2 & 0.16 &   ...\\ 
J103139-273049  & 10$^{\rm{h}}$31$^{\rm{m}}$39$^{\rm{s}}_.$4 & -27$^{\rm{d}}$30$^{\rm{m}}$46$^{\rm{s}}_.$2 & 51.0 & 3,572 &  7.4 &   0.8 &   4.9 &  0.4 &   4.5 & 0.25 &   ...\\ 
J103141-300815†  & 10$^{\rm{h}}$31$^{\rm{m}}$40$^{\rm{s}}_.$8 &  -30$^{\rm{d}}$08$^{\rm{m}}$08$^{\rm{s}}_.$2 & 74.4 & 5,205 & 12.6 &   0.7 &   9.8 &  4.3 &   5.5 & 0.31 &   ...\\ 
J103216-282359  & 10$^{\rm{h}}$32$^{\rm{m}}$15$^{\rm{s}}_.$9 & -28$^{\rm{d}}$23$^{\rm{m}}$55$^{\rm{s}}_.$7 & 89.0 & 6,227 & 15.4 &   0.5 &   8.9 &  1.1 &   7.8 & 0.31 &   ...\\ 
J103240-282058  & 10$^{\rm{h}}$32$^{\rm{m}}$40$^{\rm{s}}_.$4 & -28$^{\rm{d}}$20$^{\rm{m}}$50$^{\rm{s}}_.$9 & 56.1 & 3,926 & 12.8 &   1.4 &  10.5 &  0.9 &   9.5 & 0.25 &   ...\\ 
J103244-283639  & 10$^{\rm{h}}$32$^{\rm{m}}$44$^{\rm{s}}_.$4 & -28$^{\rm{d}}$36$^{\rm{m}}$35$^{\rm{s}}_.$1 & 51.7 & 3,621 & 20.0 &   3.4 &  21.2 &  6.4 &  14.8 & 0.24 & 36.25\\ 
J103250-301601  & 10$^{\rm{h}}$32$^{\rm{m}}$50$^{\rm{s}}_.$4 & -30$^{\rm{d}}$15$^{\rm{m}}$57$^{\rm{s}}_.$5 & 49.6 & 3,473 & 17.9 &   3.6 &  20.7 & 14.1 &   6.6 & 0.23 &   ...\\ 
J103258-274013  & 10$^{\rm{h}}$32$^{\rm{m}}$58$^{\rm{s}}_.$6 & -27$^{\rm{d}}$40$^{\rm{m}}$13$^{\rm{s}}_.$1 & 44.8 & 3,134 & 10.6 &   1.8 &   8.7 &  2.7 &   6.0 & 0.31 &   ...\\ 
J103259-273237  & 10$^{\rm{h}}$32$^{\rm{m}}$59$^{\rm{s}}_.$0 & -27$^{\rm{d}}$32$^{\rm{m}}$54$^{\rm{s}}_.$1 & 68.2 & 4,772 & 11.3 &   0.5 &   5.8 &  0.6 &   5.2 &  ... &   ...\\ 
J103335-272717  & 10$^{\rm{h}}$33$^{\rm{m}}$35$^{\rm{s}}_.$9 & -27$^{\rm{d}}$27$^{\rm{m}}$16$^{\rm{s}}_.$2 & 48.3 & 3,383 & 19.2 &   3.4 &  18.9 &  8.1 &  10.9 & 0.26 & 32.68\\ 
J103353-274945  & 10$^{\rm{h}}$33$^{\rm{m}}$53$^{\rm{s}}_.$9 & -27$^{\rm{d}}$49$^{\rm{m}}$48$^{\rm{s}}_.$1 & 39.4 & 2,756 &  9.5 &   1.6 &   5.9 &  2.5 &   3.4 & 0.10 &   ...\\ 
J103359-301003  & 10$^{\rm{h}}$33$^{\rm{m}}$58$^{\rm{s}}_.$6 & -30$^{\rm{d}}$09$^{\rm{m}}$59$^{\rm{s}}_.$8 & 49.8 & 3,483 & 19.3 &   3.1 &  18.2 &  3.8 &  14.4 & 0.39 & 11.93\\ 
J103420-265408  & 10$^{\rm{h}}$34$^{\rm{m}}$21$^{\rm{s}}_.$5 &  -26$^{\rm{d}}$54$^{\rm{m}}$07$^{\rm{s}}_.$4 & 53.3 & 3,731 & 11.5 &   1.2 &   8.0 &  1.2 &   6.8 & 0.46 &   ...\\ 
J103502-293019†  &  10$^{\rm{h}}$35$^{\rm{m}}$03$^{\rm{s}}_.$2 &  -29$^{\rm{d}}$30$^{\rm{m}}$09$^{\rm{s}}_.$5 & 53.5 & 3,744 & 12.3 &   1.2 &   8.3 &  3.8 &   4.5 &  ... &   ...\\ 
J103521-274137  & 10$^{\rm{h}}$35$^{\rm{m}}$21$^{\rm{s}}_.$8 & -27$^{\rm{d}}$41$^{\rm{m}}$38$^{\rm{s}}_.$5 & 41.2 & 2,886 &  9.4 &   1.9 &   7.5 &  4.3 &   3.2 & 0.26 &   ...\\ 
J103523-281855  & 10$^{\rm{h}}$35$^{\rm{m}}$23$^{\rm{s}}_.$3 & -28$^{\rm{d}}$18$^{\rm{m}}$50$^{\rm{s}}_.$7 & 47.1 & 3,295 & 21.7 &   6.1 &  32.0 &  7.7 &  24.3 & 0.23 &  7.07\\ 
J103541-265827  & 10$^{\rm{h}}$35$^{\rm{m}}$41$^{\rm{s}}_.$1 & -26$^{\rm{d}}$58$^{\rm{m}}$22$^{\rm{s}}_.$4 & 37.2 & 2,604 &  7.0 &   0.7 &   2.4 &  0.4 &   2.0 & 0.36 &   ...\\ 
J103546-273840  & 10$^{\rm{h}}$35$^{\rm{m}}$46$^{\rm{s}}_.$4 & -27$^{\rm{d}}$38$^{\rm{m}}$52$^{\rm{s}}_.$9 & 67.9 & 4,753 & 15.5 &   1.1 &  12.5 &  4.2 &   8.3 & 0.34 &   ...\\ 
J103602-261141†  &  10$^{\rm{h}}$36$^{\rm{m}}$02$^{\rm{s}}_.$4 & -26$^{\rm{d}}$11$^{\rm{m}}$37$^{\rm{s}}_.$7 & 46.8 & 3,274 &  6.5 &   0.6 &   2.9 &  1.6 &   1.3 & 0.18 &   ...\\ 
J103603-245430  &  10$^{\rm{h}}$36$^{\rm{m}}$03$^{\rm{s}}_.$1 & -24$^{\rm{d}}$54$^{\rm{m}}$23$^{\rm{s}}_.$8 & 56.3 & 3,944 & 14.7 &   1.6 &  11.8 &  1.7 &  10.0 & 0.38 &   ...\\ 
J103621-252235  & 10$^{\rm{h}}$36$^{\rm{m}}$21$^{\rm{s}}_.$4 & -25$^{\rm{d}}$22$^{\rm{m}}$40$^{\rm{s}}_.$6 & 57.3 & 4,011 & 12.6 &   1.4 &  11.2 &  0.6 &  10.6 & 0.39 &   ...\\ 
J103627-255957  & 10$^{\rm{h}}$36$^{\rm{m}}$26$^{\rm{s}}_.$8 & -25$^{\rm{d}}$59$^{\rm{m}}$49$^{\rm{s}}_.$9 & 46.2 & 3,235 &  5.6 &   0.3 &   1.4 &  0.6 &   0.8 &  ... &   ...\\ 
J103644-251543  & 10$^{\rm{h}}$36$^{\rm{m}}$44$^{\rm{s}}_.$2 & -25$^{\rm{d}}$15$^{\rm{m}}$52$^{\rm{s}}_.$5 & 52.4 & 3,665 & 10.7 &   1.1 &   7.4 &  1.2 &   6.3 & 0.47 &   ...\\ 
J103645-281010  & 10$^{\rm{h}}$36$^{\rm{m}}$45$^{\rm{s}}_.$7 &  -28$^{\rm{d}}$10$^{\rm{m}}$07$^{\rm{s}}_.$1 & 49.9 & 3,493 & 11.8 &   1.8 &  10.8 &  7.9 &   2.9 & 0.21 &   ...\\ 
J103646-293253  & 10$^{\rm{h}}$36$^{\rm{m}}$46$^{\rm{s}}_.$7 & -29$^{\rm{d}}$32$^{\rm{m}}$55$^{\rm{s}}_.$9 & 51.0 & 3,567 & 13.0 &   1.9 &  11.9 &  1.1 &  10.8 & 0.16 &   ...\\ 
J103651-260227  & 10$^{\rm{h}}$36$^{\rm{m}}$51$^{\rm{s}}_.$0 & -26$^{\rm{d}}$02$^{\rm{m}}$26$^{\rm{s}}_.$4 & 33.9 & 2,371 & 11.6 &   3.7 &  10.1 &  1.3 &   8.7 & 0.11 &  2.94\\ 
J103653-270311  & 10$^{\rm{h}}$36$^{\rm{m}}$53$^{\rm{s}}_.$0 & -27$^{\rm{d}}$03$^{\rm{m}}$10$^{\rm{s}}_.$6 & 50.7 & 3,549 & 14.6 &   2.3 &  14.1 &  3.9 &  10.3 & 0.15 &  7.20\\ 
J103655-265412  & 10$^{\rm{h}}$36$^{\rm{m}}$56$^{\rm{s}}_.$0 & -26$^{\rm{d}}$54$^{\rm{m}}$11$^{\rm{s}}_.$3 & 46.4 & 3,245 & 13.5 &   2.6 &  13.3 &  2.9 &  10.4 & 0.21 &  4.70\\ 
J103719-281408†  & 10$^{\rm{h}}$37$^{\rm{m}}$20$^{\rm{s}}_.$1 & -28$^{\rm{d}}$14$^{\rm{m}}$10$^{\rm{s}}_.$4 & 49.6 & 3,472 &  6.8 &   0.5 &   3.1 &  1.3 &   1.8 & 0.31 &   ...\\ 
J103721-251319  & 10$^{\rm{h}}$37$^{\rm{m}}$22$^{\rm{s}}_.$2 &  -25$^{\rm{d}}$13$^{\rm{m}}$05$^{\rm{s}}_.$8 & 15.2 & 1,068 &  1.8 &   0.3 &   0.1 &  0.1 &   0.1 &  ... &   ...\\ 
J103737-261641  & 10$^{\rm{h}}$37$^{\rm{m}}$38$^{\rm{s}}_.$3 & -26$^{\rm{d}}$16$^{\rm{m}}$37$^{\rm{s}}_.$3 & 54.3 & 3,801 & 12.7 &   1.0 &   6.9 &  1.1 &   5.8 & 0.41 &   ...\\ 
J103738-281216  & 10$^{\rm{h}}$37$^{\rm{m}}$37$^{\rm{s}}_.$0 & -28$^{\rm{d}}$12$^{\rm{m}}$22$^{\rm{s}}_.$1 & 67.4 & 4,717 &  7.9 &   0.5 &   4.8 &  4.1 &   0.8 &  ... &   ...\\ 
J103749-270715  & 10$^{\rm{h}}$37$^{\rm{m}}$49$^{\rm{s}}_.$4 & -27$^{\rm{d}}$07$^{\rm{m}}$17$^{\rm{s}}_.$0 & 35.0 & 2,446 &  8.4 &   1.6 &   4.6 &  0.4 &   4.2 & 0.13 &   ...\\ 
J103804-284333  &  10$^{\rm{h}}$38$^{\rm{m}}$04$^{\rm{s}}_.$9 & -28$^{\rm{d}}$43$^{\rm{m}}$36$^{\rm{s}}_.$2 & 63.6 & 4,451 & 10.5 &   1.1 &  10.5 &  8.5 &   2.1 & 0.32 &   ...\\ 
J103805-250537  &  10$^{\rm{h}}$38$^{\rm{m}}$06$^{\rm{s}}_.$2 & -25$^{\rm{d}}$05$^{\rm{m}}$35$^{\rm{s}}_.$0 & 56.5 & 3,951 & 12.4 &   1.1 &   8.4 &  0.4 &   8.0 & 0.40 &   ...\\ 
J103812-275607  & 10$^{\rm{h}}$38$^{\rm{m}}$11$^{\rm{s}}_.$7 & -27$^{\rm{d}}$56$^{\rm{m}}$12$^{\rm{s}}_.$0 & 40.8 & 2,856 &  5.8 &   0.3 &   1.3 &  0.6 &   0.7 & 0.39 &   ...\\ 

\end{tabular}
\end{table*}

\begin{table*}
\tiny
\contcaption{}
\centering
\begin{tabular}{cccccccccccc}
 \hline
 Name & R.A. (J2000) & Dec. (J2000) & $D$ & $V_{\rm{sys}}$ & $R_{\rm{HI}}$ & $S_{\rm{HI}}$ & $M_{\rm{total}}^{\rm{HI}}$ &  $M_{\rm{narrow}}^{\rm{HI}}$ & $M_{\rm{broad}}^{\rm{HI}}$ & $A_{\rm{map}}^{\rm{HI}}$ & $Q$\\
 &   &  & (Mpc) & (\kms)\ & (kpc) & (Jy km s$^{-1}$) & ($10^8 \ \rm{M_{\odot}}$) & ($10^8 \ \rm{M_{\odot}}$) & ($10^8 \ \rm{M_{\odot}}$) \\
    (1) & (2) & (3) & (4) & (5) & (6) & (7) & (8) & (9) & (10) & (11) & (12) \\
 \hline

J103818-285307  & 10$^{\rm{h}}$38$^{\rm{m}}$17$^{\rm{s}}_.$9 &  -28$^{\rm{d}}$53$^{\rm{m}}$07$^{\rm{s}}_.$9 & 62.5 & 4,377 & 24.6 &   5.1 &  47.3 & 13.0 &  34.2 & 0.18 &   ...\\ 
J103821-254126  & 10$^{\rm{h}}$38$^{\rm{m}}$20$^{\rm{s}}_.$8 & -25$^{\rm{d}}$41$^{\rm{m}}$23$^{\rm{s}}_.$1 & 49.0 & 3,431 & 10.8 &   0.9 &   5.3 &  3.0 &   2.2 & 0.24 &   ...\\ 
J103828-283056  & 10$^{\rm{h}}$38$^{\rm{m}}$28$^{\rm{s}}_.$6 & -28$^{\rm{d}}$30$^{\rm{m}}$50$^{\rm{s}}_.$8 & 62.2 & 4,354 & 10.5 &   1.0 &   9.1 &  4.8 &   4.3 & 0.36 &   ...\\ 
J103840-283405  & 10$^{\rm{h}}$38$^{\rm{m}}$39$^{\rm{s}}_.$7 &  -28$^{\rm{d}}$34$^{\rm{m}}$06$^{\rm{s}}_.$9 & 49.5 & 3,464 & 13.0 &   1.3 &   7.5 &  3.8 &   3.7 & 0.39 &   ...\\ 
J103841-253530  & 10$^{\rm{h}}$38$^{\rm{m}}$41$^{\rm{s}}_.$4 & -25$^{\rm{d}}$35$^{\rm{m}}$29$^{\rm{s}}_.$9 & 54.9 & 3,844 & 12.7 &   1.4 &  10.0 &  1.1 &   8.9 & 0.18 &   ...\\ 
J103842-281535  & 10$^{\rm{h}}$38$^{\rm{m}}$42$^{\rm{s}}_.$2 & -28$^{\rm{d}}$15$^{\rm{m}}$36$^{\rm{s}}_.$5 & 48.7 & 3,408 &  8.3 &   0.8 &   4.3 &  3.3 &   1.0 & 0.16 &   ...\\ 
J103902-291255  &  10$^{\rm{h}}$39$^{\rm{m}}$02$^{\rm{s}}_.$2 &  -29$^{\rm{d}}$13$^{\rm{m}}$04$^{\rm{s}}_.$7 & 45.7 & 3,198 &  8.9 &   1.1 &   5.5 &  3.0 &   2.5 & 0.55 &   ...\\ 
J103914-271511  & 10$^{\rm{h}}$39$^{\rm{m}}$14$^{\rm{s}}_.$7 & -27$^{\rm{d}}$15$^{\rm{m}}$14$^{\rm{s}}_.$7 & 64.8 & 4,536 & 10.4 &   0.5 &   4.6 &  1.1 &   3.4 & 0.32 &   ...\\ 
J103915-301757  & 10$^{\rm{h}}$39$^{\rm{m}}$14$^{\rm{s}}_.$8 & -30$^{\rm{d}}$17$^{\rm{m}}$54$^{\rm{s}}_.$1 & 53.9 & 3,774 & 32.2 &  10.7 &  73.2 & 12.7 &  60.5 & 0.35 &  7.38\\ 
J103918-265030  & 10$^{\rm{h}}$39$^{\rm{m}}$18$^{\rm{s}}_.$2 & -26$^{\rm{d}}$50$^{\rm{m}}$25$^{\rm{s}}_.$2 & 43.8 & 3,067 & 12.3 &   2.9 &  12.9 &  4.3 &   8.7 & 0.15 &   ...\\ 
J103922-293505  & 10$^{\rm{h}}$39$^{\rm{m}}$22$^{\rm{s}}_.$3 &  -29$^{\rm{d}}$35$^{\rm{m}}$06$^{\rm{s}}_.$2 & 55.4 & 3,876 & 17.1 &   3.5 &  25.0 &  6.1 &  18.9 & 0.12 &   ...\\ 
J103924-275442  & 10$^{\rm{h}}$39$^{\rm{m}}$24$^{\rm{s}}_.$7 & -27$^{\rm{d}}$54$^{\rm{m}}$42$^{\rm{s}}_.$5 & 46.1 & 3,225 & 10.0 &   1.5 &   7.4 &  1.9 &   5.5 & 0.21 &   ...\\ 
J103927-271653  & 10$^{\rm{h}}$39$^{\rm{m}}$27$^{\rm{s}}_.$5 & -27$^{\rm{d}}$16$^{\rm{m}}$49$^{\rm{s}}_.$1 & 48.1 & 3,366 &  7.5 &   0.7 &   3.8 &  2.6 &   1.2 & 0.23 &   ...\\ 
J103939-280552†  & 10$^{\rm{h}}$39$^{\rm{m}}$39$^{\rm{s}}_.$4 & -28$^{\rm{d}}$05$^{\rm{m}}$44$^{\rm{s}}_.$4 & 48.5 & 3,394 &  8.0 &   0.9 &   5.1 &  2.3 &   2.8 & 0.38 &   ...\\ 
J103958-301130  & 10$^{\rm{h}}$39$^{\rm{m}}$58$^{\rm{s}}_.$4 & -30$^{\rm{d}}$11$^{\rm{m}}$35$^{\rm{s}}_.$8 & 47.0 & 3,287 & 15.6 &   2.0 &  10.3 &  1.7 &   8.7 & 0.21 &   ...\\ 
J104000-292445  &  10$^{\rm{h}}$40$^{\rm{m}}$00$^{\rm{s}}_.$1 & -29$^{\rm{d}}$24$^{\rm{m}}$45$^{\rm{s}}_.$9 & 53.9 & 3,770 & 15.8 &   2.8 &  19.1 &  2.3 &  16.8 & 0.10 &   ...\\ 
J104004-301606  &  10$^{\rm{h}}$40$^{\rm{m}}$04$^{\rm{s}}_.$1 & -30$^{\rm{d}}$15$^{\rm{m}}$59$^{\rm{s}}_.$1 & 48.5 & 3,393 & 18.5 &   4.3 &  23.7 &  3.0 &  20.7 & 0.21 &   ...\\ 
J104016-274630  & 10$^{\rm{h}}$40$^{\rm{m}}$17$^{\rm{s}}_.$0 & -27$^{\rm{d}}$46$^{\rm{m}}$38$^{\rm{s}}_.$4 & 56.7 & 3,971 & 33.9 &  11.1 &  84.0 & 16.3 &  67.8 & 0.19 &  3.70\\ 
J104026-274853†  & 10$^{\rm{h}}$40$^{\rm{m}}$26$^{\rm{s}}_.$0 & -27$^{\rm{d}}$48$^{\rm{m}}$49$^{\rm{s}}_.$7 & 62.1 & 4,345 & 10.8 &   0.9 &   8.4 &  3.8 &   4.7 & 0.28 &   ...\\ 
J104048-244003  & 10$^{\rm{h}}$40$^{\rm{m}}$48$^{\rm{s}}_.$8 & -24$^{\rm{d}}$39$^{\rm{m}}$59$^{\rm{s}}_.$9 & 52.3 & 3,660 & 12.9 &   1.9 &  12.5 &  1.0 &  11.5 & 0.19 &   ...\\ 
J104058-274546  & 10$^{\rm{h}}$40$^{\rm{m}}$59$^{\rm{s}}_.$0 & -27$^{\rm{d}}$45$^{\rm{m}}$36$^{\rm{s}}_.$9 & 56.9 & 3,986 & 12.7 &   1.3 &   9.7 &  2.2 &   7.5 & 0.57 &   ...\\ 
J104059-270456  & 10$^{\rm{h}}$40$^{\rm{m}}$58$^{\rm{s}}_.$9 &  -27$^{\rm{d}}$05$^{\rm{m}}$02$^{\rm{s}}_.$0 & 67.7 & 4,736 & 29.2 &   6.0 &  64.5 & 14.1 &  50.4 & 0.17 &  6.95\\ 
J104100-284430†  &  10$^{\rm{h}}$41$^{\rm{m}}$00$^{\rm{s}}_.$8 & -28$^{\rm{d}}$44$^{\rm{m}}$31$^{\rm{s}}_.$3 & 53.5 & 3,748 & 11.5 &   1.4 &   9.3 &  5.5 &   3.7 & 0.33 &   ...\\ 
J104139-254049  & 10$^{\rm{h}}$41$^{\rm{m}}$39$^{\rm{s}}_.$5 & -25$^{\rm{d}}$40$^{\rm{m}}$49$^{\rm{s}}_.$5 & 55.1 & 3,855 &  9.6 &   0.9 &   6.7 &  0.6 &   6.1 & 0.13 &   ...\\ 
J104139-274639  & 10$^{\rm{h}}$41$^{\rm{m}}$39$^{\rm{s}}_.$5 & -27$^{\rm{d}}$46$^{\rm{m}}$31$^{\rm{s}}_.$4 & 62.2 & 4,354 & 10.0 &   0.9 &   8.2 &  1.5 &   6.7 & 0.44 &   ...\\ 
J104221-291748  & 10$^{\rm{h}}$42$^{\rm{m}}$22$^{\rm{s}}_.$1 & -29$^{\rm{d}}$17$^{\rm{m}}$49$^{\rm{s}}_.$9 & 63.2 & 4,426 & 14.9 &   1.7 &  16.3 &  4.6 &  11.7 & 0.17 &   ...\\ 
J104239-300357  & 10$^{\rm{h}}$42$^{\rm{m}}$39$^{\rm{s}}_.$5 & -30$^{\rm{d}}$03$^{\rm{m}}$56$^{\rm{s}}_.$6 & 35.4 & 2,477 &  5.1 &   0.8 &   2.4 &  0.3 &   2.1 & 0.17 &   ...\\ 
J104252-252014  & 10$^{\rm{h}}$42$^{\rm{m}}$53$^{\rm{s}}_.$3 & -25$^{\rm{d}}$20$^{\rm{m}}$13$^{\rm{s}}_.$7 & 91.0 & 6,371 & 12.5 &   0.5 &   9.9 &  0.9 &   9.1 & 0.33 &   ...\\ 
J104311-261500  & 10$^{\rm{h}}$43$^{\rm{m}}$11$^{\rm{s}}_.$5 & -26$^{\rm{d}}$14$^{\rm{m}}$59$^{\rm{s}}_.$3 & 64.7 & 4,532 & 30.5 &   6.0 &  59.8 & 12.0 &  47.8 & 0.10 &  8.79\\ 
J104326-251857  & 10$^{\rm{h}}$43$^{\rm{m}}$26$^{\rm{s}}_.$9 & -25$^{\rm{d}}$18$^{\rm{m}}$57$^{\rm{s}}_.$1 & 53.7 & 3,762 &  8.7 &   1.1 &   7.5 &  4.2 &   3.3 & 0.42 &   ...\\ 
J104339-285157  & 10$^{\rm{h}}$43$^{\rm{m}}$38$^{\rm{s}}_.$8 & -28$^{\rm{d}}$51$^{\rm{m}}$57$^{\rm{s}}_.$9 & 49.6 & 3,475 & 17.9 &   4.4 &  25.6 &  5.4 &  20.3 & 0.18 &  9.15\\ 
J104359-293304  & 10$^{\rm{h}}$43$^{\rm{m}}$59$^{\rm{s}}_.$1 & -29$^{\rm{d}}$32$^{\rm{m}}$59$^{\rm{s}}_.$1 & 47.8 & 3,344 &  6.7 &   0.4 &   2.4 &  0.2 &   2.2 & 0.32 &   ...\\ 
J104629-253308  & 10$^{\rm{h}}$46$^{\rm{m}}$29$^{\rm{s}}_.$5 &  -25$^{\rm{d}}$33$^{\rm{m}}$08$^{\rm{s}}_.$4 & 51.4 & 3,598 & 10.0 &   1.2 &   7.2 &  0.7 &   6.5 & 0.15 &   ...\\ 
J104824-250944  & 10$^{\rm{h}}$48$^{\rm{m}}$23$^{\rm{s}}_.$9 & -25$^{\rm{d}}$09$^{\rm{m}}$40$^{\rm{s}}_.$5 & 53.6 & 3,750 & 38.3 &   6.4 &  43.4 &  9.8 &  33.6 & 0.39 &  1.98\\ 
J104905-292232  &  10$^{\rm{h}}$49$^{\rm{m}}$05$^{\rm{s}}_.$4 & -29$^{\rm{d}}$22$^{\rm{m}}$29$^{\rm{s}}_.$5 & 61.0 & 4,273 & 20.0 &   3.8 &  33.1 &  6.1 &  27.0 & 0.22 &   ...\\ 
J105227-291155†  & 10$^{\rm{h}}$52$^{\rm{m}}$27$^{\rm{s}}_.$4 &  -29$^{\rm{d}}$12$^{\rm{m}}$02$^{\rm{s}}_.$1 & 49.2 & 3,443 &  8.8 &   0.7 &   4.2 &  2.9 &   1.2 & 0.46 &   ...\\ 
   
 \hline
 
\end{tabular}
\flushleft
 \textbf{Notes.} {The columns are (1) source name; (2) kinematic centre in R.A. units from {\sc 2dbat}; (3) kinematic centre Dec. units from {\sc 2dbat}; (4) distance to the galaxy from redshift (Mpc); (5) systemic velocity from redshift (\kms)\,; (6) radius of H{\sc i} disk from the {\sc baygaud} analysis, $R_{\rm{HI}}$ (kpc);  (7) integrated H{\sc i} intensities derived from super-profiles in Section 3 (Jy km s$^{-1}$); (8) H{\sc i} mass derived from super-profiles in Section 3 ($10^8 \ \rm{M_{\odot}}$); (9) narrow H{\sc i} mass derived from super-profiles in Section 3 ($10^8 \ \rm{M_{\odot}}$); (10) broad H{\sc i} mass derived from super-profiles in Section 3 ($10^8 \ \rm{M_{\odot}}$); (11) H{\sc i} morphological asymmetry of the galaxy ; (12) median values of the Toomre Q parameter values of the galaxy. '†' indicates the galaxy that have similar velocity widths of the narrow and broad components in the H{\sc i} suepr-profile analysis.}
 
\end{table*}
\clearpage


\begin{landscape}
\begin{table}

\tiny
\caption{H{\sc i} properties and distances of the paired galaxies in the ASKAP Hydra I cluster field.}
\label{long_hydra_pair}
\centering

\resizebox{1.3\textwidth}{!}{
\begin{tabular}{cccccccccccccc}
 \hline
 Name & R.A. (J2000) & Dec. (J2000) & $D$ & $V_{\rm{sys}}$ & $R_{\rm{HI}}$ & $S_{\rm{HI}}$ & $M_{\rm{total}}^{\rm{HI}}$ &  $M_{\rm{narrow}}^{\rm{HI}}$ & $M_{\rm{broad}}^{\rm{HI}}$ & $R_{\rm{p}}$ & $\Delta V$ & $A_{\rm{map}}^{\rm{HI}}$ & $Q$\\
 &   &  & (Mpc) & (\kms)\ & (kpc) & (Jy km s$^{-1}$) & ($10^8 \ \rm{M_{\odot}}$) & ($10^8 \ \rm{M_{\odot}}$) & ($10^8 \ \rm{M_{\odot}}$) & (kpc) & (\kms)\ \\
    (1) & (2) & (3) & (4) & (5) & (6) & (7) & (8) & (9) & (10) & (11) & (12) & (13) & (14)\\
 \hline
 
J100336-262923  & 10$^{\rm{h}}$03$^{\rm{m}}$37$^{\rm{s}}_.$4 & -26$^{\rm{d}}$29$^{\rm{m}}$39$^{\rm{s}}_.$1 & 12.9 &   901 &  4.3 &  1.8 &   0.7 &  0.2 &   0.5 &    31 &    26 &  ... &   ...\\ 
J100342-270137  & 10$^{\rm{h}}$03$^{\rm{m}}$41$^{\rm{s}}_.$9 & -27$^{\rm{d}}$01$^{\rm{m}}$39$^{\rm{s}}_.$5 & 13.7 &   957 & 15.5 & 45.7 &  20.2 &  6.8 &  13.3 &    93 &    83 & 0.09 &  2.46\\ 
J100351-263707  & 10$^{\rm{h}}$03$^{\rm{m}}$51$^{\rm{s}}_.$9 & -26$^{\rm{d}}$36$^{\rm{m}}$35$^{\rm{s}}_.$5 & 12.5 &   874 &  6.6 &  6.7 &   2.4 &  0.5 &   1.9 & 31/93 & 26/83 &  ... &   ...\\ 
J100426-282638  & 10$^{\rm{h}}$04$^{\rm{m}}$26$^{\rm{s}}_.$3 & -28$^{\rm{d}}$26$^{\rm{m}}$40$^{\rm{s}}_.$8 & 15.5 & 1,083 & 15.6 & 32.4 &  18.3 &  4.9 &  13.3 &    69 &    41 & 0.10 &  2.70\\ 
J100640-273917  & 10$^{\rm{h}}$06$^{\rm{m}}$40$^{\rm{s}}_.$2 & -27$^{\rm{d}}$39$^{\rm{m}}$19$^{\rm{s}}_.$1 & 61.0 & 4,270 & 16.3 &  2.0 &  17.7 &  0.9 &  16.9 &    75 &    24 & 0.18 &   ...\\ 
J100700-273944  &  10$^{\rm{h}}$07$^{\rm{m}}$00$^{\rm{s}}_.$8 & -27$^{\rm{d}}$39$^{\rm{m}}$41$^{\rm{s}}_.$1 & 60.6 & 4,244 & 11.3 &  1.0 &   8.8 &  5.7 &   3.1 &    75 &    24 & 0.14 &   ...\\ 
J101334-273518  & 10$^{\rm{h}}$13$^{\rm{m}}$35$^{\rm{s}}_.$0 & -27$^{\rm{d}}$35$^{\rm{m}}$23$^{\rm{s}}_.$7 & 36.4 & 2,549 &  8.4 &  2.1 &   6.6 &  3.2 &   3.4 &    47 &    49 & 0.22 &   ...\\ 
J101348-273147  & 10$^{\rm{h}}$13$^{\rm{m}}$48$^{\rm{s}}_.$8 & -27$^{\rm{d}}$31$^{\rm{m}}$42$^{\rm{s}}_.$6 & 37.1 & 2,599 &  9.0 &  1.4 &   4.7 &  0.3 &   4.4 &    47 &    49 & 0.33 &   ...\\ 
J101359-274538  & 10$^{\rm{h}}$13$^{\rm{m}}$59$^{\rm{s}}_.$5 & -27$^{\rm{d}}$45$^{\rm{m}}$30$^{\rm{s}}_.$9 & 36.6 & 2,566 &  5.1 &  0.6 &   1.8 &  0.7 &   1.1 &    90 &    50 & 0.47 &   ...\\ 
J101448-274240* & 10$^{\rm{h}}$14$^{\rm{m}}$47$^{\rm{s}}_.$7 & -27$^{\rm{d}}$42$^{\rm{m}}$40$^{\rm{s}}_.$6 & 60.5 & 4,231 & 12.4 &  1.3 &  11.0 &  4.1 &   6.9 &   ... &   ... & 0.20 &   ...\\ 
J102411-285533* & 10$^{\rm{h}}$24$^{\rm{m}}$11$^{\rm{s}}_.$5 & -28$^{\rm{d}}$55$^{\rm{m}}$40$^{\rm{s}}_.$9 & 55.6 & 3,890 & 12.3 &  1.4 &   9.9 &  5.3 &   4.6 &   ... &   ... & 0.23 &   ...\\ 
J102439-244547* & 10$^{\rm{h}}$24$^{\rm{m}}$38$^{\rm{s}}_.$5 & -24$^{\rm{d}}$45$^{\rm{m}}$47$^{\rm{s}}_.$6 & 50.0 & 3,503 & 12.7 &  1.3 &   7.7 &  2.6 &   5.1 &   ... &   ... & 0.40 &   ...\\ 
J103114-295837†  & 10$^{\rm{h}}$31$^{\rm{m}}$14$^{\rm{s}}_.$4 & -29$^{\rm{d}}$58$^{\rm{m}}$29$^{\rm{s}}_.$9 & 59.9 & 4,190 & 12.2 &  0.7 &   6.3 &  3.3 &   3.0 &    47 &    88 & 0.41 &   ...\\ 
J103124-295706  & 10$^{\rm{h}}$31$^{\rm{m}}$25$^{\rm{s}}_.$1 & -29$^{\rm{d}}$57$^{\rm{m}}$11$^{\rm{s}}_.$0 & 58.6 & 4,100 & 25.1 &  5.2 &  41.7 &  9.1 &  32.6 &    47 &    88 & 0.19 &   ...\\ 
J103241-273137  & 10$^{\rm{h}}$32$^{\rm{m}}$40$^{\rm{s}}_.$8 & -27$^{\rm{d}}$31$^{\rm{m}}$37$^{\rm{s}}_.$1 & 54.6 & 3,823 &  9.4 &  0.6 &   4.0 &  1.2 &   2.8 &    26 &   206 & 0.37 &   ...\\ 
J103248-273119  & 10$^{\rm{h}}$32$^{\rm{m}}$48$^{\rm{s}}_.$9 & -27$^{\rm{d}}$31$^{\rm{m}}$22$^{\rm{s}}_.$2 & 51.6 & 3,614 & 12.4 &  1.5 &   9.2 &  2.0 &   7.2 &    26 &   206 & 0.16 &   ...\\ 
J103436-273900  & 10$^{\rm{h}}$34$^{\rm{m}}$36$^{\rm{s}}_.$9 &  -27$^{\rm{d}}$39$^{\rm{m}}$08$^{\rm{s}}_.$8 & 43.1 & 3,017 & 14.0 &  3.1 &  13.6 &  4.2 &   9.3 &    53 &     5 &  ... & 11.81\\ 
J103455-273816  & 10$^{\rm{h}}$34$^{\rm{m}}$55$^{\rm{s}}_.$6 & -27$^{\rm{d}}$38$^{\rm{m}}$22$^{\rm{s}}_.$2 & 43.2 & 3,023 &  5.4 &  0.7 &   2.9 &  1.6 &   1.3 &    53 &     5 & 0.41 &   ...\\ 
J103459-280440  & 10$^{\rm{h}}$34$^{\rm{m}}$59$^{\rm{s}}_.$9 & -28$^{\rm{d}}$04$^{\rm{m}}$44$^{\rm{s}}_.$8 & 34.0 & 2,382 &  6.3 &  0.6 &   1.6 &  0.1 &   1.5 &    56 &    53 & 0.36 &   ...\\ 
J103507-275923  &  10$^{\rm{h}}$35$^{\rm{m}}$07$^{\rm{s}}_.$9 & -27$^{\rm{d}}$59$^{\rm{m}}$31$^{\rm{s}}_.$5 & 34.8 & 2,436 &  8.1 &  1.0 &   2.8 &  0.1 &   2.6 &    56 &    53 & 0.32 &   ...\\ 
J103521-272324* & 10$^{\rm{h}}$35$^{\rm{m}}$21$^{\rm{s}}_.$3 & -27$^{\rm{d}}$23$^{\rm{m}}$17$^{\rm{s}}_.$8 & 44.8 & 3,137 &  6.7 &  0.5 &   2.6 &  1.3 &   1.3 &   ... &   ... & 0.38 &   ...\\ 
J103522-244506  & 10$^{\rm{h}}$35$^{\rm{m}}$23$^{\rm{s}}_.$3 & -24$^{\rm{d}}$45$^{\rm{m}}$12$^{\rm{s}}_.$2 & 14.9 & 1,042 & 13.1 & 25.9 &  13.5 &  6.4 &   7.2 &    49 &    29 & 0.27 &  0.77\\ 
J103609-244856  &  10$^{\rm{h}}$36$^{\rm{m}}$09$^{\rm{s}}_.$6 & -24$^{\rm{d}}$48$^{\rm{m}}$51$^{\rm{s}}_.$1 & 14.5 & 1,012 &  4.1 &  2.8 &   1.4 &  0.1 &   1.3 &    49 &    29 & 0.23 &  2.79\\ 
J103702-273359  &  10$^{\rm{h}}$37$^{\rm{m}}$02$^{\rm{s}}_.$6 & -27$^{\rm{d}}$33$^{\rm{m}}$54$^{\rm{s}}_.$5 & 40.7 & 2,852 & 14.9 &  4.5 &  17.8 &  3.5 &  14.3 & 88/77 &  21/5 & 0.26 & 50.91\\ 
J103704-252038  &  10$^{\rm{h}}$37$^{\rm{m}}$04$^{\rm{s}}_.$5 & -25$^{\rm{d}}$20$^{\rm{m}}$44$^{\rm{s}}_.$1 & 53.0 & 3,713 & 14.5 &  1.6 &  10.3 &  2.0 &   8.3 &    53 &    96 & 0.24 &   ...\\ 
J103712-274108  & 10$^{\rm{h}}$37$^{\rm{m}}$12$^{\rm{s}}_.$5 & -27$^{\rm{d}}$41$^{\rm{m}}$13$^{\rm{s}}_.$1 & 40.4 & 2,830 &  9.9 &  1.2 &   4.7 &  0.4 &   4.3 &    88 &    21 &  ... &   ...\\ 
J103722-273235  & 10$^{\rm{h}}$37$^{\rm{m}}$22$^{\rm{s}}_.$8 & -27$^{\rm{d}}$32$^{\rm{m}}$38$^{\rm{s}}_.$8 & 39.4 & 2,754 &  6.5 &  0.6 &   2.2 &  0.7 &   1.5 &    77 &     5 &  ... &   ...\\ 
J103725-251916  & 10$^{\rm{h}}$37$^{\rm{m}}$25$^{\rm{s}}_.$0 &  -25$^{\rm{d}}$19$^{\rm{m}}$07$^{\rm{s}}_.$4 & 53.0 & 3,707 & 49.2 & 22.5 & 148.7 & 31.9 & 116.8 &    53 &    96 & 0.27 &   ...\\ 
J104513-262755* & 10$^{\rm{h}}$45$^{\rm{m}}$12$^{\rm{s}}_.$9 & -26$^{\rm{d}}$27$^{\rm{m}}$44$^{\rm{s}}_.$6 & 61.2 & 4,287 &  7.8 &  0.6 &   4.9 &  3.7 &   1.2 &   ... &   ... & 0.14 &   ...\\ 
\hline

\end{tabular}}

\flushleft
\textbf{Notes.} {The columns are (1) source name; (2) kinematic centre in R.A. units from {\sc 2dbat}; (3) kinematic centre Dec. units from {\sc 2dbat}; (4) distance to the galaxy from redshift (Mpc); (5) systemic velocity from redshift (\kms)\,; (6) radius of H{\sc i} disk from the {\sc baygaud} analysis, $R_{\rm{HI}}$ (kpc);  (7) integrated H{\sc i} intensities derived from super-profiles in Section 3 (Jy km s$^{-1}$); (8) H{\sc i} mass derived from super-profiles in Section 3 ($10^8 \ \rm{M_{\odot}}$); (9) narrow H{\sc i} mass derived from super-profiles in Section 3 ($10^8 \ \rm{M_{\odot}}$); (10) broad H{\sc i} mass derived from super-profiles in Section 3 ($10^8 \ \rm{M_{\odot}}$); (11) projected distance to the companion galaxy (kpc). For galaxies with two nearby companions, we provide two $R_{\rm{p}}$ values separated by '/'; (12) relative line-of-sight velocity to the companion galaxy (\kms)\,. For galaxies with two nearby companions, we provide two $\Delta V$ values separated by '/'; (13) H{\sc i} morphological asymmetry of the galaxy, (14) median values of the Toomre Q parameter values of the galaxy. '*' indicates the visually identified galaxy pair. '†' indicates the galaxy that have similar velocity widths of the narrow and broad components in the H{\sc i} suepr-profile analysis.}
\end{table}
\end{landscape}

\clearpage

\begin{table*}
\tiny
\caption{H{\sc i} properties and distances of the control galaxies in the ASKAP NGC 4636 field.}
\label{long_ngc_control}

\begin{tabular}{cccccccccccc}
\hline
 Name & R.A. (J2000) & Dec. (J2000) & $D$ & $V_{\rm{sys}}$ & $R_{\rm{HI}}$ & $S_{\rm{HI}}$ & $M_{\rm{total}}^{\rm{HI}}$ &  $M_{\rm{narrow}}^{\rm{HI}}$ & $M_{\rm{broad}}^{\rm{HI}}$ & $A_{\rm{map}}^{\rm{HI}}$ & $Q$\\
  &   &  & (Mpc) & (\kms)\ & (kpc) & (Jy km s$^{-1}$) & ($10^8 \ \rm{M_{\odot}}$) & ($10^8 \ \rm{M_{\odot}}$) & ($10^8 \ \rm{M_{\odot}}$) \\
    (1) & (2) & (3) & (4) & (5) & (6) & (7) & (8) & (9) & (10) & (11) & (12) \\
 \hline
 
J122708+055255  &  12$^{\rm{h}}$27$^{\rm{m}}$09$^{\rm{s}}_.$3 & +05$^{\rm{d}}$52$^{\rm{m}}$50$^{\rm{s}}_.$6 & 15.9 & 1,113 &  9.9 &  11.2 &   6.7 &  5.7 &   1.0 & 0.40 &  4.00\\ 
J122755+054312  & 12$^{\rm{h}}$27$^{\rm{m}}$55$^{\rm{s}}_.$5 & +05$^{\rm{d}}$43$^{\rm{m}}$14$^{\rm{s}}_.$8 & 32.0 & 2,243 &  5.3 &   1.2 &   2.8 &  0.4 &   2.4 & 0.19 &   ...\\ 
J122928+064623  & 12$^{\rm{h}}$29$^{\rm{m}}$28$^{\rm{s}}_.$6 & +06$^{\rm{d}}$46$^{\rm{m}}$12$^{\rm{s}}_.$0 &  7.5 &   527 &  2.3 &   3.6 &   0.5 &  0.1 &   0.4 & 0.41 &   ...\\ 
J122930+074142  & 12$^{\rm{h}}$29$^{\rm{m}}$30$^{\rm{s}}_.$5 & +07$^{\rm{d}}$41$^{\rm{m}}$46$^{\rm{s}}_.$8 & 10.7 &   748 &  3.0 &   1.5 &   0.4 &  0.1 &   0.3 & 0.22 &   ...\\ 
J122932+005020  & 12$^{\rm{h}}$29$^{\rm{m}}$32$^{\rm{s}}_.$5 & +00$^{\rm{d}}$50$^{\rm{m}}$23$^{\rm{s}}_.$3 & 31.9 & 2,235 &  8.4 &   2.9 &   7.0 &  1.0 &   6.0 & 0.19 &   ...\\ 
J122946-011745  & 12$^{\rm{h}}$29$^{\rm{m}}$46$^{\rm{s}}_.$3 & -01$^{\rm{d}}$17$^{\rm{m}}$40$^{\rm{s}}_.$8 & 32.3 & 2,259 &  6.8 &   1.8 &   4.5 &  3.9 &   0.6 & 0.25 &   ...\\ 
J123021-013813  & 12$^{\rm{h}}$30$^{\rm{m}}$21$^{\rm{s}}_.$4 & -01$^{\rm{d}}$38$^{\rm{m}}$15$^{\rm{s}}_.$3 & 30.5 & 2,132 &  5.7 &   1.3 &   2.9 &  1.5 &   1.4 & 0.25 &   ...\\ 
J123026+041450  & 12$^{\rm{h}}$30$^{\rm{m}}$26$^{\rm{s}}_.$8 & +04$^{\rm{d}}$14$^{\rm{m}}$48$^{\rm{s}}_.$1 & 34.8 & 2,437 & 16.9 &   8.7 &  24.8 &  4.7 &  20.1 & 0.12 &  8.35\\ 
J123103+014032  &  12$^{\rm{h}}$31$^{\rm{m}}$03$^{\rm{s}}_.$7 & +01$^{\rm{d}}$40$^{\rm{m}}$38$^{\rm{s}}_.$8 & 15.8 & 1,102 &  2.7 &   1.2 &   0.7 &  0.2 &   0.5 & 0.45 &   ...\\ 
J123148-025808  & 12$^{\rm{h}}$31$^{\rm{m}}$48$^{\rm{s}}_.$4 & -02$^{\rm{d}}$58$^{\rm{m}}$11$^{\rm{s}}_.$4 & 32.9 & 2,305 &  8.8 &   2.1 &   5.3 &  0.3 &   5.0 & 0.27 &   ...\\ 
J123209-010543  &  12$^{\rm{h}}$32$^{\rm{m}}$09$^{\rm{s}}_.$9 & -01$^{\rm{d}}$05$^{\rm{m}}$43$^{\rm{s}}_.$2 & 34.4 & 2,404 &  6.8 &   1.4 &   4.0 &  2.4 &   1.6 & 0.18 &   ...\\ 
J123228+002315  & 12$^{\rm{h}}$32$^{\rm{m}}$28$^{\rm{s}}_.$2 & +00$^{\rm{d}}$23$^{\rm{m}}$18$^{\rm{s}}_.$2 & 21.8 & 1,526 & 21.6 &  39.1 &  43.9 & 21.1 &  22.8 & 0.16 &  1.45\\ 
J123236+023943  & 12$^{\rm{h}}$32$^{\rm{m}}$36$^{\rm{s}}_.$6 & +02$^{\rm{d}}$39$^{\rm{m}}$41$^{\rm{s}}_.$9 & 24.7 & 1,731 & 13.8 &  10.9 &  15.7 &  5.8 &   9.8 & 0.10 &  2.53\\ 
J123244+000656  & 12$^{\rm{h}}$32$^{\rm{m}}$46$^{\rm{s}}_.$0 & +00$^{\rm{d}}$06$^{\rm{m}}$54$^{\rm{s}}_.$6 & 16.1 & 1,128 & 29.9 & 127.0 &  77.9 & 21.5 &  56.4 & 0.19 &   ...\\ 
J123246+013409  & 12$^{\rm{h}}$32$^{\rm{m}}$46$^{\rm{s}}_.$8 & +01$^{\rm{d}}$34$^{\rm{m}}$10$^{\rm{s}}_.$4 & 21.8 & 1,524 &  3.4 &   0.9 &   1.0 &  0.3 &   0.7 & 0.19 &   ...\\ 
J123257+043441  & 12$^{\rm{h}}$32$^{\rm{m}}$57$^{\rm{s}}_.$8 & +04$^{\rm{d}}$34$^{\rm{m}}$35$^{\rm{s}}_.$5 & 17.6 & 1,232 &  5.2 &   2.4 &   1.7 &  0.2 &   1.6 & 0.31 &   ...\\ 
J123308-003203  &  12$^{\rm{h}}$33$^{\rm{m}}$08$^{\rm{s}}_.$5 & -00$^{\rm{d}}$31$^{\rm{m}}$56$^{\rm{s}}_.$6 & 10.3 &   724 &  2.2 &   1.5 &   0.4 &  0.3 &   0.1 & 0.30 &   ...\\ 
J123320+013119  & 12$^{\rm{h}}$33$^{\rm{m}}$21$^{\rm{s}}_.$1 & +01$^{\rm{d}}$31$^{\rm{m}}$31$^{\rm{s}}_.$2 & 23.9 & 1,671 &  4.0 &   1.4 &   1.9 &  0.7 &   1.2 &  ... &   ...\\ 
J123329+034732  & 12$^{\rm{h}}$33$^{\rm{m}}$29$^{\rm{s}}_.$2 & +03$^{\rm{d}}$47$^{\rm{m}}$31$^{\rm{s}}_.$1 & 12.8 &   898 &  2.6 &   1.3 &   0.5 &  0.3 &   0.2 & 0.26 &   ...\\ 
J123346-023916  & 12$^{\rm{h}}$33$^{\rm{m}}$46$^{\rm{s}}_.$3 &  -02$^{\rm{d}}$39$^{\rm{m}}$03$^{\rm{s}}_.$0 & 35.5 & 2,482 & 13.4 &   5.4 &  16.1 &  5.1 &  11.0 &  ... &  1.00\\ 
J123355+033243  & 12$^{\rm{h}}$33$^{\rm{m}}$55$^{\rm{s}}_.$4 & +03$^{\rm{d}}$32$^{\rm{m}}$46$^{\rm{s}}_.$0 & 16.3 & 1,142 &  3.8 &   1.7 &   1.0 &  0.3 &   0.7 & 0.19 &   ...\\ 
J123439+070932  & 12$^{\rm{h}}$34$^{\rm{m}}$40$^{\rm{s}}_.$4 & +07$^{\rm{d}}$09$^{\rm{m}}$38$^{\rm{s}}_.$3 &  8.6 &   603 &  1.9 &   1.3 &   0.2 &  0.0 &   0.2 &  ... &   ...\\ 
J123531+062003  & 12$^{\rm{h}}$35$^{\rm{m}}$31$^{\rm{s}}_.$8 &  +06$^{\rm{d}}$20$^{\rm{m}}$02$^{\rm{s}}_.$3 & 22.9 & 1,603 &  4.0 &   0.5 &   0.7 &  0.2 &   0.4 &  ... &   ...\\ 
J123552-033632  & 12$^{\rm{h}}$35$^{\rm{m}}$52$^{\rm{s}}_.$3 & -03$^{\rm{d}}$36$^{\rm{m}}$28$^{\rm{s}}_.$0 & 38.0 & 2,660 &  7.0 &   0.6 &   2.1 &  1.3 &   0.8 & 0.45 &   ...\\ 
J123605-025225  &  12$^{\rm{h}}$36$^{\rm{m}}$05$^{\rm{s}}_.$0 & -02$^{\rm{d}}$52$^{\rm{m}}$23$^{\rm{s}}_.$2 & 36.5 & 2,558 &  5.5 &   0.5 &   1.5 &  0.6 &   0.9 & 0.24 &   ...\\ 
J123636+063715  & 12$^{\rm{h}}$36$^{\rm{m}}$37$^{\rm{s}}_.$1 & +06$^{\rm{d}}$37$^{\rm{m}}$18$^{\rm{s}}_.$8 & 15.3 & 1,073 & 10.1 &   0.8 &   0.5 &  0.3 &   0.2 & 0.17 &   ...\\ 
J123636-032344  & 12$^{\rm{h}}$36$^{\rm{m}}$36$^{\rm{s}}_.$6 & -03$^{\rm{d}}$23$^{\rm{m}}$48$^{\rm{s}}_.$3 & 17.1 & 1,196 &  2.6 &  15.8 &  10.9 &  2.9 &   8.0 & 0.30 &   ...\\ 
J123641+030621  & 12$^{\rm{h}}$36$^{\rm{m}}$42$^{\rm{s}}_.$1 & +03$^{\rm{d}}$06$^{\rm{m}}$31$^{\rm{s}}_.$2 & 20.5 & 1,436 & 10.0 &   6.4 &   6.3 &  2.9 &   3.5 & 0.33 &   ...\\ 
J123651-003515  & 12$^{\rm{h}}$36$^{\rm{m}}$51$^{\rm{s}}_.$8 & -00$^{\rm{d}}$35$^{\rm{m}}$15$^{\rm{s}}_.$3 & 36.8 & 2,573 &  6.7 &   0.6 &   1.9 &  1.0 &   0.9 & 0.28 &   ...\\ 
J123656+004118  & 12$^{\rm{h}}$36$^{\rm{m}}$56$^{\rm{s}}_.$4 & +00$^{\rm{d}}$41$^{\rm{m}}$13$^{\rm{s}}_.$5 & 36.4 & 2,544 &  4.7 &   3.0 &   9.3 &  1.1 &   8.2 & 0.35 &   ...\\ 
J123703+065533  &  12$^{\rm{h}}$37$^{\rm{m}}$03$^{\rm{s}}_.$3 & +06$^{\rm{d}}$55$^{\rm{m}}$39$^{\rm{s}}_.$0 & 23.4 & 1,637 & 12.2 &   0.7 &   0.8 &  0.1 &   0.7 & 0.26 &  2.71\\ 
J123729+044504  & 12$^{\rm{h}}$37$^{\rm{m}}$29$^{\rm{s}}_.$5 & +04$^{\rm{d}}$44$^{\rm{m}}$59$^{\rm{s}}_.$1 & 23.4 & 1,638 &  5.6 &   6.8 &   8.8 &  1.9 &   6.9 & 0.35 &   ...\\ 
J123746-024651  & 12$^{\rm{h}}$37$^{\rm{m}}$46$^{\rm{s}}_.$3 & -02$^{\rm{d}}$46$^{\rm{m}}$38$^{\rm{s}}_.$1 & 15.6 & 1,093 &  2.1 &   1.3 &   0.8 &  0.2 &   0.5 &  ... &   ...\\ 
J123748-012038  & 12$^{\rm{h}}$37$^{\rm{m}}$48$^{\rm{s}}_.$2 & -01$^{\rm{d}}$20$^{\rm{m}}$45$^{\rm{s}}_.$1 & 22.1 & 1,547 &  2.8 &   0.4 &   0.5 &  0.0 &   0.4 & 0.37 &   ...\\ 
J123803-021549  &  12$^{\rm{h}}$38$^{\rm{m}}$03$^{\rm{s}}_.$1 & -02$^{\rm{d}}$15$^{\rm{m}}$49$^{\rm{s}}_.$7 & 36.8 & 2,574 & 12.6 &   0.4 &   1.2 &  0.1 &   1.2 & 0.19 &   ...\\ 
J123827+041912  & 12$^{\rm{h}}$38$^{\rm{m}}$28$^{\rm{s}}_.$8 &  +04$^{\rm{d}}$19$^{\rm{m}}$00$^{\rm{s}}_.$4 & 11.6 &   812 &  2.9 &   3.8 &   1.2 &  0.8 &   0.5 &  ... & 52.94*\\ 
J123905-020044  &  12$^{\rm{h}}$39$^{\rm{m}}$05$^{\rm{s}}_.$9 & -02$^{\rm{d}}$00$^{\rm{m}}$46$^{\rm{s}}_.$7 & 37.5 & 2,627 &  7.7 &   1.5 &   5.0 &  3.1 &   1.9 & 0.30 &   ...\\ 
J123912+060038  & 12$^{\rm{h}}$39$^{\rm{m}}$12$^{\rm{s}}_.$5 & +06$^{\rm{d}}$00$^{\rm{m}}$43$^{\rm{s}}_.$8 & 34.7 & 2,432 & 11.3 &   1.0 &   3.0 &  1.8 &   1.2 & 0.21 & 14.97\\ 
J123917-003149  & 12$^{\rm{h}}$39$^{\rm{m}}$18$^{\rm{s}}_.$4 & -00$^{\rm{d}}$31$^{\rm{m}}$51$^{\rm{s}}_.$8 & 15.3 & 1,072 & 32.6 &   2.9 &   1.6 &  0.6 &   1.1 & 0.15 &   ...\\ 
J123922+045608  & 12$^{\rm{h}}$39$^{\rm{m}}$21$^{\rm{s}}_.$9 &  +04$^{\rm{d}}$56$^{\rm{m}}$07$^{\rm{s}}_.$3 & 23.1 & 1,614 &  3.7 & 174.8 & 219.2 & 88.2 & 131.0 & 0.25 &   ...\\ 
J123949+014016  & 12$^{\rm{h}}$39$^{\rm{m}}$50$^{\rm{s}}_.$3 & +01$^{\rm{d}}$40$^{\rm{m}}$17$^{\rm{s}}_.$3 & 17.6 & 1,228 &  4.6 &   0.9 &   0.6 &  0.3 &   0.3 & 0.19 &   ...\\ 
J123955-000822  & 12$^{\rm{h}}$39$^{\rm{m}}$54$^{\rm{s}}_.$9 & -00$^{\rm{d}}$08$^{\rm{m}}$16$^{\rm{s}}_.$5 & 43.8 & 3,062 & 12.5 &   1.9 &   8.7 &  2.5 &   6.2 & 0.31 &  7.56\\ 
J124002-010258  &  12$^{\rm{h}}$40$^{\rm{m}}$02$^{\rm{s}}_.$9 & -01$^{\rm{d}}$02$^{\rm{m}}$58$^{\rm{s}}_.$1 & 22.8 & 1,595 &  4.1 &   2.8 &   3.4 &  1.8 &   1.7 & 0.15 &   ...\\ 
J124002-033623  &  12$^{\rm{h}}$40$^{\rm{m}}$01$^{\rm{s}}_.$1 & -03$^{\rm{d}}$36$^{\rm{m}}$26$^{\rm{s}}_.$8 & 36.4 & 2,546 & 14.9 &   1.3 &   4.2 &  1.8 &   2.3 & 0.56 &  2.24\\ 
J124009+065029  &  12$^{\rm{h}}$40$^{\rm{m}}$09$^{\rm{s}}_.$6 & +06$^{\rm{d}}$50$^{\rm{m}}$19$^{\rm{s}}_.$2 & 14.3 & 1,003 &  3.4 &   0.5 &   0.2 &  0.1 &   0.1 &  ... &   ...\\ 
J124009-002103  &  12$^{\rm{h}}$40$^{\rm{m}}$09$^{\rm{s}}_.$1 &  -00$^{\rm{d}}$21$^{\rm{m}}$00$^{\rm{s}}_.$9 & 24.4 & 1,706 &  3.4 &   5.0 &   7.0 &  2.8 &   4.2 & 0.17 &   ...\\ 
J124111+012435  & 12$^{\rm{h}}$41$^{\rm{m}}$11$^{\rm{s}}_.$4 & +01$^{\rm{d}}$24$^{\rm{m}}$46$^{\rm{s}}_.$4 & 24.2 & 1,696 & 10.3 &   0.7 &   1.0 &  0.5 &   0.5 & 0.35 &  8.45\\ 
J124123-030327  & 12$^{\rm{h}}$41$^{\rm{m}}$23$^{\rm{s}}_.$1 & -03$^{\rm{d}}$03$^{\rm{m}}$29$^{\rm{s}}_.$1 & 20.7 & 1,448 &  5.5 &   6.2 &   6.3 &  0.8 &   5.4 & 0.19 &   ...\\ 
J124128-031517  & 12$^{\rm{h}}$41$^{\rm{m}}$28$^{\rm{s}}_.$0 & -03$^{\rm{d}}$15$^{\rm{m}}$27$^{\rm{s}}_.$5 & 25.7 & 1,800 &  4.0 &   2.7 &   4.2 &  0.2 &   4.0 &  ... &   ...\\ 
J124139+015014  & 12$^{\rm{h}}$41$^{\rm{m}}$39$^{\rm{s}}_.$4 & +01$^{\rm{d}}$50$^{\rm{m}}$15$^{\rm{s}}_.$2 & 21.8 & 1,528 &  2.3 &   0.6 &   0.7 &  0.4 &   0.4 & 0.21 &   ...\\ 
J124218+054425  & 12$^{\rm{h}}$42$^{\rm{m}}$18$^{\rm{s}}_.$2 & +05$^{\rm{d}}$44$^{\rm{m}}$27$^{\rm{s}}_.$1 & 13.9 &   974 &  3.0 &   0.4 &   0.2 &  0.0 &   0.2 & 0.19 &   ...\\ 
J124228+010549  & 12$^{\rm{h}}$42$^{\rm{m}}$28$^{\rm{s}}_.$4 & +01$^{\rm{d}}$05$^{\rm{m}}$50$^{\rm{s}}_.$1 & 13.8 &   964 &  1.7 &   1.7 &   0.8 &  0.4 &   0.3 & 0.21 &   ...\\ 
J124231+035729  & 12$^{\rm{h}}$42$^{\rm{m}}$31$^{\rm{s}}_.$1 & +03$^{\rm{d}}$57$^{\rm{m}}$29$^{\rm{s}}_.$2 & 10.6 &   743 &  4.6 &   0.4 &   0.1 &  0.0 &   0.1 & 0.15 &  6.30\\ 
J124232-012111  & 12$^{\rm{h}}$42$^{\rm{m}}$32$^{\rm{s}}_.$9 &  -01$^{\rm{d}}$21$^{\rm{m}}$07$^{\rm{s}}_.$5 & 15.8 & 1,109 & 13.2 &   5.6 &   3.3 &  0.4 &   2.9 &  ... &  2.11\\ 
J124244+072019  & 12$^{\rm{h}}$42$^{\rm{m}}$44$^{\rm{s}}_.$6 & +07$^{\rm{d}}$20$^{\rm{m}}$23$^{\rm{s}}_.$7 & 34.4 & 2,407 &  9.0 &  23.8 &  66.4 & 25.9 &  40.5 & 0.38 &  1.73\\ 
J124307+073914  &  12$^{\rm{h}}$43$^{\rm{m}}$07$^{\rm{s}}_.$8 & +07$^{\rm{d}}$39$^{\rm{m}}$23$^{\rm{s}}_.$9 & 18.7 & 1,309 &  6.1 &   1.5 &   1.2 &  0.1 &   1.1 &  ... &   ...\\ 
J124317-003841  & 12$^{\rm{h}}$43$^{\rm{m}}$17$^{\rm{s}}_.$9 & -00$^{\rm{d}}$38$^{\rm{m}}$41$^{\rm{s}}_.$5 & 37.9 & 2,651 & 16.4 &   1.7 &   5.6 &  3.4 &   2.2 & 0.17 & 12.86\\ 
J124322+054557  & 12$^{\rm{h}}$43$^{\rm{m}}$22$^{\rm{s}}_.$9 & +05$^{\rm{d}}$45$^{\rm{m}}$52$^{\rm{s}}_.$0 & 25.1 & 1,754 &  3.6 &   6.5 &   9.7 &  3.3 &   6.4 & 0.27 &   ...\\ 
J124350-003344  & 12$^{\rm{h}}$43$^{\rm{m}}$51$^{\rm{s}}_.$0 & -00$^{\rm{d}}$33$^{\rm{m}}$42$^{\rm{s}}_.$5 & 37.6 & 2,633 & 29.9 &   0.5 &   1.7 &  1.0 &   0.8 & 0.19 &  3.08\\ 
J124428+002815  & 12$^{\rm{h}}$44$^{\rm{m}}$28$^{\rm{s}}_.$7 &  +00$^{\rm{d}}$28$^{\rm{m}}$04$^{\rm{s}}_.$0 & 16.9 & 1,183 & 10.1 &  21.4 &  14.4 &  3.4 &  11.0 & 0.17 &  2.38\\ 
J124433-021916  & 12$^{\rm{h}}$44$^{\rm{m}}$33$^{\rm{s}}_.$5 & -02$^{\rm{d}}$19$^{\rm{m}}$16$^{\rm{s}}_.$5 & 22.7 & 1,589 & 13.5 &  12.5 &  15.2 &  5.1 &  10.1 & 0.13 &  1.37\\ 
J124541-020154  & 12$^{\rm{h}}$45$^{\rm{m}}$41$^{\rm{s}}_.$0 &  -02$^{\rm{d}}$02$^{\rm{m}}$09$^{\rm{s}}_.$3 & 23.2 & 1,626 &  3.5 &  12.7 &  16.2 &  4.6 &  11.5 &  ... &   ...\\ 
J124601+042248  &  12$^{\rm{h}}$46$^{\rm{m}}$01$^{\rm{s}}_.$2 &  +04$^{\rm{d}}$23$^{\rm{m}}$03$^{\rm{s}}_.$4 &  9.3 &   649 &  1.9 &   4.0 &   0.8 &  0.0 &   0.8 &  ... &   ...\\ 
J124645+055723  & 12$^{\rm{h}}$46$^{\rm{m}}$45$^{\rm{s}}_.$8 & +05$^{\rm{d}}$57$^{\rm{m}}$19$^{\rm{s}}_.$0 & 11.9 &   833 &  6.9 &   0.7 &   0.2 &  0.1 &   0.2 & 0.19 &   ...\\ 
J124701-013444  &  12$^{\rm{h}}$47$^{\rm{m}}$01$^{\rm{s}}_.$0 & -01$^{\rm{d}}$34$^{\rm{m}}$44$^{\rm{s}}_.$7 & 38.9 & 2,724 & 11.5 &   1.4 &   5.1 &  0.3 &   4.8 & 0.17 &   ...\\ 
J124813-032006  & 12$^{\rm{h}}$48$^{\rm{m}}$12$^{\rm{s}}_.$0 & -03$^{\rm{d}}$19$^{\rm{m}}$49$^{\rm{s}}_.$2 & 16.1 & 1,125 &  6.5 &  10.5 &   6.4 &  2.0 &   4.4 &  ... &   ...\\ 
J125038+012749  & 12$^{\rm{h}}$50$^{\rm{m}}$39$^{\rm{s}}_.$0 & +01$^{\rm{d}}$27$^{\rm{m}}$51$^{\rm{s}}_.$0 & 18.2 & 1,276 &  7.1 &   2.4 &   1.9 &  0.5 &   1.4 & 0.21 & 14.29\\ 

\hline
\end{tabular}

\flushleft
 \textbf{Notes.} {The columns are (1) source name; (2) kinematic centre in R.A. units from {\sc 2dbat}; (3) kinematic centre Dec. units from {\sc 2dbat}; (4) distance to the galaxy from redshift (Mpc); (5) systemic velocity from redshift (\kms)\,; (6) radius of H{\sc i} disk from the {\sc baygaud} analysis, $R_{\rm{HI}}$ (kpc);  (7) integrated H{\sc i} intensities derived from super-profiles in Section 3 (Jy km s$^{-1}$); (8) H{\sc i} mass derived from super-profiles in Section 3 ($10^8 \ \rm{M_{\odot}}$); (9) narrow H{\sc i} mass derived from super-profiles in Section 3 ($10^8 \ \rm{M_{\odot}}$); (10) broad H{\sc i} mass derived from super-profiles in Section 3 ($10^8 \ \rm{M_{\odot}}$); (11) H{\sc i} morphological asymmetry of the galaxy, (12) median values of the Toomre Q parameter values of the galaxy ('*': {\sc moment1} velocity field is used for the rotation curve analysis).}
\end{table*}

\begin{landscape}
\begin{table}
\tiny
\caption{H{\sc i} properties and distances of the paired galaxies in the ASKAP NGC 4636 field.}
\label{long_ngc_pair}

\resizebox{1.3\textwidth}{!}{
\begin{tabular}{cccccccccccccc}
 \hline
 Name & R.A. (J2000) & Dec. (J2000) & $D$ & $V_{\rm{sys}}$ & $R_{\rm{HI}}$ & $S_{\rm{HI}}$ & $M_{\rm{total}}^{\rm{HI}}$ &  $M_{\rm{narrow}}^{\rm{HI}}$ & $M_{\rm{broad}}^{\rm{HI}}$ & $R_{\rm{p}}$ & $\Delta V$ & $A_{\rm{map}}^{\rm{HI}}$ & $Q$\\
  &   &  & (Mpc) & (\kms)\ & (kpc) & (Jy km s$^{-1}$) & ($10^8 \ \rm{M_{\odot}}$) & ($10^8 \ \rm{M_{\odot}}$) & ($10^8 \ \rm{M_{\odot}}$) & (kpc) & (\kms)\ \\
    (1) & (2) & (3) & (4) & (5) & (6) & (7) & (8) & (9) & (10) & (11) & (12) & (13) & (14)\\
 \hline
 
J122710+071549  & 12$^{\rm{h}}$27$^{\rm{m}}$11$^{\rm{s}}_.$3 & +07$^{\rm{d}}$15$^{\rm{m}}$49$^{\rm{s}}_.$2 & 13.3 &   929 & 11.3 &  21.1 &   8.8 &  4.7 &   4.0 & 86 & 68 & 0.20 &   ...\\ 
J122729+073841  & 12$^{\rm{h}}$27$^{\rm{m}}$29$^{\rm{s}}_.$4 & +07$^{\rm{d}}$38$^{\rm{m}}$48$^{\rm{s}}_.$4 & 12.3 &   860 &  4.6 &   3.5 &   1.3 &  0.8 &   0.5 & 86 & 68 & 0.36 &   ...\\ 
J123422+021914  & 12$^{\rm{h}}$34$^{\rm{m}}$21$^{\rm{s}}_.$8 & +02$^{\rm{d}}$19$^{\rm{m}}$25$^{\rm{s}}_.$3 & 25.4 & 1,779 &  7.5 &   3.5 &   5.4 &  0.2 &   5.2 & 58 & 19 &  ... &   ...\\ 
J124508-002747  &  12$^{\rm{h}}$45$^{\rm{m}}$08$^{\rm{s}}_.$4 & -00$^{\rm{d}}$27$^{\rm{m}}$47$^{\rm{s}}_.$4 & 21.9 & 1,534 & 34.3 &  80.5 &  91.2 & 30.0 &  61.2 & 47 & 84 & 0.17 &  9.51\\ 
J124531-003203  & 12$^{\rm{h}}$45$^{\rm{m}}$31$^{\rm{s}}_.$9 & -00$^{\rm{d}}$32$^{\rm{m}}$10$^{\rm{s}}_.$4 & 23.1 & 1,619 & 14.3 &  12.7 &  16.1 &  6.1 &  10.0 & 50/47 & 55/84 & 0.25 &  3.50\\ 
J124548-002600  & 12$^{\rm{h}}$45$^{\rm{m}}$48$^{\rm{s}}_.$5 & -00$^{\rm{d}}$25$^{\rm{m}}$58$^{\rm{s}}_.$8 & 23.9 & 1,676 &  5.6 &   0.9 &   1.2 &  0.9 &   0.3 & 50 & 55 & 0.37 &   ...\\ 
J124747+042017  & 12$^{\rm{h}}$47$^{\rm{m}}$46$^{\rm{s}}_.$6 & +04$^{\rm{d}}$20$^{\rm{m}}$16$^{\rm{s}}_.$1 & 14.1 &   986 & 12.6 &  27.4 &  12.8 &  5.7 &   7.2 & 22 & 36 & 0.24 &  0.28\\ 
J124800+042609  & 12$^{\rm{h}}$47$^{\rm{m}}$59$^{\rm{s}}_.$5 & +04$^{\rm{d}}$26$^{\rm{m}}$16$^{\rm{s}}_.$0 & 14.6 & 1,023 &  3.4 &   1.2 &   0.6 &  0.3 &   0.3 & 27 & 36 &  ... &   ...\\ 

\hline

\end{tabular}}

\flushleft
\textbf{Notes.} {The columns are (1) source name; (2) kinematic centre in R.A. units from {\sc 2dbat}; (3) kinematic centre Dec. units from {\sc 2dbat}; (4) distance to the galaxy from redshift (Mpc); (5) systemic velocity from redshift (\kms)\,; (6) radius of H{\sc i} disk from the {\sc baygaud} analysis, $R_{\rm{HI}}$ (kpc);  (7) integrated H{\sc i} intensities derived from super-profiles in Section 3 (Jy km s$^{-1}$); (8) H{\sc i} mass derived from super-profiles in Section 3 ($10^8 \ \rm{M_{\odot}}$); (9) narrow H{\sc i} mass derived from super-profiles in Section 3 ($10^8 \ \rm{M_{\odot}}$); (10) broad H{\sc i} mass derived from super-profiles in Section 3 ($10^8 \ \rm{M_{\odot}}$); (11) projected distance to the companion galaxy (kpc). For galaxies with two nearby companions, we provide two $R_{\rm{p}}$ values separated by '/'; (12) relative line-of-sight velocity to the companion galaxy (\kms)\,. For galaxies with two nearby companions, we provide two $\Delta V$ values separated by '/'; (13) H{\sc i} morphological asymmetry of the galaxy; (14) median values of the Toomre Q parameter of the galaxy.}

\end{table}
\end{landscape}
\clearpage

\begin{table*}
\tiny
\caption{{H{\sc i} properties and distances of the control galaxies in the ASKAP Norma cluster field.}}
\label{long_norma_control}

\begin{tabular}{cccccccccccc}
 \hline
 Name & R.A. (J2000) & Dec. (J2000) & $D$ & $V_{\rm{sys}}$ & $R_{\rm{HI}}$ & $S_{\rm{HI}}$ & $M_{\rm{total}}^{\rm{HI}}$ &  $M_{\rm{narrow}}^{\rm{HI}}$ & $M_{\rm{broad}}^{\rm{HI}}$ & $A_{\rm{map}}^{\rm{HI}}$ & $Q$\\
 &   &  & (Mpc) & (\kms)\ & (kpc) & (Jy km s$^{-1}$) & ($10^8 \ \rm{M_{\odot}}$) & ($10^8 \ \rm{M_{\odot}}$) & ($10^8 \ \rm{M_{\odot}}$) \\
    (1) & (2) & (3) & (4) & (5) & (6) & (7) & (8) & (9) & (10) & (11) & (12) \\
 \hline
 
J163435-620248  & 16$^{\rm{h}}$34$^{\rm{m}}$34$^{\rm{s}}_.$7 & -62$^{\rm{d}}$02$^{\rm{m}}$45$^{\rm{s}}_.$3 &  63.6 & 4,450 & 28.4 &   5.2 &  49.4 & 10.8 &  38.5 & 0.14 &  3.14\\ 
J163452-603705  & 16$^{\rm{h}}$34$^{\rm{m}}$52$^{\rm{s}}_.$8 &  -60$^{\rm{d}}$37$^{\rm{m}}$07$^{\rm{s}}_.$8 &  46.7 & 3,266 & 32.5 &  15.1 &  77.5 & 18.3 &  59.2 & 0.15 &  0.03\\ 
J163518-581311  & 16$^{\rm{h}}$35$^{\rm{m}}$19$^{\rm{s}}_.$1 &  -58$^{\rm{d}}$13$^{\rm{m}}$09$^{\rm{s}}_.$0 &  22.3 & 1,562 & 13.4 &  10.6 &  12.4 &  1.3 &  11.2 & 0.24 &  4.55\\ 
J163600-611510  &  16$^{\rm{h}}$36$^{\rm{m}}$00$^{\rm{s}}_.$4 & -61$^{\rm{d}}$15$^{\rm{m}}$11$^{\rm{s}}_.$2 &  75.8 & 5,307 & 16.3 &   1.3 &  17.4 & 11.9 &   5.5 & 0.32 &   ...\\ 
J163644-573547  & 16$^{\rm{h}}$36$^{\rm{m}}$42$^{\rm{s}}_.$4 &  -57$^{\rm{d}}$36$^{\rm{m}}$02$^{\rm{s}}_.$5 &  11.6 &   812 &  2.1 &   0.7 &   0.2 &  0.0 &   0.2 &  ... &   ...\\ 
J163650-601632  & 16$^{\rm{h}}$36$^{\rm{m}}$52$^{\rm{s}}_.$2 & -60$^{\rm{d}}$16$^{\rm{m}}$35$^{\rm{s}}_.$0 &  48.6 & 3,400 & 30.4 &  12.9 &  71.6 & 25.5 &  46.1 & 0.18 &  2.35\\ 
J163841-570554  & 16$^{\rm{h}}$38$^{\rm{m}}$39$^{\rm{s}}_.$3 & -57$^{\rm{d}}$05$^{\rm{m}}$53$^{\rm{s}}_.$0 &  72.3 & 5,060 & 22.5 &   2.3 &  28.5 &  4.3 &  24.2 &  ... &   ...\\ 
J163951-602257  & 16$^{\rm{h}}$39$^{\rm{m}}$50$^{\rm{s}}_.$4 & -60$^{\rm{d}}$23$^{\rm{m}}$04$^{\rm{s}}_.$3 &  81.8 & 5,729 & 17.8 &   1.1 &  17.9 &  4.5 &  13.4 & 0.46 &   ...\\ 
J164052-602341  & 16$^{\rm{h}}$40$^{\rm{m}}$52$^{\rm{s}}_.$5 & -60$^{\rm{d}}$23$^{\rm{m}}$39$^{\rm{s}}_.$6 &  75.2 & 5,268 & 43.2 &  10.5 & 140.0 & 36.0 & 104.0 & 0.11 &  5.16\\ 
J164107-605902  &  16$^{\rm{h}}$41$^{\rm{m}}$06$^{\rm{s}}_.$6 & -60$^{\rm{d}}$58$^{\rm{m}}$55$^{\rm{s}}_.$3 &  49.8 & 3,484 & 15.8 &   3.5 &  20.4 &  5.1 &  15.2 & 0.31 &   ...\\ 
J164113-610322  & 16$^{\rm{h}}$41$^{\rm{m}}$12$^{\rm{s}}_.$2 & -61$^{\rm{d}}$03$^{\rm{m}}$25$^{\rm{s}}_.$5 &  16.6 & 1,162 &  3.0 &   0.7 &   0.5 &  0.3 &   0.2 & 0.42 &   ...\\ 
J164130-573717  & 16$^{\rm{h}}$41$^{\rm{m}}$31$^{\rm{s}}_.$4 & -57$^{\rm{d}}$37$^{\rm{m}}$25$^{\rm{s}}_.$1 &  70.9 & 4,962 & 13.6 &   0.9 &  10.4 &  3.8 &   6.6 & 0.46 &   ...\\ 
J164157-622910  & 16$^{\rm{h}}$41$^{\rm{m}}$57$^{\rm{s}}_.$1 &  -62$^{\rm{d}}$29$^{\rm{m}}$09$^{\rm{s}}_.$8 &  62.7 & 4,387 & 22.4 &   3.5 &  32.3 &  1.9 &  30.3 & 0.27 &   ...\\ 
J164206-613441  &  16$^{\rm{h}}$42$^{\rm{m}}$07$^{\rm{s}}_.$0 & -61$^{\rm{d}}$34$^{\rm{m}}$46$^{\rm{s}}_.$2 &  74.9 & 5,244 & 22.1 &   2.8 &  37.0 &  9.4 &  27.6 & 0.17 &   ...\\ 
J164219-595904  & 16$^{\rm{h}}$42$^{\rm{m}}$18$^{\rm{s}}_.$9 &  -59$^{\rm{d}}$59$^{\rm{m}}$04$^{\rm{s}}_.$4 &  78.3 & 5,484 & 19.7 &   1.6 &  22.6 & 12.1 &  10.5 & 0.34 &   ...\\ 
J164416-582056  & 16$^{\rm{h}}$44$^{\rm{m}}$16$^{\rm{s}}_.$9 & -58$^{\rm{d}}$20$^{\rm{m}}$55$^{\rm{s}}_.$5 &  75.5 & 5,282 & 19.3 &   1.6 &  20.9 &  5.6 &  15.3 & 0.20 &   ...\\ 
J164437-605103  & 16$^{\rm{h}}$44$^{\rm{m}}$37$^{\rm{s}}_.$7 &  -60$^{\rm{d}}$51$^{\rm{m}}$01$^{\rm{s}}_.$6 &  75.5 & 5,287 & 19.2 &   1.5 &  20.0 &  1.7 &  18.3 & 0.24 &   ...\\ 
J164508-611614  &  16$^{\rm{h}}$45$^{\rm{m}}$08$^{\rm{s}}_.$7 & -61$^{\rm{d}}$16$^{\rm{m}}$10$^{\rm{s}}_.$2 &  52.6 & 3,683 & 14.3 &   2.4 &  15.4 &  5.0 &  10.4 & 0.27 &   ...\\ 
J164619-614832  & 16$^{\rm{h}}$46$^{\rm{m}}$21$^{\rm{s}}_.$2 & -61$^{\rm{d}}$48$^{\rm{m}}$34$^{\rm{s}}_.$3 &  45.0 & 3,147 & 10.4 &   1.2 &   5.9 &  1.3 &   4.6 & 0.40 &   ...\\ 
J164749-613936  & 16$^{\rm{h}}$47$^{\rm{m}}$48$^{\rm{s}}_.$9 & -61$^{\rm{d}}$39$^{\rm{m}}$35$^{\rm{s}}_.$9 &  75.0 & 5,250 & 25.9 &   3.6 &  48.0 & 21.8 &  26.2 & 0.15 &   ...\\ 
J164826-623040  & 16$^{\rm{h}}$48$^{\rm{m}}$25$^{\rm{s}}_.$9 & -62$^{\rm{d}}$30$^{\rm{m}}$35$^{\rm{s}}_.$4 &  63.3 & 4,429 & 22.0 &   2.5 &  23.6 &  2.4 &  21.3 & 0.22 &   ...\\ 
J164848-614143  & 16$^{\rm{h}}$48$^{\rm{m}}$49$^{\rm{s}}_.$9 & -61$^{\rm{d}}$41$^{\rm{m}}$47$^{\rm{s}}_.$4 &  73.1 & 5,117 & 15.6 &   1.1 &  14.1 &  7.6 &   6.5 & 0.36 &   ...\\ 
J164859-621337  & 16$^{\rm{h}}$48$^{\rm{m}}$59$^{\rm{s}}_.$6 & -62$^{\rm{d}}$13$^{\rm{m}}$24$^{\rm{s}}_.$0 &  64.2 & 4,496 & 15.0 &   1.2 &  11.6 &  4.3 &   7.2 &  ... &   ...\\ 
J164934-600157  & 16$^{\rm{h}}$49$^{\rm{m}}$34$^{\rm{s}}_.$4 &  -60$^{\rm{d}}$02$^{\rm{m}}$06$^{\rm{s}}_.$6 &  67.4 & 4,718 & 13.5 &   0.7 &   7.9 &  6.6 &   1.3 & 0.48 & 14.17*\\ 
J164950-623318  & 16$^{\rm{h}}$49$^{\rm{m}}$49$^{\rm{s}}_.$8 & -62$^{\rm{d}}$33$^{\rm{m}}$22$^{\rm{s}}_.$4 &  61.2 & 4,286 & 22.2 &   3.0 &  26.5 &  9.7 &  16.8 & 0.28 &   ...\\ 
J165030-614932  & 16$^{\rm{h}}$50$^{\rm{m}}$29$^{\rm{s}}_.$2 & -61$^{\rm{d}}$49$^{\rm{m}}$18$^{\rm{s}}_.$0 &  61.5 & 4,308 & 17.4 &   1.6 &  13.9 &  7.0 &   6.9 & 0.46 &   ...\\ 
J165040-622852  & 16$^{\rm{h}}$50$^{\rm{m}}$41$^{\rm{s}}_.$5 & -62$^{\rm{d}}$28$^{\rm{m}}$50$^{\rm{s}}_.$0 &  47.7 & 3,339 & 10.7 &   1.3 &   6.8 &  2.2 &   4.5 & 0.43 &   ...\\ 
J165101-604809  &  16$^{\rm{h}}$51$^{\rm{m}}$05$^{\rm{s}}_.$7 & -60$^{\rm{d}}$47$^{\rm{m}}$58$^{\rm{s}}_.$7 &  46.9 & 3,282 & 27.8 &   4.5 &  23.5 & 10.5 &  13.0 &  ... &   ...\\ 
J165103-600109  &  16$^{\rm{h}}$51$^{\rm{m}}$03$^{\rm{s}}_.$5 &  -60$^{\rm{d}}$01$^{\rm{m}}$06$^{\rm{s}}_.$5 &  82.9 & 5,804 & 27.5 &   3.0 &  48.2 & 14.4 &  33.8 & 0.17 &  4.19\\ 
J165105-585918  &  16$^{\rm{h}}$51$^{\rm{m}}$06$^{\rm{s}}_.$2 & -58$^{\rm{d}}$59$^{\rm{m}}$22$^{\rm{s}}_.$3 &  22.4 & 1,565 & 18.4 &  35.2 &  41.5 & 22.8 &  18.6 & 0.22 &   ...\\ 
J165131-603514  & 16$^{\rm{h}}$51$^{\rm{m}}$31$^{\rm{s}}_.$5 &  -60$^{\rm{d}}$35$^{\rm{m}}$05$^{\rm{s}}_.$9 &  47.7 & 3,342 & 11.2 &   1.5 &   8.0 &  3.0 &   5.0 & 0.38 &   ...\\ 
J165248-585644  & 16$^{\rm{h}}$52$^{\rm{m}}$48$^{\rm{s}}_.$9 & -58$^{\rm{d}}$56$^{\rm{m}}$50$^{\rm{s}}_.$8 &  41.2 & 2,885 &  8.8 &   0.8 &   3.4 &  0.5 &   2.9 & 0.28 &   ...\\ 
J165248-591309  & 16$^{\rm{h}}$52$^{\rm{m}}$46$^{\rm{s}}_.$1 & -59$^{\rm{d}}$12$^{\rm{m}}$55$^{\rm{s}}_.$1 &  21.1 & 1,476 & 31.0 &  58.3 &  61.1 & 19.9 &  41.2 &  ... &  6.90\\ 
J165249-574157  & 16$^{\rm{h}}$52$^{\rm{m}}$48$^{\rm{s}}_.$9 &  -57$^{\rm{d}}$42$^{\rm{m}}$03$^{\rm{s}}_.$4 &  41.6 & 2,909 &  5.7 &   0.6 &   2.6 &  0.8 &   1.8 & 0.28 &   ...\\ 
J165331-595759  & 16$^{\rm{h}}$53$^{\rm{m}}$32$^{\rm{s}}_.$4 &  -59$^{\rm{d}}$58$^{\rm{m}}$01$^{\rm{s}}_.$9 &  66.6 & 4,664 & 15.9 &   1.8 &  18.9 &  4.0 &  14.9 & 0.24 &   ...\\ 
J165348-590530  & 16$^{\rm{h}}$53$^{\rm{m}}$48$^{\rm{s}}_.$8 & -59$^{\rm{d}}$05$^{\rm{m}}$32$^{\rm{s}}_.$1 &  23.6 & 1,650 &  8.8 &   4.4 &   5.7 &  2.2 &   3.5 & 0.19 &  1.27\\ 
J165429-592515  & 16$^{\rm{h}}$54$^{\rm{m}}$28$^{\rm{s}}_.$9 & -59$^{\rm{d}}$25$^{\rm{m}}$13$^{\rm{s}}_.$6 &  21.6 & 1,513 &  6.3 &   2.9 &   3.2 &  2.4 &   0.8 & 0.19 &  2.20\\ 
J165437-602755  & 16$^{\rm{h}}$54$^{\rm{m}}$36$^{\rm{s}}_.$7 & -60$^{\rm{d}}$27$^{\rm{m}}$52$^{\rm{s}}_.$2 &  15.0 & 1,048 &  2.1 &   0.4 &   0.2 &  0.1 &   0.1 & 0.38 &   ...\\ 
J165439-605730  & 16$^{\rm{h}}$54$^{\rm{m}}$38$^{\rm{s}}_.$9 & -60$^{\rm{d}}$57$^{\rm{m}}$35$^{\rm{s}}_.$6 &  44.4 & 3,108 &  9.6 &   1.6 &   7.4 &  3.0 &   4.5 & 0.33 &   ...\\ 
J165455-610856†  & 16$^{\rm{h}}$54$^{\rm{m}}$55$^{\rm{s}}_.$1 & -61$^{\rm{d}}$08$^{\rm{m}}$53$^{\rm{s}}_.$0 &  62.8 & 4,396 &  9.5 &   0.7 &   6.6 &  3.5 &   3.0 & 0.27 &   ...\\ 
J165520-592615  & 16$^{\rm{h}}$55$^{\rm{m}}$19$^{\rm{s}}_.$4 & -59$^{\rm{d}}$26$^{\rm{m}}$13$^{\rm{s}}_.$6 &  70.3 & 4,922 & 14.4 &   1.3 &  15.4 &  8.1 &   7.3 & 0.21 &   ...\\ 
J165531-570010  & 16$^{\rm{h}}$55$^{\rm{m}}$31$^{\rm{s}}_.$2 & -57$^{\rm{d}}$00$^{\rm{m}}$32$^{\rm{s}}_.$3 & 116.3 & 8,139 & 18.1 &   0.5 &  15.3 &  0.7 &  14.6 &  ... &   ...\\ 
J165545-590931  & 16$^{\rm{h}}$55$^{\rm{m}}$44$^{\rm{s}}_.$9 & -59$^{\rm{d}}$09$^{\rm{m}}$36$^{\rm{s}}_.$1 &  38.3 & 2,681 &  3.8 &   0.2 &   0.9 &  0.4 &   0.5 & 0.40 &   ...\\ 
J165559-592439  & 16$^{\rm{h}}$55$^{\rm{m}}$59$^{\rm{s}}_.$3 & -59$^{\rm{d}}$24$^{\rm{m}}$39$^{\rm{s}}_.$9 &  68.5 & 4,794 & 18.4 &   2.1 &  22.8 & 10.0 &  12.8 & 0.24 & 12.84\\ 
J165610-592620  & 16$^{\rm{h}}$56$^{\rm{m}}$11$^{\rm{s}}_.$9 & -59$^{\rm{d}}$26$^{\rm{m}}$37$^{\rm{s}}_.$0 &  36.9 & 2,584 &  9.1 &   1.1 &   3.4 &  0.1 &   3.3 &  ... &   ...\\ 
J165644-622413  & 16$^{\rm{h}}$56$^{\rm{m}}$44$^{\rm{s}}_.$6 &  -62$^{\rm{d}}$24$^{\rm{m}}$07$^{\rm{s}}_.$6 &  72.8 & 5,099 & 33.1 &   5.8 &  72.2 & 16.5 &  55.7 & 0.15 &  6.52\\ 
J165645-620555  & 16$^{\rm{h}}$56$^{\rm{m}}$45$^{\rm{s}}_.$4 & -62$^{\rm{d}}$05$^{\rm{m}}$55$^{\rm{s}}_.$2 & 128.2 & 8,971 & 20.9 &   0.5 &  18.8 &  2.1 &  16.7 & 0.23 &   ...\\ 
J165724-595238  & 16$^{\rm{h}}$57$^{\rm{m}}$22$^{\rm{s}}_.$8 & -59$^{\rm{d}}$52$^{\rm{m}}$39$^{\rm{s}}_.$2 &  45.7 & 3,200 & 20.7 &   7.0 &  34.6 & 13.5 &  21.1 & 0.23 &  2.67\\ 
J165758-624336  & 16$^{\rm{h}}$57$^{\rm{m}}$58$^{\rm{s}}_.$4 & -62$^{\rm{d}}$43$^{\rm{m}}$33$^{\rm{s}}_.$9 &  71.8 & 5,023 & 34.6 &   5.9 &  71.3 & 24.9 &  46.4 & 0.17 &  6.44\\ 
J165804-605308  &  16$^{\rm{h}}$58$^{\rm{m}}$05$^{\rm{s}}_.$7 & -60$^{\rm{d}}$52$^{\rm{m}}$46$^{\rm{s}}_.$3 &  14.8 & 1,037 & 10.1 &  15.4 &   8.0 &  5.0 &   3.0 & 0.39 &  1.11\\ 
J165840-582909  & 16$^{\rm{h}}$58$^{\rm{m}}$41$^{\rm{s}}_.$1 & -58$^{\rm{d}}$29$^{\rm{m}}$10$^{\rm{s}}_.$7 &  87.5 & 6,127 & 23.4 &   1.6 &  28.1 &  8.0 &  20.1 & 0.24 &   ...\\ 
J165847-612501  & 16$^{\rm{h}}$58$^{\rm{m}}$48$^{\rm{s}}_.$1 & -61$^{\rm{d}}$24$^{\rm{m}}$55$^{\rm{s}}_.$7 &  67.7 & 4,738 & 14.1 &   1.5 &  16.0 &  5.6 &  10.4 & 0.36 &   ...\\ 
J165850-611725  & 16$^{\rm{h}}$58$^{\rm{m}}$53$^{\rm{s}}_.$5 & -61$^{\rm{d}}$17$^{\rm{m}}$39$^{\rm{s}}_.$2 &  71.2 & 4,981 & 18.1 &   1.7 &  20.2 &  3.3 &  16.9 &  ... &   ...\\ 
J165901-601241  &        ... &         ... &  16.3 & 1,143 & 34.6 & 150.9 &  95.0 & 34.9 &  60.1 &  ... &   ...\\ 
J165934-623256  & 16$^{\rm{h}}$59$^{\rm{m}}$36$^{\rm{s}}_.$5 &  -62$^{\rm{d}}$32$^{\rm{m}}$43$^{\rm{s}}_.$6 &  69.5 & 4,866 & 21.6 &   2.5 &  28.1 &  7.6 &  20.5 &  ... &   ...\\ 
J165949-584125  &        ... &         ... &  68.1 & 4,768 & 33.8 &   8.5 &  92.6 & 20.7 &  71.9 &  ... &   ...\\ 
J165952-573929  & 16$^{\rm{h}}$59$^{\rm{m}}$52$^{\rm{s}}_.$7 & -57$^{\rm{d}}$39$^{\rm{m}}$43$^{\rm{s}}_.$6 & 109.1 & 7,635 & 16.8 &   0.7 &  19.6 &  5.2 &  14.4 & 0.52 &   ...\\ 
J170039-584218  & 17$^{\rm{h}}$00$^{\rm{m}}$41$^{\rm{s}}_.$3 & -58$^{\rm{d}}$42$^{\rm{m}}$22$^{\rm{s}}_.$2 &  89.0 & 6,227 & 17.4 &   1.3 &  23.5 &  5.6 &  17.9 &  ... & 14.11\\ 
J170103-582824  &  17$^{\rm{h}}$01$^{\rm{m}}$03$^{\rm{s}}_.$7 & -58$^{\rm{d}}$28$^{\rm{m}}$18$^{\rm{s}}_.$5 & 111.9 & 7,835 & 28.6 &   1.6 &  48.3 & 18.3 &  30.0 & 0.22 &   ...\\ 
J170149-603901  & 17$^{\rm{h}}$01$^{\rm{m}}$48$^{\rm{s}}_.$9 & -60$^{\rm{d}}$38$^{\rm{m}}$57$^{\rm{s}}_.$0 &  70.0 & 4,899 & 18.0 &   1.4 &  16.6 &  1.6 &  15.1 & 0.33 & 11.81\\ 
J170232-584322  & 17$^{\rm{h}}$02$^{\rm{m}}$31$^{\rm{s}}_.$7 & -58$^{\rm{d}}$43$^{\rm{m}}$15$^{\rm{s}}_.$4 &  87.6 & 6,130 & 22.3 &   1.3 &  23.2 &  8.3 &  14.9 & 0.24 &   ...\\ 
J170248-602617  & 17$^{\rm{h}}$02$^{\rm{m}}$48$^{\rm{s}}_.$8 & -60$^{\rm{d}}$26$^{\rm{m}}$27$^{\rm{s}}_.$0 &  15.4 & 1,080 &  2.8 &   0.5 &   0.3 &  0.1 &   0.2 & 0.36 &   ...\\ 
J170303-614604  &  17$^{\rm{h}}$03$^{\rm{m}}$03$^{\rm{s}}_.$4 &  -61$^{\rm{d}}$46$^{\rm{m}}$03$^{\rm{s}}_.$6 &  38.9 & 2,725 & 26.9 &  16.4 &  58.5 & 17.6 &  40.9 & 0.17 &  1.64\\ 
J170349-595038  & 17$^{\rm{h}}$03$^{\rm{m}}$49$^{\rm{s}}_.$1 & -59$^{\rm{d}}$50$^{\rm{m}}$43$^{\rm{s}}_.$2 &  84.3 & 5,899 & 20.0 &   1.8 &  29.6 &  9.4 &  20.2 & 0.27 &   ...\\ 
J170546-614959  & 17$^{\rm{h}}$05$^{\rm{m}}$45$^{\rm{s}}_.$1 &  -61$^{\rm{d}}$50$^{\rm{m}}$00$^{\rm{s}}_.$6 &  15.9 & 1,116 &  2.5 &   0.8 &   0.5 &  0.2 &   0.3 & 0.29 &   ...\\ 
J170610-590352  & 17$^{\rm{h}}$06$^{\rm{m}}$10$^{\rm{s}}_.$7 & -59$^{\rm{d}}$03$^{\rm{m}}$52$^{\rm{s}}_.$6 &  88.3 & 6,183 & 18.6 &   0.8 &  15.3 & 10.6 &   4.7 & 0.20 &   ...\\ 
J170642-604341  & 17$^{\rm{h}}$06$^{\rm{m}}$42$^{\rm{s}}_.$0 & -60$^{\rm{d}}$43$^{\rm{m}}$41$^{\rm{s}}_.$8 & 114.7 & 8,026 & 21.5 &   1.2 &  36.4 & 20.0 &  16.4 & 0.14 &   ...\\ 
J170700-600654  & 17$^{\rm{h}}$06$^{\rm{m}}$59$^{\rm{s}}_.$3 &  -60$^{\rm{d}}$07$^{\rm{m}}$01$^{\rm{s}}_.$9 &  61.0 & 4,271 & 14.9 &   1.8 &  16.2 &  9.4 &   6.7 &  ... &   ...\\ 
J170747-595059  & 17$^{\rm{h}}$07$^{\rm{m}}$47$^{\rm{s}}_.$3 &  -59$^{\rm{d}}$51$^{\rm{m}}$01$^{\rm{s}}_.$3 &  67.0 & 4,690 & 26.6 &   6.3 &  66.5 & 10.9 &  55.6 & 0.15 &   ...\\ 
J170842-592322  & 17$^{\rm{h}}$08$^{\rm{m}}$43$^{\rm{s}}_.$3 & -59$^{\rm{d}}$23$^{\rm{m}}$17$^{\rm{s}}_.$5 &  16.1 & 1,131 &  2.0 &   0.5 &   0.3 &  0.1 &   0.2 & 0.24 &   ...\\ 
J170915-613637  & 17$^{\rm{h}}$09$^{\rm{m}}$16$^{\rm{s}}_.$6 & -61$^{\rm{d}}$36$^{\rm{m}}$39$^{\rm{s}}_.$0 &  60.4 & 4,227 & 27.2 &   5.0 &  43.3 & 14.0 &  29.4 & 0.27 &  4.80\\ 
J170937-602429  & 17$^{\rm{h}}$09$^{\rm{m}}$35$^{\rm{s}}_.$9 & -60$^{\rm{d}}$24$^{\rm{m}}$19$^{\rm{s}}_.$5 &  67.4 & 4,720 & 13.7 &   1.2 &  12.3 &  3.0 &   9.3 & 0.42 &   ...\\ 
J171015-615505  & 17$^{\rm{h}}$10$^{\rm{m}}$15$^{\rm{s}}_.$6 &  -61$^{\rm{d}}$55$^{\rm{m}}$04$^{\rm{s}}_.$9 &  68.9 & 4,824 & 31.5 &   6.9 &  77.2 & 17.2 &  60.0 & 0.20 &  2.94\\ 
J171040-605237  & 17$^{\rm{h}}$10$^{\rm{m}}$40$^{\rm{s}}_.$5 & -60$^{\rm{d}}$52$^{\rm{m}}$33$^{\rm{s}}_.$9 &  20.2 & 1,415 &  2.8 &   0.5 &   0.5 &  0.2 &   0.3 & 0.16 &   ...\\ 

\end{tabular}
\end{table*}
\clearpage

\begin{table*}
\tiny
\contcaption{}
\label{long_norma_control}
\centering
\begin{tabular}{cccccccccccc}
 \hline
 Name & R.A. (J2000) & Dec. (J2000) & $D$ & $V_{\rm{sys}}$ & $R_{\rm{HI}}$ & $S_{\rm{HI}}$ & $M_{\rm{total}}^{\rm{HI}}$ &  $M_{\rm{narrow}}^{\rm{HI}}$ & $M_{\rm{broad}}^{\rm{HI}}$ & $A_{\rm{map}}^{\rm{HI}}$ & $Q$\\
 &   &  & (Mpc) & (\kms)\ & (kpc) & (Jy km s$^{-1}$) & ($10^8 \ \rm{M_{\odot}}$) & ($10^8 \ \rm{M_{\odot}}$) & ($10^8 \ \rm{M_{\odot}}$) \\
    (1) & (2) & (3) & (4) & (5) & (6) & (7) & (8) & (9) & (10) & (11) & (12) \\
 \hline
 
J171120-603454  & 17$^{\rm{h}}$11$^{\rm{m}}$20$^{\rm{s}}_.$8 & -60$^{\rm{d}}$34$^{\rm{m}}$55$^{\rm{s}}_.$4 &  71.0 & 4,968 & 13.0 &   0.9 &  11.1 &  4.1 &   7.0 & 0.23 &   ...\\ 
J171134-610802  & 17$^{\rm{h}}$11$^{\rm{m}}$34$^{\rm{s}}_.$4 &  -61$^{\rm{d}}$08$^{\rm{m}}$08$^{\rm{s}}_.$4 &  38.1 & 2,665 &  6.2 &   0.9 &   3.2 &  0.3 &   2.8 & 0.25 &   ...\\ 
J171309-603124  & 17$^{\rm{h}}$13$^{\rm{m}}$10$^{\rm{s}}_.$0 & -60$^{\rm{d}}$31$^{\rm{m}}$28$^{\rm{s}}_.$6 &  69.1 & 4,837 & 19.9 &   1.4 &  15.7 &  8.7 &   7.0 & 0.39 &   ...\\ 
J171312-612310  & 17$^{\rm{h}}$13$^{\rm{m}}$10$^{\rm{s}}_.$4 & -61$^{\rm{d}}$23$^{\rm{m}}$13$^{\rm{s}}_.$6 & 115.8 & 8,108 & 23.5 &   1.6 &  49.9 & 22.7 &  27.2 & 0.27 &   ...\\ 
J171338-590621  & 17$^{\rm{h}}$13$^{\rm{m}}$39$^{\rm{s}}_.$7 & -59$^{\rm{d}}$06$^{\rm{m}}$24$^{\rm{s}}_.$8 &  50.3 & 3,521 & 33.5 &  11.4 &  67.8 & 23.4 &  44.4 & 0.22 &  1.23\\ 
J171339-615822  & 17$^{\rm{h}}$13$^{\rm{m}}$39$^{\rm{s}}_.$8 & -61$^{\rm{d}}$58$^{\rm{m}}$19$^{\rm{s}}_.$2 &  64.0 & 4,480 & 14.3 &   1.5 &  14.1 &  3.8 &  10.3 & 0.31 &   ...\\ 
J171530-601427  & 17$^{\rm{h}}$15$^{\rm{m}}$30$^{\rm{s}}_.$9 & -60$^{\rm{d}}$14$^{\rm{m}}$35$^{\rm{s}}_.$1 &  83.3 & 5,829 & 12.4 &   0.5 &   8.8 &  7.3 &   1.5 &  ... &   ...\\ 
J171538-602409  & 17$^{\rm{h}}$15$^{\rm{m}}$32$^{\rm{s}}_.$3 & -60$^{\rm{d}}$23$^{\rm{m}}$54$^{\rm{s}}_.$3 &  83.0 & 5,813 & 19.5 &   1.5 &  23.8 & 11.4 &  12.4 &  ... &   ...\\ 
J171558-590923  & 17$^{\rm{h}}$15$^{\rm{m}}$57$^{\rm{s}}_.$9 & -59$^{\rm{d}}$09$^{\rm{m}}$23$^{\rm{s}}_.$9 &  50.0 & 3,503 & 11.1 &   1.5 &   8.8 &  4.7 &   4.1 & 0.39 &   ...\\ 
J171804-575135  &  17$^{\rm{h}}$18$^{\rm{m}}$07$^{\rm{s}}_.$1 & -57$^{\rm{d}}$51$^{\rm{m}}$35$^{\rm{s}}_.$1 &  15.5 & 1,084 & 20.6 &  52.9 &  29.9 &  7.9 &  22.0 & 0.25 &  1.20\\ 
J171850-601104  & 17$^{\rm{h}}$18$^{\rm{m}}$51$^{\rm{s}}_.$6 &  -60$^{\rm{d}}$11$^{\rm{m}}$00$^{\rm{s}}_.$2 &  65.9 & 4,615 & 16.8 &   1.7 &  17.6 & 10.5 &   7.1 & 0.43 &   ...\\ 
J171855-613632  & 17$^{\rm{h}}$18$^{\rm{m}}$55$^{\rm{s}}_.$6 & -61$^{\rm{d}}$36$^{\rm{m}}$32$^{\rm{s}}_.$5 & 107.3 & 7,509 & 35.5 &   3.0 &  80.2 & 26.8 &  53.4 & 0.18 &   ...\\ 
J171924-615016  & 17$^{\rm{h}}$19$^{\rm{m}}$25$^{\rm{s}}_.$2 & -61$^{\rm{d}}$50$^{\rm{m}}$10$^{\rm{s}}_.$4 & 104.8 & 7,333 & 28.7 &   1.7 &  44.6 & 10.7 &  34.0 & 0.23 &   ...\\ 
J172007-600928  &  17$^{\rm{h}}$20$^{\rm{m}}$08$^{\rm{s}}_.$0 & -60$^{\rm{d}}$09$^{\rm{m}}$26$^{\rm{s}}_.$1 &  64.5 & 4,513 & 36.7 &   8.9 &  87.8 & 12.9 &  74.9 & 0.14 &  5.32\\ 
 
 \hline
 
\end{tabular}
\flushleft
 \textbf{Notes.} {The columns are (1) source name; (2) kinematic centre in R.A. units from {\sc 2dbat}; (3) kinematic centre Dec. units from {\sc 2dbat}; (4) distance to the galaxy from redshift (Mpc); (5) systemic velocity from redshift (\kms)\,; (6) radius of H{\sc i} disk from the {\sc baygaud} analysis, $R_{\rm{HI}}$ (kpc);  (7) integrated H{\sc i} intensities derived from super-profiles in Section 3 (Jy km s$^{-1}$); (8) H{\sc i} mass derived from super-profiles in Section 3 ($10^8 \ \rm{M_{\odot}}$); (9) narrow H{\sc i} mass derived from super-profiles in Section 3 ($10^8 \ \rm{M_{\odot}}$); (10) broad H{\sc i} mass derived from super-profiles in Section 3 ($10^8 \ \rm{M_{\odot}}$); (11) H{\sc i} morphological asymmetry of the galaxy; (12) median values of the Toomre Q parameter values of the galaxy ('*': {\sc moment1} velocity field is used for the rotation curve analysis). '†' indicates the galaxy that have similar velocity widths of the narrow and broad components in the H{\sc i} suepr-profile analysis.}

\end{table*}
\clearpage

\begin{landscape}
\begin{table}
\tiny
\caption{H{\sc i} properties and distances of the paired galaxies in the ASKAP Norma cluster field.}
\label{long_norma_pair}

\resizebox{1.3\textwidth}{!}{
\begin{tabular}{cccccccccccccc}
 \hline
 Name & R.A. (J2000) & Dec. (J2000) & $D$ & $V_{\rm{sys}}$ & $R_{\rm{HI}}$ & $S_{\rm{HI}}$ & $M_{\rm{total}}^{\rm{HI}}$ &  $M_{\rm{narrow}}^{\rm{HI}}$ & $M_{\rm{broad}}^{\rm{HI}}$ & $R_{\rm{p}}$ & $\Delta V$ & $A_{\rm{map}}^{\rm{HI}}$ & $Q$\\
 &   &  & (Mpc) & (\kms)\ & (kpc) & (Jy km s$^{-1}$) & ($10^8 \ \rm{M_{\odot}}$) & ($10^8 \ \rm{M_{\odot}}$) & ($10^8 \ \rm{M_{\odot}}$) & (kpc) & (\kms)\ \\
    (1) & (2) & (3) & (4) & (5) & (6) & (7) & (8) & (9) & (10) & (11) & (12) & (13) & (14)\\
 \hline
 
J163754-564907  & 16$^{\rm{h}}$37$^{\rm{m}}$55$^{\rm{s}}_.$7 & -56$^{\rm{d}}$48$^{\rm{m}}$59$^{\rm{s}}_.$2 &  21.1 & 1,481 &  4.8 &   1.0 &   1.0 &  0.5 &   0.5 &  76 &  14 &  ... &   ...\\ 
J163834-601517  & 16$^{\rm{h}}$38$^{\rm{m}}$34$^{\rm{s}}_.$2 & -60$^{\rm{d}}$15$^{\rm{m}}$14$^{\rm{s}}_.$2 &  49.2 & 3,446 &  9.7 &   1.5 &   8.4 &  2.4 &   6.0 &  91 &  91 & 0.23 &   ...\\ 
J163924-565221  & 16$^{\rm{h}}$39$^{\rm{m}}$24$^{\rm{s}}_.$7 & -56$^{\rm{d}}$52$^{\rm{m}}$18$^{\rm{s}}_.$4 &  20.9 & 1,467 & 15.0 &  17.1 &  17.7 &  5.4 &  12.3 &  76 &  14 & 0.17 &  1.15\\ 
J164249-610527* & 16$^{\rm{h}}$42$^{\rm{m}}$49$^{\rm{s}}_.$7 & -61$^{\rm{d}}$05$^{\rm{m}}$24$^{\rm{s}}_.$9 &  63.8 & 4,463 & 22.3 &   4.1 &  39.0 &  3.7 &  35.3 & ... & ... & 0.13 &   ...\\ 
J164739-570817  & 16$^{\rm{h}}$47$^{\rm{m}}$38$^{\rm{s}}_.$7 & -57$^{\rm{d}}$08$^{\rm{m}}$13$^{\rm{s}}_.$1 &  11.5 &   804 &  3.9 &   4.3 &   1.3 &  0.2 &   1.2 &  62 &  40 & 0.35 &   ...\\ 
J164740-600851* & 16$^{\rm{h}}$47$^{\rm{m}}$40$^{\rm{s}}_.$4 & -60$^{\rm{d}}$08$^{\rm{m}}$54$^{\rm{s}}_.$7 &  46.6 & 3,261 & 27.2 &  12.5 &  64.1 & 11.6 &  52.5 & ... & ... & 0.16 &  4.09\\ 
J165419-615311* & 16$^{\rm{h}}$54$^{\rm{m}}$19$^{\rm{s}}_.$9 &  -61$^{\rm{d}}$53$^{\rm{m}}$03$^{\rm{s}}_.$3 &  64.6 & 4,522 & 20.3 &   3.2 &  31.8 &  4.1 &  27.7 & ... & ... & 0.20 &   ...\\ 
J165731-583412* & 16$^{\rm{h}}$57$^{\rm{m}}$30$^{\rm{s}}_.$2 & -58$^{\rm{d}}$34$^{\rm{m}}$28$^{\rm{s}}_.$7 &  39.9 & 2,791 & 23.3 &  12.6 &  47.3 & 27.7 &  19.6 & ... & ... & 0.24 &   ...\\ 
J170550-620939  & 17$^{\rm{h}}$05$^{\rm{m}}$54$^{\rm{s}}_.$9 & -62$^{\rm{d}}$09$^{\rm{m}}$50$^{\rm{s}}_.$5 &  21.4 & 1,494 &  4.5 &   0.8 &   0.9 &  0.3 &   0.6 &  58 &  14 &  ... &   ...\\ 
J170659-620502  & 17$^{\rm{h}}$06$^{\rm{m}}$59$^{\rm{s}}_.$8 &  -62$^{\rm{d}}$05$^{\rm{m}}$05$^{\rm{s}}_.$0 &  21.6 & 1,509 & 26.7 &  52.2 &  57.2 & 15.5 &  41.7 &  58 &  14 &  ... &  3.35\\ 
J171500-601723* &        ... &         ... &  84.4 & 5,907 & 24.7 &   1.9 &  31.2 &  1.6 &  29.7 & ... & ... &  ... &   ...\\ 

\hline

\end{tabular}}

\flushleft
\textbf{Notes.} {The columns are (1) source name; (2) kinematic centre in R.A. units from {\sc 2dbat}; (3) kinematic centre Dec. units from {\sc 2dbat}; (4) distance to the galaxy from redshift (Mpc); (5) systemic velocity from redshift (\kms)\,; (6) radius of H{\sc i} disk from the {\sc baygaud} analysis, $R_{\rm{HI}}$ (kpc);  (7) integrated H{\sc i} intensities derived from super-profiles in Section 3 (Jy km s$^{-1}$); (8) H{\sc i} mass derived from super-profiles in Section 3 ($10^8 \ \rm{M_{\odot}}$); (9) narrow H{\sc i} mass derived from super-profiles in Section 3 ($10^8 \ \rm{M_{\odot}}$); (10) broad H{\sc i} mass derived from super-profiles in Section 3 ($10^8 \ \rm{M_{\odot}}$); (11) projected distance to the companion galaxy (kpc); (12) relative line-of-sight velocity to the companion galaxy (\kms)\,; (13) H{\sc i} morphological asymmetry of the galaxy; (14) median values of the Toomre Q parameter of the galaxy. '*' indicates the visually identified galaxy pair.}

\end{table}
\end{landscape}

\clearpage


\begin{figure*}
    \centering
    \includegraphics[width=8.5cm]{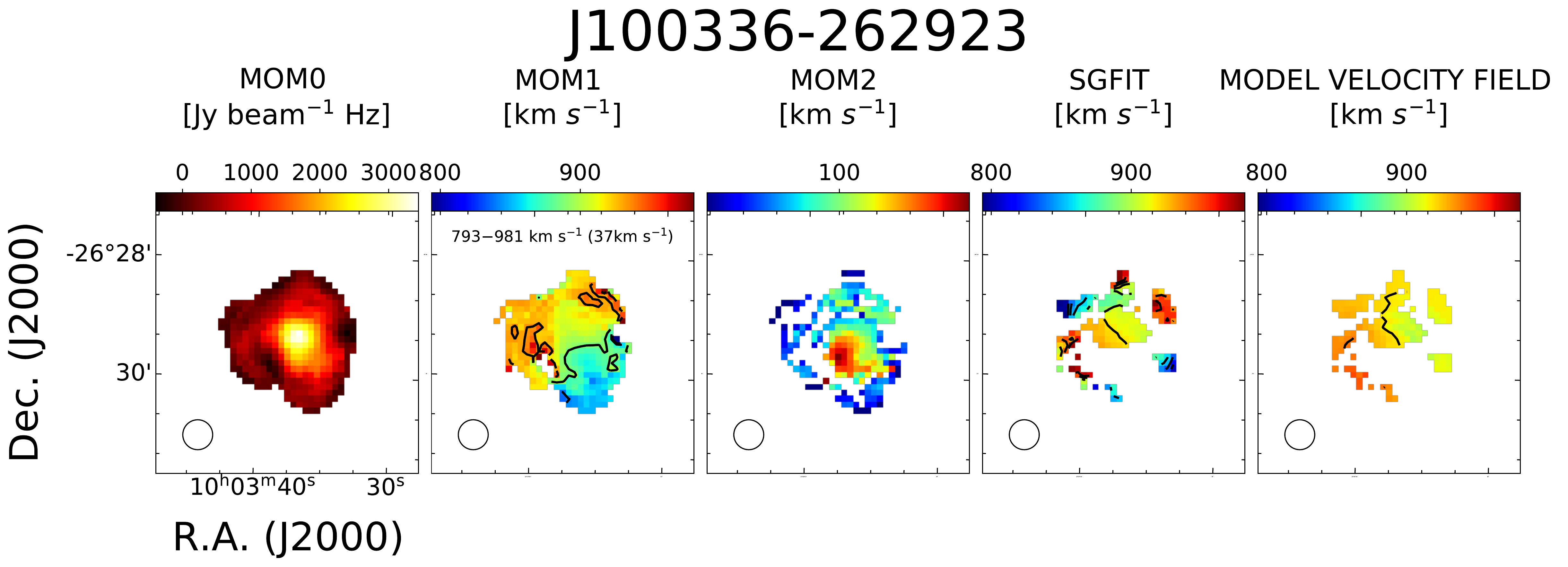}
    \vspace{0.5cm}
    \hspace{0.5cm}
    \includegraphics[width=8.5cm]{Figure/Hydra_J100342-270137.pdf}
    \includegraphics[width=8.5cm]{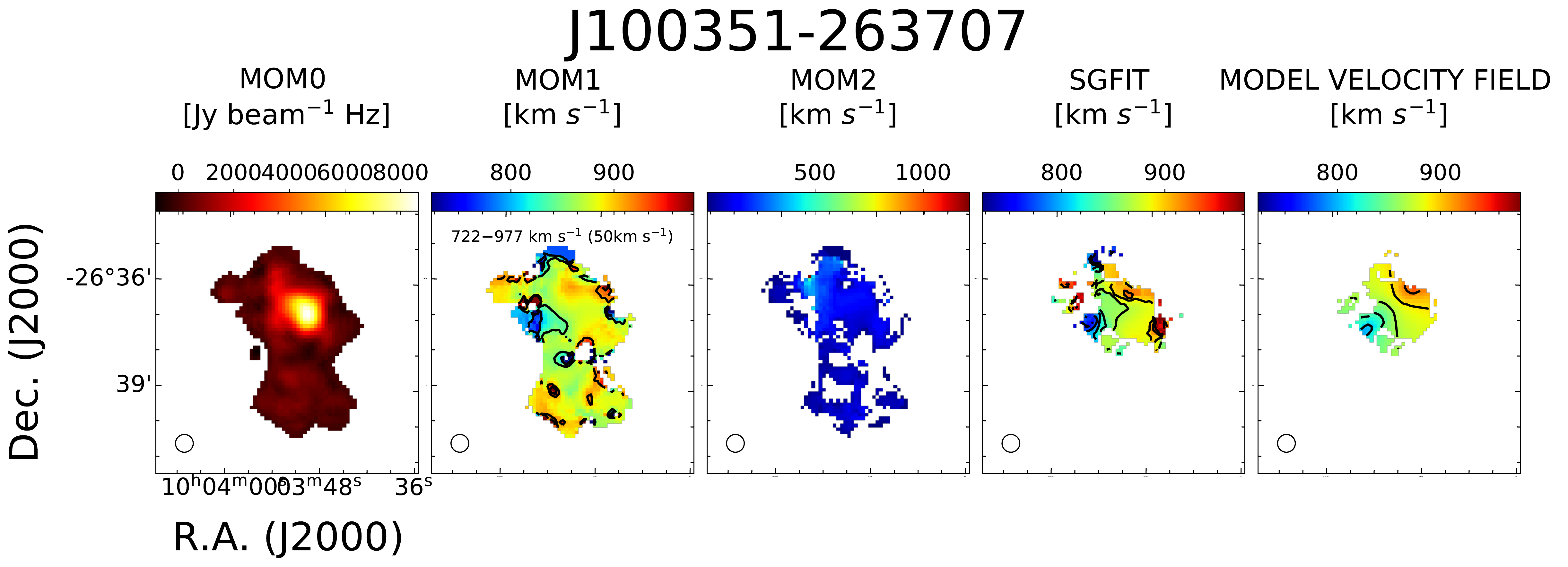}
    \vspace{0.5cm}
    \hspace{0.5cm}
    \includegraphics[width=8.5cm]{Figure/Hydra_J100426-282638.pdf}
    \includegraphics[width=8.5cm]{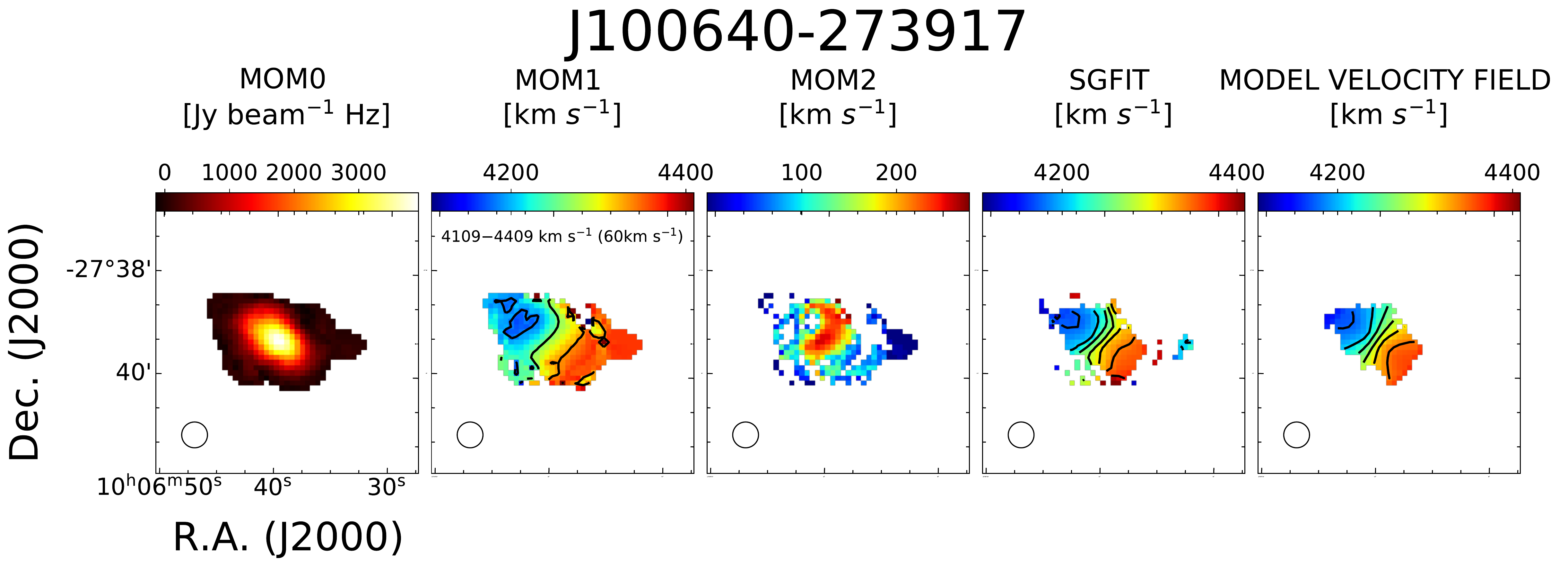}
    \vspace{0.5cm}
    \hspace{0.5cm}
    \includegraphics[width=8.5cm]{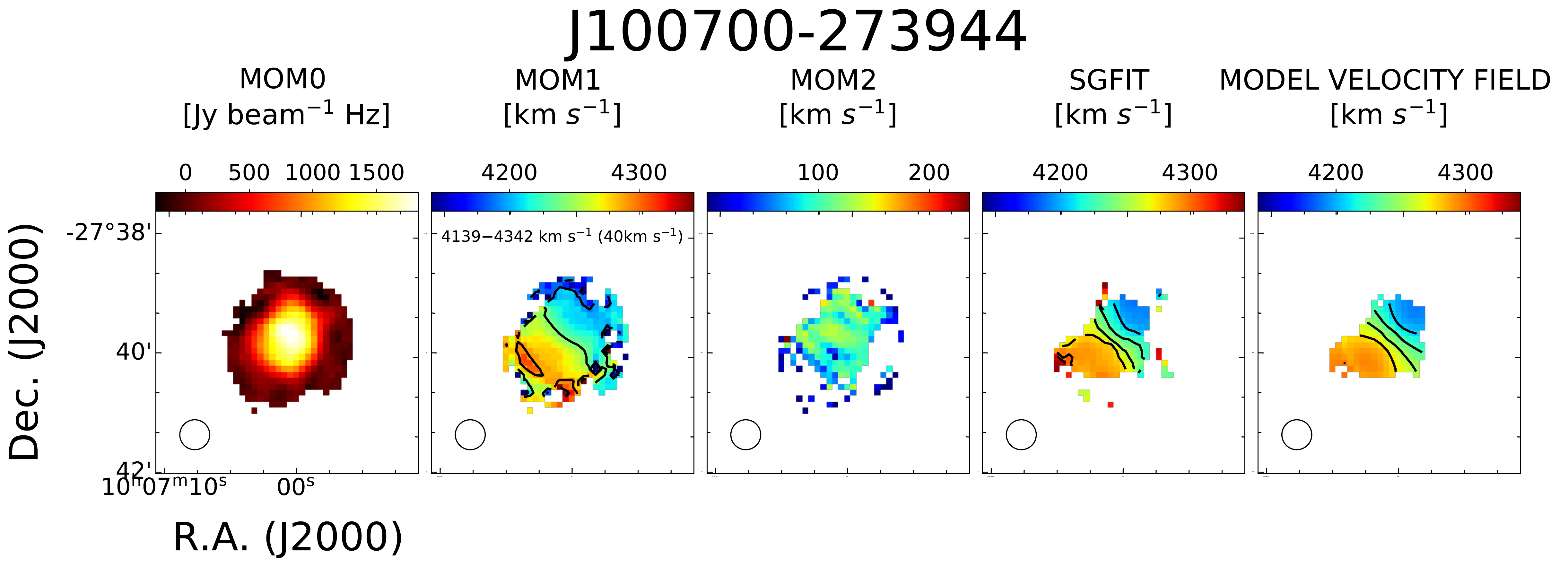}
    \includegraphics[width=8.5cm]{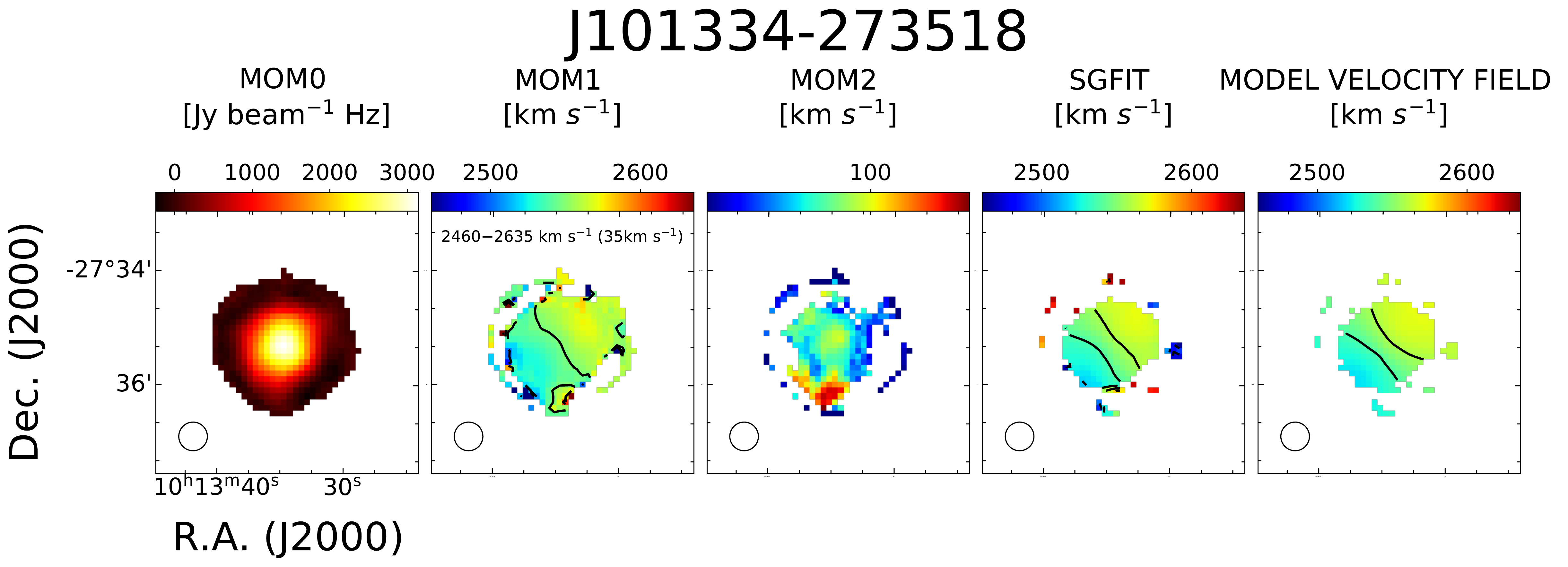}
    \vspace{0.5cm}
    \hspace{0.5cm}
    \includegraphics[width=8.5cm]{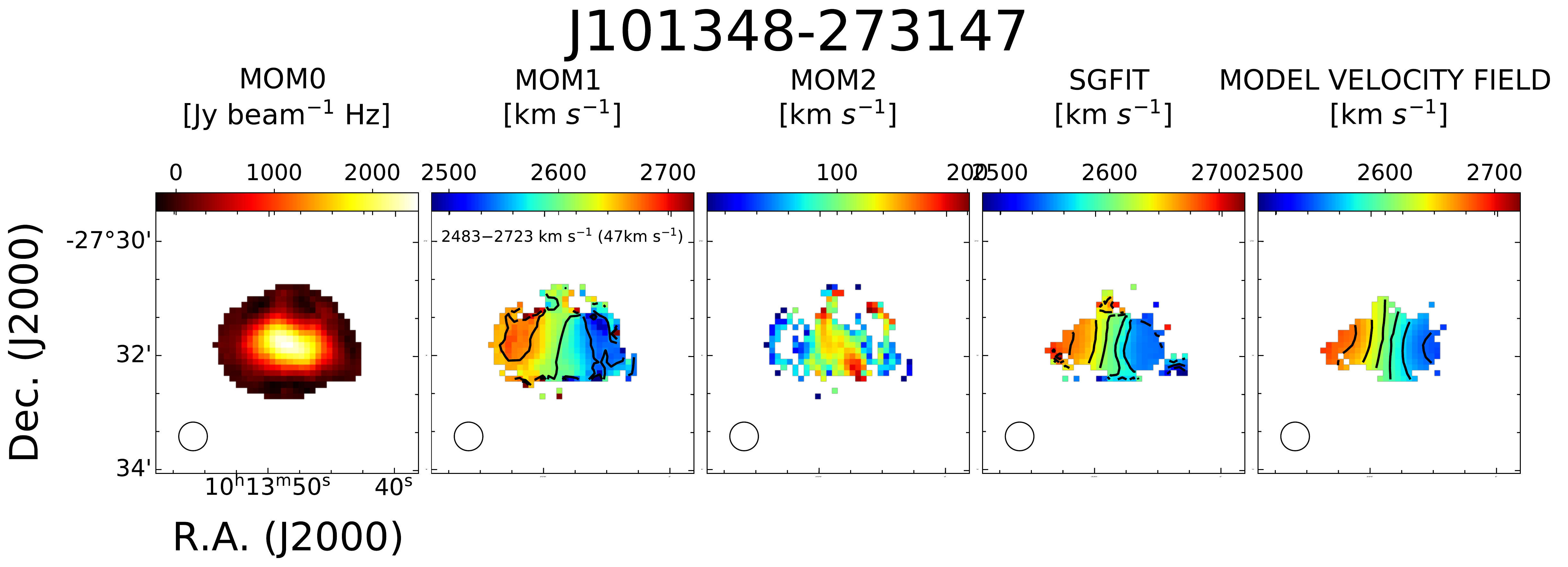}
    \includegraphics[width=8.5cm]{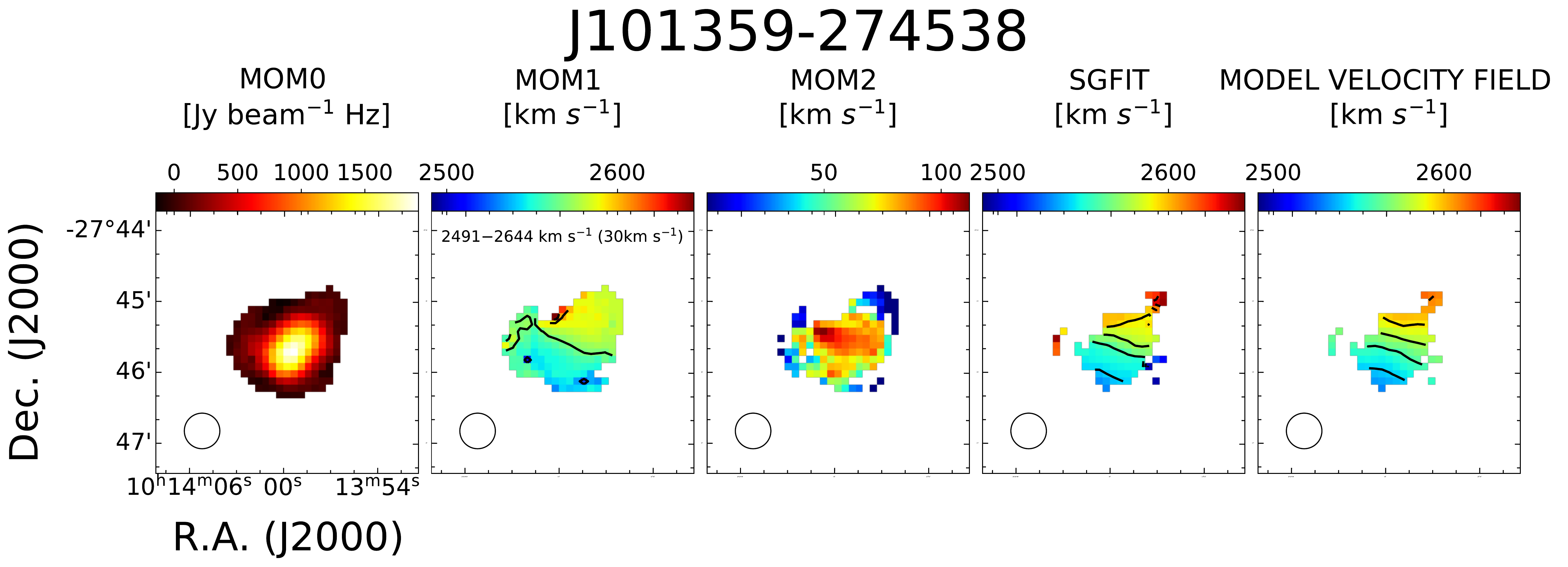}
    \vspace{0.5cm}
    \hspace{0.5cm}
    \includegraphics[width=8.5cm]{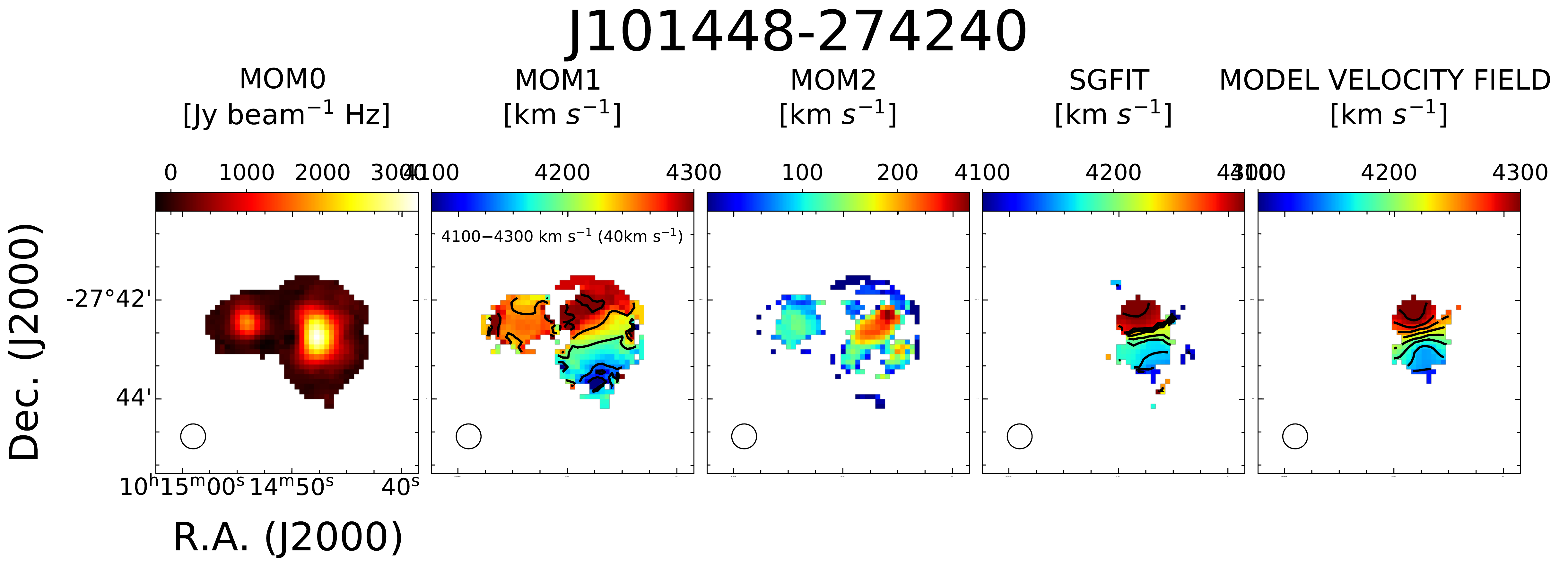}
    \includegraphics[width=8.5cm]{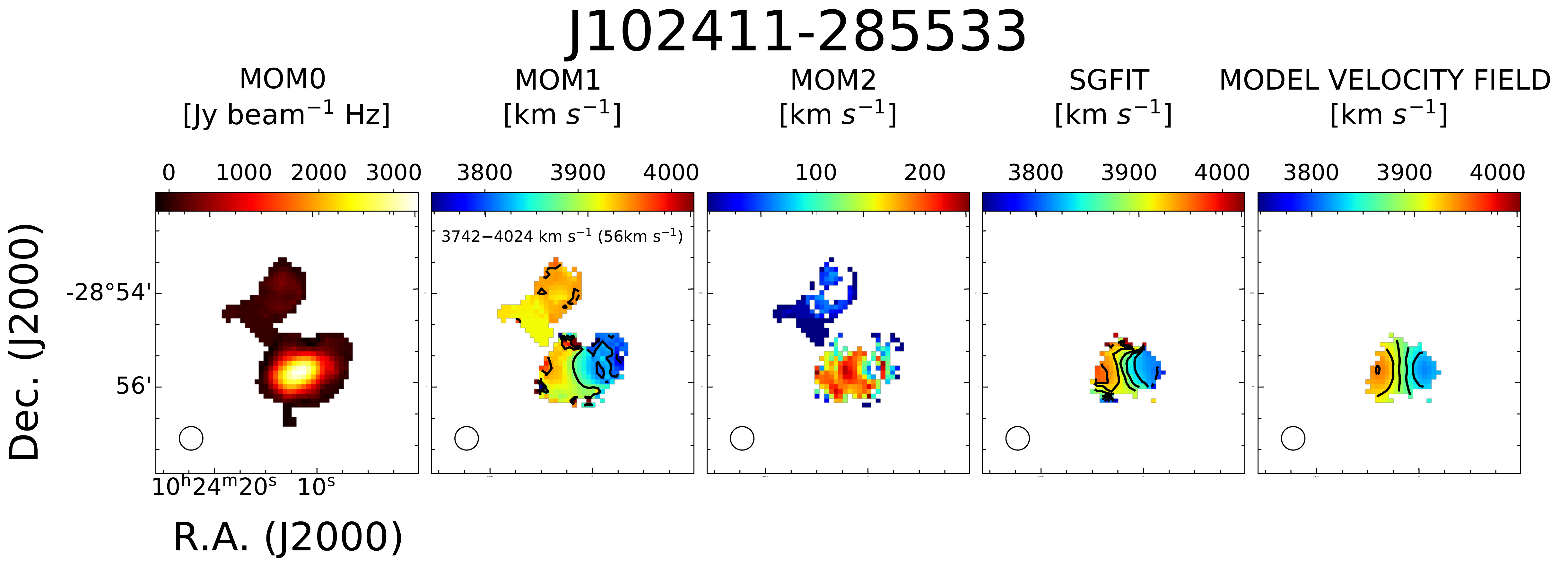}
    \vspace{0.5cm}
    \hspace{0.5cm}
    \includegraphics[width=8.5cm]{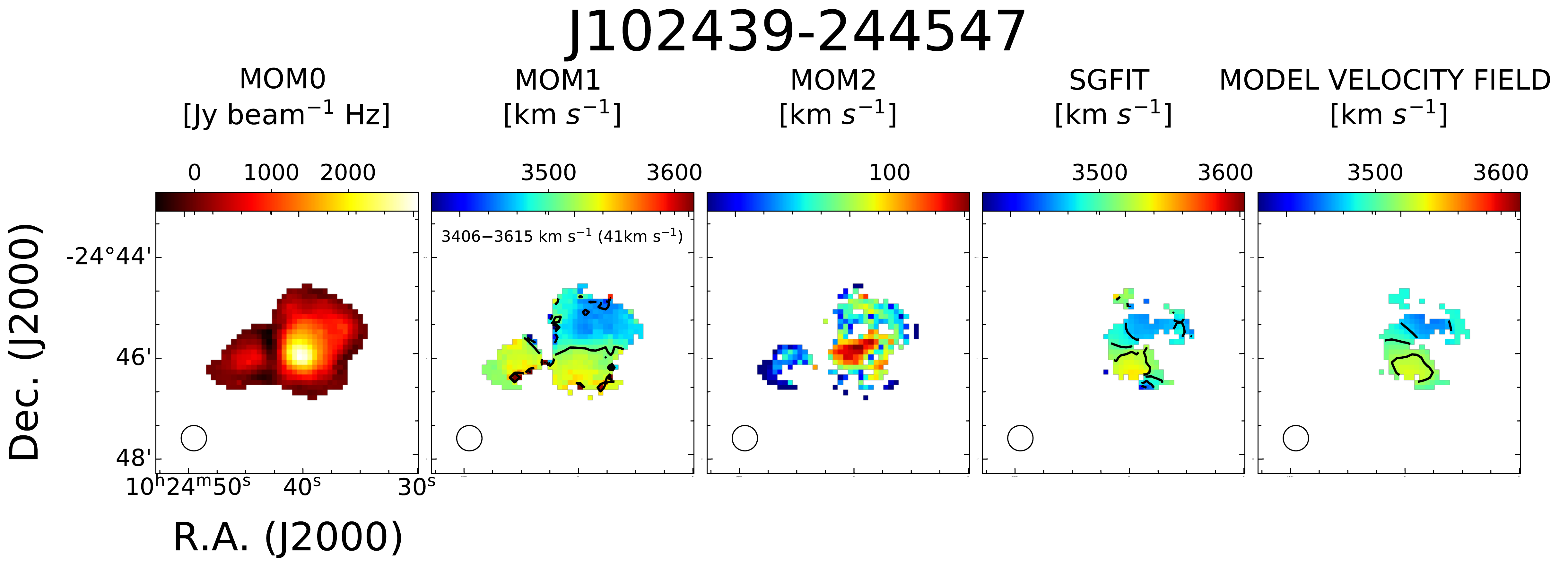}
    \caption{H{\sc i} images of the galaxy pairs (ordered by the WALLABY ID) in the ASKAP Hydra I cluster field: ‘MOM0’ (SoFiA2 integrated intensity map, {\sc moment0}), ‘MOM1’ (SoFiA2 velocity field map , {\sc moment1}), ‘MOM2’ (SoFiA2, velocity dispersion map, {\sc moment2}), SGFIT ({\sc baygaud} Single Gaussian fitting (SGfit) velocity field map), and ‘MODEL VELOCITY FIELD’ ({\sc 2dbat} model velocity field map). The contour levels for each velocity field are denoted in the second panel of each figure. The ASKAP beam is shown as an ellipse on the bottom-left corner of each panel.}\label{figA1}
   
\end{figure*}

\begin{figure*}
\ContinuedFloat
    \centering
    \includegraphics[width=8.5cm]{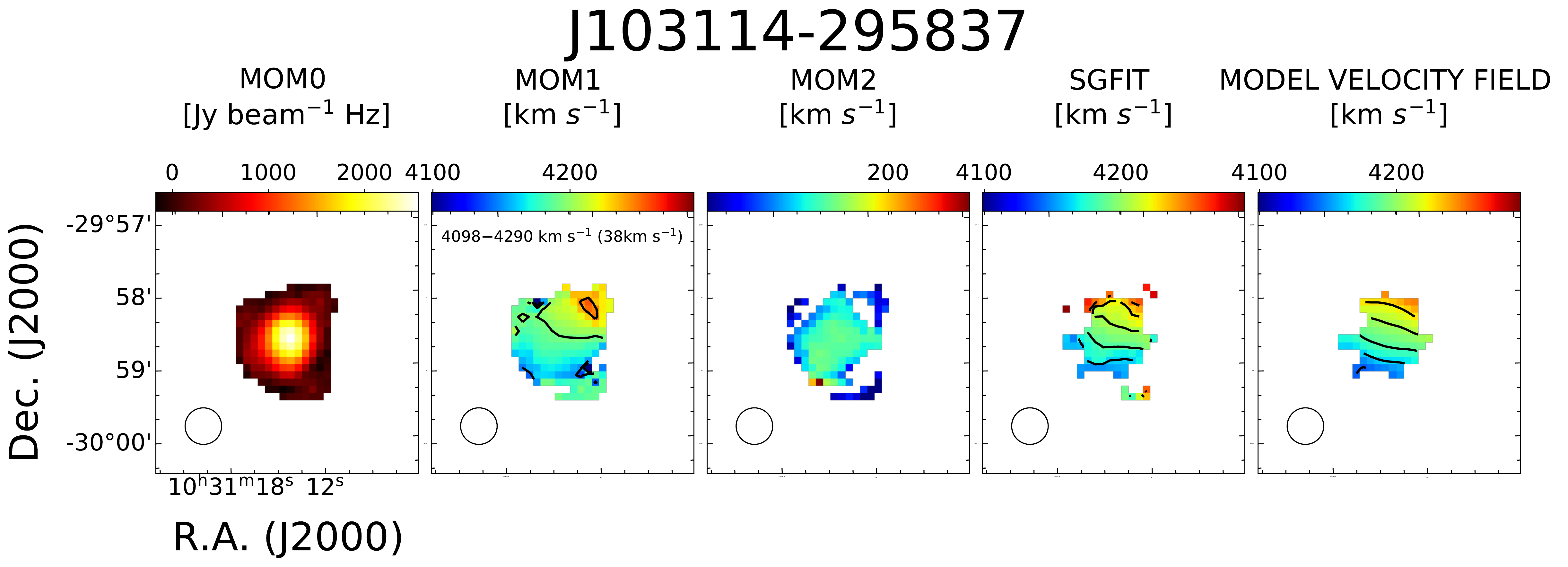}
    \vspace{0.5cm}
    \hspace{0.5cm}
    \includegraphics[width=8.5cm]{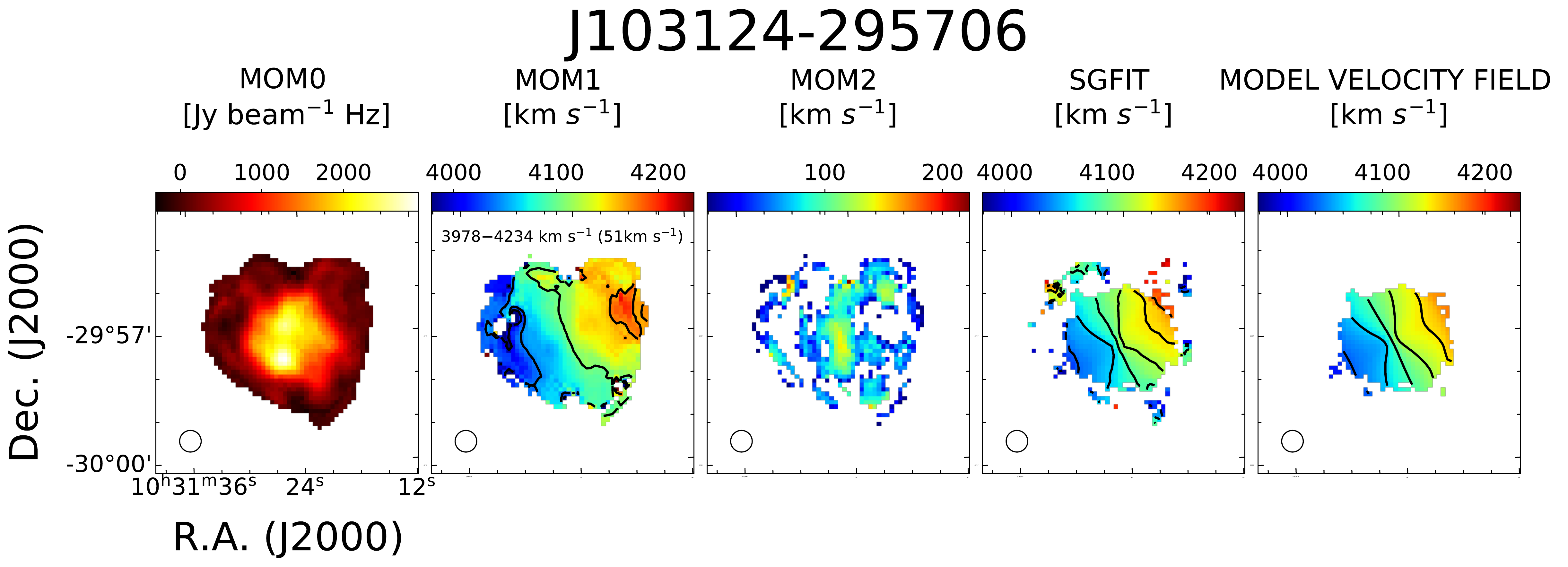}
    \includegraphics[width=8.5cm]{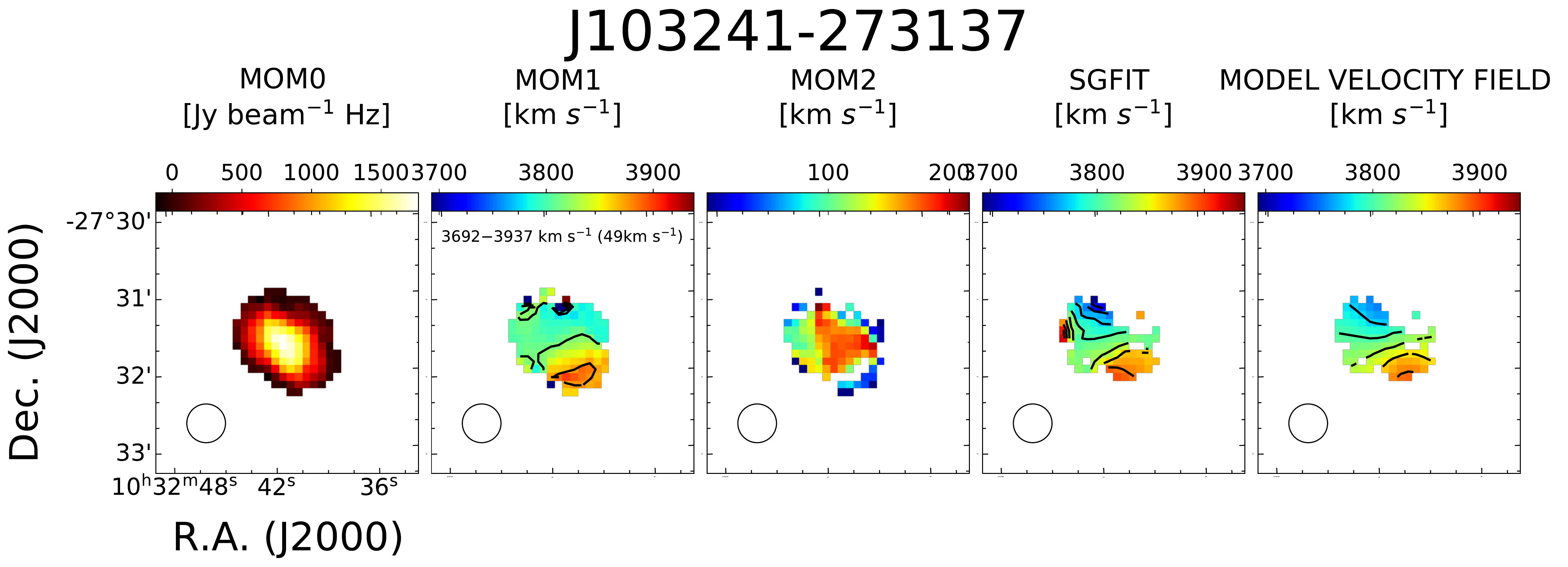}
    \vspace{0.5cm}
    \hspace{0.5cm}
    \includegraphics[width=8.5cm]{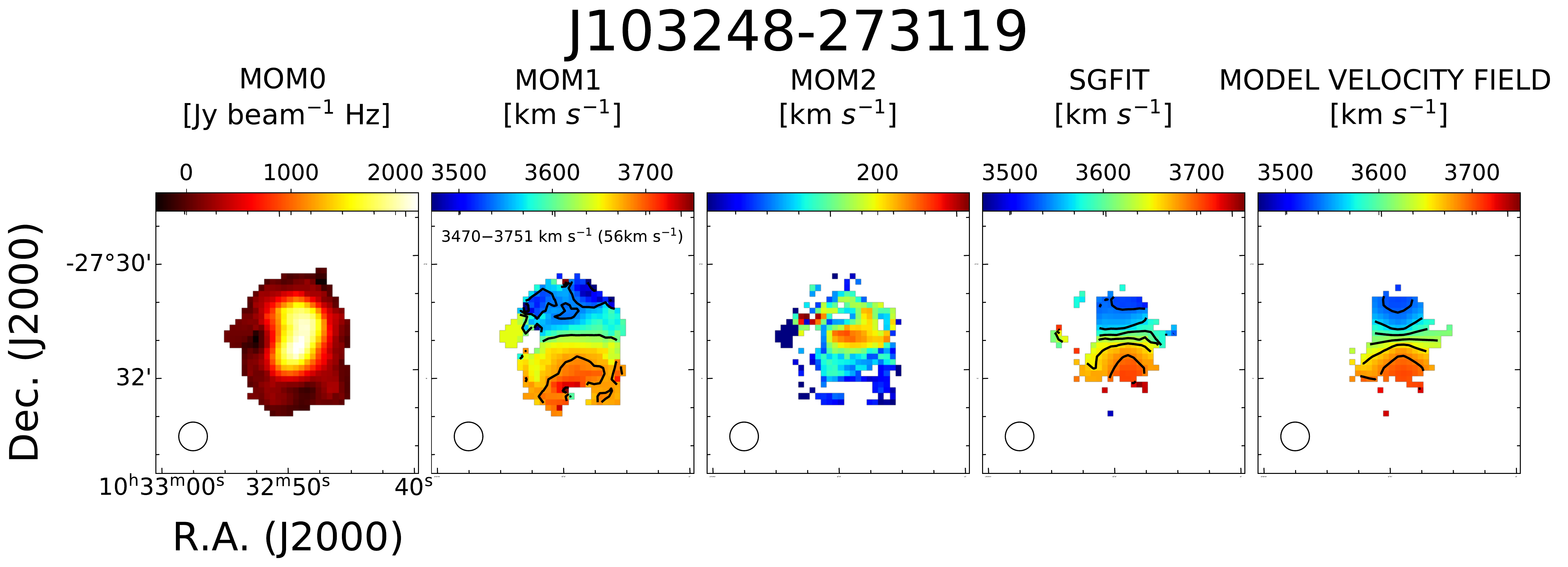}
    \includegraphics[width=8.5cm]{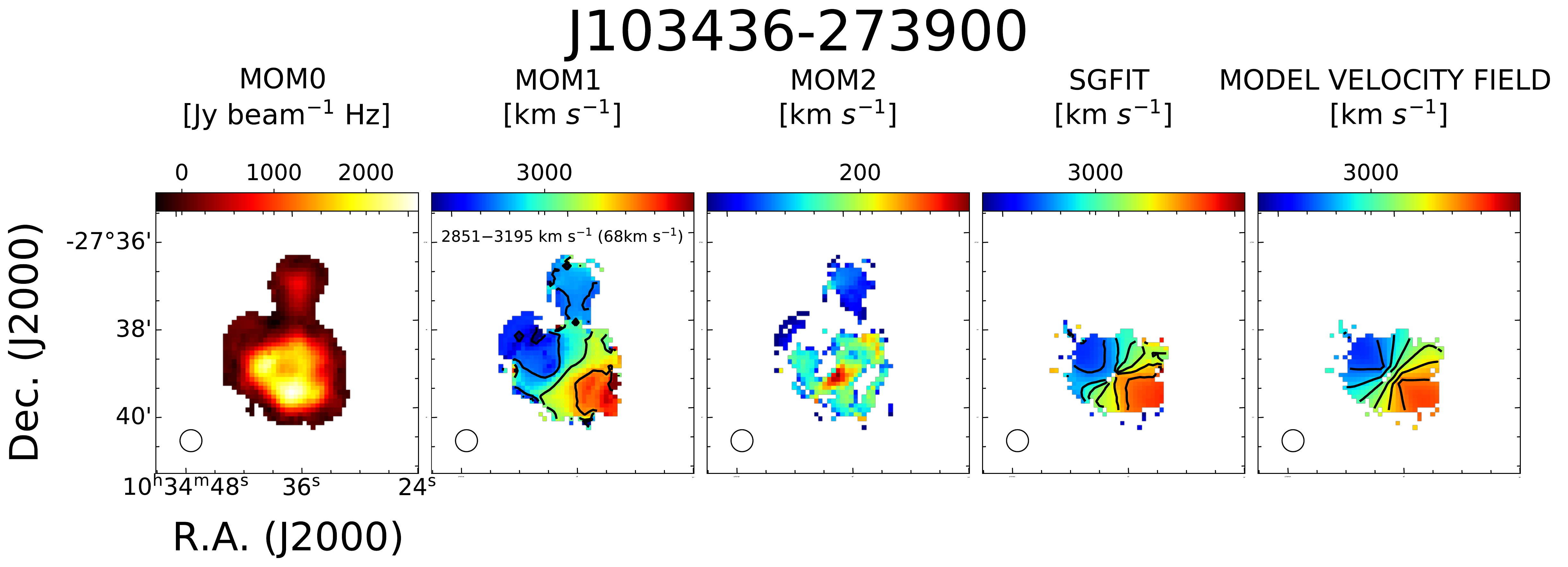}
    \vspace{0.5cm}
    \hspace{0.5cm}
    \includegraphics[width=8.5cm]{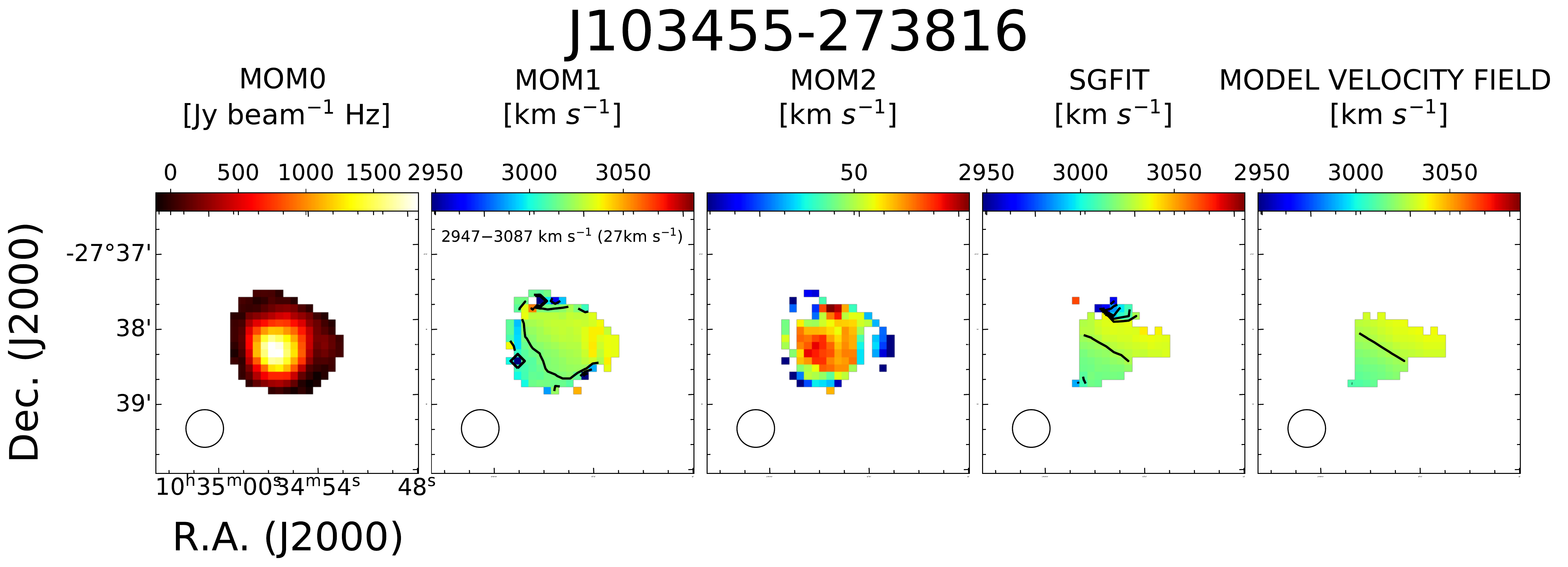}
    \includegraphics[width=8.5cm]{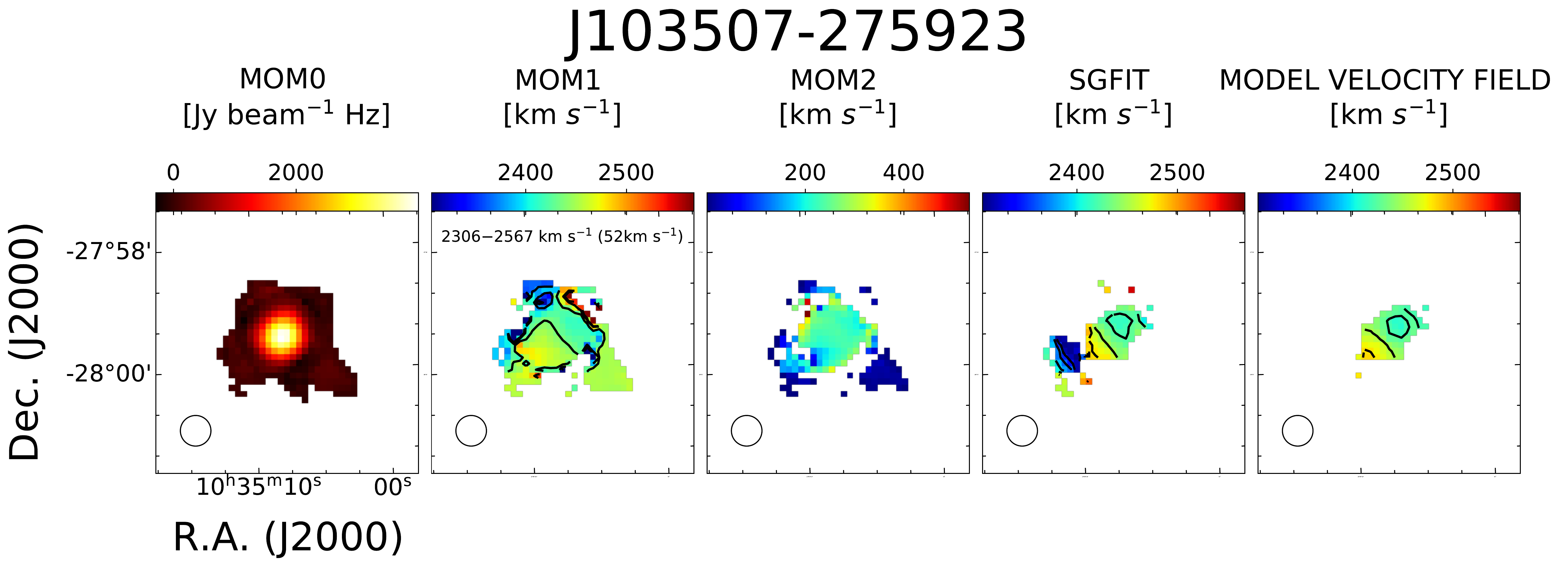}
    \vspace{0.5cm}
    \hspace{0.5cm}
    \includegraphics[width=8.5cm]{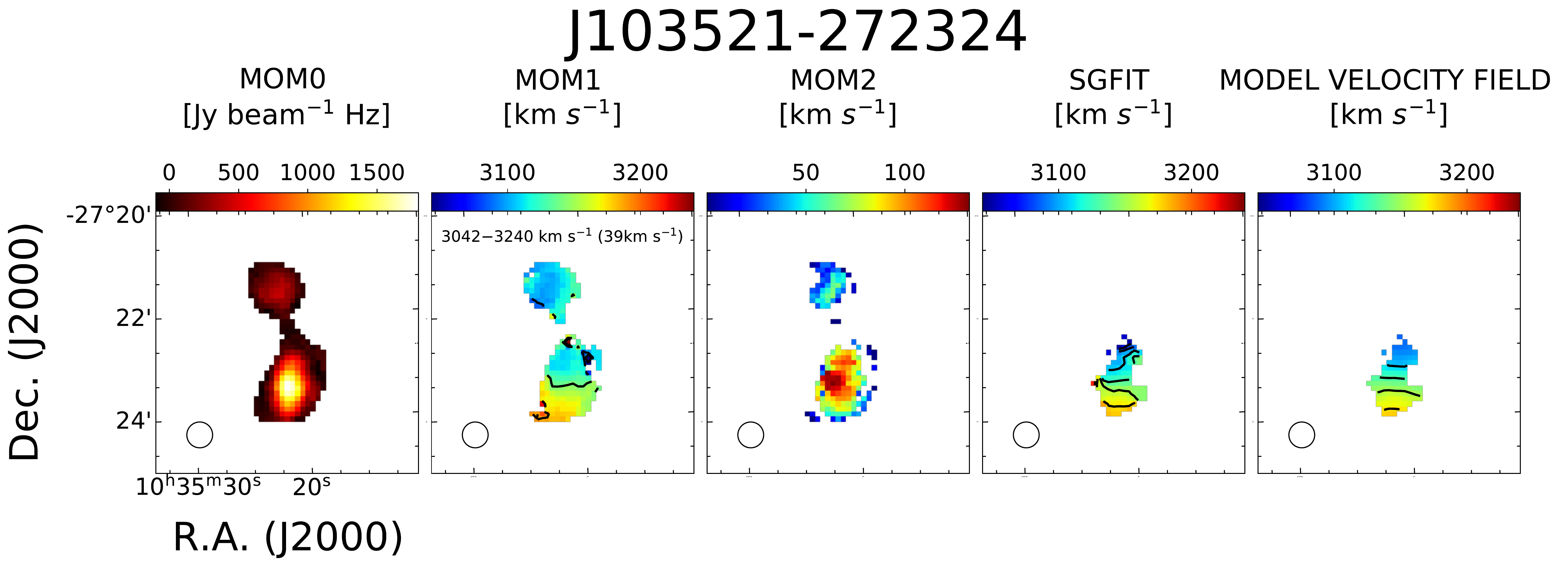}
    \includegraphics[width=8.5cm]{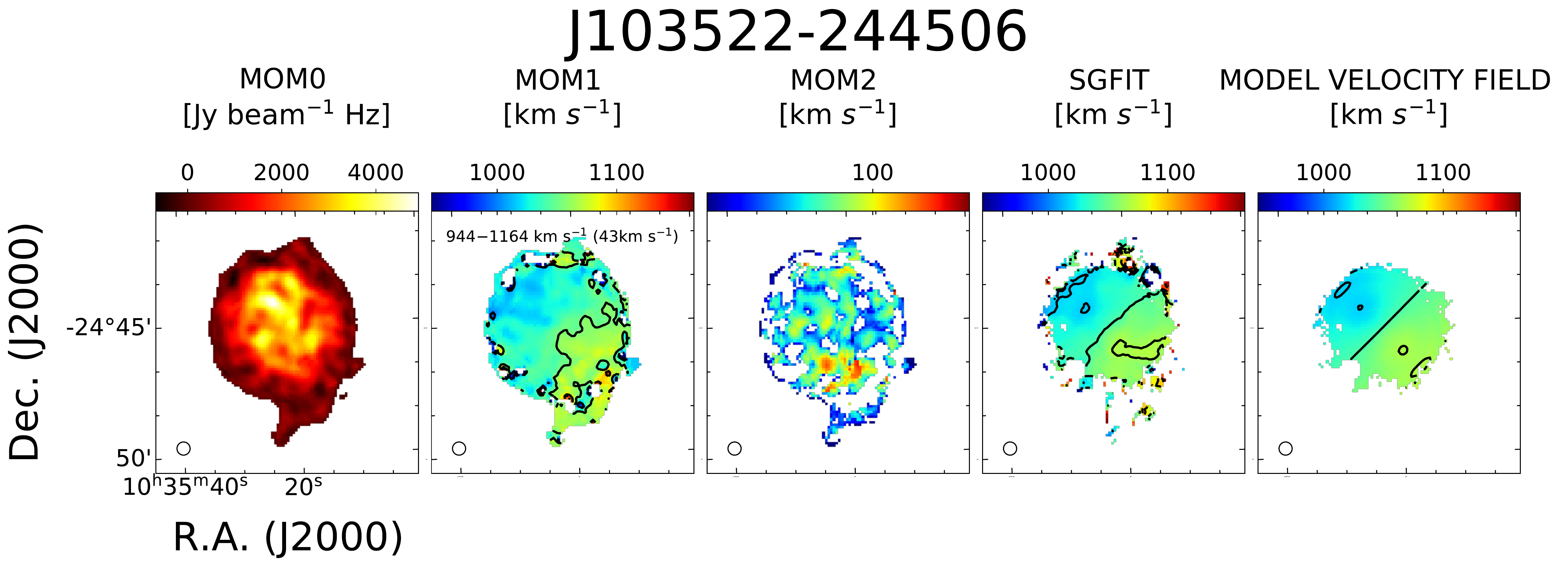}
    \vspace{0.5cm}
    \hspace{0.5cm}
    \includegraphics[width=8.5cm]{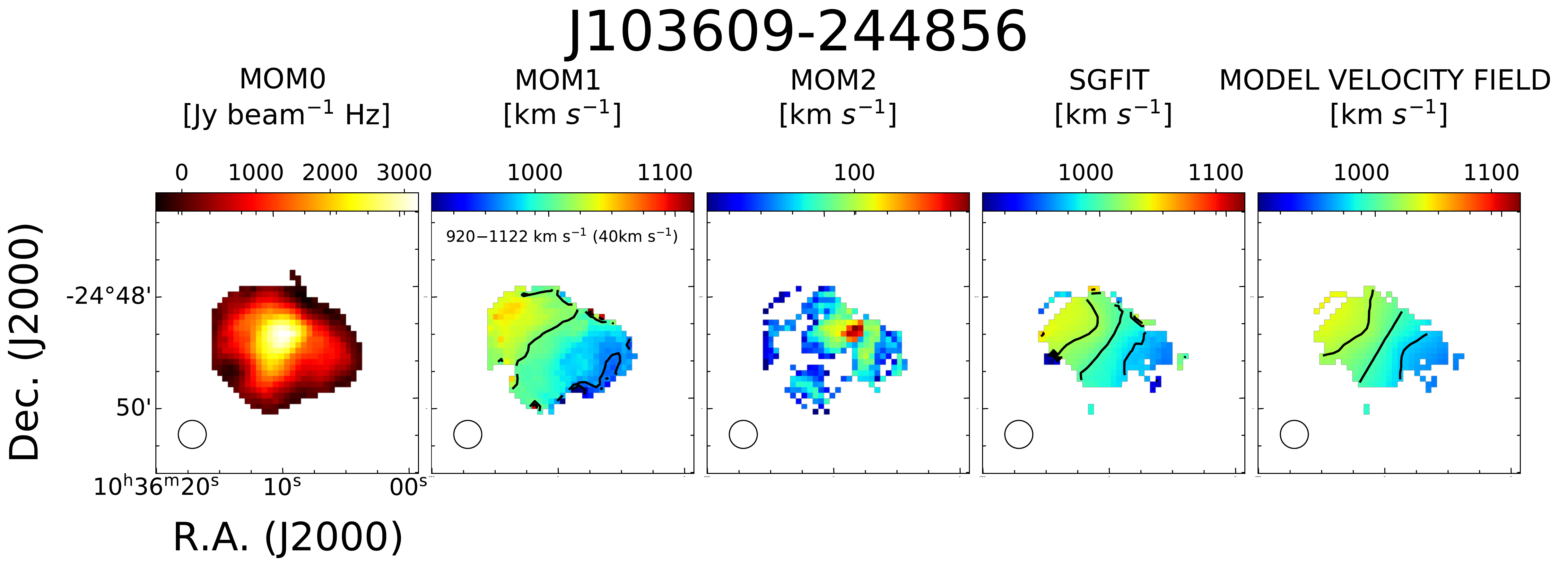}
    \includegraphics[width=8.5cm]{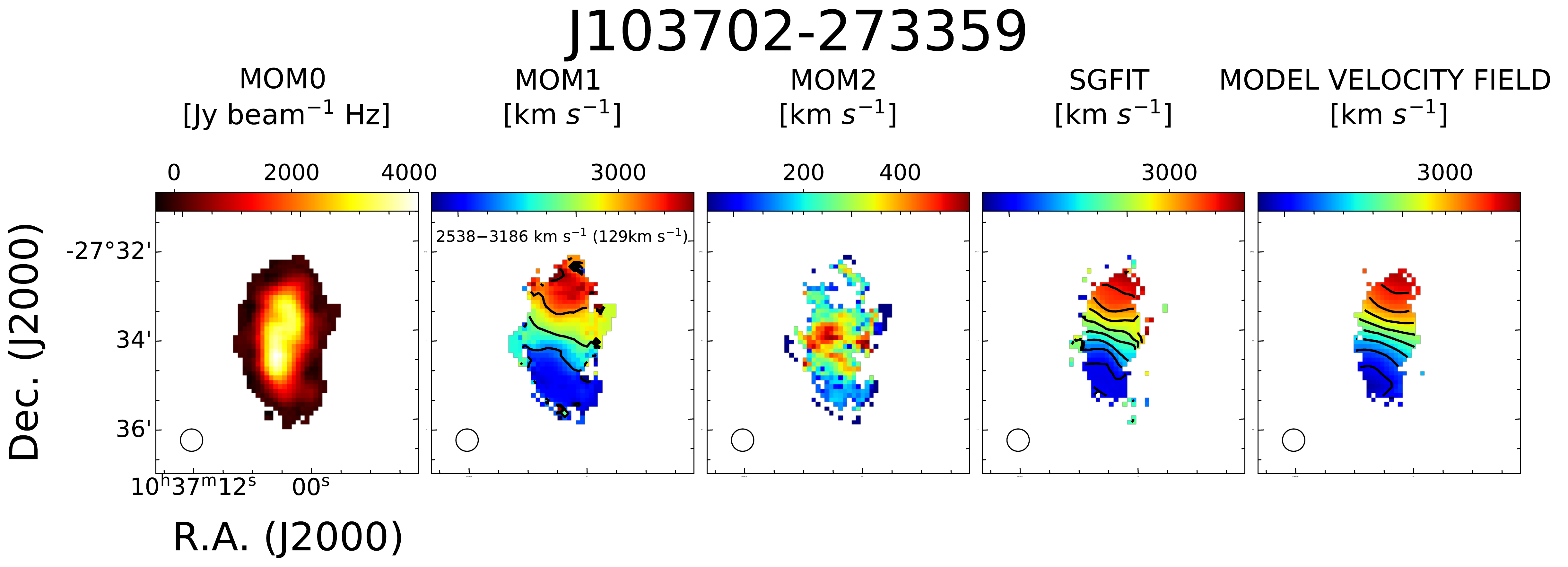}
    \vspace{0.5cm}
    \hspace{0.5cm}
    \includegraphics[width=8.5cm]{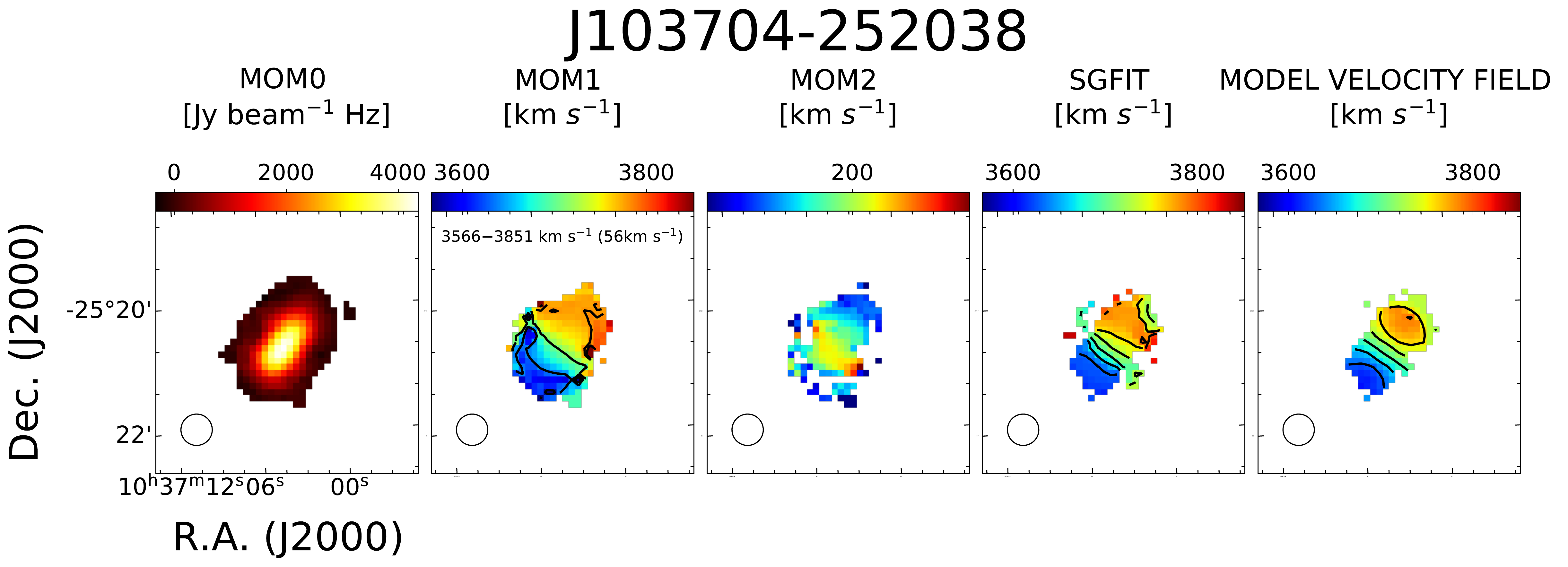}
    
    \caption{(continued)}
    \label{fig:my_label}
    \end{figure*}
    
\begin{figure*}
\ContinuedFloat
    \includegraphics[width=8.5cm]{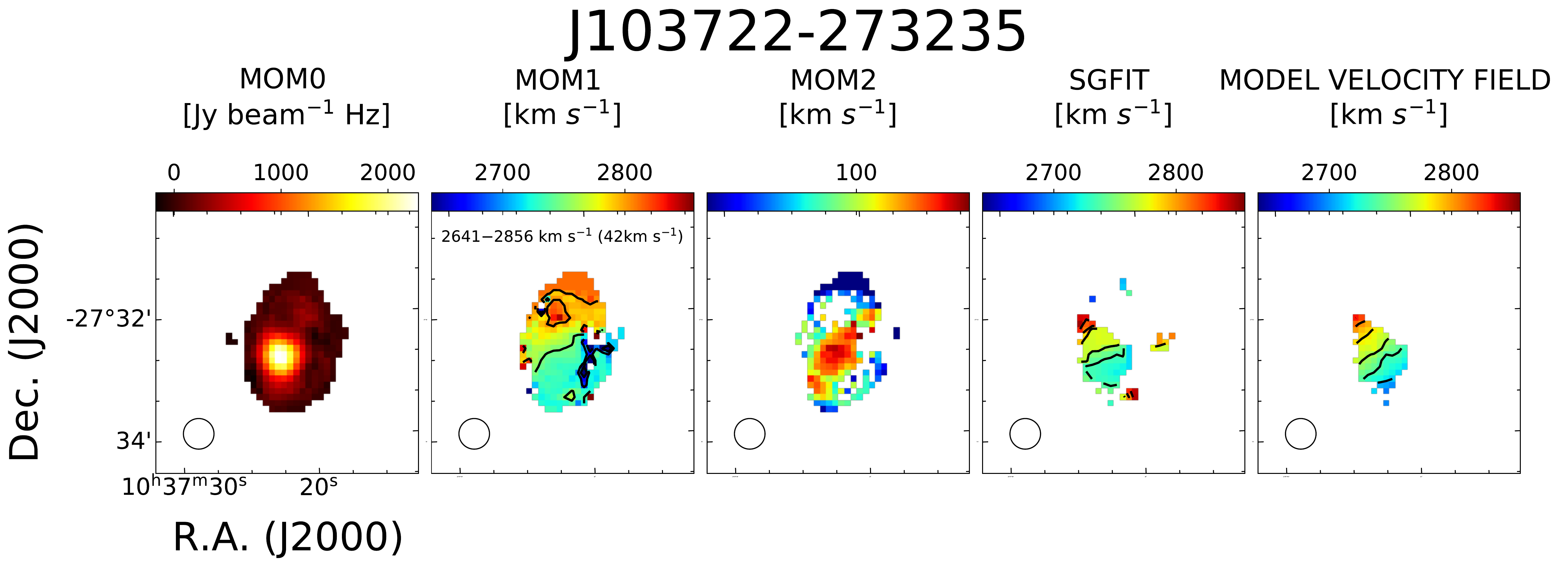}
    \vspace{0.5cm}
    \hspace{0.5cm}
    \includegraphics[width=8.5cm]{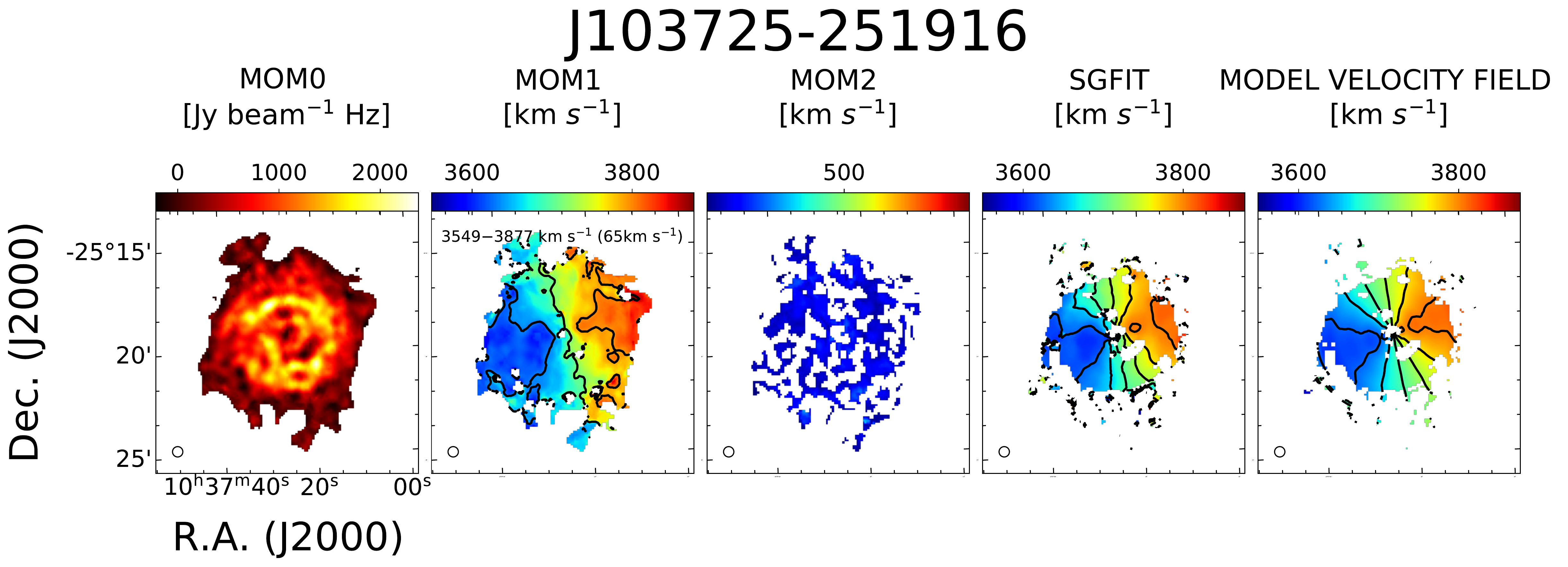}
    \includegraphics[width=8.5cm]{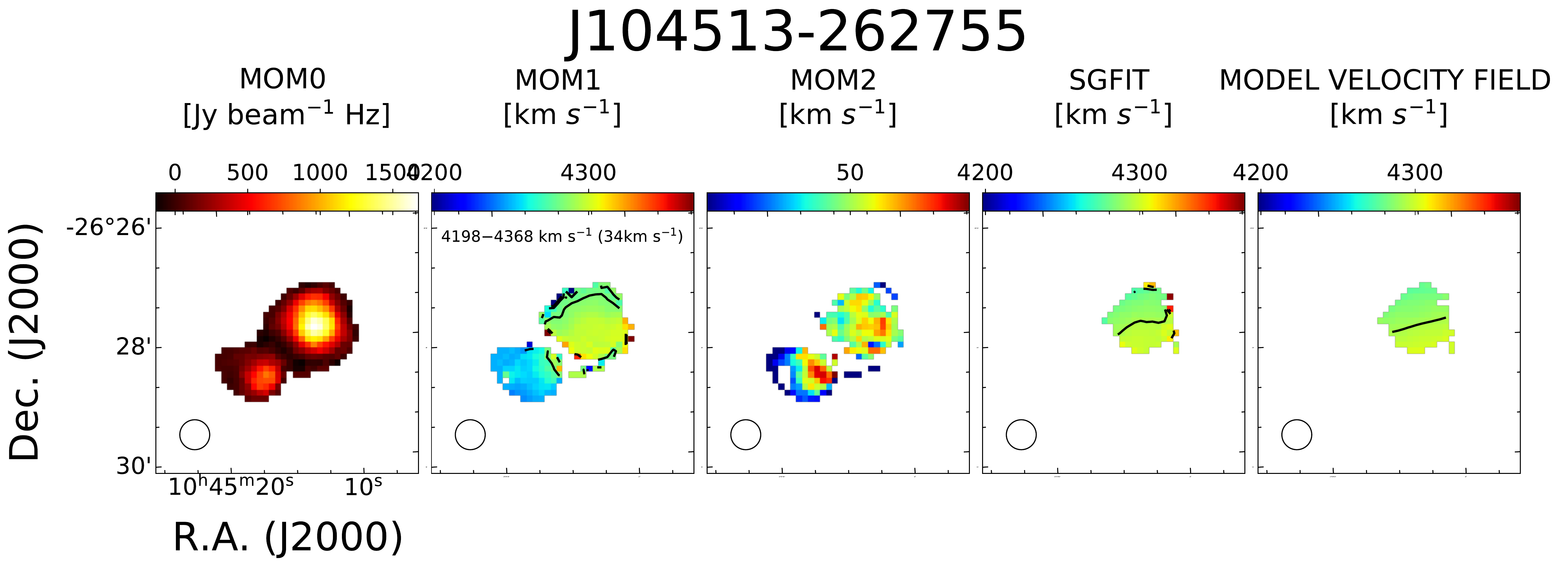}
    \vspace{0.5cm}
    \hspace{0.5cm}
    \includegraphics[width=8.5cm]{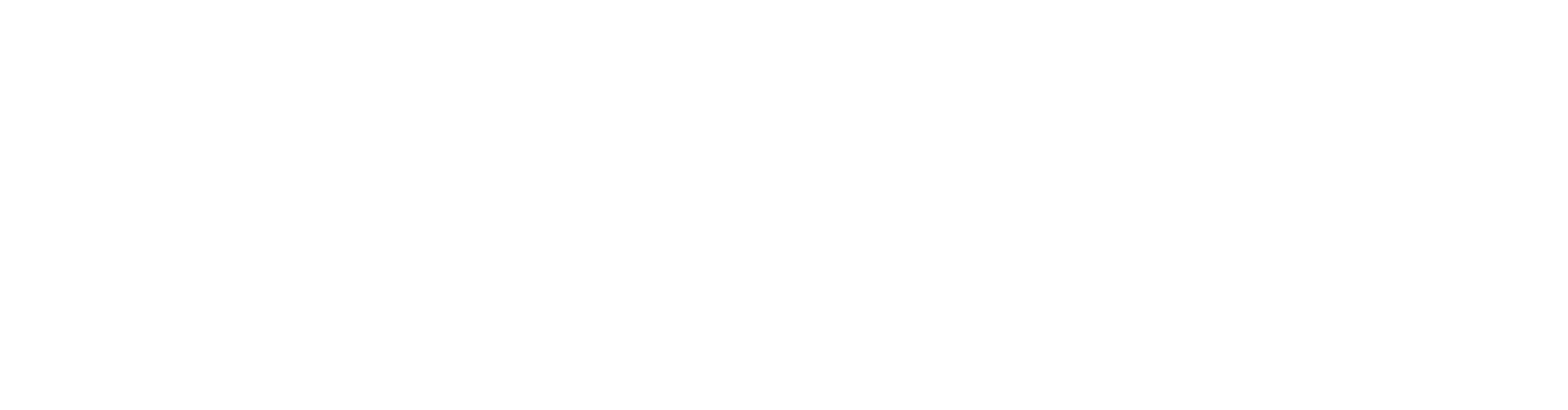}
    \caption{(continued)}
    \end{figure*}

\begin{figure*}
    \includegraphics[width=8.5cm]{Figure/NGC4636_J122710+071549.pdf}
    \vspace{0.5cm}
    \hspace{0.5cm}
    \includegraphics[width=8.5cm]{Figure/NGC4636_J122729+073841.pdf}
    \includegraphics[width=8.5cm]{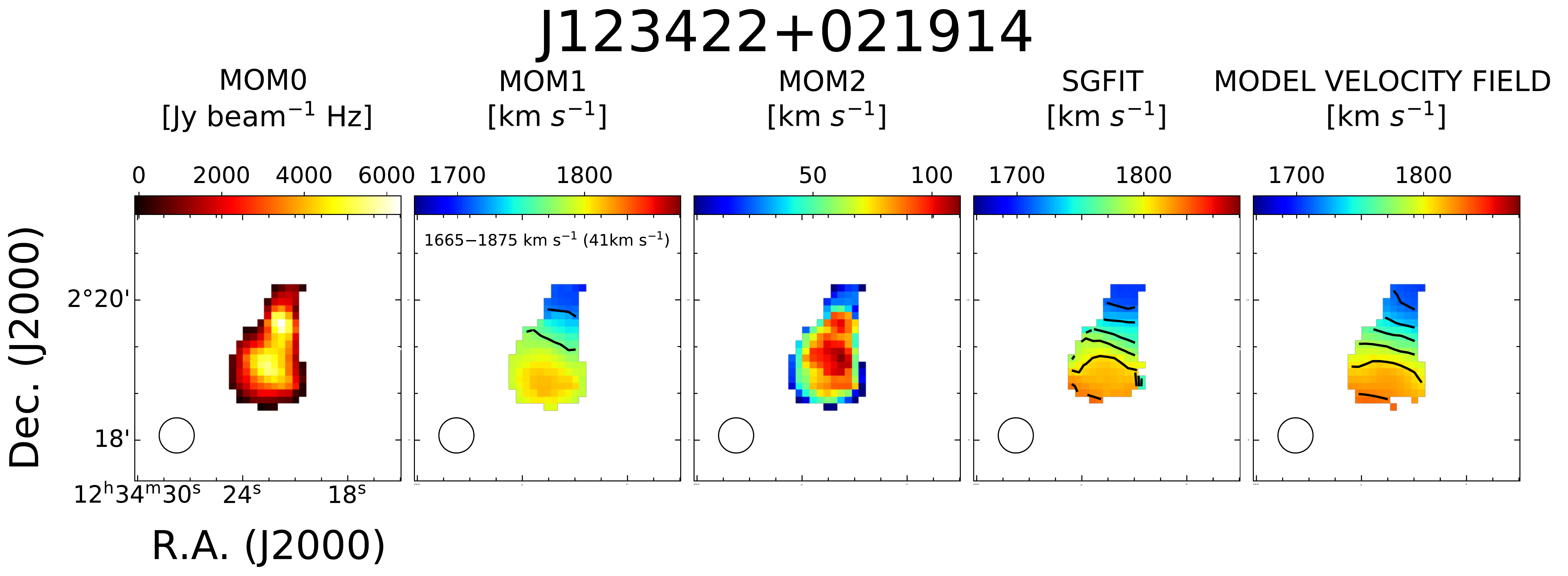}
    \vspace{0.5cm}
    \hspace{0.5cm}
    \includegraphics[width=8.5cm]{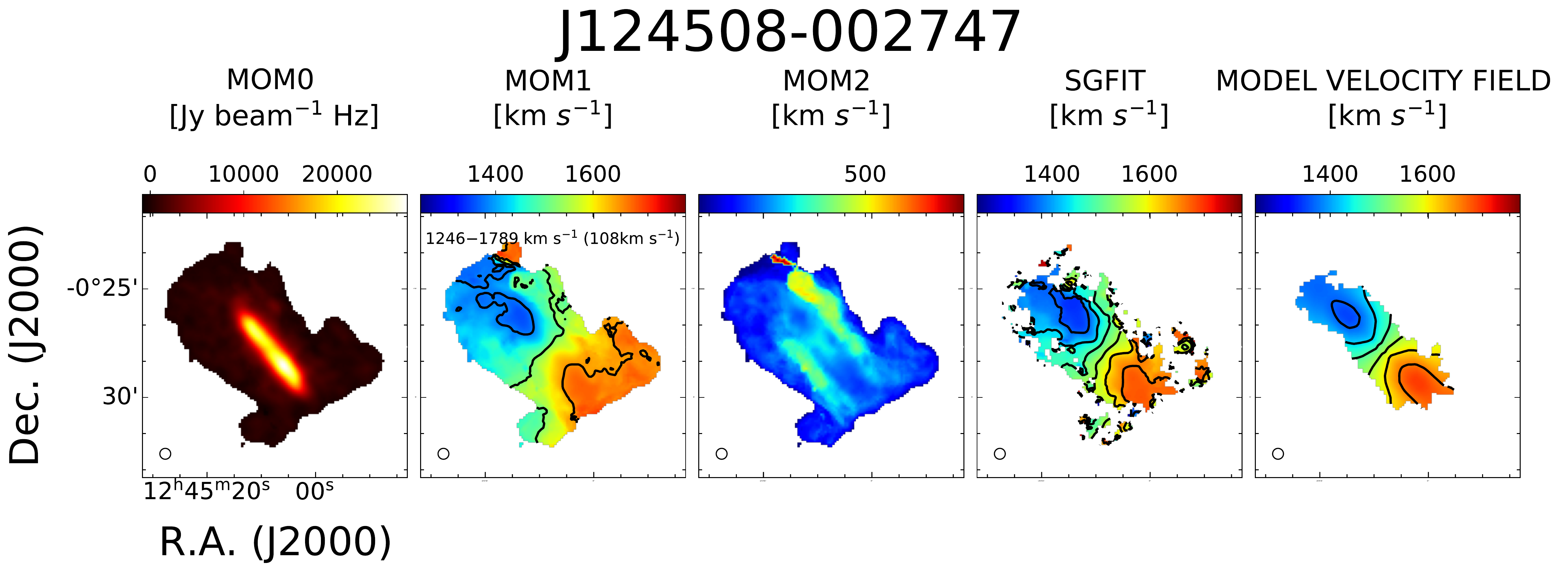}
    \includegraphics[width=8.5cm]{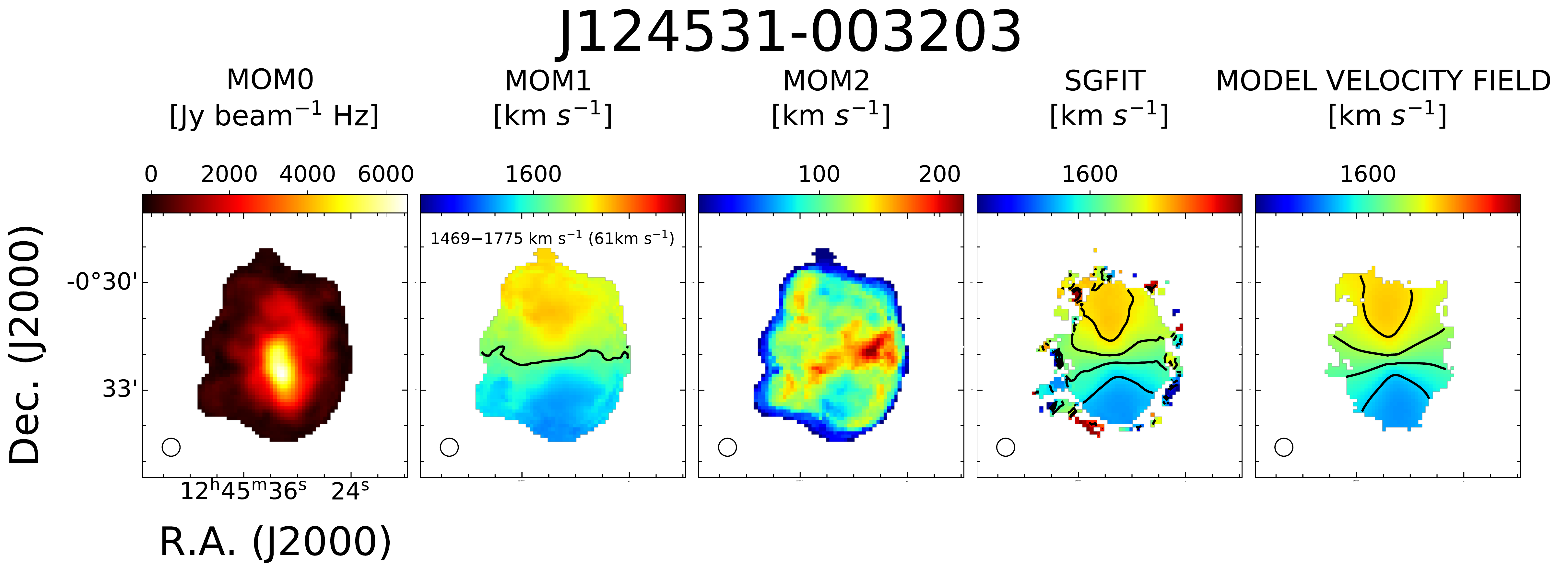}
    \vspace{0.5cm}
    \hspace{0.5cm}
    \includegraphics[width=8.5cm]{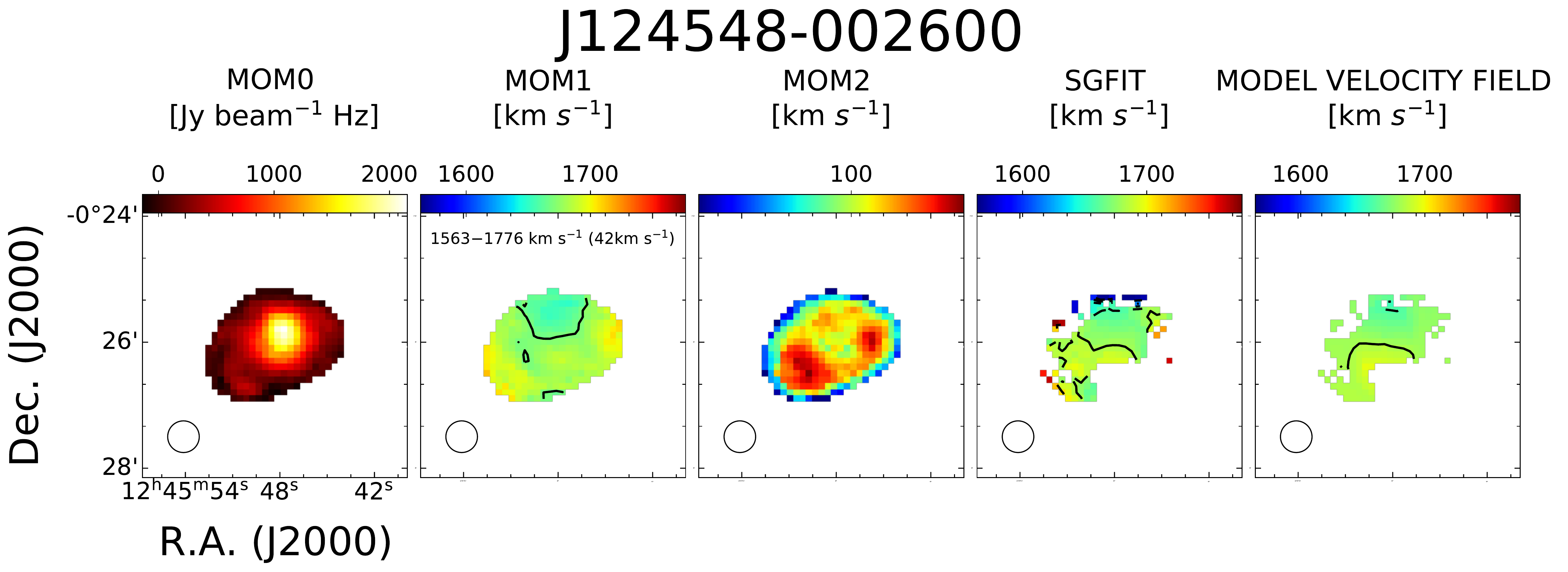}
    \includegraphics[width=8.5cm]{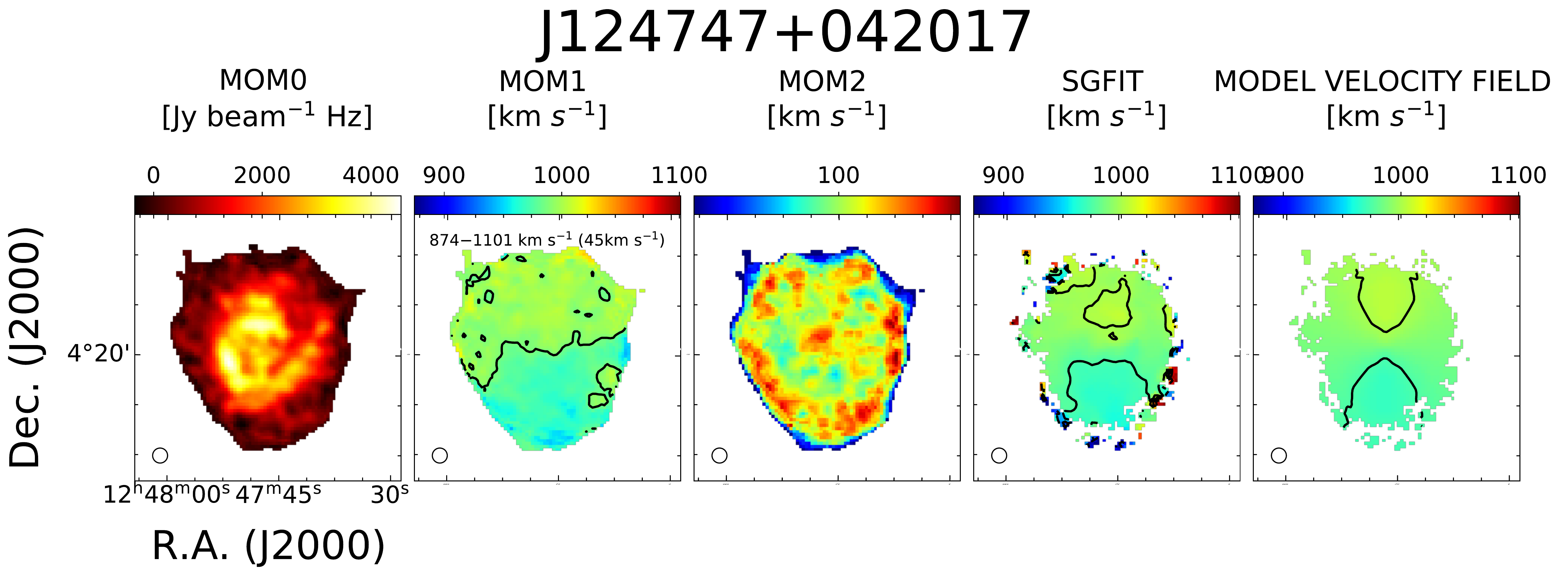}
    \vspace{0.5cm}
    \hspace{0.5cm}
    \includegraphics[width=8.5cm]{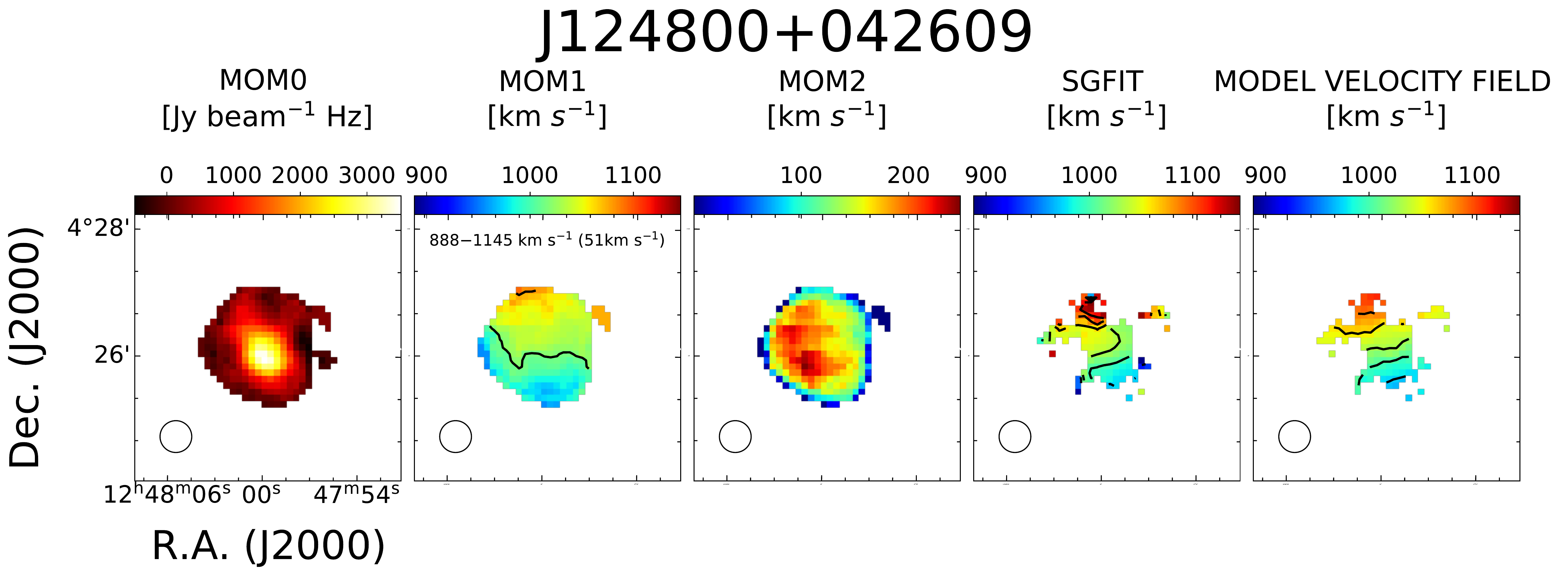}
    \caption{H{\sc i} images of the galaxy pairs (ordered by the WALLABY ID) in the ASKAP NGC 4636 field: ‘MOM0’ (SoFiA2 integrated intensity map, {\sc moment0}), ‘MOM1’ (SoFiA2 velocity field map , {\sc moment1}), ‘MOM2’ (SoFiA2, velocity dispersion map, {\sc moment2}), SGFIT ({\sc baygaud} Single Gaussian fitting (SGfit) velocity field map), and ‘MODEL VELOCITY FIELD’ ({\sc 2dbat} model velocity field map). The contour levels for each velocity field are denoted in the second panel of each figure. The ASKAP beam is shown as an ellipse on the bottom-left corner of each panel.}
    \label{fig:my_label}\label{figA2}
\end{figure*}

\begin{figure*}
    \includegraphics[width=8.5cm]{Figure/Norma_J163754-564907.pdf}
    \vspace{0.5cm}
    \hspace{0.5cm}
    \includegraphics[width=8.5cm]{Figure/Norma_J163834-601517.pdf}
    \includegraphics[width=8.5cm]{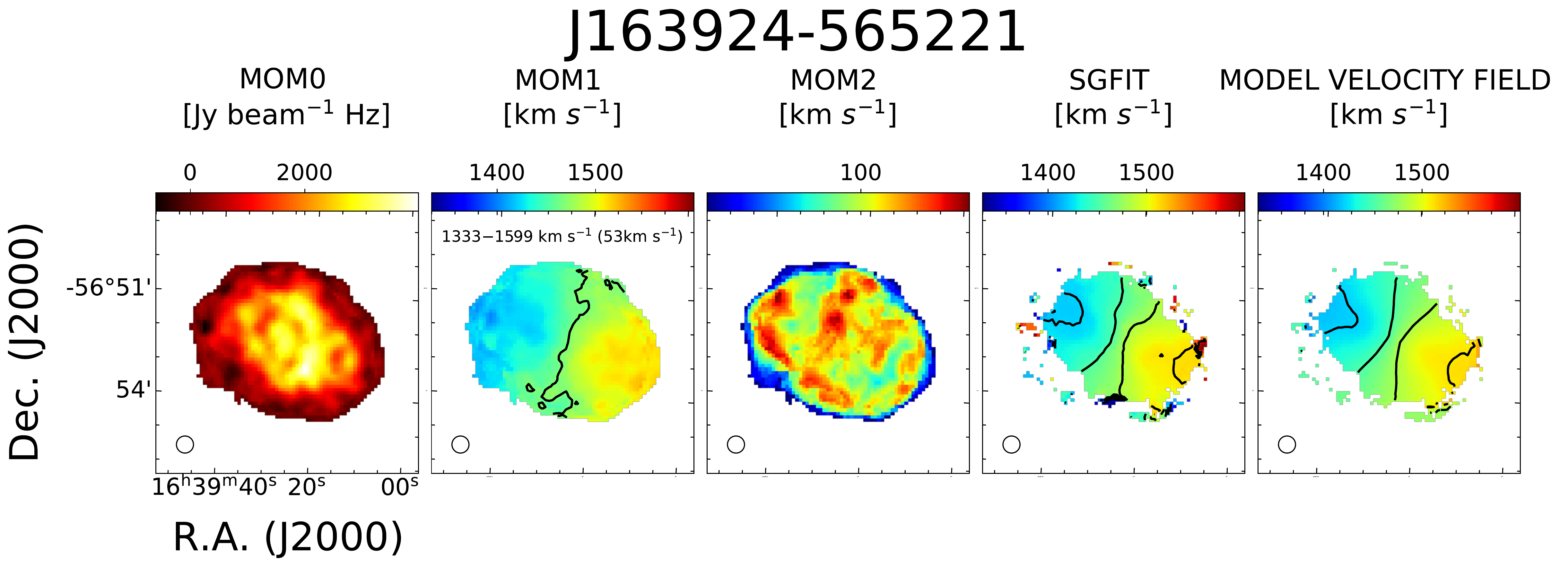}
    \vspace{0.5cm}
    \hspace{0.5cm}
    \includegraphics[width=8.5cm]{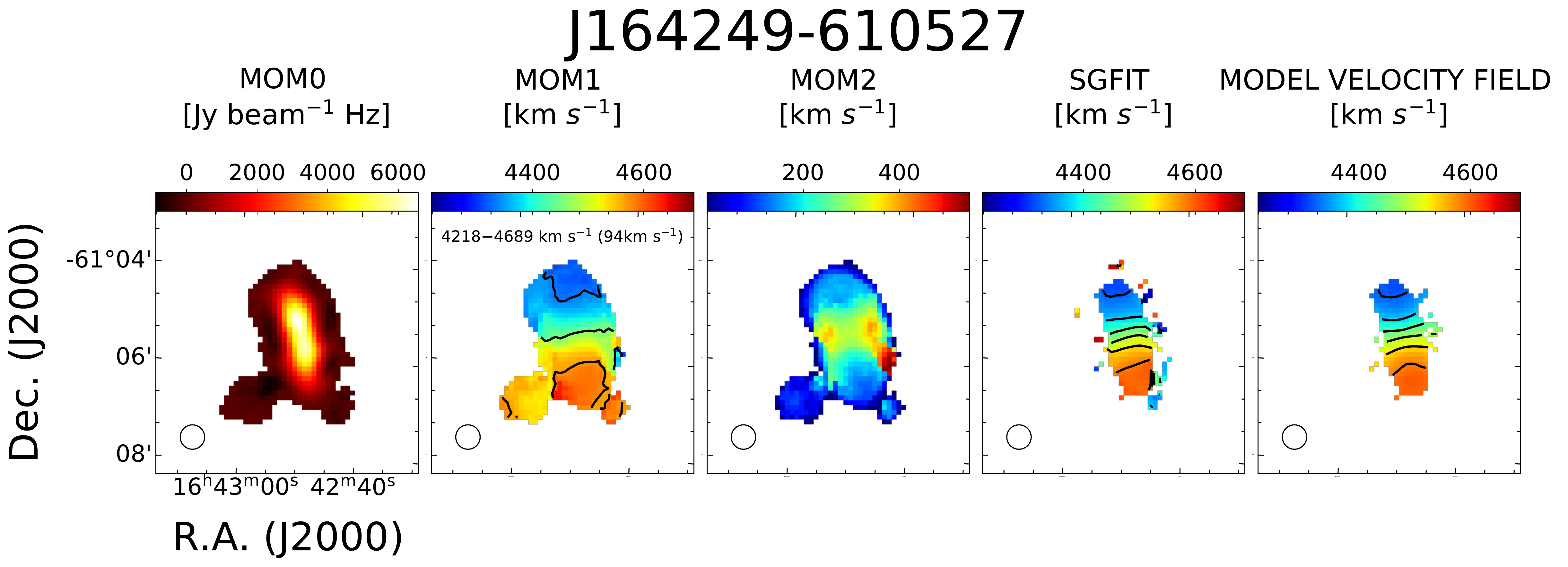}
    \includegraphics[width=8.5cm]{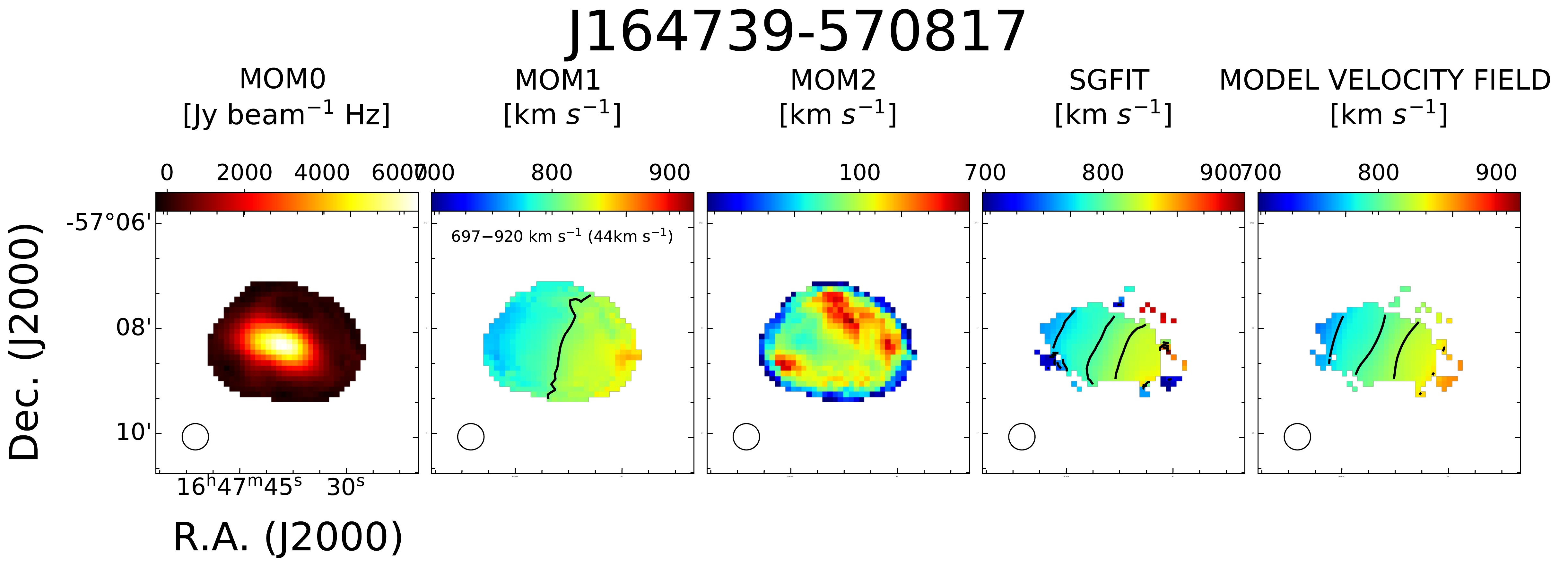}
    \vspace{0.5cm}
    \hspace{0.5cm}
    \includegraphics[width=8.5cm]{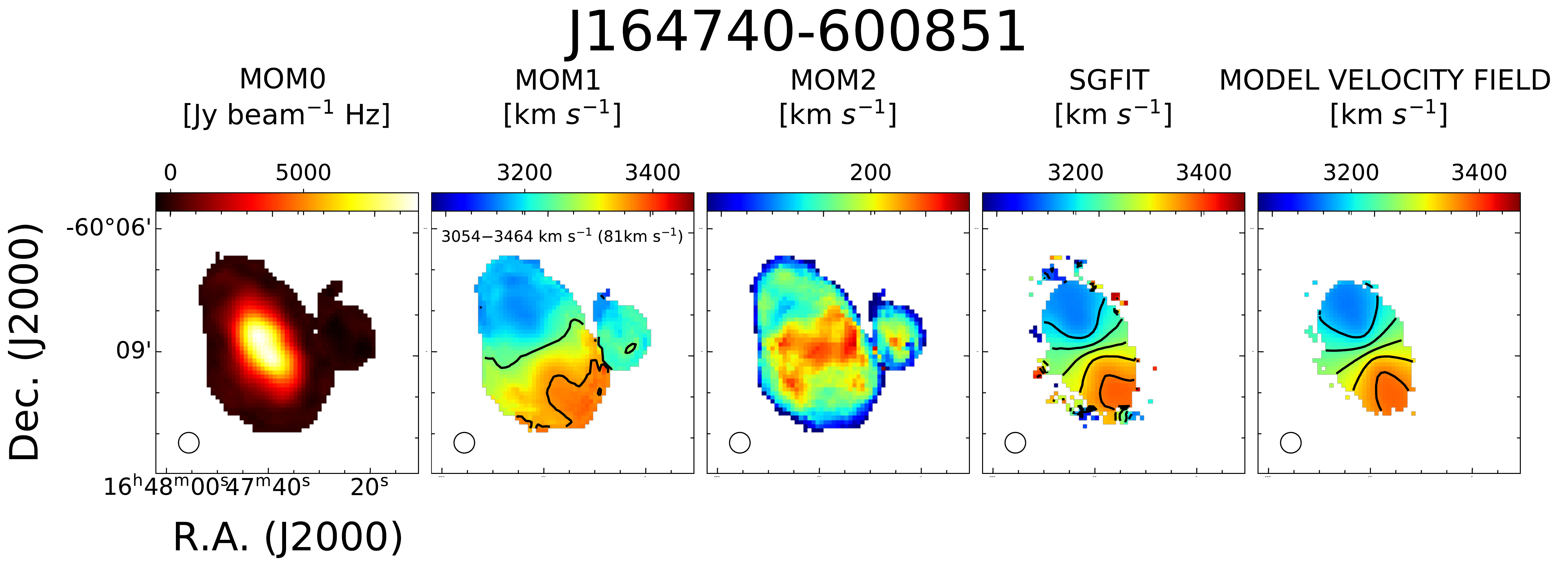}
    \includegraphics[width=8.5cm]{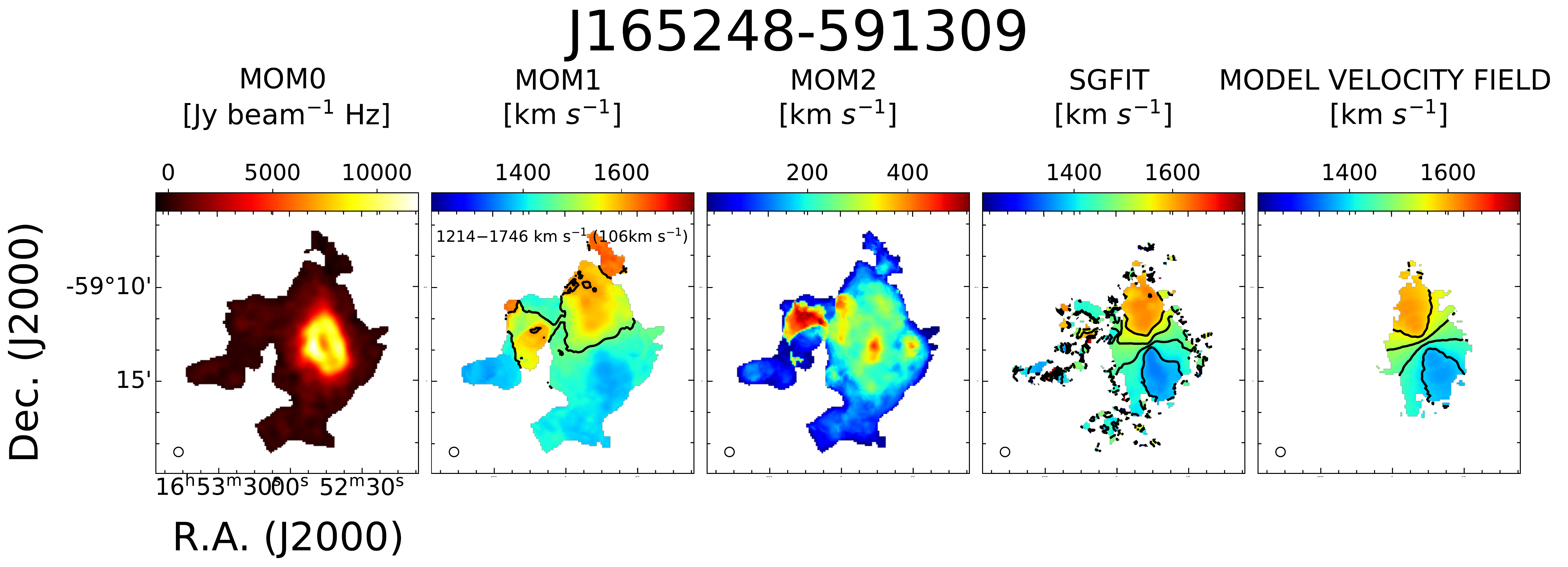}
    \vspace{0.5cm}
    \hspace{0.5cm}
    \includegraphics[width=8.5cm]{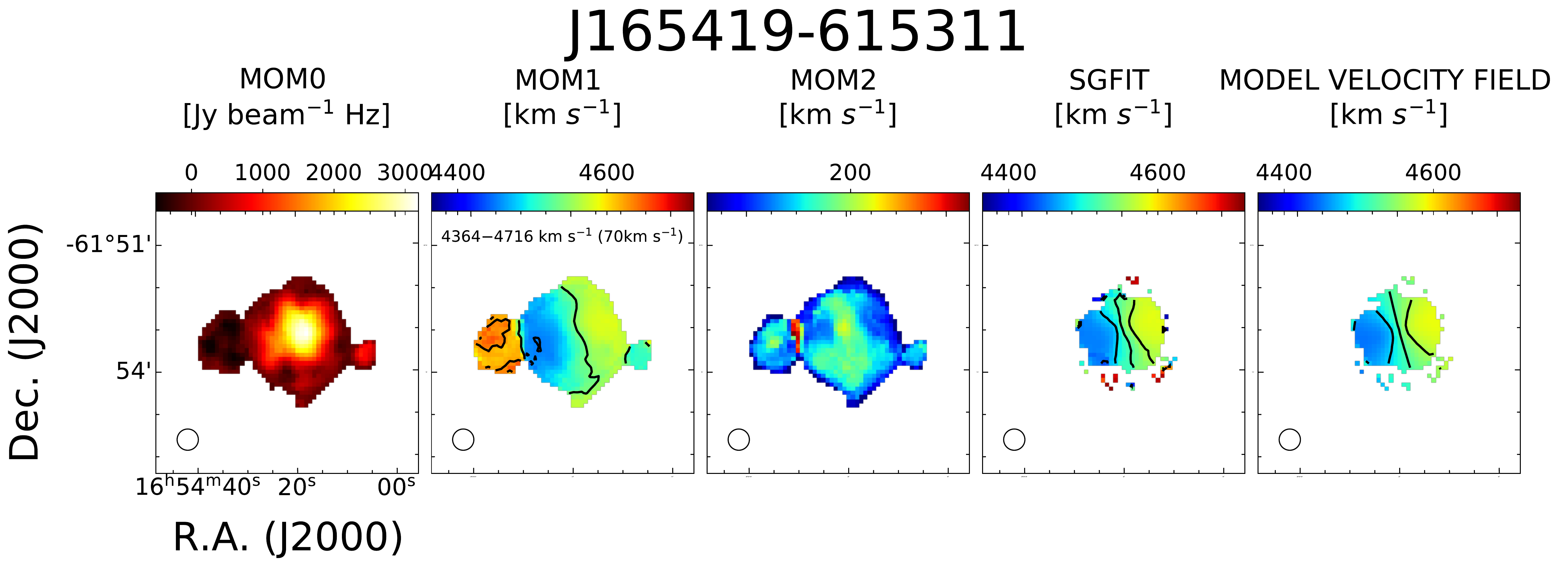}
    \includegraphics[width=8.5cm]{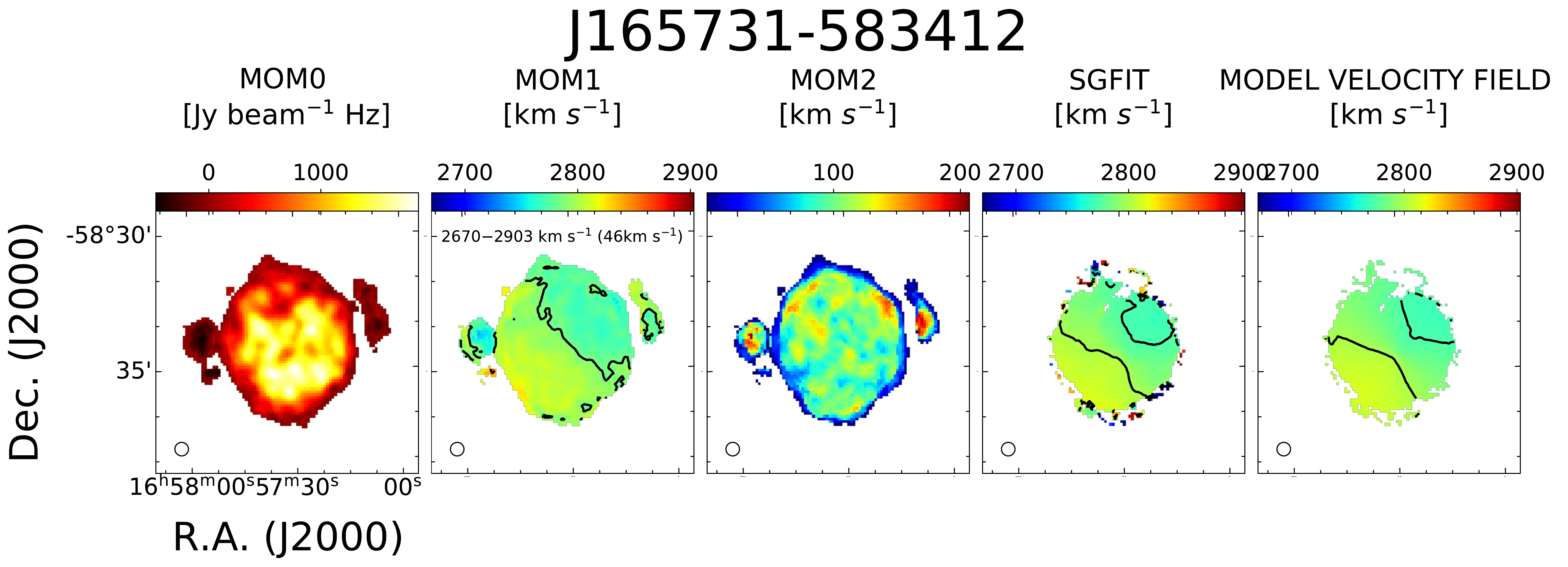}
    \vspace{0.5cm}
    \hspace{0.5cm}
    \includegraphics[width=8.5cm]{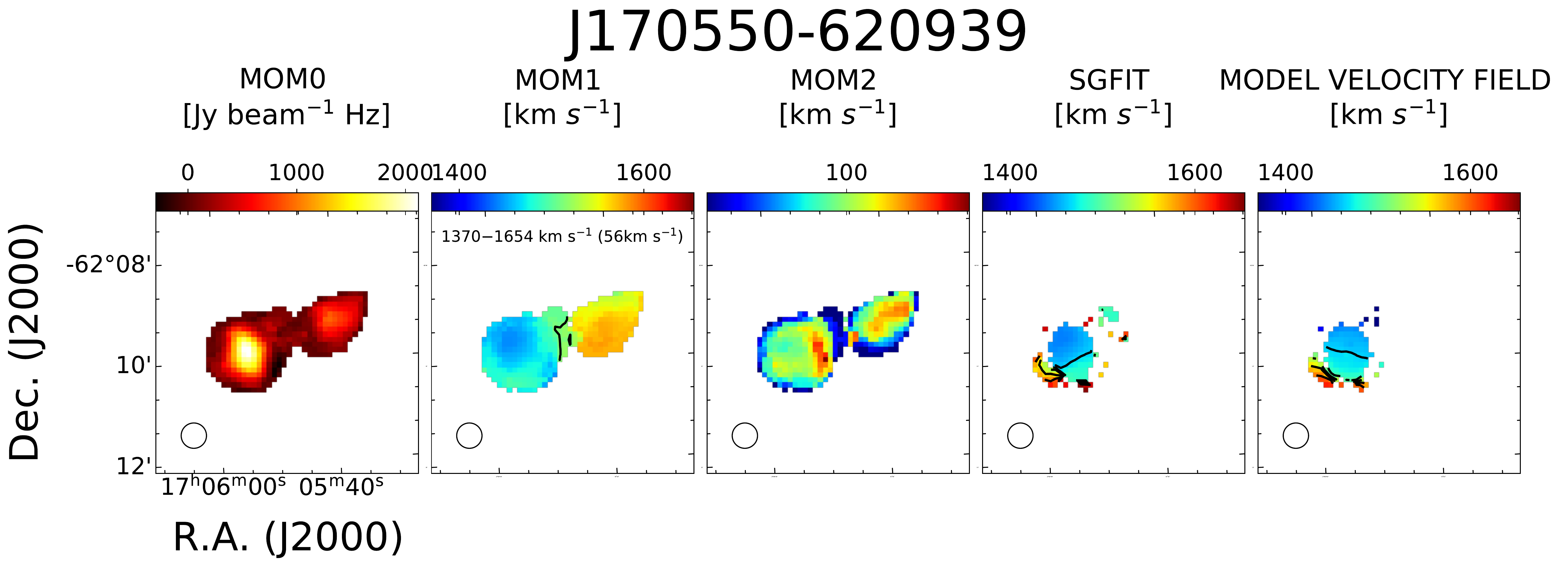}
    \includegraphics[width=8.5cm]{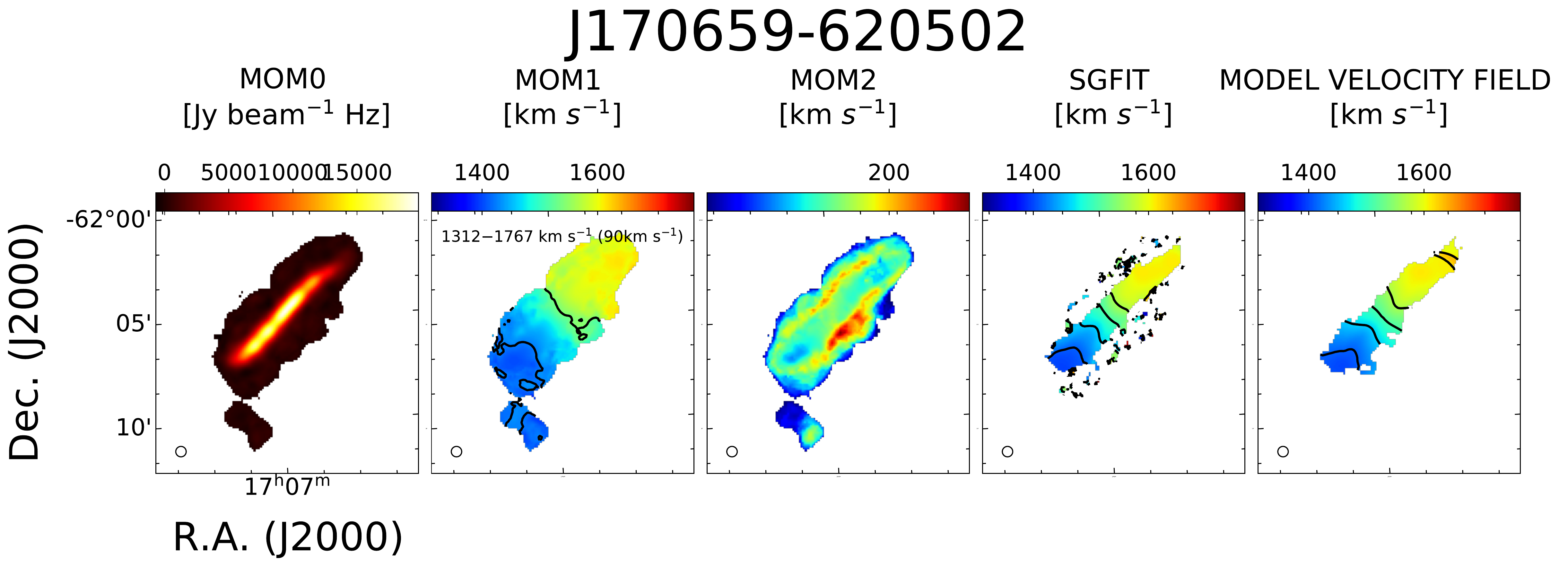}
    \vspace{0.5cm}
    \hspace{0.5cm}
    \includegraphics[width=8.5cm]{Figure/white_multiplot.pdf}
    \caption{H{\sc i} images of the galaxy pairs (ordered by the WALLABY ID) in the ASKAP Norma cluster field: ‘MOM0’ (SoFiA2 integrated intensity map, {\sc moment0}), ‘MOM1’ (SoFiA2 velocity field map , {\sc moment1}), ‘MOM2’ (SoFiA2, velocity dispersion map, {\sc moment2}), SGFIT ({\sc baygaud} Single Gaussian fitting (SGfit) velocity field map), and ‘MODEL VELOCITY FIELD’ ({\sc 2dbat} model velocity field map). The contour levels for each velocity field are denoted in the second panel of each figure. The ASKAP beam is shown as an ellipse on the bottom-left corner of each panel.}\label{figA3}
    \end{figure*}


\begin{figure*}
    \centering
    \includegraphics[width=0.45\textwidth]{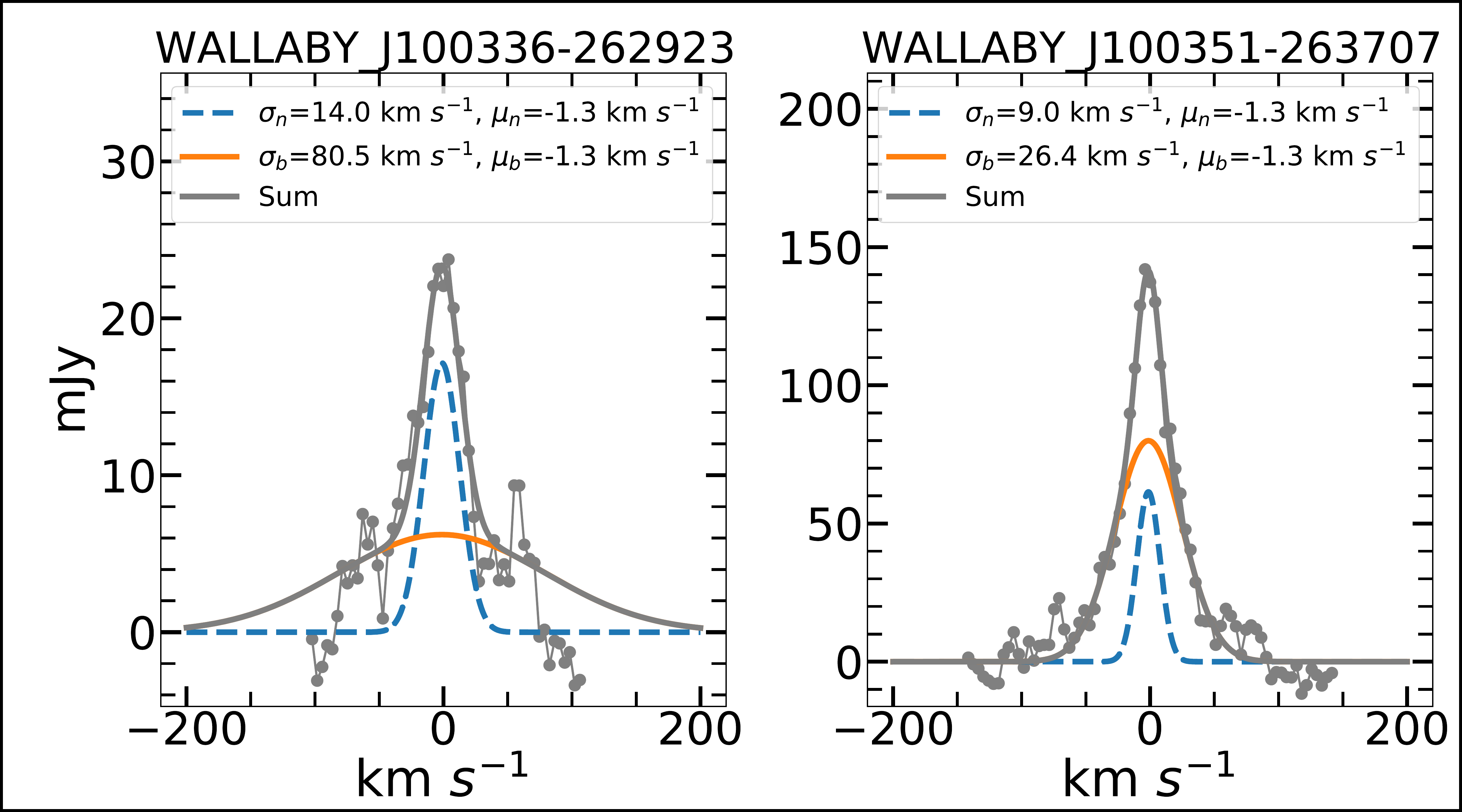}
    \hspace{0.3cm}
    \vspace{0.5cm}
    \includegraphics[width=0.45\textwidth]{Figure/WALLABY_J100342-270137_superprofile.pdf}
    \vspace{0.5cm}
    \includegraphics[width=0.45\textwidth]{Figure/WALLABY_J100426-282638_superprofile.pdf}
    \hspace{0.3cm}
    \includegraphics[width=0.45\textwidth]{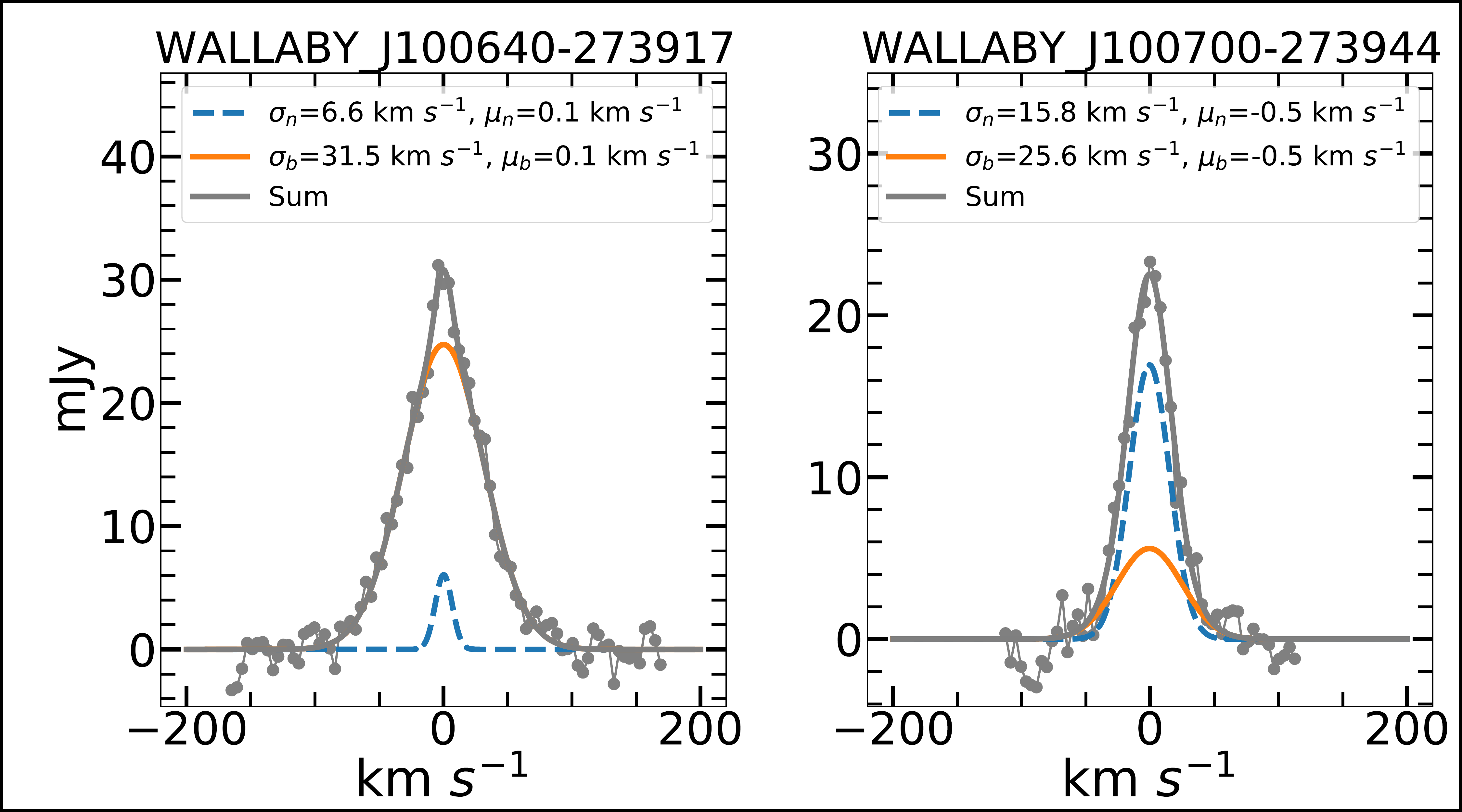}
    \vspace{0.5cm}
    \includegraphics[width=0.45\textwidth]{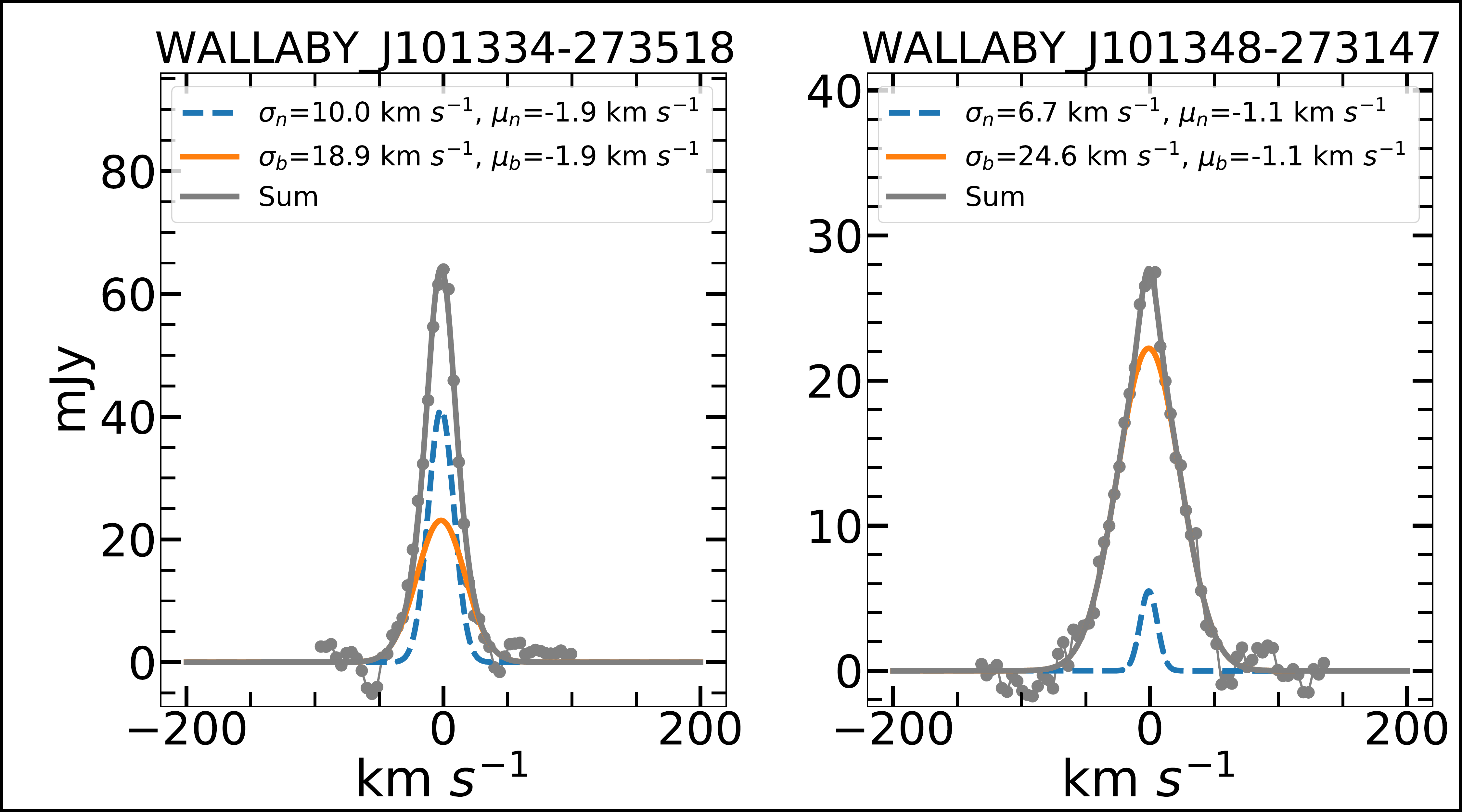}
    \hspace{0.3cm}
    \includegraphics[width=0.45\textwidth]{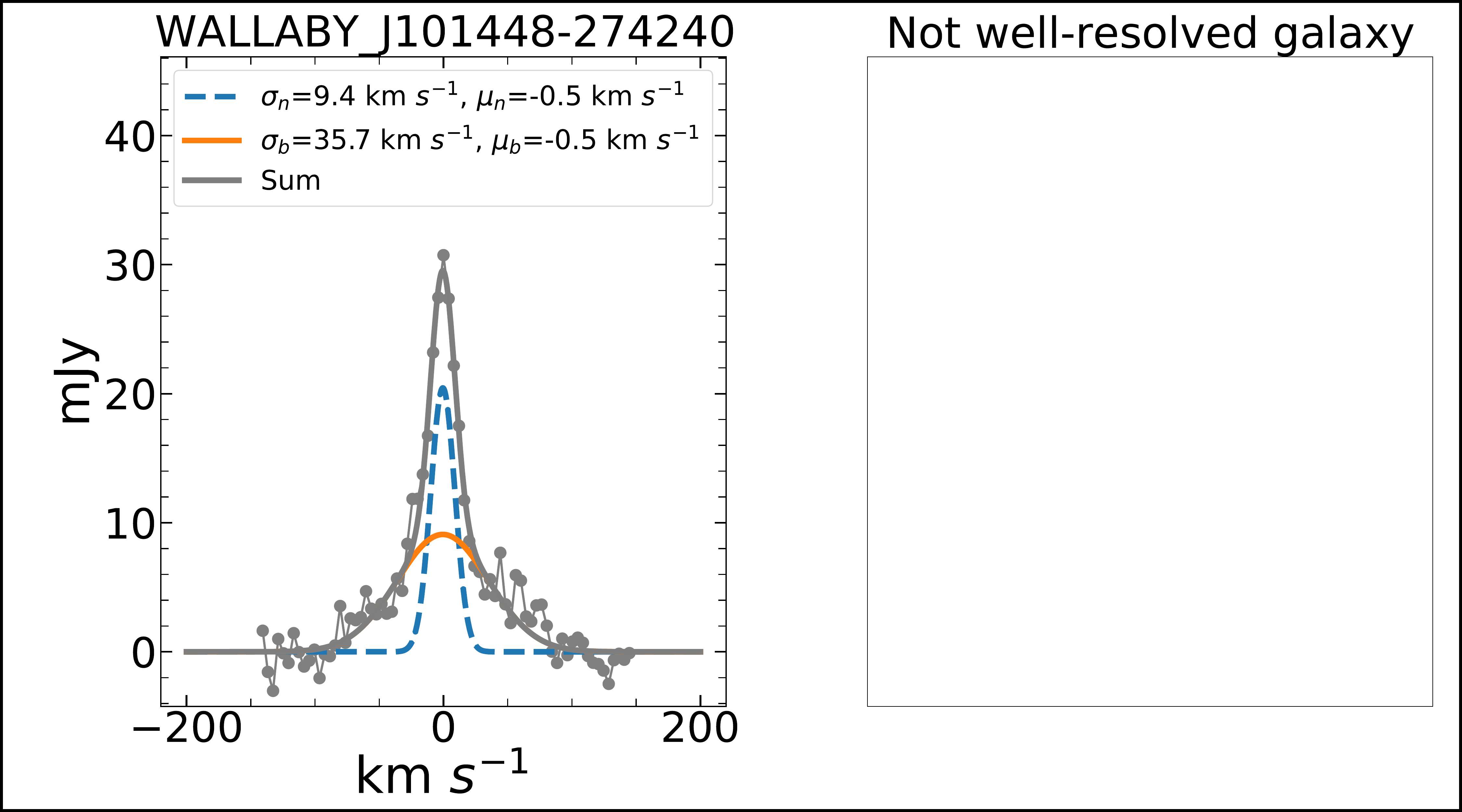}
    \vspace{0.5cm}
    \includegraphics[width=0.45\textwidth]{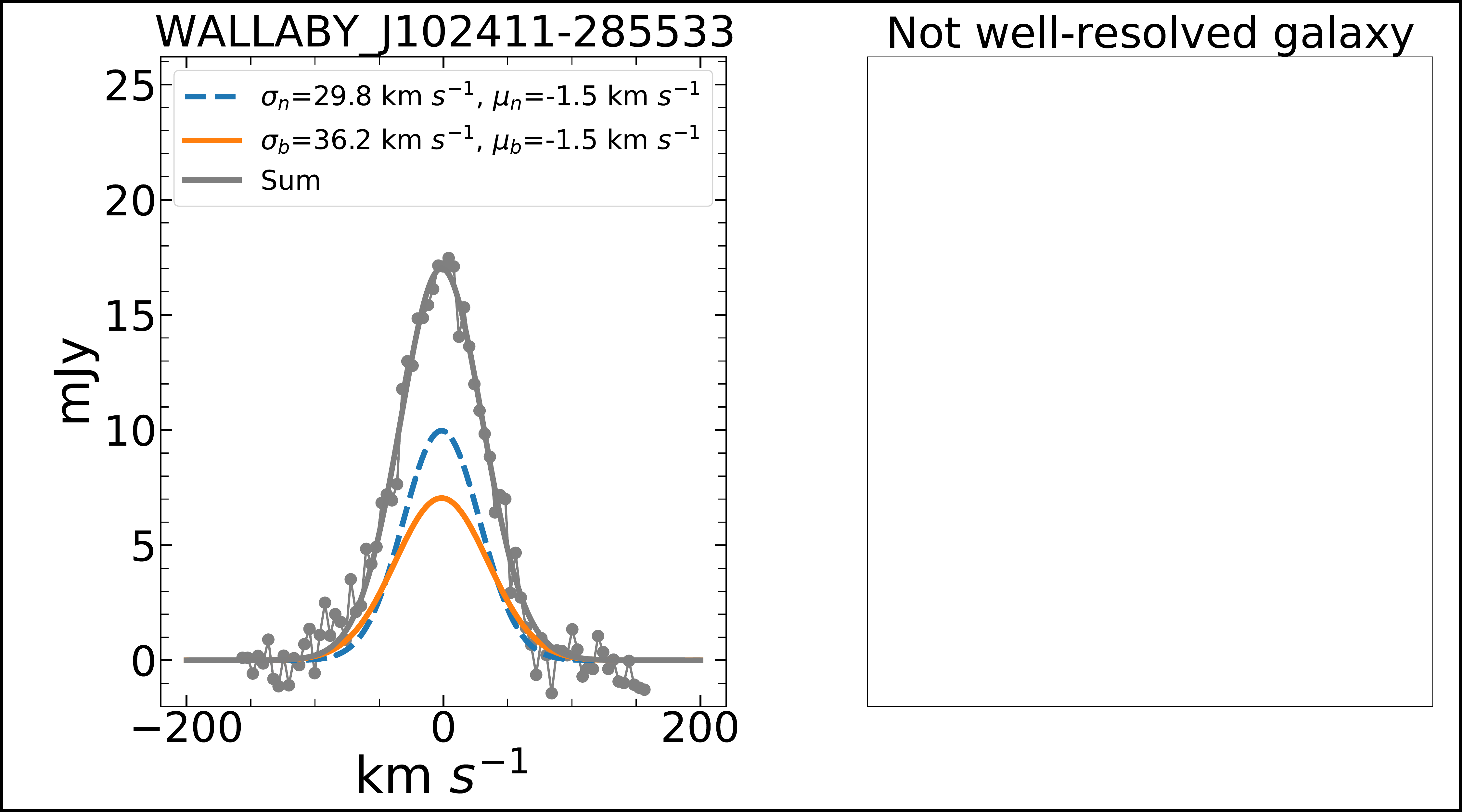}
    \hspace{0.3cm}
    \includegraphics[width=0.45\textwidth]{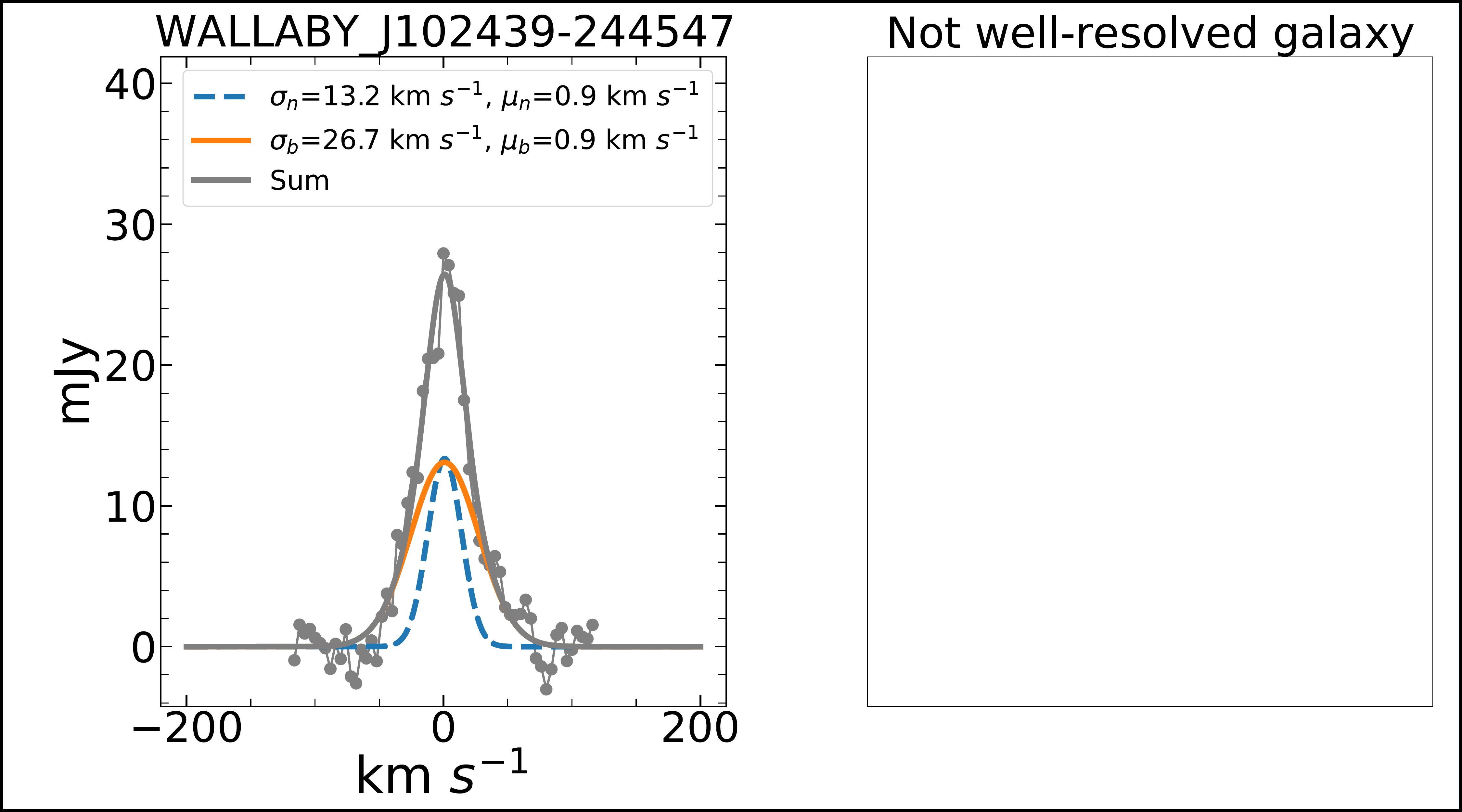}
    \vspace{0.5cm}
    \caption{{\sc baygaud} based H{\sc i} super-profiles of the galaxy pairs (ordered by the WALLABY ID) in the ASKAP Hydra I cluster field. The grey circles show the stacked fluxes at the corresponding channels. The blue dashed and orange solid lines represent the narrow and broad components decomposed from the double Gaussian fitting to the super-profiles. Their velocity dispersions ($\sigma$) and centroid velocities ($\mu$) are labeled on the top-right corner of each panel. H{\sc i} super-profiles of galaxies which are poorly resolved are blanked.}
    \label{figA4}
\end{figure*}

\begin{figure*} 
\ContinuedFloat
    \includegraphics[width=0.45\textwidth]{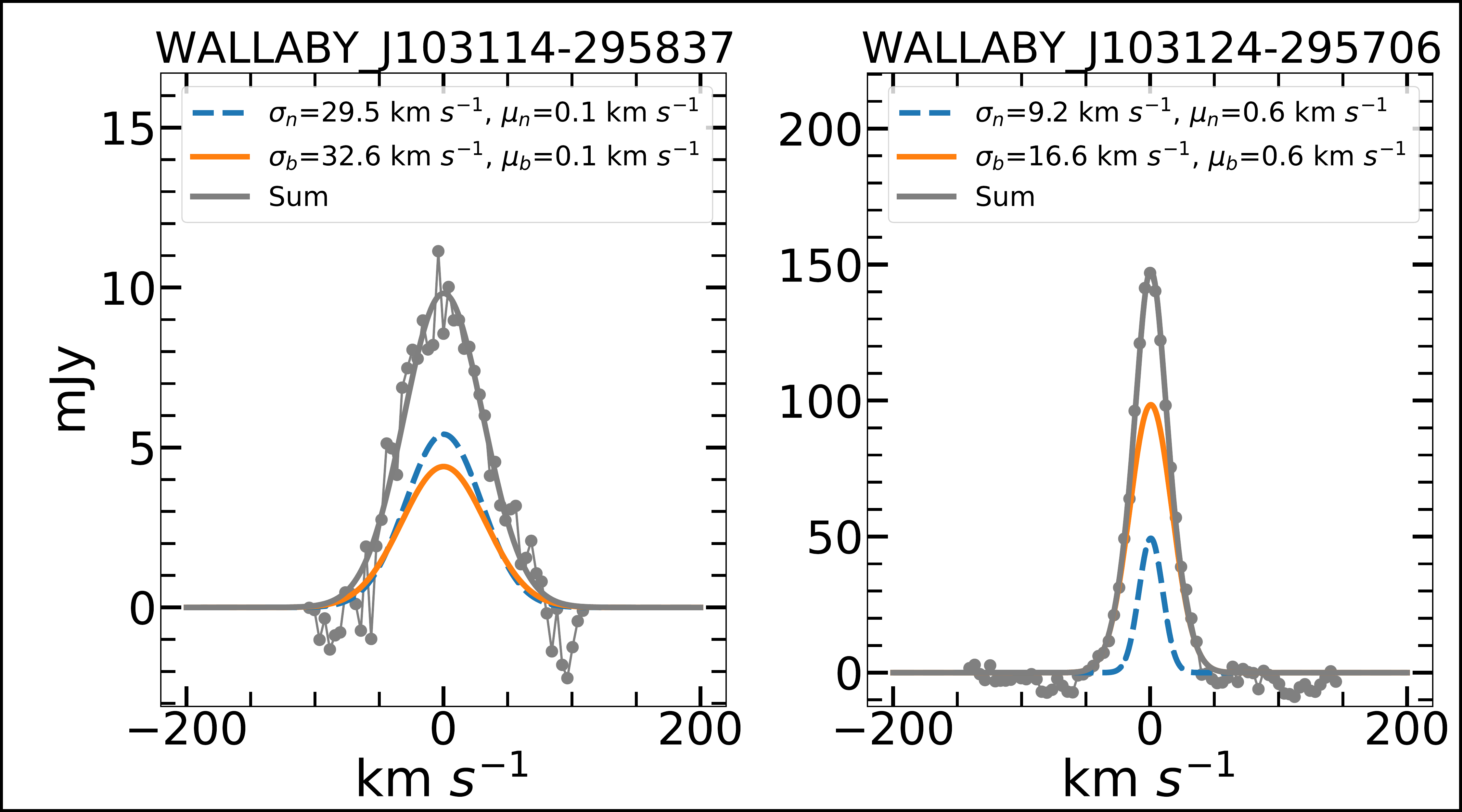}
    \hspace{0.3cm}
    \vspace{0.5cm}
    \includegraphics[width=0.45\textwidth]{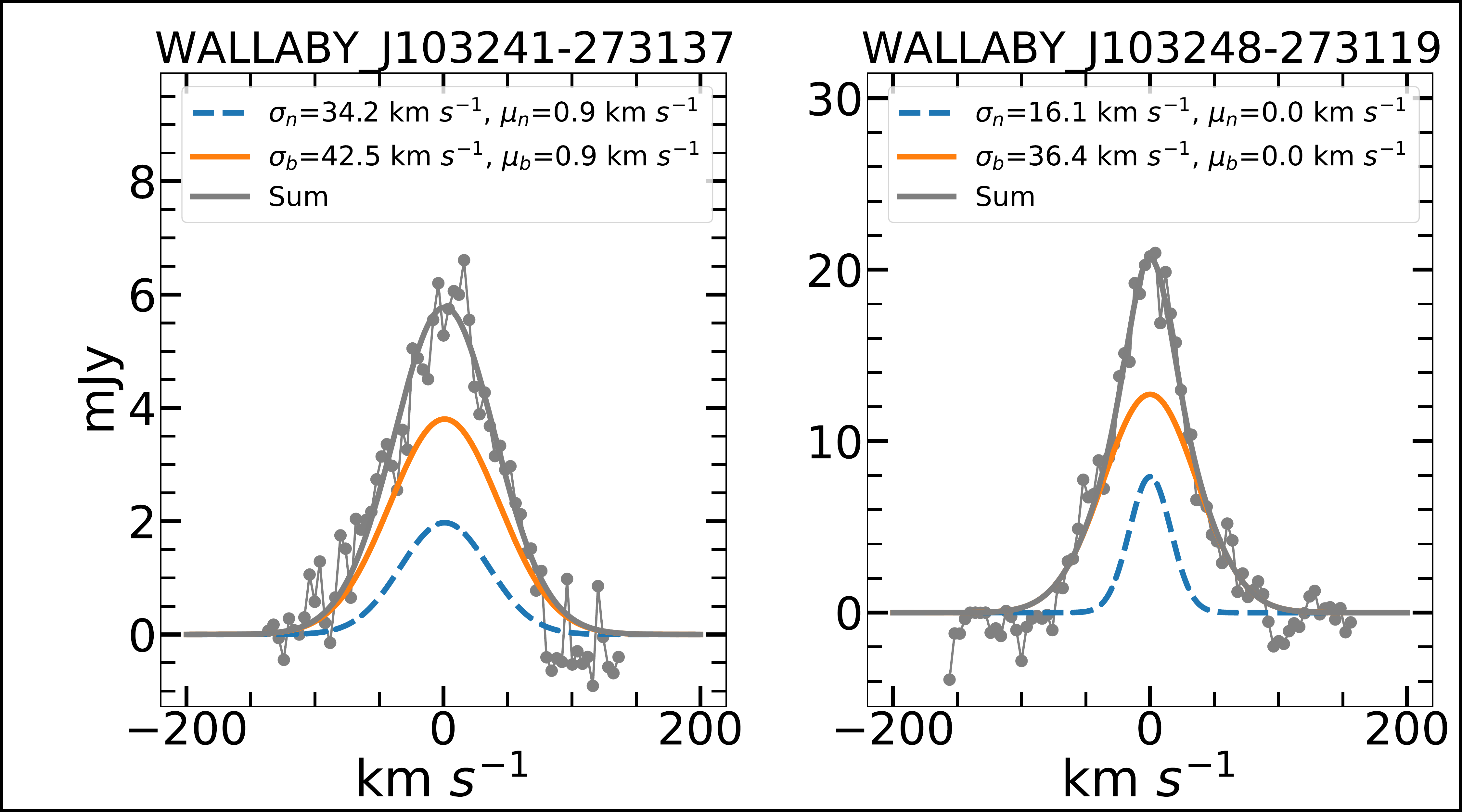}
    \vspace{0.5cm}
    \includegraphics[width=0.45\textwidth]{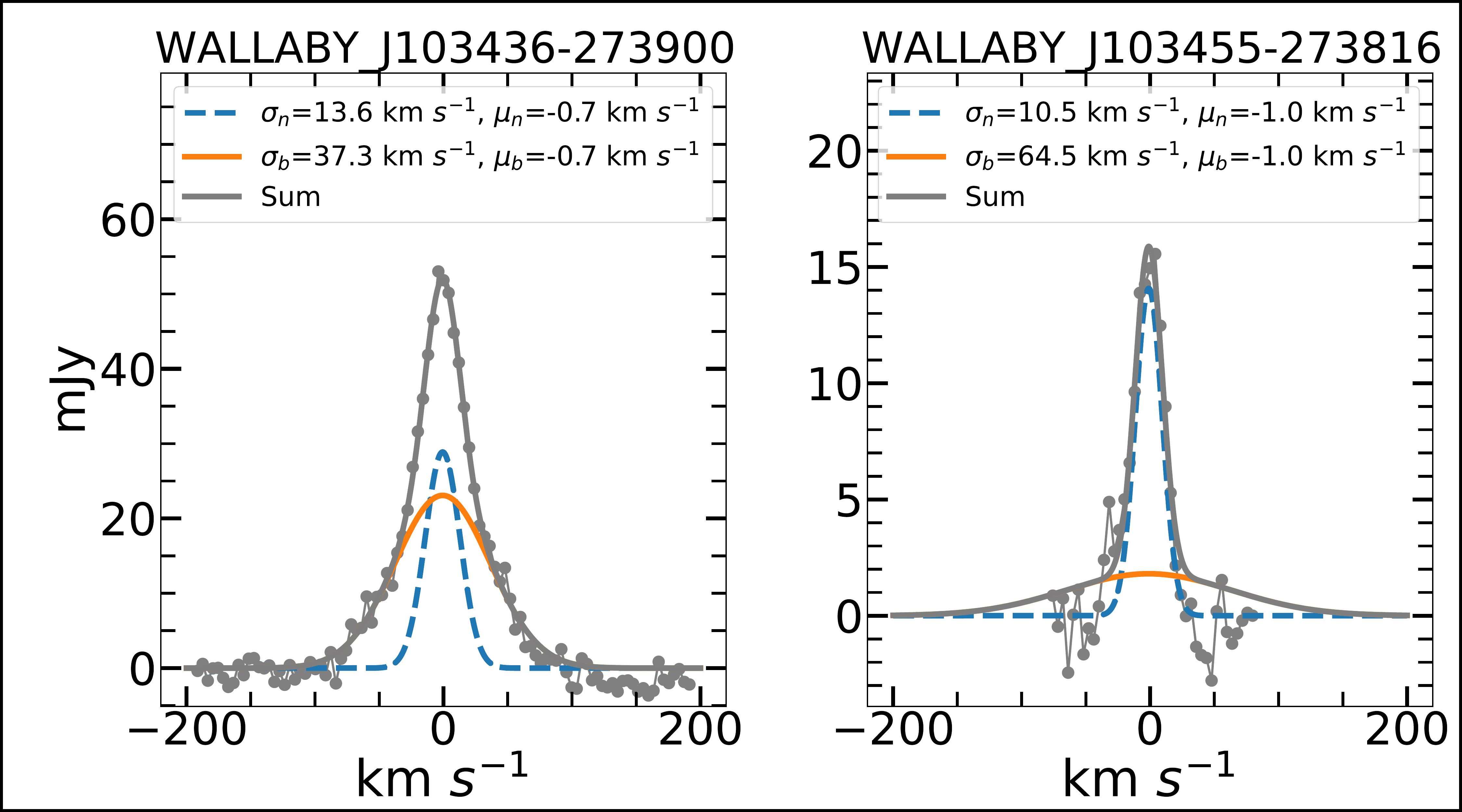}
    \hspace{0.3cm}
    \includegraphics[width=0.45\textwidth]{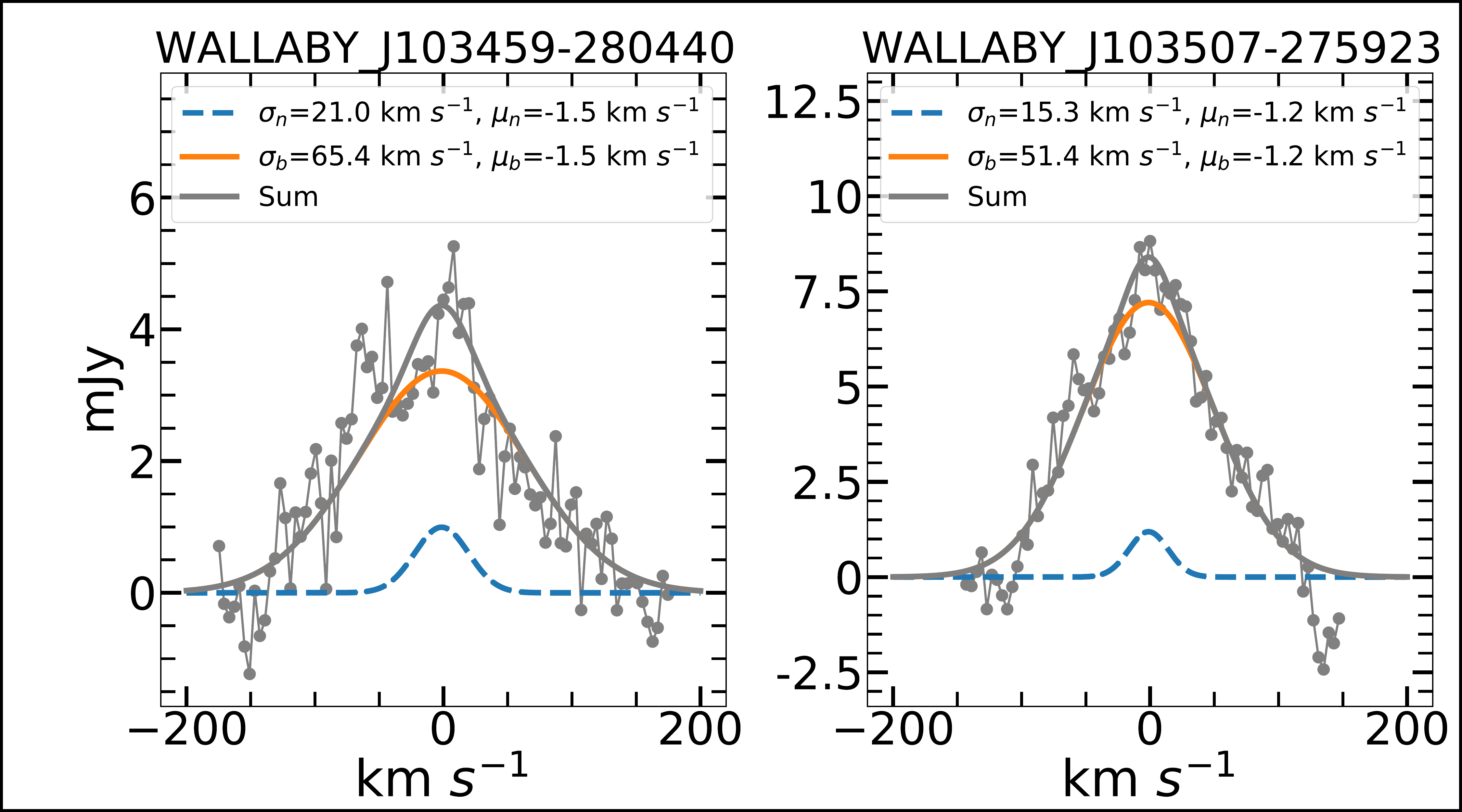}
    \vspace{0.5cm}
    \includegraphics[width=0.45\textwidth]{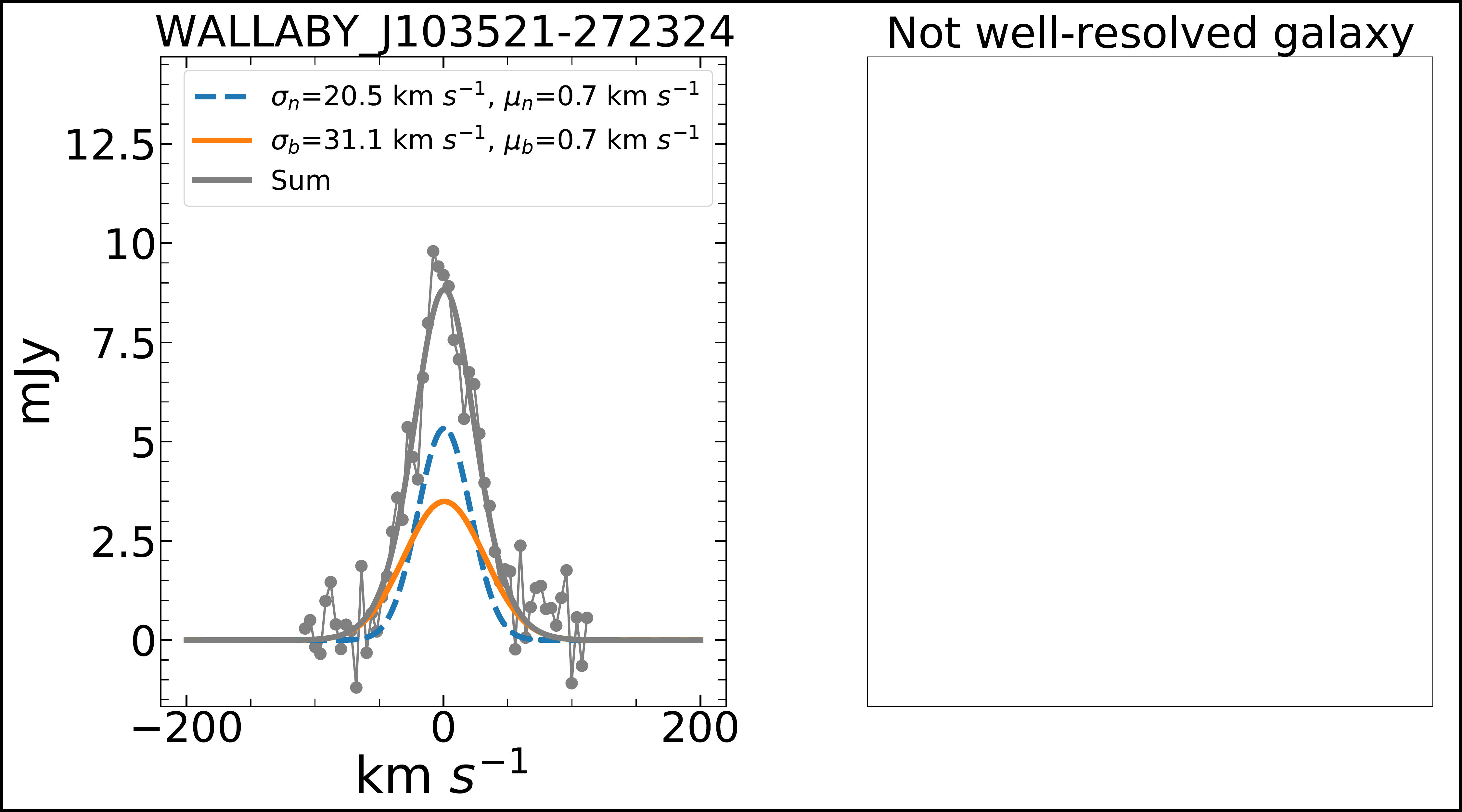}
    \hspace{0.3cm}
    \includegraphics[width=0.45\textwidth]{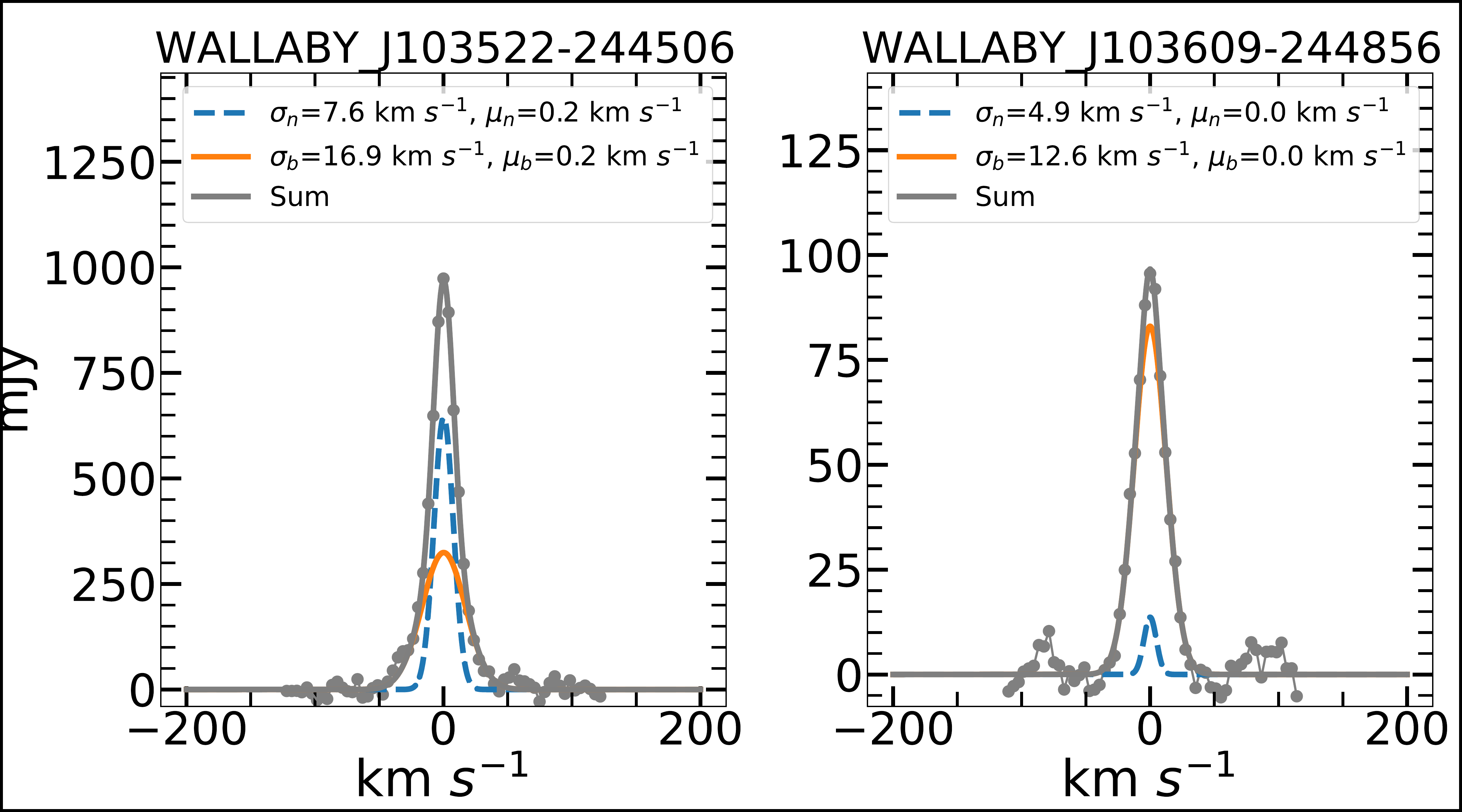}
    \vspace{0.5cm}
    \includegraphics[width=0.45\textwidth]{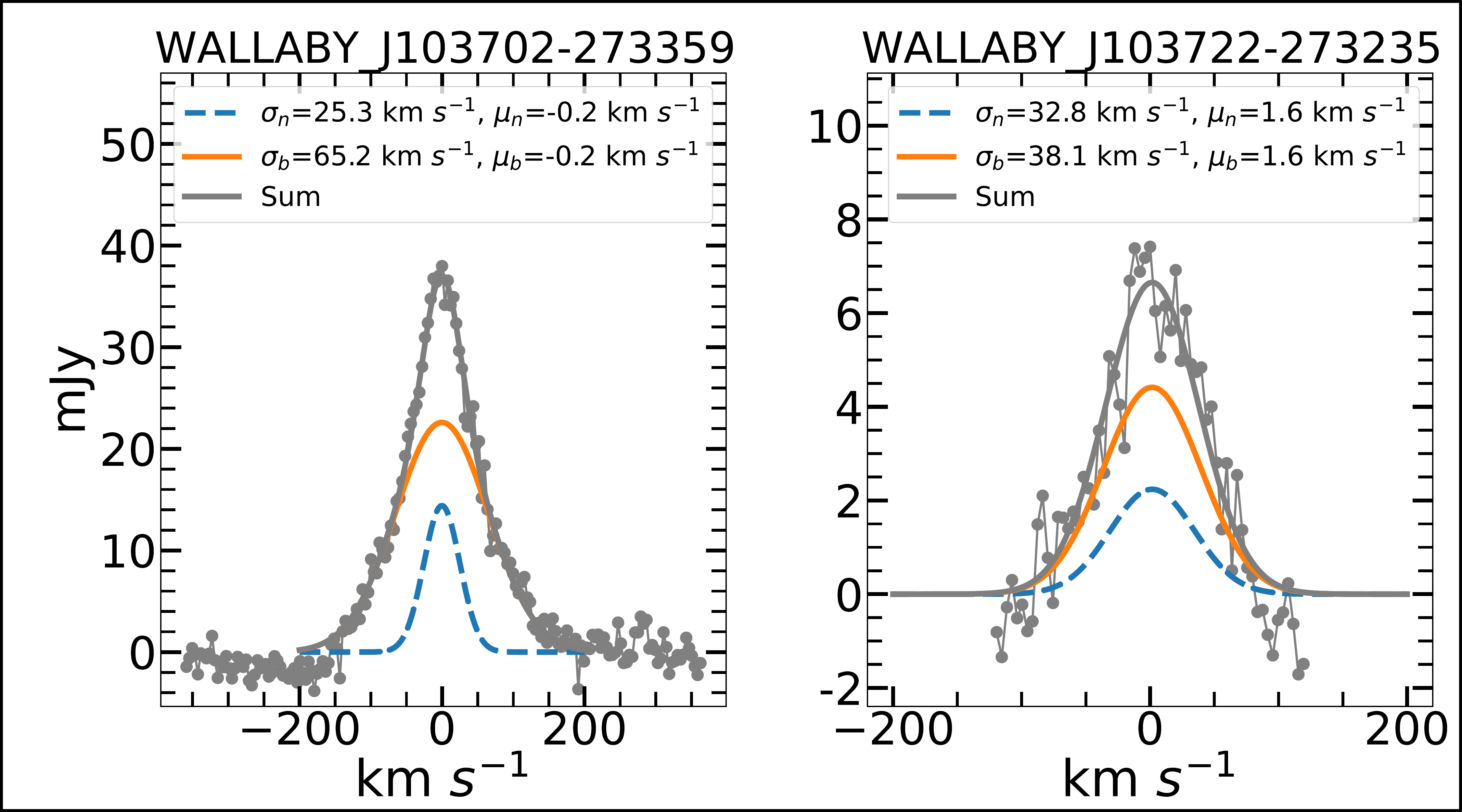}
    \hspace{0.3cm}
    \includegraphics[width=0.45\textwidth]{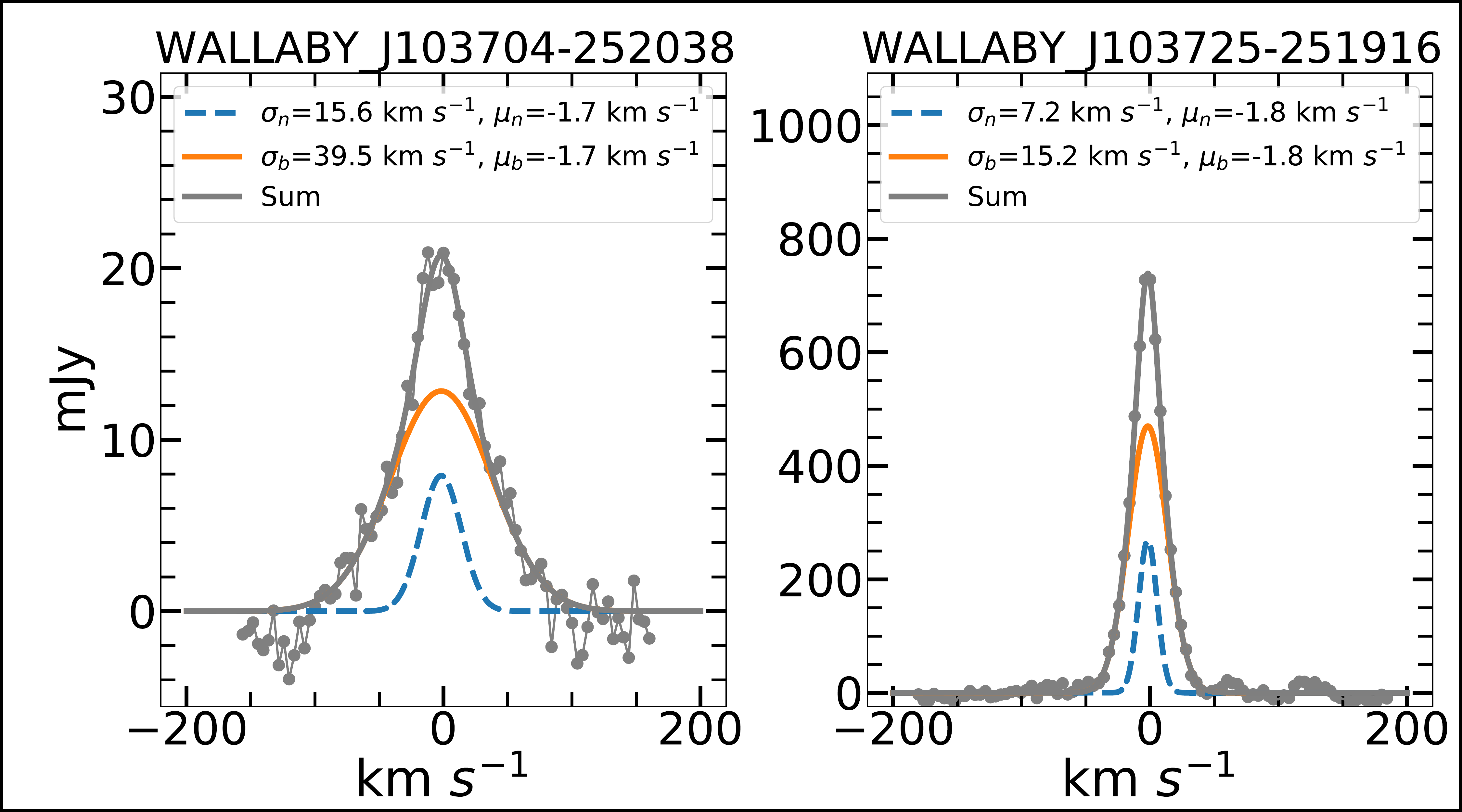}
    \vspace{0.5cm}
    \caption{(continued)}
\end{figure*}

\begin{figure*} 
\ContinuedFloat
    \includegraphics[width=0.45\textwidth]{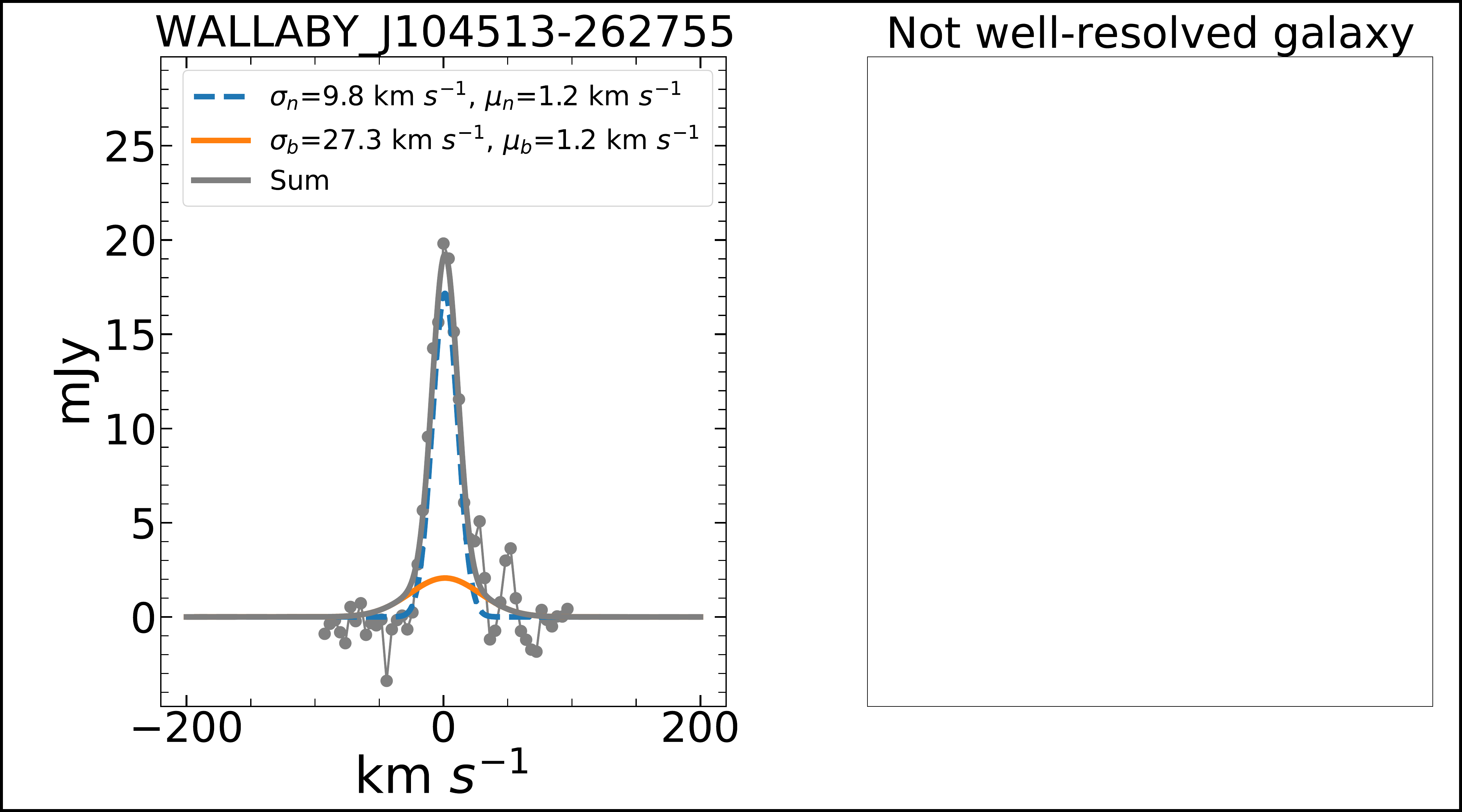}
    \hspace{0.3cm}
    \vspace{0.5cm}
    \includegraphics[width=0.45\textwidth]{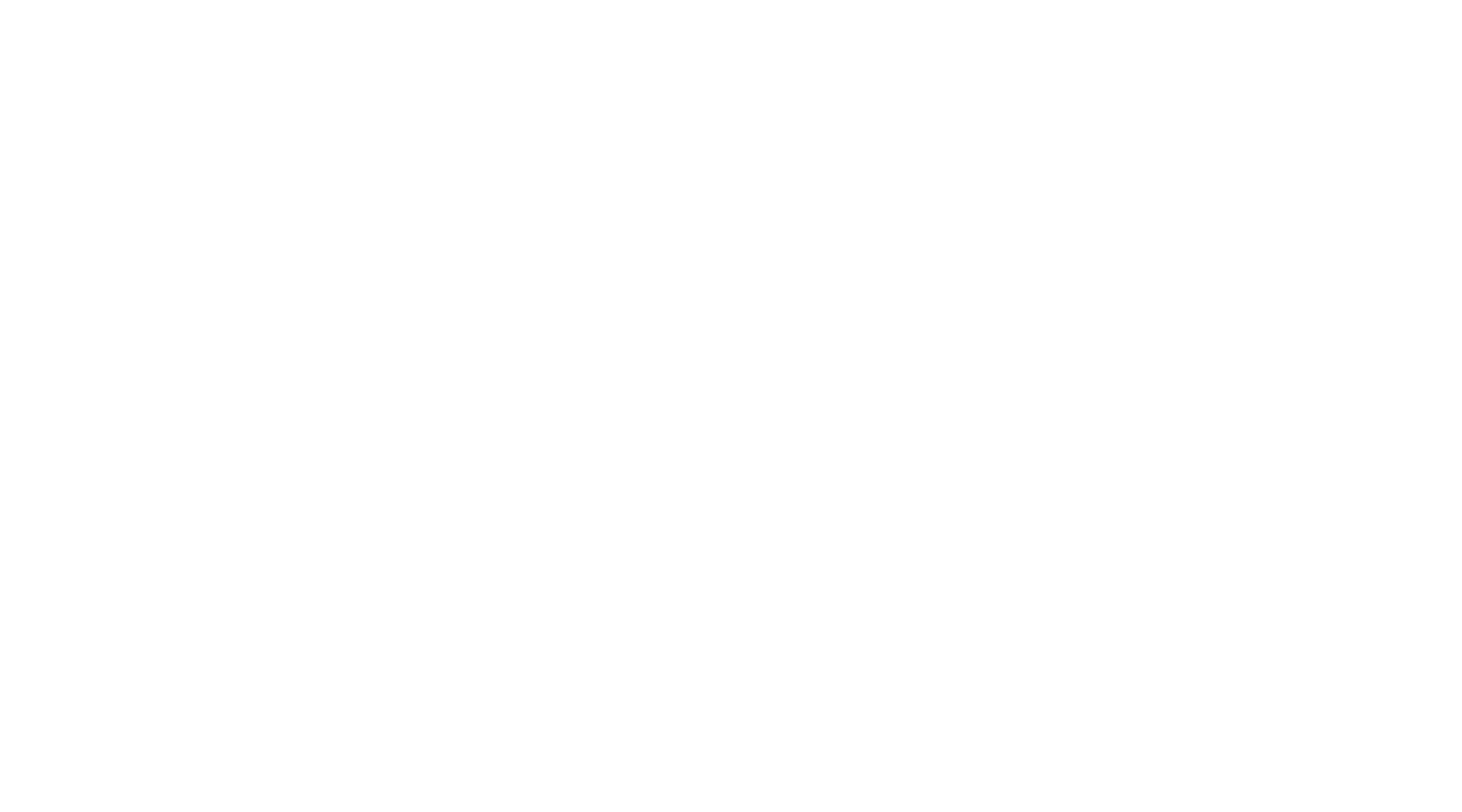}
    \caption{(continued)}
\end{figure*}

\begin{figure*}  
    \includegraphics[width=0.45\textwidth]{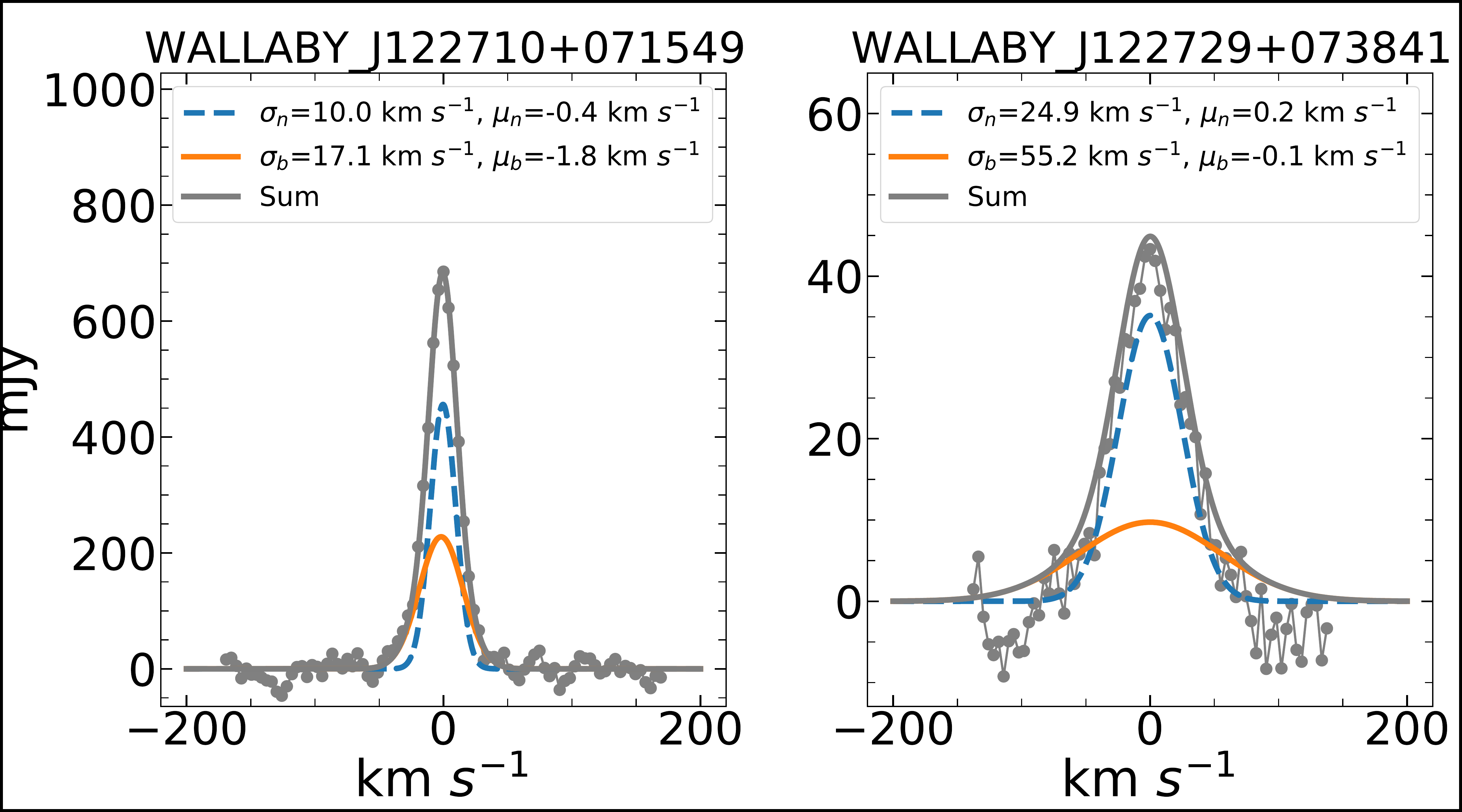}
    \hspace{0.3cm}
    \vspace{0.5cm}
    \includegraphics[width=0.45\textwidth]{Figure/WALLABY_J123422+021914_superprofile.pdf}
    \vspace{0.5cm}
    \includegraphics[width=0.45\textwidth]{Figure/WALLABY_J124508-002747_superprofile.pdf}
    \hspace{0.3cm}
    \includegraphics[width=0.45\textwidth]{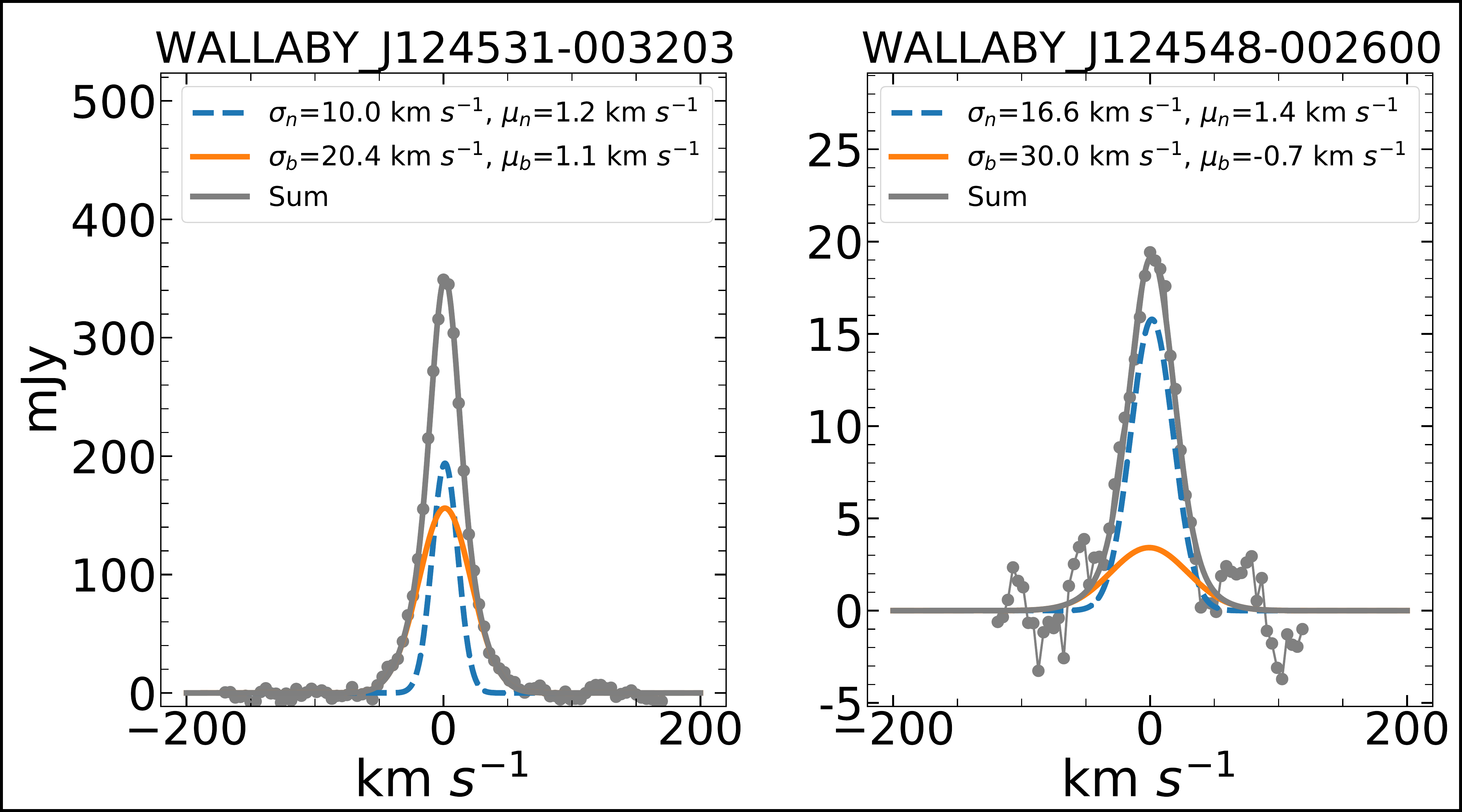}
    \vspace{0.5cm}
    \includegraphics[width=0.45\textwidth]{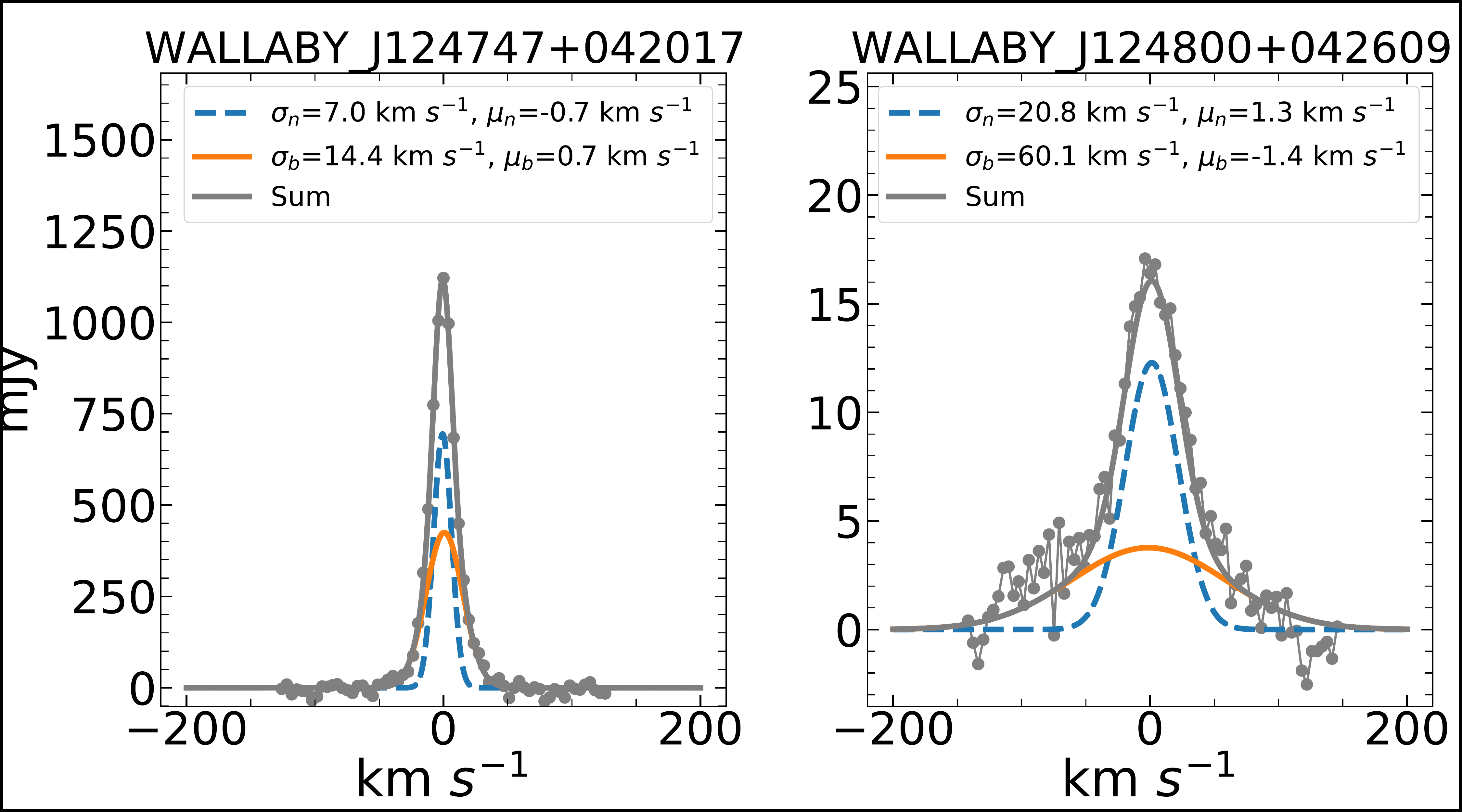}
    \hspace{0.3cm}
    \includegraphics[width=0.45\textwidth]{Figure/white.pdf}
    \vspace{0.5cm}
    \caption{{\sc baygaud} based H{\sc i} super-profiles of the galaxy pairs (ordered by the WALLABY ID) in the ASKAP NGC 4636 group field. The grey circles show the stacked fluxes at the corresponding channels. The blue dashed and orange solid lines represent the narrow and broad components decomposed from the double Gaussian fitting to the super-profiles. Their velocity dispersions ($\sigma$) and centroid velocities ($\mu$) are labeled on the top-right corner of each panel. H{\sc i} super-profiles of galaxies which are poorly resolved are blanked.}
    \label{figA5}
    \end{figure*}
 
\begin{figure*}
    \includegraphics[width=0.45\textwidth]{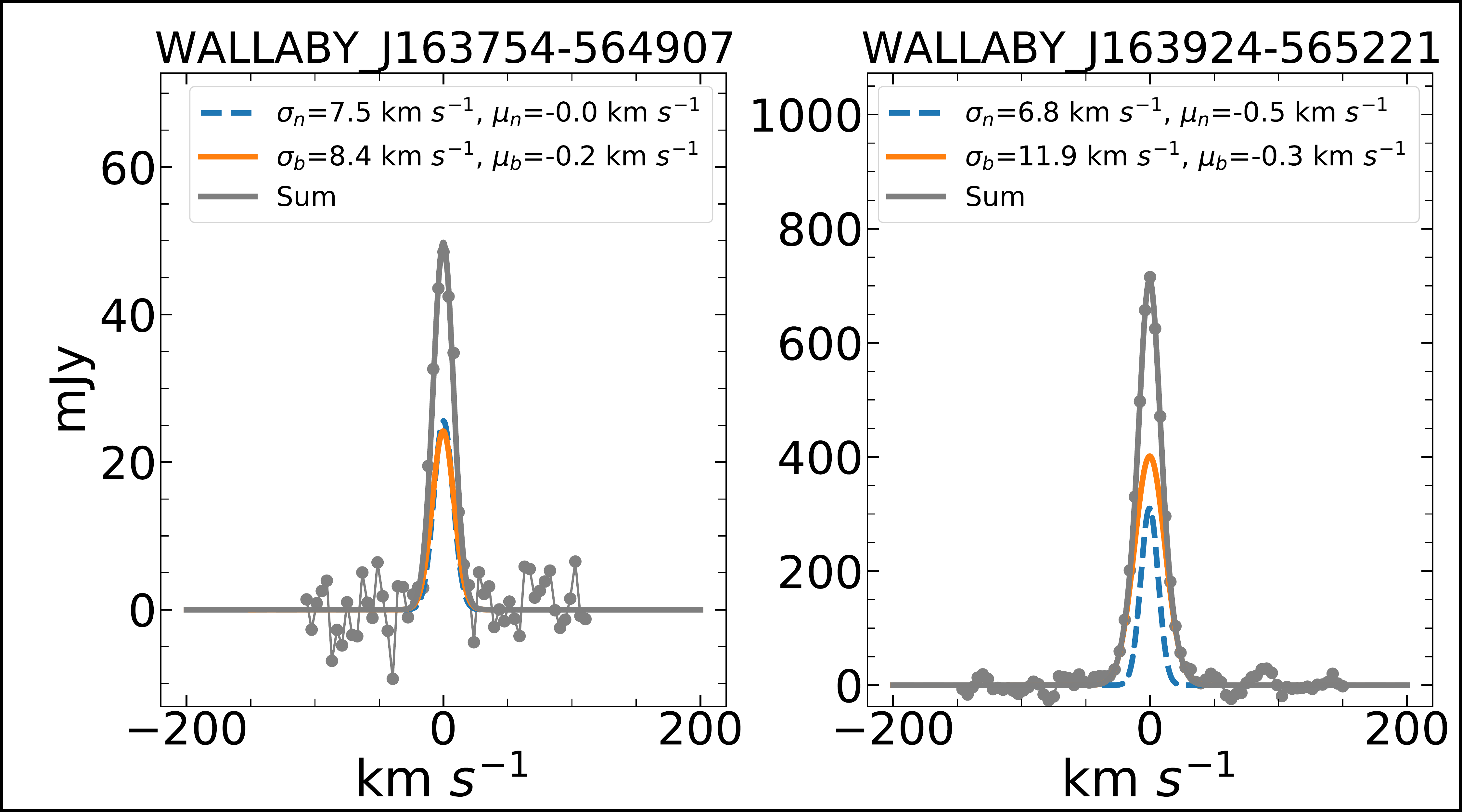}
    \hspace{0.3cm}
    \vspace{0.5cm}
    \includegraphics[width=0.45\textwidth]{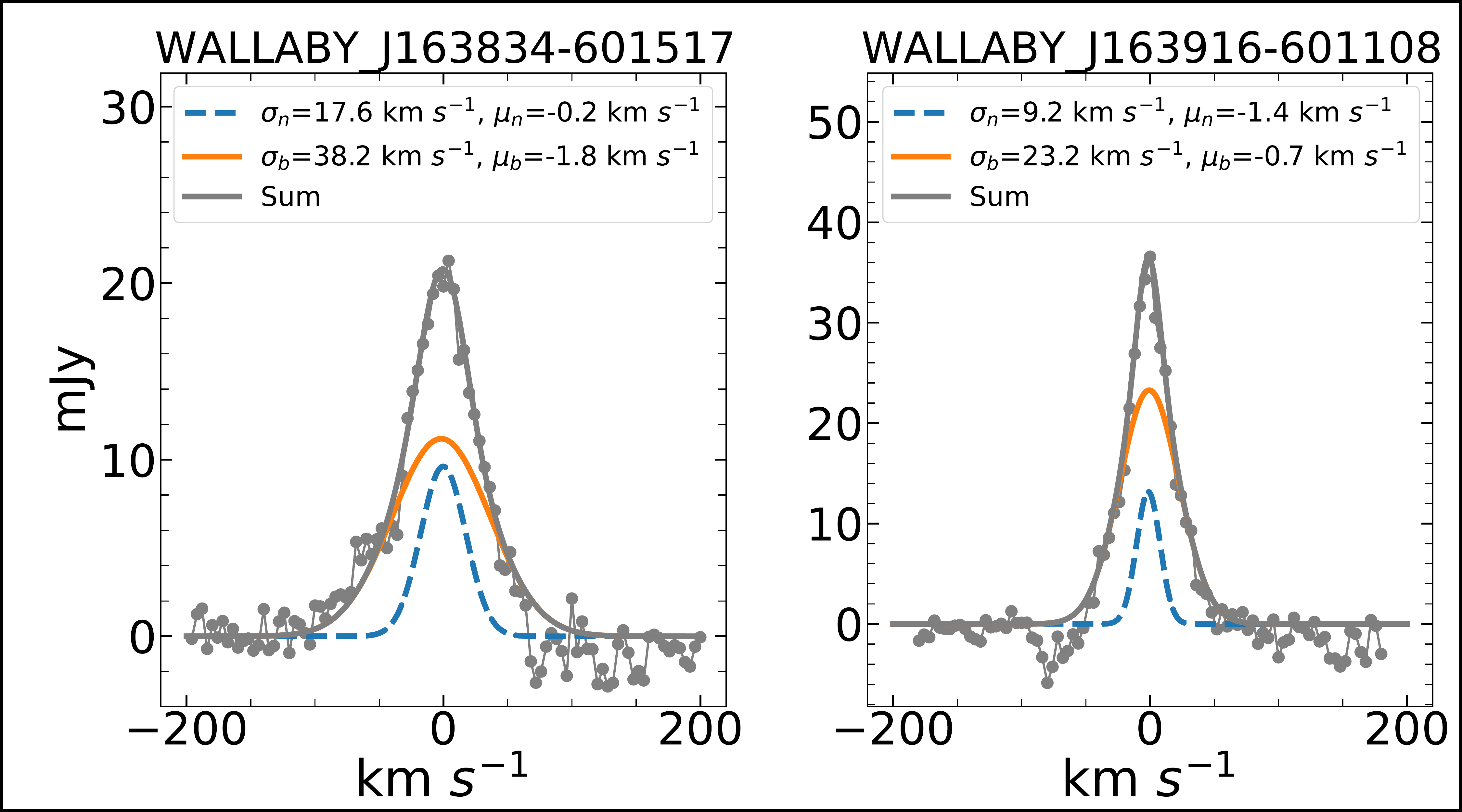}
    \vspace{0.5cm}
    \includegraphics[width=0.45\textwidth]{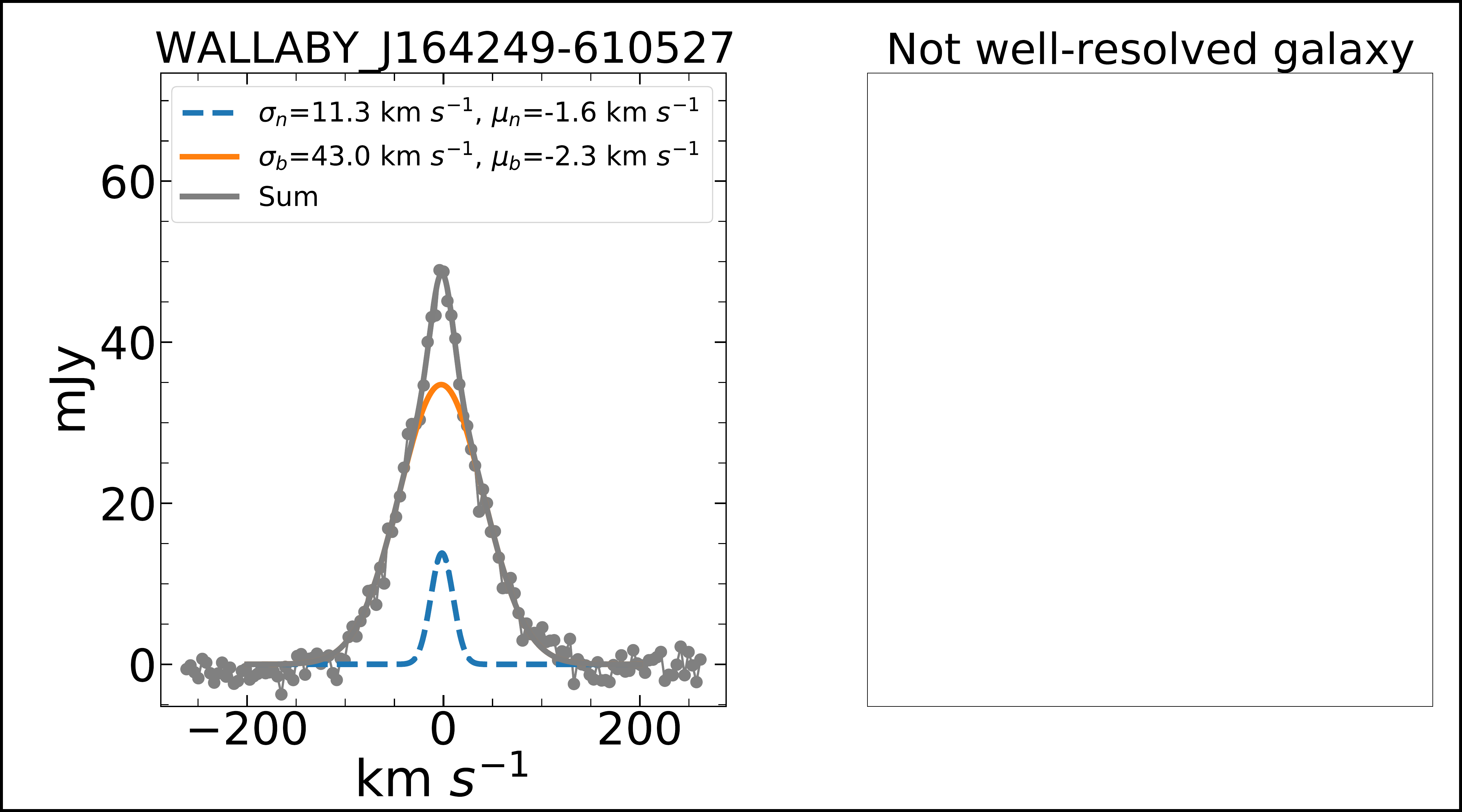}
    \hspace{0.3cm}
    \includegraphics[width=0.45\textwidth]{Figure/WALLABY_J164720-572629_superprofile.pdf}
    \vspace{0.5cm}
    \includegraphics[width=0.45\textwidth]{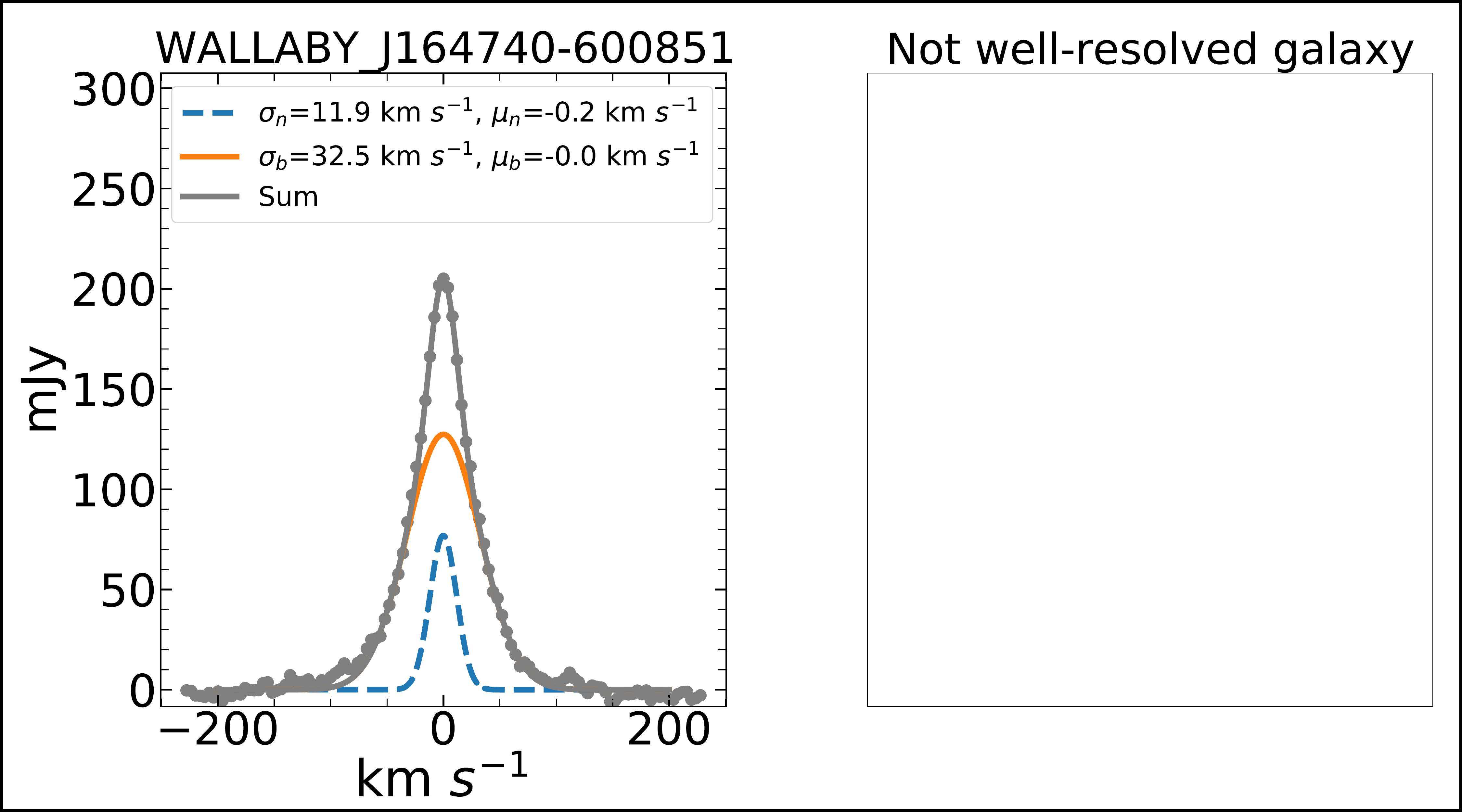}
    \hspace{0.3cm}
    \includegraphics[width=0.45\textwidth]{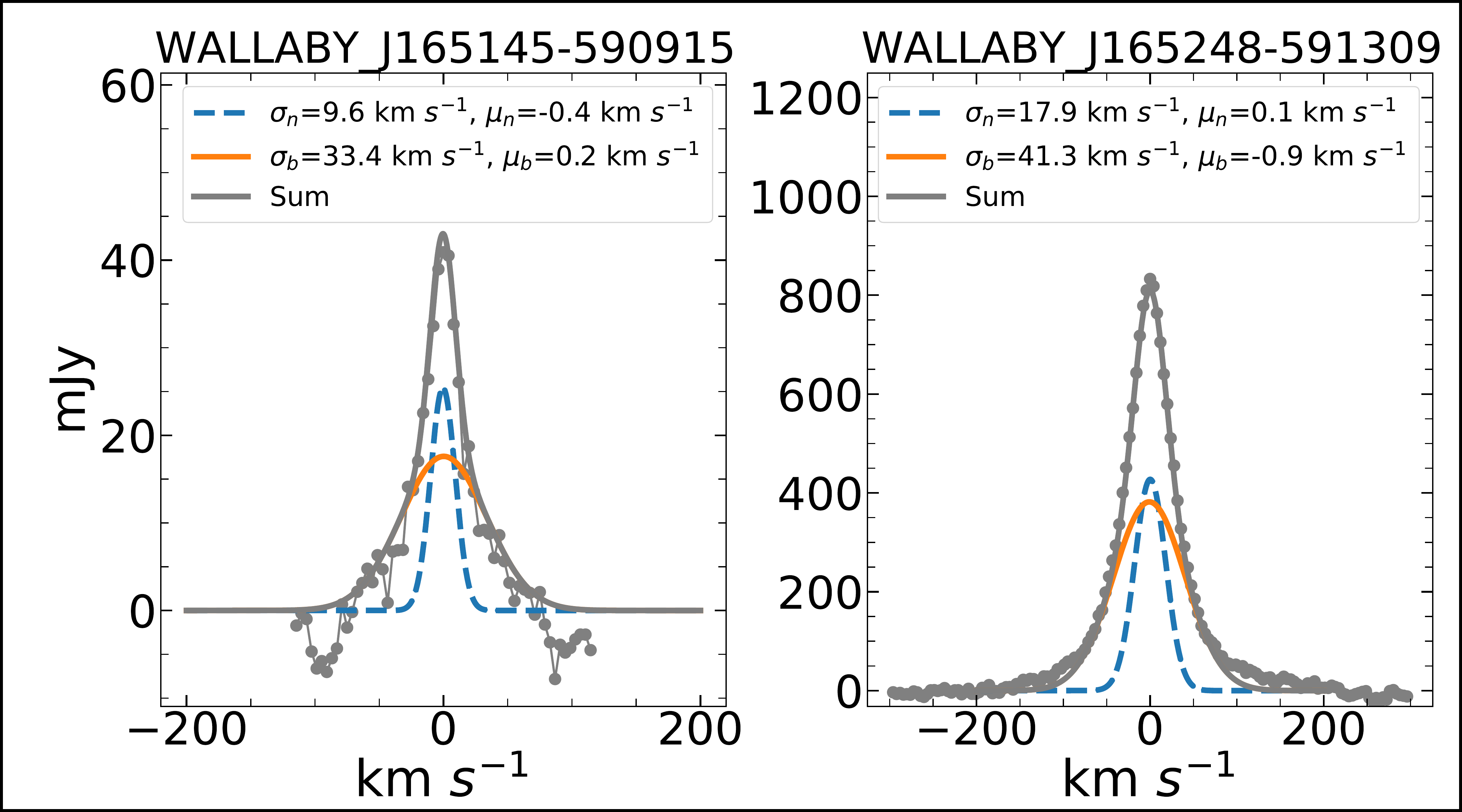}
    \vspace{0.5cm}
    \includegraphics[width=0.45\textwidth]{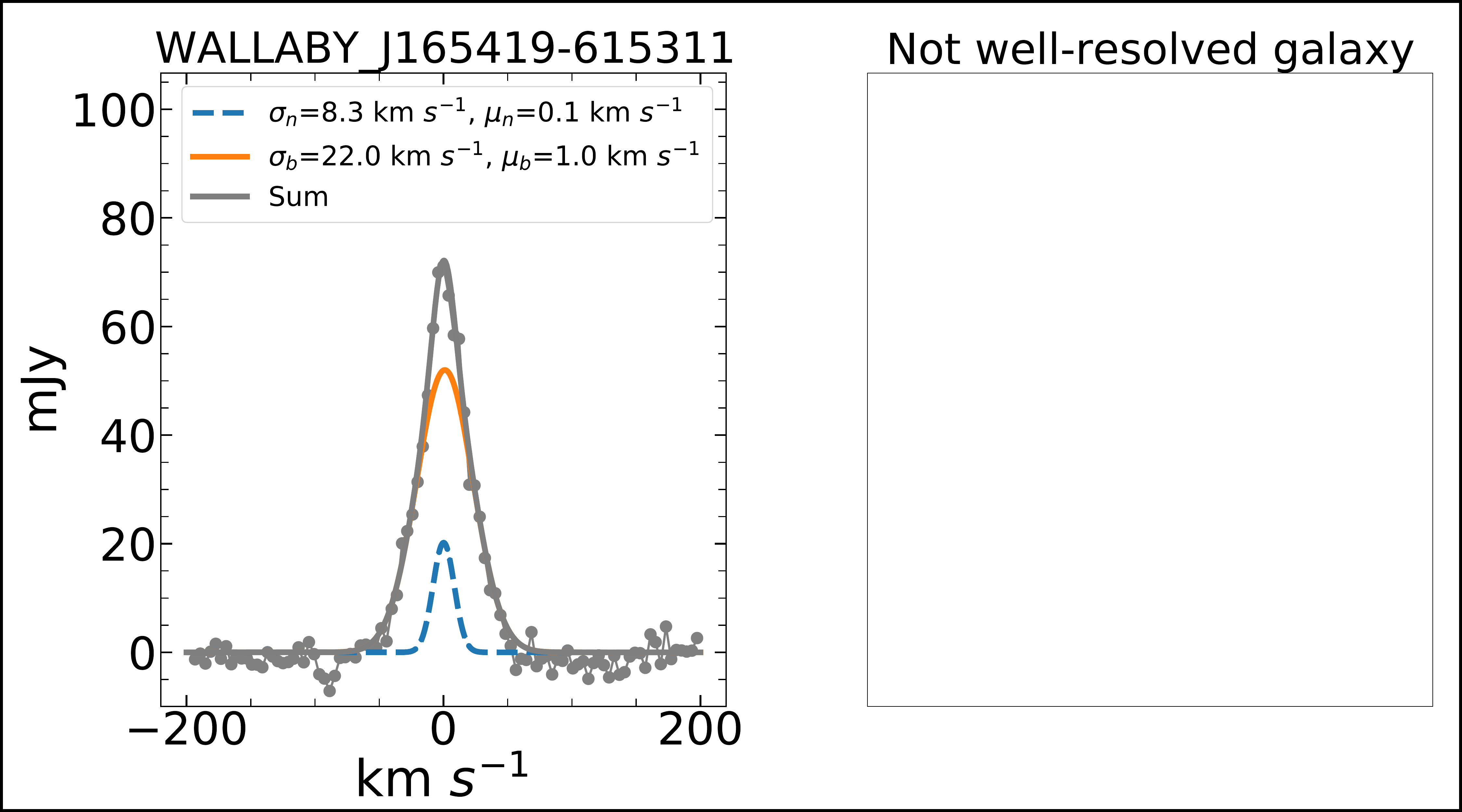}
    \hspace{0.3cm}
    \includegraphics[width=0.45\textwidth]{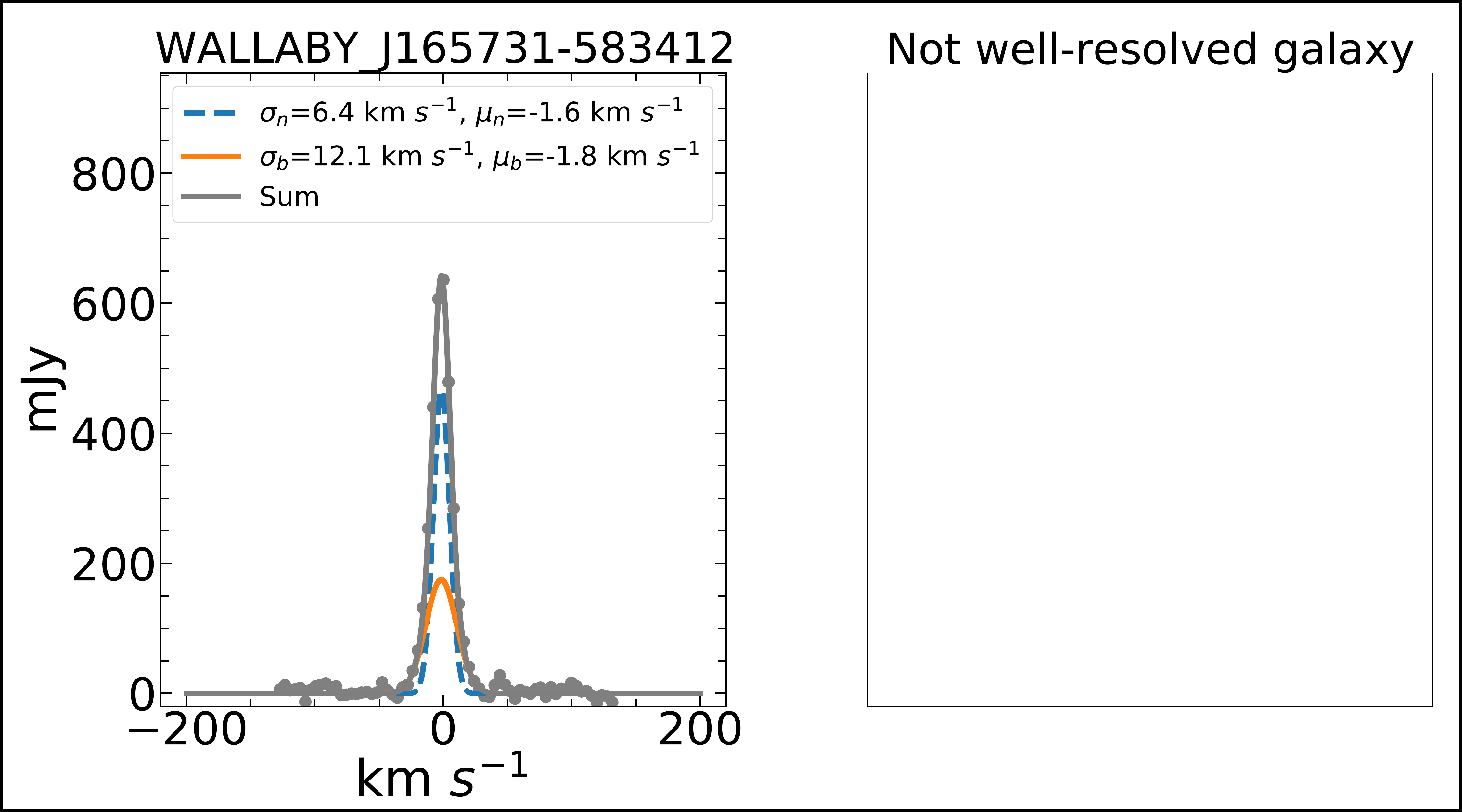}
    \vspace{0.5cm}
    \caption{{\sc baygaud} based H{\sc i} super-profiles of the galaxy pairs (ordered by the WALLABY ID) in the ASKAP Norma cluster field. The grey circles show the stacked fluxes at the corresponding channels. The blue dashed and orange solid lines represent the narrow and broad components decomposed from the double Gaussian fitting to the super-profiles. Their velocity dispersions ($\sigma$) and centroid velocities ($\mu$) are labeled on the top-right corner of each panel. H{\sc i} super-profiles of galaxies which are poorly resolved are blanked.}
    \label{figA6}
\end{figure*}

\begin{figure*}
\ContinuedFloat
    \includegraphics[width=0.45\textwidth]{Figure/WALLABY_J170550-620939_superprofile.pdf}
    \hspace{0.3cm}
    \includegraphics[width=0.45\textwidth]{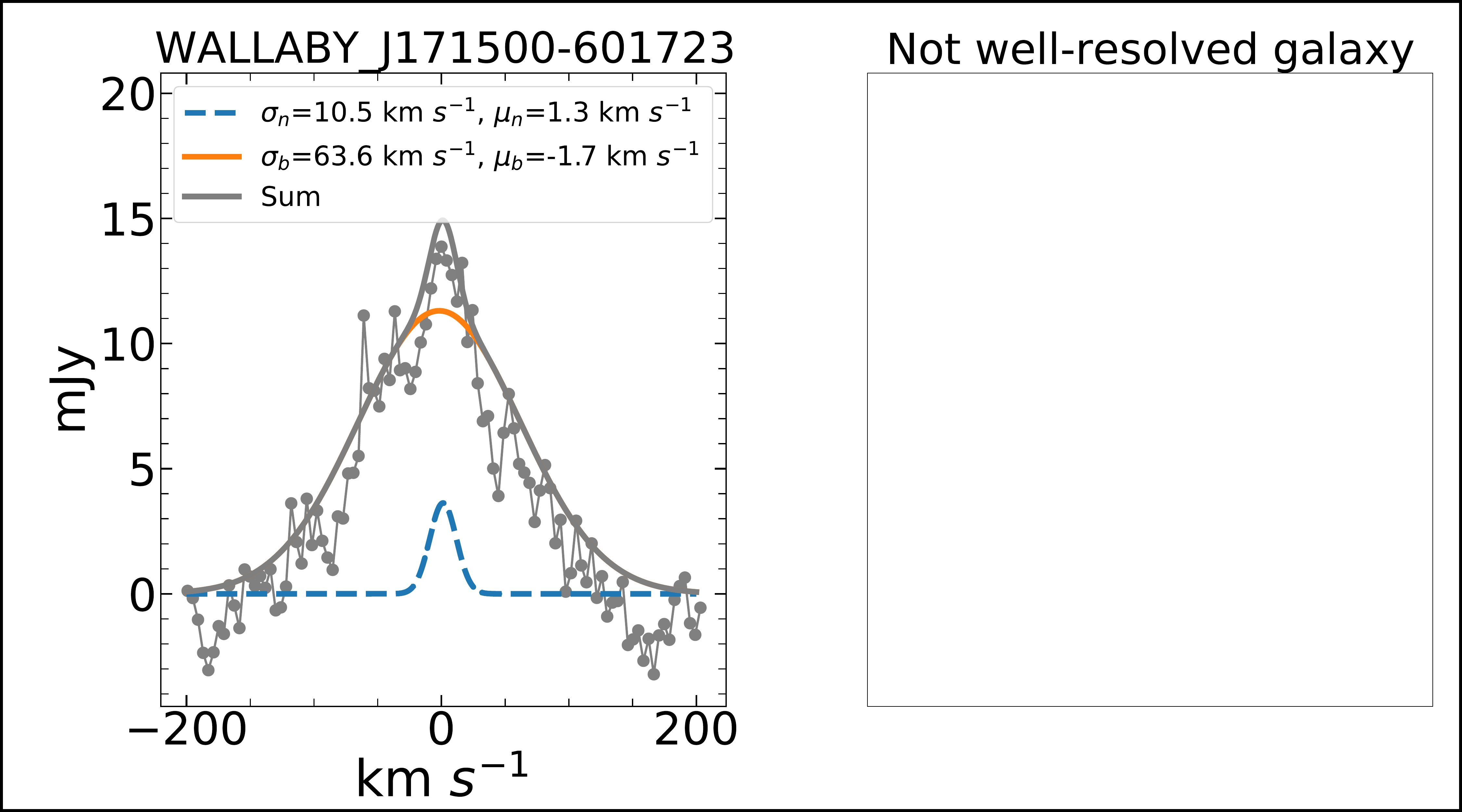}
    \caption{(continued)}
\end{figure*}




\bsp	
\label{lastpage}

\end{document}